\documentclass[floatfix,twocolumn,showpacs,preprintnumbers,amsmath,amssymb,pra,superscriptaddress,longbibliography]{revtex4-1}
\usepackage{color}
\usepackage[usenames,dvipsnames,svgnames,table]{xcolor}
\usepackage[colorlinks=true,linkcolor=blue,urlcolor=blue,citecolor=blue]{hyperref}
\usepackage{mathtools}
\usepackage{graphicx}
\usepackage{dcolumn}
\usepackage{array}
\usepackage{lipsum}
\usepackage{bm}
\usepackage{subfigure}
\usepackage{amssymb}
\usepackage{multirow}
\usepackage{tabularx}
\usepackage{amsmath}
\usepackage{braket}
\usepackage{csquotes}
\graphicspath{{plots/}}
 \usepackage{lipsum}
\usepackage{mathrsfs}
\usepackage{MnSymbol}
	
\newcommand{\beq}{\begin{equation}}
\newcommand{\eeq}{\end{equation}}
\newcommand{\bea}{\begin{eqnarray}}
\newcommand{\eea}{\end{eqnarray}}

\begin{document}
\title{Taylor series perspective on \emph{ab initio} path integral Monte Carlo simulations with Fermi-Dirac statistics}

\author{Tobias Dornheim}
\email{t.dornheim@hzdr.de}
\affiliation{Center for Advanced Systems Understanding (CASUS), D-02826 G\"orlitz, Germany}
\affiliation{Helmholtz-Zentrum Dresden-Rossendorf (HZDR), D-01328 Dresden, Germany}

\author{Alexander Benedix Robles}
\affiliation{Center for Advanced Systems Understanding (CASUS), D-02826 G\"orlitz, Germany}
\affiliation{Helmholtz-Zentrum Dresden-Rossendorf (HZDR), D-01328 Dresden, Germany}
\affiliation{Technische  Universit\"at  Dresden,  D-01062  Dresden,  Germany}

\author{Paul Hamann}
\affiliation{Center for Advanced Systems Understanding (CASUS), D-02826 G\"orlitz, Germany}
\affiliation{Helmholtz-Zentrum Dresden-Rossendorf (HZDR), D-01328 Dresden, Germany}
\affiliation{Institut f\"ur Physik, Universit\"at Rostock, D-18057 Rostock, Germany}

\author{Thomas M.~Chuna}
\affiliation{Center for Advanced Systems Understanding (CASUS), D-02826 G\"orlitz, Germany}
\affiliation{Helmholtz-Zentrum Dresden-Rossendorf (HZDR), D-01328 Dresden, Germany}

\author{Pontus~Svensson}
\affiliation{Center for Advanced Systems Understanding (CASUS), D-02826 G\"orlitz, Germany}
\affiliation{Helmholtz-Zentrum Dresden-Rossendorf (HZDR), D-01328 Dresden, Germany}

\author{Sebastian Schwalbe}

\author{Zhandos A.~Moldabekov}
\affiliation{Center for Advanced Systems Understanding (CASUS), D-02826 G\"orlitz, Germany}
\affiliation{Helmholtz-Zentrum Dresden-Rossendorf (HZDR), D-01328 Dresden, Germany}

\author{Panagiotis Tolias}
\affiliation{Space and Plasma Physics, Royal Institute of Technology (KTH), Stockholm, SE-100 44, Sweden}

\author{Jan Vorberger}
\affiliation{Helmholtz-Zentrum Dresden-Rossendorf (HZDR), D-01328 Dresden, Germany}

\begin{abstract}
The fermion sign problem constitutes a fundamental computational bottleneck across a plethora of research fields in physics, quantum chemistry and related disciplines. Recently, it has been suggested to alleviate the sign problem in \emph{ab initio} path integral Molecular Dynamics and path integral Monte Carlo (PIMC) calculations based on the simulation of fictitious identical particles that are represented by a continuous quantum statistics variable $\xi$ [\textit{J.~Chem.~Phys.}~\textbf{157}, 094112 (2022)]. This idea facilitated a host of applications including the interpretation of an x-ray scattering experiment with strongly compressed beryllium at the National Ignition Facility [\textit{Nature Commun.}~\textbf{16}, 5103 (2025)]. In the present work, we express the original isothermal $\xi$-extrapolation method as a special case of a truncated Taylor series expansion around the $\xi=0$ limit of distinguishable particles. We derive new PIMC estimators that allow us to evaluate the Taylor coefficients up to arbitrary order and we carry out extensive new PIMC simulations of the warm dense electron gas to systematically analyze the sign problem from this new perspective. This gives us important insights into the applicability of the  $\xi$-extrapolation method for different levels of quantum degeneracy in terms of the Taylor series radius of convergence. Moreover, the direct PIMC evaluation of the $\xi$-derivatives, in principle, removes the necessity for simulations at different values of $\xi$ and can facilitate more efficient simulations that are designed to maximize compute time in those regions of the full permutation space that contribute most to the final Taylor estimate of the fermionic expectation value of interest. 
\end{abstract}
\maketitle

\section{Introduction\label{sec:introduction}}

For over half a century, \emph{ab initio} quantum Monte Carlo (QMC) methods~\cite{anderson2007quantum,Foulkes_RMP_2001,cep,Pollet_2012,Booth_Nature_2013} have been employed with great success to study a broad range of quantum many-body systems in both the zero-temperature limit~\cite{Ceperley_Alder_PRL_1980,anderson2007quantum,Foulkes_RMP_2001,Booth_JCP_2009,Booth_Nature_2013,Pfau_PRR_2020} and thermal equilibrium~\cite{cep,PhysRevE.73.056703,Ceperley_PRL_1992,Dornheim_POP_2017,Schoof_PRL_2015,Blunt_PRB_2014,Malone_JCP_2015,Driver_PRL_2012,boninsegni1,Saccani_Supersolid_PRL_2012,kawashima2004recent,Joonho_JCP_2021,Filinov_PRA_2012}. The basic idea is usually to cast the expectation value of a given observable of interest $\hat{O}$ in terms of a high-dimensional (with $D\sim10^{3-7}$ dimensions not being unusual) integral. The associated curse of dimensionality is then avoided with Monte Carlo importance sampling integration, often using a special purpose implementation of the celebrated Metropolis algorithm~\cite{metropolis}. At finite temperatures, a particularly successful method is offered by the \emph{ab initio} path integral Monte Carlo (PIMC) approach~\cite{Pollock_PRB_1984,Berne_JCP_1982,Takahashi_Imada_PIMC_1984}, which is based on the well-known \emph{classical isomorphism}~\cite{Chandler_JCP_1981}, where the quantum many-body system of interest is mapped onto an effectively classical system of interacting ring polymers.
Having originally been developed for the simulation of ultracold helium~\cite{Jordan_PR_1968,Fosdick_PR_1966}, PIMC has since given deep insights into important phenomena such as superfluidity~\cite{cep,Sindzingre_PRL_1989,Filinov_PRL_2010,Dornheim_PRB_2015,Pollock_PRB_1987}, collective and single-particle excitations~\cite{Filinov_PRA_2012,Ferre_PRB_2016,Dornheim_SciRep_2022,dornheim_dynamic,Saccani_Supersolid_PRL_2012,Vitali_PRB_2010,Boninsegni_maximum_entropy}, as well as crystallization~\cite{Clark_PRL_2009,Filinov_PRL_2001,PhysRevB.51.2723,Bhattacharya2016}.

A decisive factor regarding the efficiency of the PIMC approach concerns the quantum statistics that are obeyed by the simulated particles. For bosons and boltzmannons (i.e., hypothetical distinguishable quantum particles), all contributions to the partition function are strictly non-negative. Thus, the corresponding path integral configuration space is ideally suited for Metropolis Monte Carlo sampling, and quasi-exact (within the given statistical Monte Carlo error bars) simulations of up to $N\sim10^4$ particles have been reported~\cite{boninsegni1,boninsegni2}. The situation is dramatically changed for fermions, where the anti-symmetry of the thermal density matrix under the exchange of particle coordinates induces sign changes in the respective contributions to both the partition function and the observables. The resulting cancellation of positive and negative terms is the root cause of the notorious \emph{fermion sign problem}~\cite{troyer,Loh_PRB_1990,dornheim_sign_problem}, which manifests itself as an exponential increase in the compute time required to attain a specific level of accuracy upon increasing the system size $N$ or decreasing the temperature $T$. In practice, the sign problem limits direct PIMC simulations of fermions to moderate system sizes ($N\sim10^0-10^2$) and weak to moderate levels of quantum degeneracy~\cite{dornheim_sign_problem,Dornheim_JPA_2021}. This is unfortunate, as degenerate Fermi systems potentially offer a wealth of interesting physics, such as the BCS-BEC transition in ultracold $^3$He~\cite{Kagan_2019}, roton-type collective modes in Fermi liquids~\cite{Godfrin2012,Dornheim_Nature_2022,Hamann_PRR_2023,Chuna_JCP_2025,Trigger} and the formation of Wigner molecules or crystals at low density~\cite{Bhattacharya2016,Azadi_PRB_2022,Drummond_PRB_Wigner_2004,PhysRevLett.82.3320,Filinov_PRL_2001}. A particularly important application is given by so-called \emph{warm dense matter}~\cite{vorberger2025roadmapwarmdensematter}, an extreme state that naturally occurs in a host of compact astrophysical objects~\cite{wdm_book,drake2018high,Guillot2018,SAUMON20221}, which is also highly relevant for material science~\cite{Kraus2016,Kraus2017,Lazicki2021} and inertial confinement fusion experiments~\cite{hu_ICF,Batani_roadmap,Hurricane_RevModPhys_2023,AbuShawareb_PRL_2024}. This regime is characterized by the complex interplay of Coulomb correlations, thermal excitations and partial electron degeneracy, making holistic descriptions using state-of-the-art \emph{ab initio} methods such as PIMC paramount~\cite{new_POP,Dornheim_review,Bonitz_POP_2024}.

The rather pressing need to understand interacting quantum many-fermion systems has sparked a surge of new developments in corresponding finite-temperature QMC methodologies~\cite{Schoof_CPP_2015,Groth_PRB_2016,Chin_PRE_2015,Dornheim_NJP_2015,Malone_JCP_2015,Malone_PRL_2016,Blunt_PRB_2014,Rubenstein_JCTC_2020,Joonho_JCP_2021,Brown_PRL_2013,Driver_PRL_2012,Xiong_JCP_2022,Xiong_PRE_2024,Hirshberg_JCP_2020,Dornheim_JCP_2020,Dornheim_JCP_xi_2023,Yilmaz_JCP_2020,Dornheim_JCP_2024,Dornheim_NatComm_2025,Filinov_PRE_2020,FILINOV2025130542,Xiong:2025dwx,dornheim2025reweightingestimatorabinitio}, see also Refs.~\cite{Dornheim_review,Bonitz_POP_2024,vorberger2025roadmapwarmdensematter} and references therein. A particularly interesting approach is offered by the simulation of fictitious identical particles guided by the continuous quantum statistics variable $\xi$~\cite{Xiong_JCP_2022,Dornheim_JCP_xi_2023,Xiong_PRE_2023,Xiong_PRE_2024,Dornheim_JCP_2024,Dornheim_JPCL_2024,Dornheim_NatComm_2025,Dornheim_MRE_2024,schwalbe2025staticlineardensityresponse,Dornheim_JCTC_2025,svensson2025acceleratedfreeenergyestimation,morresi2025studyuniformelectrongas,dornheim2025reweightingestimatorabinitio,Morresi_PRB_2025,Yang_Entropy_2025}, where the cases of $\xi=-1$, $\xi=0$ and $\xi=1$ correspond to the physically relevant cases of Fermi-Dirac, Maxwell-Boltzmann and Bose-Einstein statistics, respectively. Specifically, Xiong and Xiong~\cite{Xiong_JCP_2022} have suggested to carry out simulations in the sign-problem free domain of $\xi\geq0$ and then extrapolate to the fermionic limit of $\xi=-1$ using an empirical quadratic ansatz. Subsequent studies~\cite{Dornheim_JCP_xi_2023,Dornheim_JCP_2024} have substantiated the high quality of the $\xi$-extrapolation method for weak to moderate levels of quantum degeneracy. Importantly, when applicable, this method completely removes the fermionic cancellation problem, facilitating simulations of up to $N=1000$ electrons~\cite{Dornheim_JPCL_2024} as well as the first direct comparison between \emph{ab initio} PIMC simulations and an x-ray scattering measurement on warm dense beryllium performed at the National Ignition Facility (NIF) in Livermore~\cite{Dornheim_Nature_2022,Dornheim_POP_2025,schwalbe2025staticlineardensityresponse}. Moreover, complete access to the full set of observables (including dynamic information as encoded into the variety of imaginary-time correlation functions~\cite{Dornheim_JCP_ITCF_2021,Boninsegni1996,Berne_JCP_1983,Dornheim_MRE_2023,Dornheim_PTR_2023}) is retained, in contrast to other methods (e.g., those relying on nodal restrictions~\cite{Ceperley1991}).

In the present work, we report a generalized Taylor series perspective on PIMC simulations of fictitious identical particles, which contains the original, empirical $\xi$-extrapolation technique as a special case. To this end, we derive new estimators that allow us to evaluate the Taylor extrapolation to the fermionic limit of $\xi=-1$ (and any other value of $\xi$) up to arbitrary order from a single PIMC simulation at any reasonable reference value $\xi_\textnormal{ref}>0$. Choosing the boltzmannon case of $\xi=0$ as the basis for the Taylor series, we empirically verify that the physically meaningful cases of $\xi=1$ and $\xi=-1$ are indeed within its radius of convergence for all studied cases, and that the radius appears to approach unity from above upon decreasing the temperature. Crucially, access to the full Taylor series allows us to rigorously assess the convergence with the number of Taylor coefficients and, in this way, to choose appropriate levels of truncation. This, in turn, nicely verifies the quadratic empirical extrapolation that has been employed in previous works~\cite{Xiong_JCP_2022,Dornheim_JCP_xi_2023,Dornheim_JCP_2024,Dornheim_MRE_2024}. As a practical example, we consider the warm dense uniform electron gas (UEG)~\cite{review,loos,quantum_theory}---the quantum version of the classical one-component plasma---which has attracted a surge of interest over the last decade or so~\cite{Brown_PRL_2013,Brown_PRB_2013,ksdt,Karasiev_status_2019,dornheim_prl,groth_prl,review,Malone_PRL_2016,Joonho_JCP_2021,Hou_PRB_2022,dornheim_ML,dornheim_dynamic,Dornheim_PRL_2020,Dornheim_PRL_2020_ESA,Hunger_PRE_2021,tanaka_hnc,Tanaka_CPP_2017,arora,Tolias_JCP_2021,Tolias_JCP_2023,Dornheim_PRB_2024,Tolias_PRB_2024,stls2} owing to its relevance for a variety of warm dense matter applications, including thermal density functional theory simulations~\cite{karasiev_importance,kushal,Karasiev_PRL_2018,Karasiev_PRB_2022,moldabekov2024density,Moldabekov_JCTC_2024,Sjostrom_PRB_2014}. Nevertheless, our methodology and conclusions are very general and expected to be relevant for PIMC simulations of a gamut of interacting Fermi-Dirac systems, including ultracold atoms and real warm dense matter systems comprised of both electrons and ions.

The paper is organized as follows: In Sec.~\ref{sec:theory}, we introduce the required theoretical background, starting with the UEG model~(\ref{sec:UEG}), a brief discussion of PIMC (\ref{sec:PIMC}) and the simulation of fictitious identical particles (\ref{sec:fictitious}). This is followed by the introduction of the new Taylor series perspective (\ref{sec:Taylor}) and the new PIMC estimators for the $\xi$-derivatives (\ref{sec:derivatives}). Sec.~\ref{sec:results} contains the detailed presentation of our simulation results, including the analysis of the convergence radii (\ref{sec:general}), the study of $\xi$-derivatives (\ref{sec:derivatives_results}) and the investigation of the practical convergence of the Taylor series at different conditions (\ref{sec:taylor_results}). The paper is concluded by a summary and outlook in Sec.~\ref{sec:outlook}.

\section{Theory\label{sec:theory}}

We assume Hartree atomic units throughout this work.

\subsection{Uniform electron gas model\label{sec:UEG}}

The UEG, also known as \emph{jellium}, constitutes the quantum version of the classical one-component plasma~\cite{Baus_Hansen_OCP,OCP_bridge_2022}. In the context of this work, we simulate $N=N^\uparrow+N^\downarrow$ \emph{spin-unpolarized} electrons in standard periodic boundary conditions, with $N^\uparrow$ and $N^\downarrow$ the equal numbers of spin-up and spin-down electrons. Following Fraser \textit{et al.}~\cite{Fraser_PRB_1996}, we express the corresponding $N$-body Hamiltonian as
\begin{eqnarray}\label{eq:Hamiltonian}
    \hat{H}_\textnormal{E} = -\frac{1}{2}\sum_{l=1}^N \nabla_l^2 + \sum_{l<k}^N \phi_\textnormal{E}(\hat{\mathbf{r}}_l,\hat{\mathbf{r}}_k) + \frac{N\xi_\textnormal{M}}{2}\ ,
\end{eqnarray}
where $\xi_\textnormal{M}=-2.837297(3/4\pi)^{1/3}N^{-1/3}r_s^{-1}$ is the Madelung constant~\cite{Schoof_PRL_2015}.
Here $\phi_\textnormal{E}(\mathbf{r}_a,\mathbf{r}_b)$ corresponds to the usual Ewald pair potential and already includes the interaction between the electrons and the homogeneous neutralizing positive background; other potentials have been explored in the literature~\cite{Fukuda2012,Dornheim_PRE_2025,Yakub2005,Yakub_JCP_2003,Filinov_PRE_2020,Filinov_PRE_2023,svensson2025acceleratedfreeenergyestimation,Demyanov_2022}, but using the standard Ewald method allows for direct comparisons with other PIMC variants, most notably configuration PIMC~\cite{Schoof_CPP_2015,Schoof_PRL_2015,Groth_PRB_2016,Dornheim_PRB_2016,review} at high densities.

The unpolarized UEG is conveniently characterized by two dimensionless parameters~\cite{Ott2018}: the density parameter $r_s=(3/4\pi n)^{1/3}=d/a_\textnormal{B}$ is given by the Wigner-Seitz radius in units of the Bohr radius and serves as the quantum coupling parameter; the reduced degeneracy temperature $\Theta=T/E_\textnormal{F}$ is given by the ratio of the thermal energy to the Fermi energy and constitutes a straightforward measure for the degree of quantum degeneracy (with $\Theta\ll1$ and $\Theta\gg1$ the fully degenerate and semi-classical limits, respectively). In the warm dense matter regime, which is of considerable interest for astrophysics, planetary science, material science and inertial fusion energy~\cite{vorberger2025roadmapwarmdensematter}, we have $r_s\sim\Theta\sim1$, indicating a complex interplay of thermal, quantum and Coulomb coupling effects.

Originally developed as a model for conduction electrons in simple metals~\cite{mahan1990many,quantum_theory}, the UEG has emerged as the archetypal model system of interacting electrons with a plethora of practical applications. For example, highly accurate parametrizations of UEG properties such as the exchange--correlation (free) energy~\cite{vwn,Perdew_Wang,Perdew_Zunger_PRB_1981,ksdt,groth_prl} and linear density response function~\cite{cdop,Utsumi_PRA_1982,farid,dornheim_ML,Dornheim_PRB_ESA_2021}, based on state-of-the-art QMC simulations~\cite{Ceperley_Alder_PRL_1980,Ortiz_PRB_1994,Ortiz_PRL_1999,Spink_PRB_2013,dornheim_prl,review,dynamic_folgepaper,moroni,moroni2}, constitute important input for other calculations, most notably density functional theory simulations of real materials~\cite{vorberger2025roadmapwarmdensematter,Bonitz_POP_2024,Jones_RMP_2015}.

In the warm dense matter regime, the extensive body of thermal QMC simulations includes results for the energy, free energy and chemical potential~\cite{Brown_PRL_2013,Schoof_CPP_2015,Schoof_PRL_2015,Dornheim_JCP_2015,Groth_PRB_2016,Dornheim_PRB_2016,dornheim_prl,Yilmaz_JCP_2020,Malone_JCP_2015,Malone_PRL_2016,Joonho_JCP_2021,Dornheim_PRB_2025,Dornheim_PRR_2025,Dornheim_JCTC_2025,Dornheim_PRB_ChemPot_2025}, momentum distribution~\cite{Militzer_Pollock_PRL_2002,Dornheim_PRB_nk_2021,Dornheim_PRE_2021,Hunger_PRE_2021}, linear~\cite{dornheim_ML,dynamic_folgepaper,Dornheim_PRL_2020_ESA,dornheim_electron_liquid,dornheim_HEDP,Dornheim_HEDP_2022,Dornheim_PRB_2024,Hou_PRB_2022} and non-linear~\cite{dornheim_pre,groth_jcp,Dornheim_PRL_2020,Dornheim_PRR_2021,Dornheim_JPSJ_2021,Dornheim_CPP_2021,Dornheim_CPP_2022,Dornheim_JCP_ITCF_2021,Tolias_EPL_2023} density response, dynamic structure factor~\cite{dornheim_dynamic,Chuna_JCP_2025,Dornheim_PRE_2020,Chuna_PRB_2025,robles2025pylitreformulationimplementationanalytic} and related dynamic properties~\cite{Hamann_CPP_2020,Hamann_PRB_2020}.

\subsection{Path integral Monte Carlo\label{sec:PIMC}}

In the following, we give a concise introduction to certain aspects of the PIMC method that they are relevant for the present investigation. More detailed introductions to PIMC can be found, e.g., in Refs.~\cite{cep,Pollock_PRB_1984,Takahashi_Imada_PIMC_1984,boninsegni1,Marienhagen_JCP_2025}.

Throughout this work, we consider $N$ particles in a cubic simulation cell of volume $\Omega=L^3$ in the canonical ensemble, meaning that $N$, $\Omega$ and the inverse temperature $\beta=1/T$ are fixed.
On the most abstract level, the path integral form of the canonical partition function $Z$ can be expressed as
\begin{eqnarray}\label{eq:Z_generic}
    Z_\xi(\beta,N,\Omega) = \sumint\textnormal{d}\mathbf{X}\ \underbrace{W(\mathbf{X})\xi^{N_\textnormal{pp}(\mathbf{X})}}_{W_\xi(\mathbf{X})}\ ,
\end{eqnarray}
where the meta-variable $\mathbf{X}=(\mathbf{R}_0,\dots,\mathbf{R}_{P-1})^T$ contains the coordinates of all particles on all $P$ imaginary time slices, with $\mathbf{R}_\alpha = (\mathbf{r}_1,\dots,\mathbf{r}_N)^T$ being the $N$-particle coordinate on time slice $\alpha$ at $\tau=-i\alpha\epsilon$ where $\epsilon=\beta/P$ is the imaginary time step.
We note that it holds $\mathbf{R}_0\equiv\mathbf{R}_P$; the paths are, hence, closed, giving rise to their interpretation as ring polymers~\cite{Chandler_JCP_1981}, with $W(\mathbf{X})$ being their associated configuration weight containing both kinetic and potential terms; it is a function that can straightforwardly be evaluated in practice.
The notation $\sumint\textnormal{d}\mathbf{X}$ in Eq.~(\ref{eq:Z_generic}) implies that we have to integrate over all $3NP$ dimensions of the meta-variable $\mathbf{X}$ and, in addition, also includes the sum over all possible permutations of particle coordinates of particles of the same spin orientation (i.e., spin-up and spin-down electrons for the unpolarized UEG). Finally, the aforementioned variable $\xi$ takes into account quantum statistics, with $\xi=-1,0,1$ corresponding to fermions, boltzmannons, and bosons, respectively, and $N_\textnormal{pp}(\mathbf{X})$ being the corresponding number of pair permutations. For the purposes of the present work, any $\xi\in\mathbb{R}$ constitutes a valid option. The basic idea of the PIMC method is to use a modern implementation of the Metropolis algorithm~\cite{boninsegni1,Dornheim_PRB_nk_2021,mezza} to stochastically generate a Markov chain $\{\mathbf{X}\}_i$ that is distributed according to $P_\xi(\mathbf{X})=W_\xi(\mathbf{X})/Z_\xi$; we note that the arguments $(\beta,N,\Omega)$ will be suppressed for simplicity throughout the remainder of this work.
The corresponding Monte Carlo estimate for the thermodynamic equilibrium expectation value of an arbitrary observable $\hat{O}$ is
\begin{eqnarray}\label{eq:expectation_value}
    \braket{\hat O}_\xi = \frac{1}{Z_\xi}\sumint\textnormal{d}\mathbf{X}\ W_\xi(\mathbf{X}) O(\mathbf{X)}\ ,
\end{eqnarray}
where $O(\mathbf{X})$ denotes the corresponding Monte Carlo estimator. 

Unfortunately, both $P_\xi(\mathbf{X})$ and $W_\xi(\mathbf{X})$ can be either positive or negative for $\xi<0$, which is respectively the case when $N_\textnormal{pp}$ is even or odd. This precludes their interpretation as a probability distribution, requiring a further intermediate step. As a practical workaround, we sample path configurations $\mathbf{X}$ according to the modified probability distribution $P'_\xi(\mathbf{X}) = |W_\xi(\mathbf{X})|/Z'_\xi = P_{|\xi|}(\mathbf{X})$, and the exact signful expectation value is then given by
\begin{eqnarray}\label{eq:ratio}
    \braket{\hat O}_{\xi<0} = \frac{\braket{\hat{O}\hat{S}}_{|\xi|}}{\braket{\hat{S}}_{|\xi|}} \ ,
\end{eqnarray}
where the numerator and denominator are computed as a Monte Carlo average with respect to the absolute value of $\xi$.
We note that $S\equiv \braket{\hat{S}}_{|\xi|}$, with $S(\mathbf{X})=|W_\xi(\mathbf{X})|/W_\xi(\mathbf{X})$, is known as the \emph{average sign} in the literature and constitutes a direct measure for the degree of cancellations within the PIMC simulation for fermionic observables. Fermionic PIMC simulations are generally feasible for $S\gtrsim 10^{-2}-10^{-3}$~\cite{dornheim_sign_problem}.

\subsection{Fictitious identical particles and re-weighting\label{sec:fictitious}}

In their seminal work, Xiong \& Xiong~\cite{Xiong_JCP_2022} have proposed to carry out path integral molecular dynamics simulations~\cite{Hirshberg_JCP_2020,Hirshberg_PNAS_2019} for continuous non-negative values of $\xi$ and then subsequently extrapolate to the fermionic limit of $\xi=-1$ using the empirical parabolic ansatz
\begin{eqnarray}\label{eq:extrapolation}
    O(\xi) = a_O + b_O\xi + c_O\xi^2\ ,
\end{eqnarray}
with $a_O$, $b_O$, $c_O$ free fit parameters. Usually, simulations have been performed on a grid with $\xi\in[0,1]$ although alternative intervals $\xi\in[\xi_\textnormal{min},\xi_\textnormal{max}]$ are certainly possible.
While choosing $\xi_\textnormal{min}>0$ makes no practical sense, it seems worthwhile to attempt to choose $\xi_\textnormal{min}$ as close to $\xi=-1$ as possible to explicitly capture more of the fermionic sector; this can be possible even in situations where the full fermion sign problem is rather severe as the average sign increases exponentially upon decreasing $|\xi|$ in the fermionic sector of $\xi<0$~\cite{Dornheim_JCP_xi_2023,Dornheim_JCTC_2025}.
Similarly, one might choose $\xi_\textnormal{max}$ larger or smaller than unity. A more systematic answer to this question can be given in terms of the radius of convergence of the Taylor series, see Eqs.~(\ref{eq:radius_root}),(\ref{eq:radius_ratio}) below.

Clearly, PIMC simulations for $\xi\geq0$ are not subject to the exponential computation bottleneck inherent to the fermion sign problem. Having to perform $N_\xi=5-20$ independent PIMC simulations for different values of $\xi\geq0$, thus appears to be a reasonable trade-off to extend the capabilities of standard PIMC to larger system sizes. At the same time, simulations of $N\sim\mathcal{O}\left(10^2\right)$ particles on $P\sim\mathcal{O}\left(10^2\right)$ imaginary-time slices can be computationally demanding even without the sign problem. To avoid this obstacle, Dornheim \textit{et al.}~\cite{dornheim2025reweightingestimatorabinitio} have recently presented a re-weighting estimator, which, in principle allows the evaluation of the full $\xi$-dependence from a single PIMC simulation at an arbitrary reference value $\xi_\textnormal{ref}$. Specifically, we can express the expectation value of an observable $\hat{O}$ at any $\xi$ in terms of its expectation values at $\xi_\textnormal{ref}$,
\begin{widetext}
\begin{eqnarray}
\braket{\hat{O}}_\xi &=& \frac{Z_{\xi_\textnormal{ref}}}{Z_\xi}  \frac{1}{Z_{\xi_\textnormal{ref}}} \sum_{\sigma\in S_N}\int\textnormal{d}\mathbf{X}\ W_{\xi_\textnormal{ref}}(\mathbf{X}) \underbrace{ \frac{W_\xi(\mathbf{X})}{W_{\xi_\textnormal{ref}}(\mathbf{X})} 
O(\mathbf{X}) }_{O_{\xi_\textnormal{ref},\xi}(\mathbf{X})} = \frac{Z_{\xi_\textnormal{ref}}}{Z_\xi} \braket{\hat{O}_{\xi_\textnormal{ref},\xi}}_{\xi_\textnormal{ref}} \ , \\
\frac{Z_\xi}{Z_{\xi_\textnormal{ref}}} &=& \frac{1}{Z_{\xi_\textnormal{ref}}} \sum_{\sigma\in S_N}\int\textnormal{d}\mathbf{X}\ W_{\xi_\textnormal{ref}}(\mathbf{X}) \frac{W_\xi(\mathbf{X})}{W_{\xi_\textnormal{ref}}(\mathbf{X})} = \left< \frac{W_\xi(\mathbf{X})}{W_{\xi_\textnormal{ref}}(\mathbf{X})} \right>_{\xi_\textnormal{ref}} \ , \label{eq:Down_Nola}
\end{eqnarray}
\end{widetext}
leading to the estimator
\begin{eqnarray}
\braket{\hat O}_\xi = \frac{ \braket{\hat{O}_{\xi_\textnormal{ref},\xi}}_{\xi_\textnormal{ref}} }{ \left< \frac{W_\xi(\mathbf{X})}{W_{\xi_\textnormal{ref}}(\mathbf{X})} \right>_{\xi_\textnormal{ref}} } \ , \label{eq:final}
\end{eqnarray}
with the ratio of configuration weights given by
\begin{eqnarray}\label{eq:re}
    \frac{W_\xi(\mathbf{X})}{W_{\xi_\textnormal{ref}}(\mathbf{X})} = \left(\frac{\xi}{\xi_\textnormal{ref}}\right)^{N_\textnormal{pp}}\ .
\end{eqnarray}
In essence, Eq.~(\ref{eq:final}) constitutes a re-weighting estimator, where measurements of the estimator $O(\mathbf{X})$ are weighted by Eq.~(\ref{eq:re}) and then re-normalized by Eq.~(\ref{eq:Down_Nola}).

\subsection{Taylor series and extrapolation\label{sec:Taylor}}

Within the reasonable assumption that the $O(\xi)$ expectation value of an observable $\hat{O}$ constitutes a smooth, continuous and infinitely differentiable function with respect to $\xi$ at a reference real number $\xi_\textnormal{T}$, as it has been verified empirically in recent PIMC investigations~\cite{Xiong_JCP_2022,Dornheim_JCP_xi_2023,dornheim2025reweightingestimatorabinitio}, we can express $O(\xi)$ as a Taylor series around the reference value $\xi_\textnormal{T}$,
\begin{eqnarray}\label{eq:taylor_full}
    O(\xi)=O_{\infty,\xi_\textnormal{T}}(\xi) = \left. \sum_{\nu=0}^\infty \frac{(\xi-\xi_\textnormal{T})^\nu}{\nu !} \frac{\partial^\nu}{\partial\xi^\nu} O(\xi) \right|_{\xi=\xi_\textnormal{T}}\ .
\end{eqnarray}
In practice, the Taylor series would need to be truncated at a polynomial degree $p$,
\begin{eqnarray}\label{eq:taylor_truncated}
    O_{p,\xi_\textnormal{T}}(\xi) = \left. \sum_{\nu=0}^{p} \frac{(\xi-\xi_\textnormal{T})^\nu}{\nu !} \frac{\partial^\nu}{\partial\xi^\nu} O(\xi) \right|_{\xi=\xi_\textnormal{T}}\ .
\end{eqnarray}
It is apparent that the standard $\xi$-extrapolation method, based on the empirical parabolic fit formula Eq.~(\ref{eq:extrapolation}), corresponds to the special case of $O_{2,0}(\xi)$.

Heuristically, it makes sense to choose $\xi_\textnormal{T}=0$, which minimizes the extrapolation interval from the bosonic sector to the fermionic limit of interest. Then, the truncated version $O_{p,0}(\xi)$ is automatically expressed as a canonical polynomial,
\begin{eqnarray}\label{eq:canonical_polynomial}
    O_{p,0}(\xi) = \sum_{\nu=0}^p c_\nu \xi^\nu\ ,
\end{eqnarray}
which, among other things, lends itself for a fit to PIMC results for $O(\xi)$ in the sign-problem free domain of $\xi\geq0$. Naively, it might seem beneficial to consider a $\xi$-interval with $\xi_\textnormal{max}\gg1$ in order to extract as many coefficients $c_\nu$ as possible. In this regard, however, it is important to recall the a-priori unknown radius of convergence $r$ of Eq.~(\ref{eq:taylor_full}).  In general, a Taylor series converges within the radius $r$ if one of two limits exists~\cite{apostol_book}
\begin{eqnarray}\label{eq:radius_root}
    r&=&\lim_{\nu\to\infty}\frac{1}{\sqrt[\nu]{|c_{\nu}|}}\ , \\
    r&=&\lim_{\nu\to\infty}\left|\frac{c_{\nu}}{c_{\nu+1}} \right|\ . \label{eq:radius_ratio}
\end{eqnarray}
In practice, the root test of Eq.~(\ref{eq:radius_root}) will be considered throughout this investigation, since the ratio test of Eq.~(\ref{eq:radius_ratio}) is rendered ineffective by the oscillating nature of the $c_\nu$ coefficients at strong degeneracy. A polynomial fit according to Eqs.~(\ref{eq:taylor_truncated}) and (\ref{eq:canonical_polynomial}) outside of $\xi\in[0,r)$ (where we again limit ourselves to the bosonic sector) will thus not yield any information that is pertinent to the attempted extrapolation to $\xi=-1$. In fact, it is not obvious to the present authors that the circle of convergence of $O_{p,0}(\xi)$ must include the physical limits of $\xi=\pm1$. However, we empirically verify this based on extensive new PIMC results in Sec.~\ref{sec:results} below.


\subsection{PIMC estimation of $\xi$-derivatives\label{sec:derivatives}}

Aiming to derive estimators for the $\xi-$derivatives of $\braket{\hat O}_\xi$, we introduce the convenient short-hand notation
\begin{eqnarray}\label{eq:O_Z_ratio}
    \braket{\hat O}_\xi = \frac{1}{Z_\xi} \sumint\textnormal{d}\mathbf{X}\ W(\mathbf{X}) O(\mathbf{X}) \xi^{N_\textnormal{pp}} = \frac{O_\xi}{Z_\xi}\ ,
\end{eqnarray}
where $O_\xi$ denotes the un-normalized $\hat{O}$ expectation value at $\xi$. We first consider the case of $\xi\neq0$, for which
\begin{eqnarray}\label{eq:Oxi1}
    \frac{\partial O_{\xi}}{\partial \xi} &=& \sumint W(\mathbf{X}) \xi^{N_\textnormal{pp}(\mathbf{X})} \frac{O(\mathbf{X})N_\textnormal{pp}(\mathbf{X})}{\xi}\ , \\
    \frac{\partial Z_{\xi}}{\partial \xi} &=& \sumint W(\mathbf{X}) \xi^{N_\textnormal{pp}(\mathbf{X})} \frac{N_\textnormal{pp}(\mathbf{X})}{\xi}\ . \label{eq:Zxi1}
\end{eqnarray}
Combining Eqs.~(\ref{eq:Oxi1}), (\ref{eq:Zxi1}) with the quotient rule yields
\begin{eqnarray}\label{eq:derivative}
    \frac{\partial \braket{\hat{O}}_{\xi}}{\partial\xi} = \frac{\braket{\hat{O}\hat{N}_\textnormal{pp}}_\xi - \braket{\hat{O}}_\xi \braket{\hat{N}_\textnormal{pp}}_\xi}{\xi}\ .
\end{eqnarray}
Evidently, Eq.~(\ref{eq:derivative}) can directly be used to compute the $\xi$-derivatives of any expectation value, including $\braket{\hat{O}\hat{N}_\textnormal{pp}}_\xi$ and $\braket{\hat{N}_\textnormal{pp}}_\xi$ that are needed for the computation of the higher-order derivatives. In this way, Eq.~(\ref{eq:derivative}) serves as a generator of derivatives of arbitrary order. We find
\begin{widetext}
\begin{eqnarray}\label{eq:derivative2}
    \frac{\partial^2\braket{\hat{O}}_{\xi}}{\partial\xi^2} &=& \frac{1}{\xi^2}
    \Bigg\{
    \braket{\hat{O}}_\xi
    \left(\braket{\hat{N}_{pp}}_\xi+2\braket{\hat{N}_{pp}}^2_\xi-\braket{\hat{N}^2_{pp}}_\xi\right)
    -\braket{\hat{O}\hat{N}_{pp}}_\xi
    \left(2\braket{\hat{N}_{pp}}_\xi+1\right)
    +\braket{\hat{O}\hat{N}^2_{pp}}_\xi
    \Bigg\}\ . \\ \label{eq:derivative3}
    \frac{\partial^3\braket{\hat{O}}_{\xi}}{\partial\xi^3} &=& \frac{1}{\xi^3}
    \Bigg\{
    \braket{\hat{O}}_\xi
    \left(2\braket{\hat{N}_{pp}}_\xi+6\braket{\hat{N}_{pp}}^2_\xi-3\braket{\hat{N}^2_{pp}}_\xi\right.
    \left.+6\braket{\hat{N}_{pp}}^3_\xi
    -6\braket{\hat{N}_{pp}}_\xi\braket{\hat{N}^2_{pp}}_\xi+\braket{\hat{N}^3_{pp}}_\xi\right)
    \\\nonumber & & -\braket{\hat{O}\hat{N}_{pp}}_\xi
    \left(2+6\braket{\hat{N}_{pp}}_\xi+6\braket{\hat{N}_{pp}}^2_\xi-3\braket{\hat{N}^2_{pp}}_\xi\right)
    +\braket{\hat{O}\hat{N}^2_{pp}}_\xi
    3\left(1+\braket{\hat{N}_{pp}}_\xi\right)
    -\braket{\hat{O}\hat{N}^3_{pp}}_\xi
    \Bigg\}\,.
\end{eqnarray}
\end{widetext}


We then consider the case where $\xi=0$, which is of the highest interest for the Taylor extrapolation to the fermionic limit of $\xi=-1$ discussed above. It is straightforward to obtain
\begin{eqnarray}\label{eq:Z_derivatives}
 \frac{\partial^\nu}{\partial\xi^\nu}Z_\xi\bigg|_{\xi=0} &=& \nu!\ \sumint_{N_\textnormal{pp}\equiv\nu}\textnormal{d}\mathbf{X}\ W(\mathbf{X}) =: Z(\nu)\ \nu! 
\end{eqnarray}
\begin{eqnarray}\label{eq:O_derivatives}
    \frac{\partial^\nu}{\partial\xi^\nu}O_\xi\bigg|_{\xi=0} &=& \nu!\ \sumint_{N_\textnormal{pp}\equiv\nu}\textnormal{d}\mathbf{X}\ W(\mathbf{X}) O(\mathbf{X})\\  &=& Z(\nu)\braket{\hat{O}}_{N_\textnormal{pp}\equiv\nu}\ \nu!\ , \nonumber
\end{eqnarray}
where $\sumint_{N_\textnormal{pp}\equiv\nu}\textnormal{d}\mathbf{X}$ implies that only path-integral configurations with $N_\textnormal{pp}=\nu$ contribute.
Eqs.~(\ref{eq:O_Z_ratio}) lends itself to an implementation of the generalized quotient rule~\cite{comtet_book}
\begin{widetext}
    \begin{eqnarray}\label{eq:quotient_rule}
      \frac{\partial^\nu}{\partial\xi^\nu} \left(\frac{O_\xi}{Z_\xi}\right) = \sum_{\kappa=0}^\nu \binom{\nu}{\kappa} O_\xi^{(\kappa)} \frac{(-1)^{\nu-\kappa}}{[Z_\xi]^{\nu-\kappa+1}} B_{\nu-\kappa}\left( Z_\xi^{(1)}, Z_\xi^{(2)},\dots,Z_\xi^{(\nu-\kappa)} \right) \quad ,
    \end{eqnarray}
\end{widetext}
with $B_{\nu-\kappa}(\dots)$ the complete Bell polynomials. In practice, we can estimate $Z(\nu)$ and $\braket{\hat{O}}_{N_\textnormal{pp}\equiv\nu}$, up to a proportionality factor that cancels in view of Eq.~(\ref{eq:quotient_rule}), by only measuring observables in our PIMC simulation when $N_\textnormal{pp}=\nu$,
\begin{eqnarray}\label{eq:wroclaw}
    \frac{Z(\nu)}{Z_\textnormal{tot}} &=& \left< \hat{\delta}_{N_\textnormal{pp},\nu} \right>_\textnormal{tot} \\ \nonumber
    \braket{\hat{O}}_{N_\textnormal{pp}\equiv\nu} &=& \frac{Z_\textnormal{tot}}{Z(\nu)}\left< \hat{O}\hat{\delta}_{N_\textnormal{pp},\nu}  \right>_\textnormal{tot} \ ,
\end{eqnarray}
with $\braket{\dots}_\textnormal{tot}$ and $Z_\textnormal{tot}$ indicating the full configuration space that can contain paths with any value of $N_\textnormal{pp}$.
For the first three orders, Eq.~(\ref{eq:quotient_rule}) gives
\begin{widetext}
\begin{eqnarray}\label{eq:I1}
    \frac{\partial\braket{\hat O}_\xi}{\partial\xi}\Bigg|_{\xi=0} &=& \frac{Z_{N_\textnormal{pp}\equiv1}}{Z_{N_\textnormal{pp}\equiv0}}\Big\{ \braket{\hat O}_{N_\textnormal{pp}\equiv1} - \braket{\hat O}_{N_\textnormal{pp}\equiv0} \Big\} \\\label{eq:I2}
    \frac{\partial^2\braket{\hat O}_\xi}{\partial\xi^2}\Bigg|_{\xi=0} &=& 2\Bigg\{ \frac{Z_{N_\textnormal{pp}\equiv2}}{Z_{N_\textnormal{pp}\equiv0}}\Big( \braket{\hat O}_{N_\textnormal{pp}\equiv2}-\braket{\hat O}_{N_\textnormal{pp}\equiv0}\Big) -  \left(\frac{Z_{N_\textnormal{pp}\equiv1}}{Z_{N_\textnormal{pp}\equiv0}}\right)^2\Big( \braket{\hat O}_{N_\textnormal{pp}\equiv1}-\braket{\hat O}_{N_\textnormal{pp}\equiv0}\Big)\Bigg\} \\ \label{eq:I3}
    \frac{\partial^3\braket{\hat O}_\xi}{\partial\xi^3}\Bigg|_{\xi=0} &=& 6 \Bigg\{ \frac{Z_{N_\textnormal{pp}\equiv3}}{Z_{N_\textnormal{pp}\equiv0}}\Big( \braket{\hat{O}}_{N_\textnormal{pp}\equiv3} - \braket{\hat{O}}_{N_\textnormal{pp}\equiv0}\Big) + \left(\frac{Z_{N_\textnormal{pp}\equiv1}}{Z_{N_\textnormal{pp}\equiv0}}\right)^3\Big(\braket{\hat{O}}_{N_\textnormal{pp}\equiv1}-\braket{\hat{O}}_{N_\textnormal{pp}\equiv0}\Big) \Bigg\} \\\nonumber 
    & & - \left(  \frac{Z_{N_\textnormal{pp}\equiv2}}{Z_{N_\textnormal{pp}\equiv0}}\right)\left( \frac{Z_{N_\textnormal{pp}\equiv1}}{Z_{N_\textnormal{pp}\equiv0}} \right)\Bigg\{ 6 \braket{\hat{O}}_{N_\textnormal{pp}\equiv2}  + 6\braket{\hat{O}}_{N_\textnormal{pp}\equiv1} - 12 \braket{\hat{O}}_{N_\textnormal{pp}\equiv0}\Bigg\}\ . 
\end{eqnarray}
\end{widetext}

We conclude the theoretical background with a concise summary of the structure of the resulting Taylor extrapolation around $\xi_\textnormal{T}=0$ to the fermionic limit of $\xi=-1$. First, evaluating the Taylor series up to an order $p$ involves path configurations with up to $N_\textnormal{pp}=p$ pair permutations. These path configurations can, in general, be comprised of a variety of combinations of permutation cycles and topologies~\cite{Dornheim_permutation_cycles,DuBois,Vorontsov_PRA_1993}. For a spin-unpolarized system, the maximum possible number of pair exchanges is given by $N_\textnormal{pp}^\textnormal{max}=2(N/2-1)$, nevertheless, derivatives of the order $\nu>N_\textnormal{pp}^\textnormal{max}$ do not vanish, since terms with $N_\textnormal{pp}<\nu$ contribute to Eq.~(\ref{eq:quotient_rule}). Finally, we note that the evaluation of Eqs.~(\ref{eq:Z_derivatives},\ref{eq:O_derivatives}), which is the only PIMC expectation value required to evaluate the Taylor series, does not involve any fermionic cancellation and is, thus, formally sign-problem free. Unfortunately, the generalized quotient rule Eq.~(\ref{eq:quotient_rule}) does involve cancellations between these expectation values [this can be discerned already in Eqs.~(\ref{eq:I1}-\ref{eq:I3})].
As we shall see in Sec.~\ref{sec:results}, this re-introduces the original fermion sign problem in the limit of $p\to\infty$. Yet, a reasonable and often well justified truncation of the Taylor series at a finite $p\lesssim3$ can produce reliable results without the full cancellation problem as we will demonstrate empirically in the following.

\section{Results\label{sec:results}}

We use a canonical adaption~\cite{Dornheim_PRB_nk_2021} of the worm algorithm by Boninsegni \emph{et al.}~\cite{boninsegni1,boninsegni2} as it has been implemented into the open-source \texttt{ISHTAR} code~\cite{ISHTAR}.
A repository containing all PIMC results is available online~\cite{repo}. CPIMC results have been obtained using the open-source Julia implementation \texttt{CPIMC.jl} \cite{cpimc.jl}.

\subsection{General results: dependence on $\xi$ and radius of convergence\label{sec:general}}

\begin{figure*}
    \centering
    \includegraphics[width=0.44\linewidth]{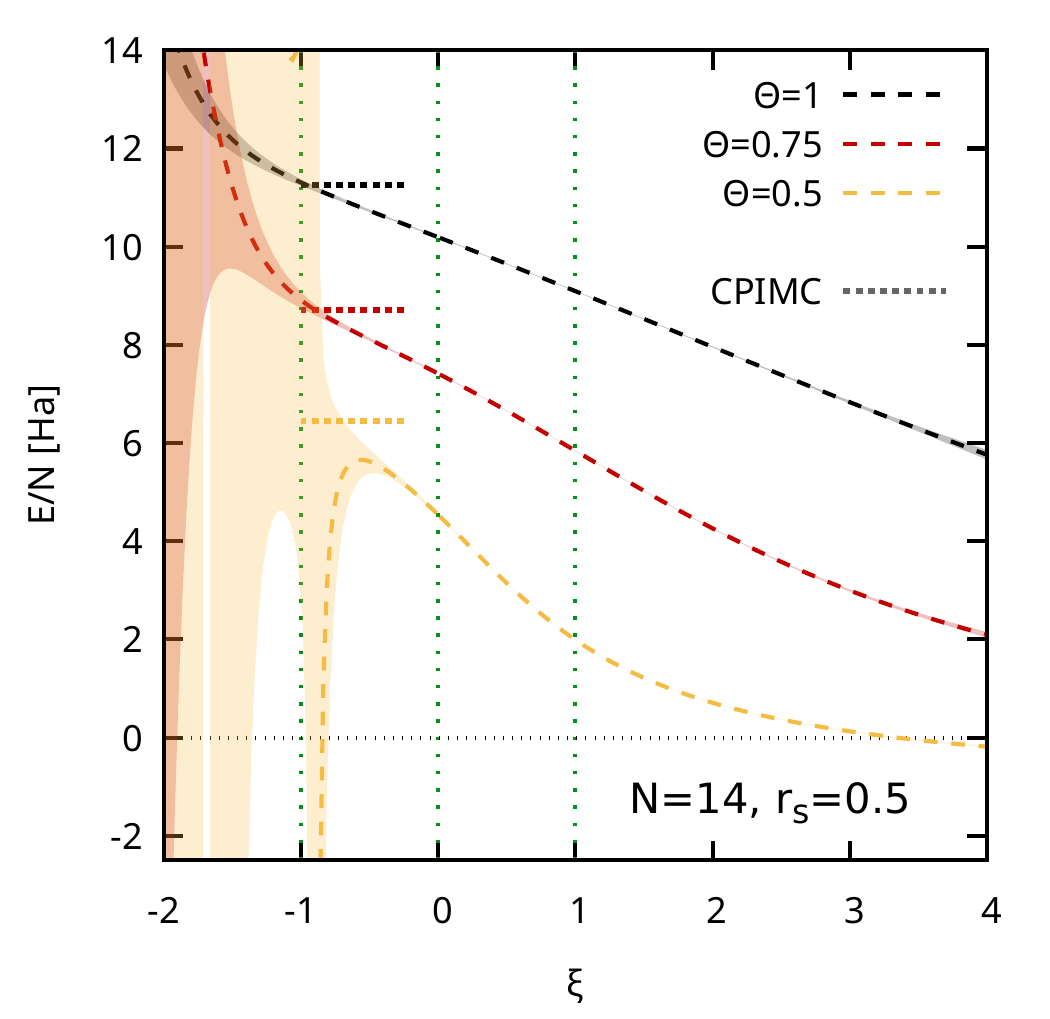}\includegraphics[width=0.44\linewidth]{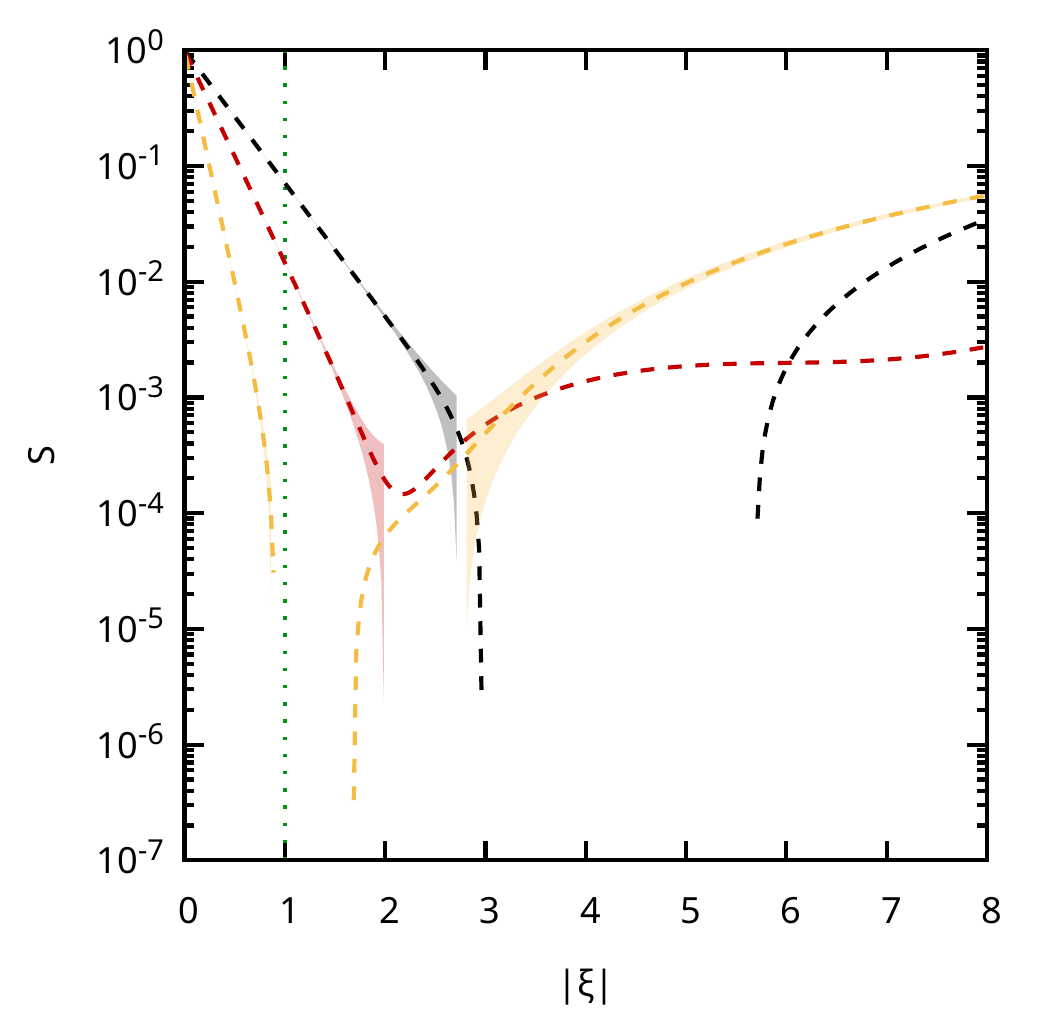}
    \caption{\label{fig:compare_UEG_N14_rs0p5_theta} Dependence of the total energy (per particle) [left] and the average sign $S$ [right] on the fictitious quantum statistics variable $\xi$ for the unpolarized UEG at $r_s=0.5$ with $N=14$ electrons. Results are shown for $\Theta=1$ (black), $\Theta=0.75$ (red), $\Theta=0.5$ (gold). The shaded areas indicate the corresponding uncertainty, the horizontal dotted lines correspond to highly accurate CPIMC reference results for $\xi=-1$, and the vertical dotted green lines indicate the physically meaningful cases of $\xi=-1$ (Fermi-Dirac), $\xi=0$ (Maxwell-Boltzmann) and $\xi=1$ (Bose-Einstein). For each quantum degeneracy, the entire depicted $\xi$-dependence has been obtained from a single PIMC simulation at a reference value of $\xi_\textnormal{ref}=1$ using the re-weighting estimator Eq.~(\ref{eq:final})~\cite{dornheim2025reweightingestimatorabinitio}.}
\end{figure*}

As a starting point, we investigate the dependence of different observables on the quantum statistics variable $\xi$ for the UEG at $r_s=0.5$ in Fig.~\ref{fig:compare_UEG_N14_rs0p5_theta}. These conditions can be realized in compression experiments, e.g., at the NIF~\cite{Tilo_Nature_2023,Moses_NIF}, and play an important role in inertial fusion energy applications~\cite{vorberger2025roadmapwarmdensematter}. Furthermore, due to the relatively weak coupling, highly accurate CPIMC~\cite{Groth_PRB_2016,Schoof_PRL_2015} reference results are available as a rigorous benchmark for any attempted Taylor series extrapolation.
Finally, these conditions were also considered by Dornheim \emph{et al.}~\cite{Dornheim_JCP_xi_2023}, who found that the original $\xi$-extrapolation guided by Eq.~(\ref{eq:extrapolation}) breaks down for $\Theta=0.5$. Below, we will analyze this breakdown from the perspective of the full Taylor series expansion introduced in Sec.~\ref{sec:Taylor}.

All results shown in Fig.~\ref{fig:compare_UEG_N14_rs0p5_theta} have been obtained for $N=14$ particles, and the black, red, and golden curves correspond to $\Theta=1$, $\Theta=0.75$, and $\Theta=0.5$, respectively. Specifically, we use the re-weighting estimator from Eq.~(\ref{eq:final}) with a reference value of $\xi_\textnormal{ref}=1$~\cite{dornheim2025reweightingestimatorabinitio}.
The shaded areas represent the statistical uncertainty that we estimate from the ratio estimate given by Hatano~\cite{hatano1994data}; we have compared these error estimates with a standard jackknife procedure~\cite{berg2004markov} for selected cases and found them to be indistinguishable. The vertical dotted green lines indicate the physically relevant cases of fermions, boltzmannons, and bosons.

The left panel shows results for the $\xi$-dependence of the total energy per particle, and the horizontal dotted lines represent quasi-exact CPIMC data (with the associated statistical errors being smaller than the width of the lines). For $\Theta=1$ and $\Theta=0.75$, the direct PIMC method is capable of providing accurate results for the fermionic limit of $\xi=-1$, which nicely agree with the CPIMC results within the error intervals. In stark contrast, the error interval diverges for $\xi\to-1$ for $\Theta=0.5$.
This is a direct consequence of the average sign $S$ vanishing within the given Monte Carlo error bars, see the right panel of Fig.~\ref{fig:compare_UEG_N14_rs0p5_theta}.
At a first glance, the dependence of $S$ on $\xi$ over the full depicted $\xi$-range appears to be counter-intuitive and non-trivial. Starting at $|\xi|=0$, we find the usual monotonic decay of the sign that appears to follow the expected exponential behavior~\cite{Dornheim_JCTC_2025} for all three temperatures. At some point, the sign vanishes within error bars; unfortunately, this happens before the fermionic limit of $\xi=-1$ for $\Theta=0.5$ and for somewhat larger $|\xi|$ for the two higher values of $\Theta$. For even larger $|\xi|$, the sign starts to increase again. This is a direct consequence of the function of the $\xi$-term as an $N_\textnormal{pp}$-dependent weight in the partition function, cf.~Eq.~(\ref{eq:Z_generic}). For $|\xi|\gg1$, configurations with large $N_\textnormal{pp}$ are actually being favored in the PIMC generated Markov chain, eventually leading to an exclusive sampling of paths with the maximum possible number $N_\textnormal{pp}^\textnormal{max}$. In this limit (which is always associated with a positive sign for spin-unpolarized systems), there is no more cancellation between adjacent permutation sectors, and the simulation will be sign-problem free for both positive and negative values of $\xi$. 

Returning to the $\xi$-dependence of the total energy per particle in the left panel of Fig.~\ref{fig:compare_UEG_N14_rs0p5_theta}, we observe an increasingly steep dependence of $E(\xi)$ on $\xi$ for $\xi\in[-1,1]$ with decreasing temperatures, reflecting the increasing dissimilarity between the corresponding bosonic and fermionic limits as the impact of quantum statistics becomes more important. Moreover, the dependence on $\xi$ appears to be almost linear in this range for $\Theta=1$ and still relatively simple for $\Theta=0.75$.
In contrast, the exact CPIMC result for $\xi=-1$ implies the presence of an inflection point around $\xi\approx-0.7$ for $\Theta=0.5$, thus resulting in a substantially more complicated dependence of $E$ on $\xi$. In the context of the present work, a crucial question is thus if it is even theoretically possible to use a Taylor series to go beyond the inflection point and reach $\xi=-1$ even in the limit of $p\to\infty$.

\begin{figure}
    \centering
    \includegraphics[width=0.44\textwidth]{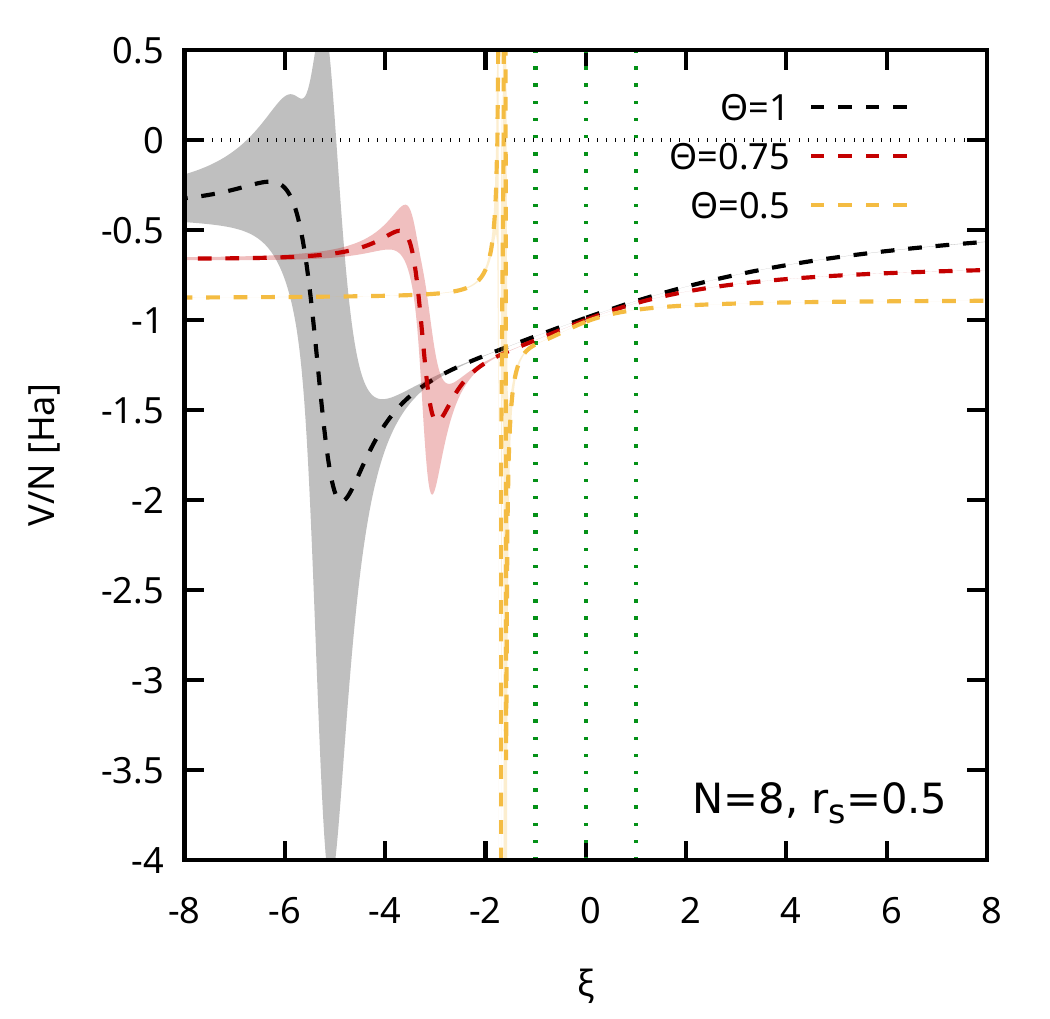}\\\vspace{-0.00005cm}\includegraphics[width=0.44\textwidth]{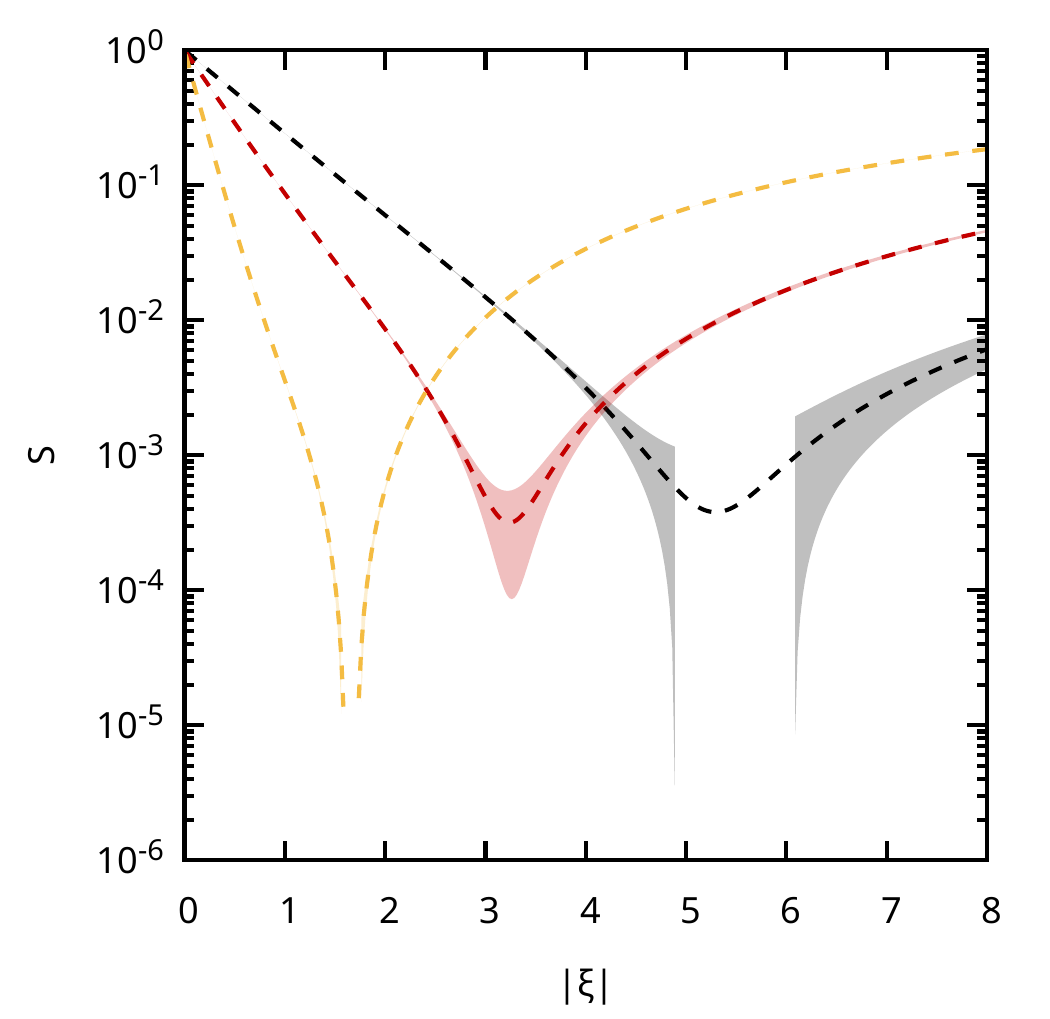}
    \caption{\label{fig:temperature_in_poland} Dependence of the interaction energy per particle [top] and the average sign [bottom] on the fictitious quantum statistics variable $\xi$ for the unpolarized UEG with $N=8$ particles at $r_s=0.5$ and $\Theta=1$ (black), $\Theta=0.75$ (red), $\Theta=0.5$ (gold).}
\end{figure}

\begin{figure}
    \centering
\includegraphics[width=0.44\textwidth]{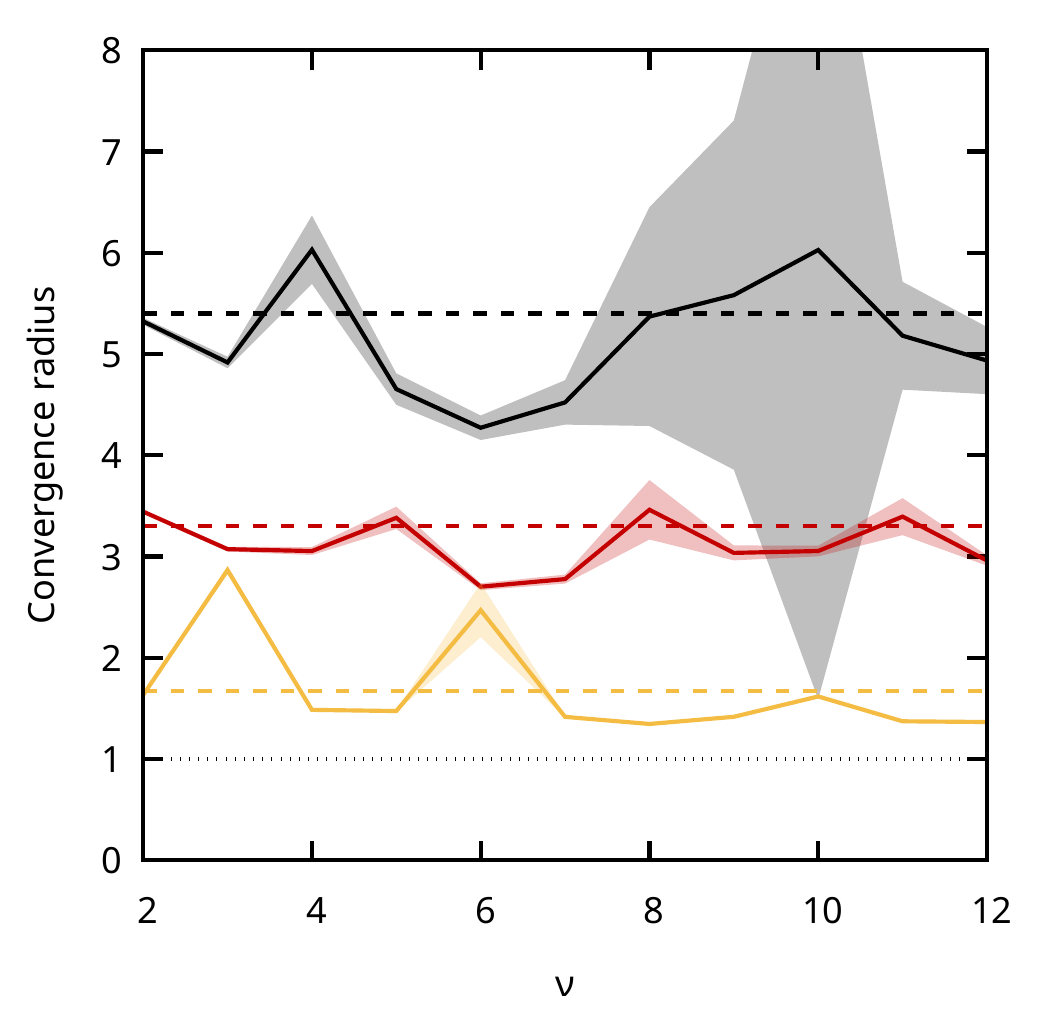}
    \caption{\label{fig:temperature_in_poland2} Dependence of the radius of convergence estimate of the Taylor series around $\xi_\textnormal{T}=0$ on the polynomial order $\nu$ [cf.~Eq.~(\ref{eq:radius_root})] for the unpolarized UEG with $N=8$ particles at $r_s=0.5$ and $\Theta=1$ (black), $\Theta=0.75$ (red), $\Theta=0.5$ (gold). The horizontal dashed lines indicate the position of the "pseudo-poles" shown in the top panel of Fig.~\ref{fig:temperature_in_poland}. The error estimates (shaded areas) have been obtained using a standard jackknife resampling scheme~\cite{berg2004markov}.
    }
\end{figure}

To answer this question, we have repeated the above simulations for $N=8$ electrons, for which the fermionic limit of $\xi=-1$ can be reached within the limit of feasible computational effort, and we find an average sign of $S=0.00359(2)$ for $\Theta=0.5$.
The results are shown in Fig.~\ref{fig:temperature_in_poland}, and the top panel shows the interaction energy per particle as a function of $\xi$ with the usual color code.
Equivalent conclusions can be drawn from the total energy and kinetic energy, but the interaction energy is particularly convenient for our purposes as its Monte Carlo estimator has an intrinsically lower variance~\cite{Janke_JCP_1997}. Interestingly, we find a pole-like phenomenon for all three curves for different values of $|\xi|$. These "pseudo-divergencies" can be directly traced back to the average sign vanishing within error bars, see the bottom panel of Fig.~\ref{fig:temperature_in_poland}.
For completeness, we note that the existence of an actual pole, or, equivalently, of a truly vanishing sign fundamentally cannot be resolved with PIMC: the only conclusion that we can draw is that the average sign is definitely smaller than our statistical error bars; resolving an actual pole with $S\equiv0$ would, by definition, require an infinite amount of compute time. Since the average sign in direct PIMC is just the ratio of fermionic and bosonic partition functions, $S=Z_\xi/Z_{|\xi|}$~\cite{dornheim_sign_problem}, a vanishing sign would be connected with a zero in $Z_\xi$ as $Z_{|\xi|}$ is always larger than zero.
This phenomenon has recently been explored by He \emph{et al.}~\cite{he2025revisitingfermionsignproblem}. In particular, these authors extended the domain of definition of $\xi$ from the real axis to the complex plane and studied the distribution of the complex roots of $Z_{\xi}$, which translate to complex poles of $\langle\hat{O}\rangle_{\xi}$. When these zeroes approach the fermionic sector $\xi\in[-1,0]$ of the real axis, any polynomial extrapolation is bound to fail. In particular, in the zero-temperature limit, these authors report the occurrence of such $Z_\xi$ zeroes for $\xi=-1,-1/2,-1/3,\dots,-1/(N-1)$, thus fundamentally preventing bosonic to fermionic extrapolations along the real axis. This conclusion is somewhat expected as it has been known that PIMC cannot directly reach the limit of $\beta\to\infty$~\cite{krauth2006statistical}. Fortunately, we do not find any zeros in $Z_\xi$ for $|\xi|\leq1$. At the same time, our PIMC simulations clearly show that the observed "pseudo-poles" / "pseudo-zeros" move closer to $|\xi|=1$ with decreasing $\Theta$ for the investigated temperatures. Despite the fact that we constrained ourselves along the real axis, this is qualitatively consistent with Ref.~\cite{he2025revisitingfermionsignproblem}.

In Fig.~\ref{fig:temperature_in_poland2}, we show the radius of convergence $r$ computed via Eq.~(\ref{eq:radius_root}) as a function of the degree of the coefficients $\nu$ (solid lines) evaluated from the same PIMC simulations as Fig.~\ref{fig:temperature_in_poland} for the expansion point $\xi_\textnormal{T}=0$. Interestingly, these curves are in very good agreement with the position of the "pseudo-poles", which are included as the horizontal dashed lines. Importantly, the radius of convergence includes the physically relevant limit of $|\xi|=1$ in all cases, leaving the door open to a controlled Taylor extrapolation to the fermionic limit.

\begin{figure}
    \centering
    \includegraphics[width=0.44\textwidth]{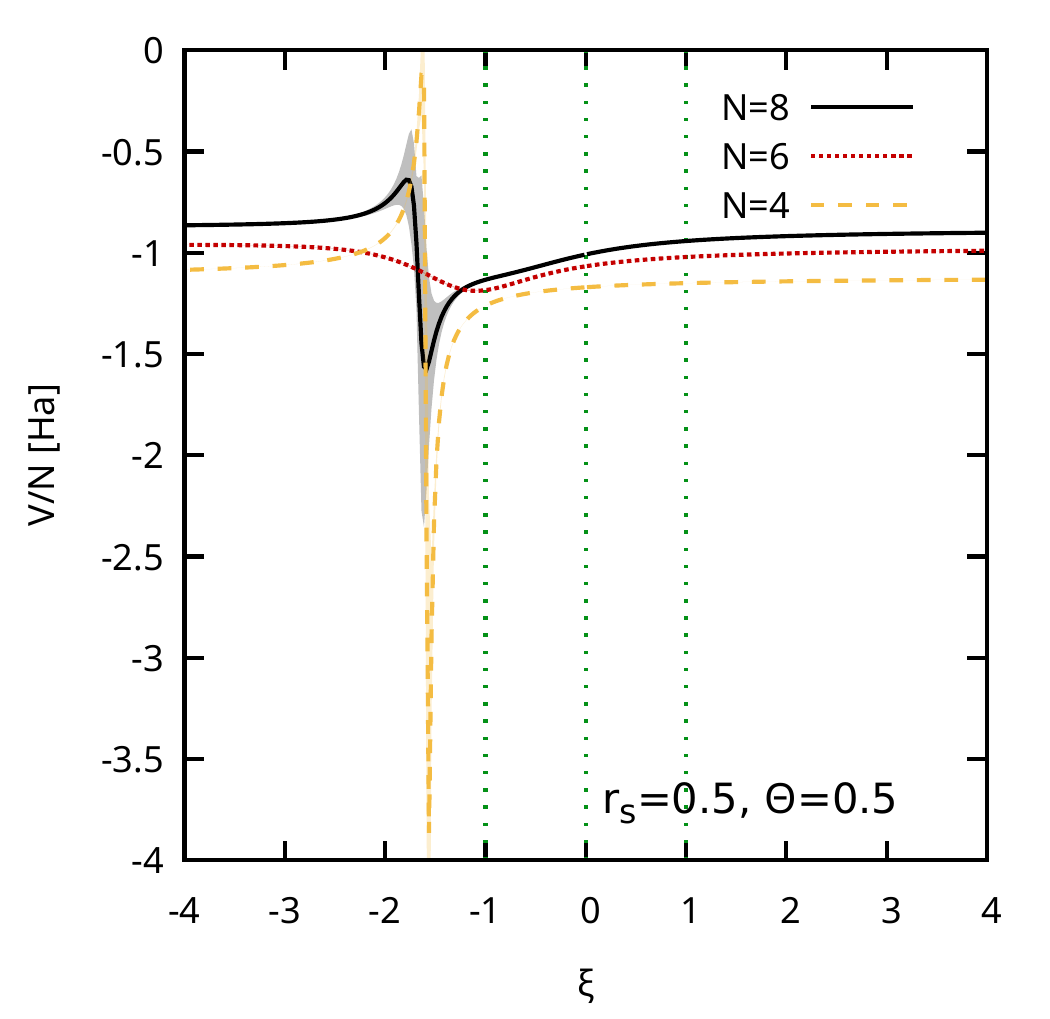}\\\vspace*{-0.005cm}\includegraphics[width=0.44\textwidth]{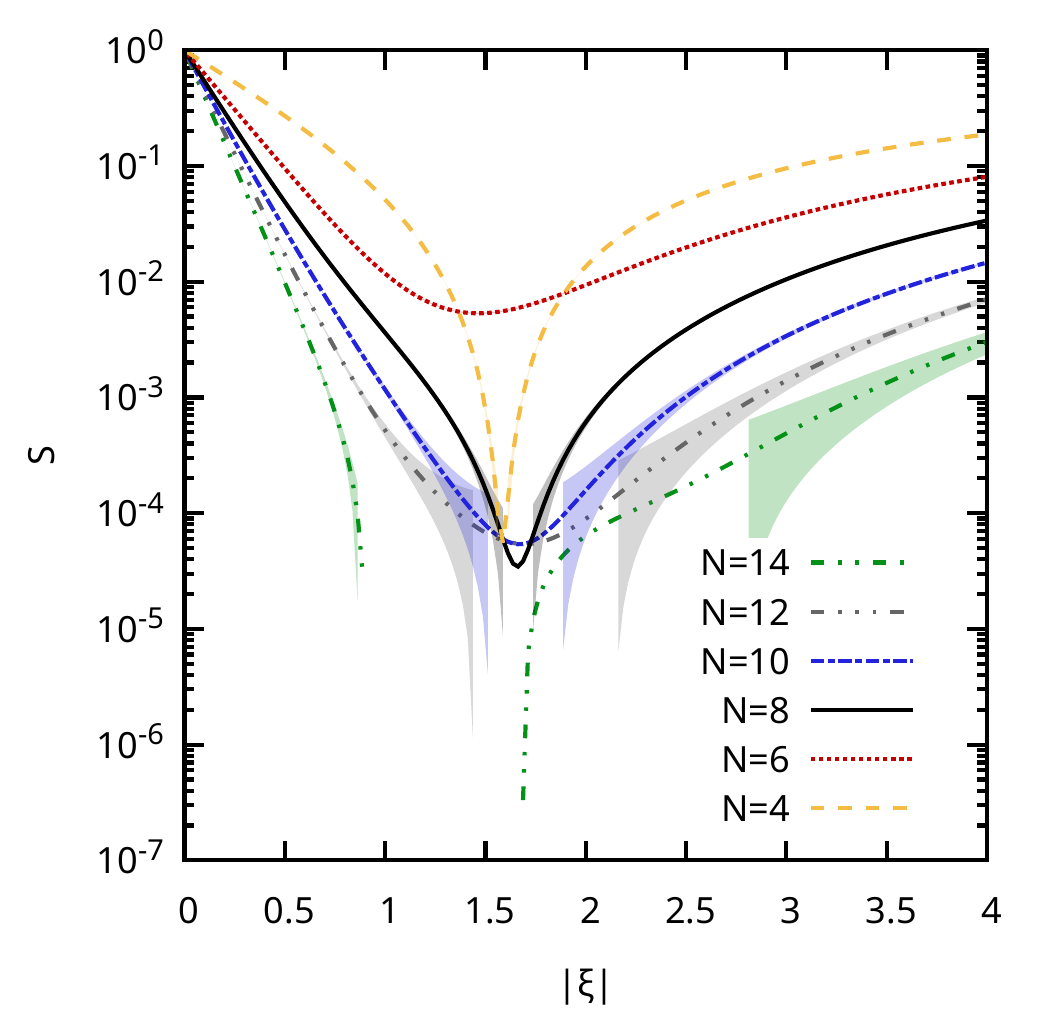}
    \caption{\label{fig:poland} Dependence of the total energy (per particle) [top] and average sign $S$ [bottom] on the fictitious quantum statistics variable $\xi$ for the unpolarized UEG at $r_s=0.5$ and $\Theta=0.5$. Results for $N=8$ (solid black), $N=6$ (dotted red), and $N=4$ (dashed golden). For the average sign, results are also shown for $N=14$ (dashed-double-dotted green), $N=12$ (dashed-triple-dotted grey) and $N=10$ (dash-dotted blue). Shaded areas indicate the statistical uncertainty.}
\end{figure}

\begin{figure}
    \centering
\includegraphics[width=0.44\textwidth]{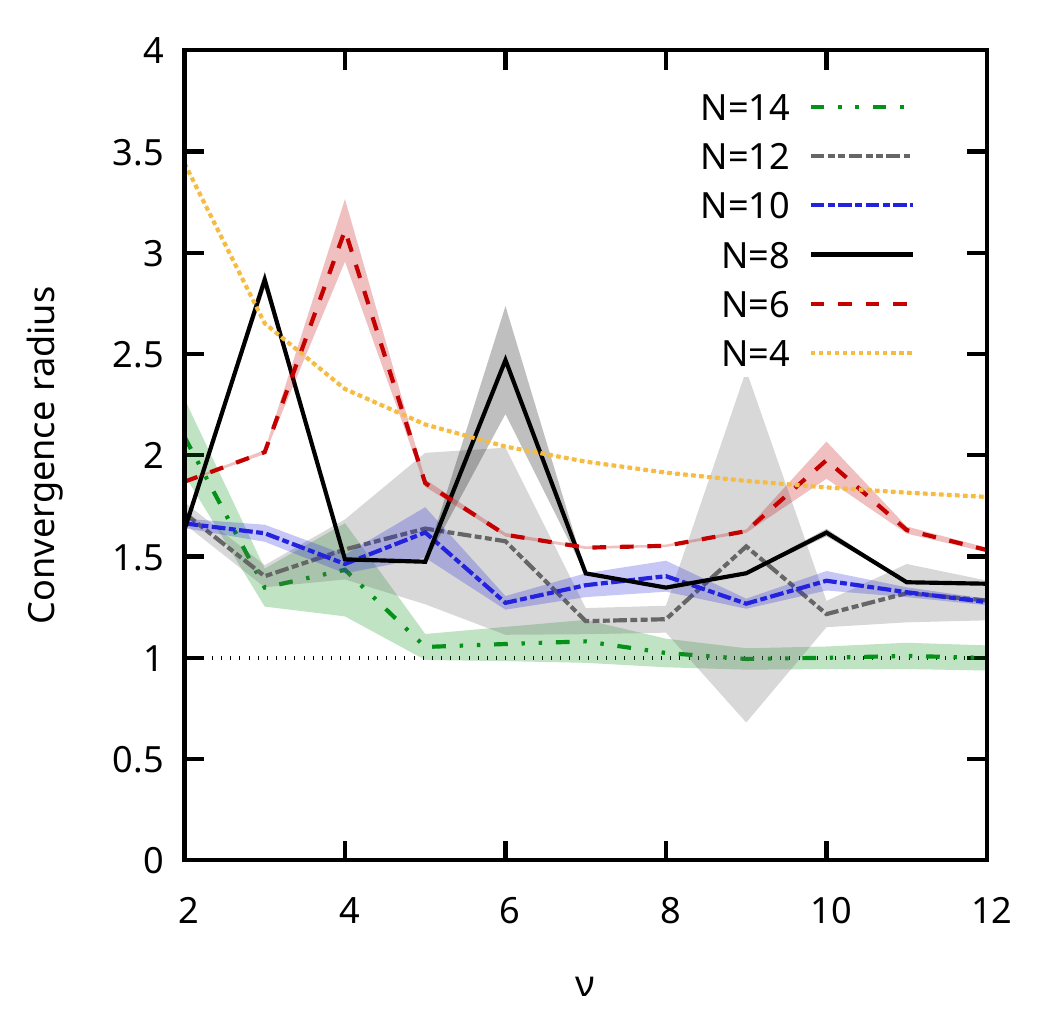}
    \caption{\label{fig:poland2} Dependence of the radius of convergence estimate of the Taylor series around $\xi_\textnormal{T}=0$ on the polynomial order $\nu$ [cf.~Eq.~(\ref{eq:radius_root})] for the unpolarized UEG at $r_s=0.5$ and $\Theta=0.5$. Results for $N=14$ (dashed-double-dotted green), $N=12$ (dashed-triple-dotted grey), $N=10$ (dash-dotted blue), $N=8$ (solid black), $N=6$ (dotted red), and $N=4$ (dashed golden).}
\end{figure}

\begin{figure}
    \centering
    \includegraphics[width=0.44\textwidth]{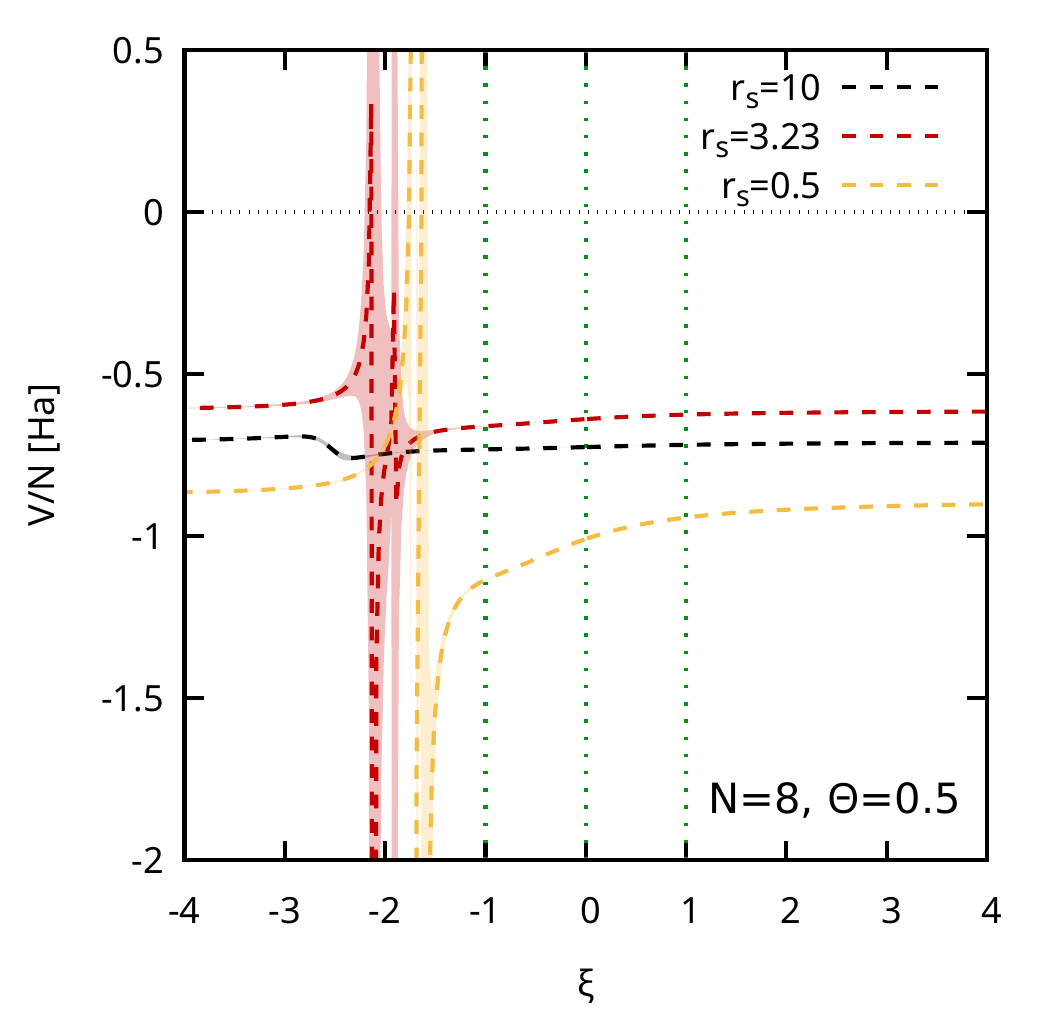}\\\vspace*{-0.005cm}\includegraphics[width=0.44\textwidth]{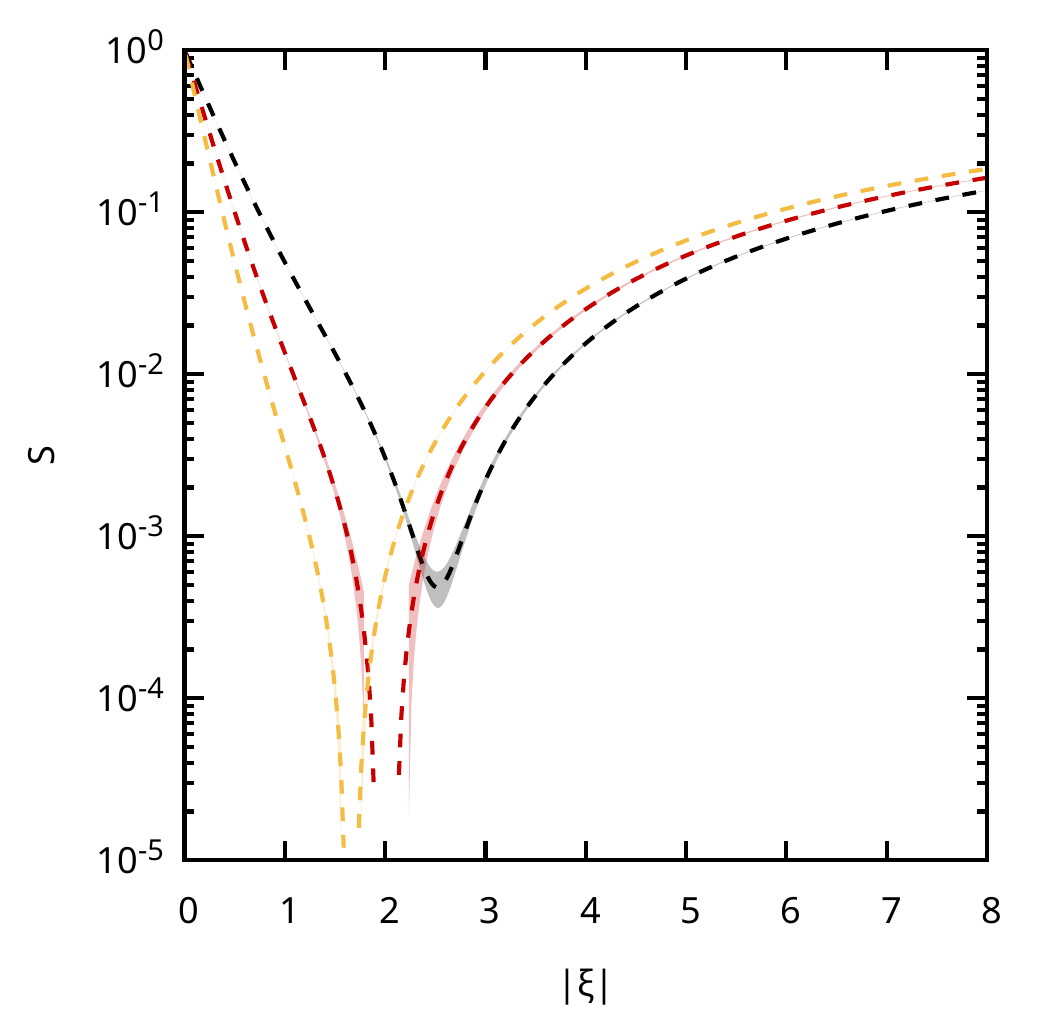}
    \caption{\label{fig:poland_rs} Dependence of the interaction energy per particle [top] and the average sign [bottom] on the fictitious quantum statistics variable $\xi$ for the unpolarized UEG with $N=8$ particles at $\Theta=0.5$. Results for $r_s=10$ (dashed black), $r_s=3.23$ (dashed red), and $r_s=0.5$ (dashed golden).}
\end{figure}

\begin{figure}
    \centering
\includegraphics[width=0.44\textwidth]{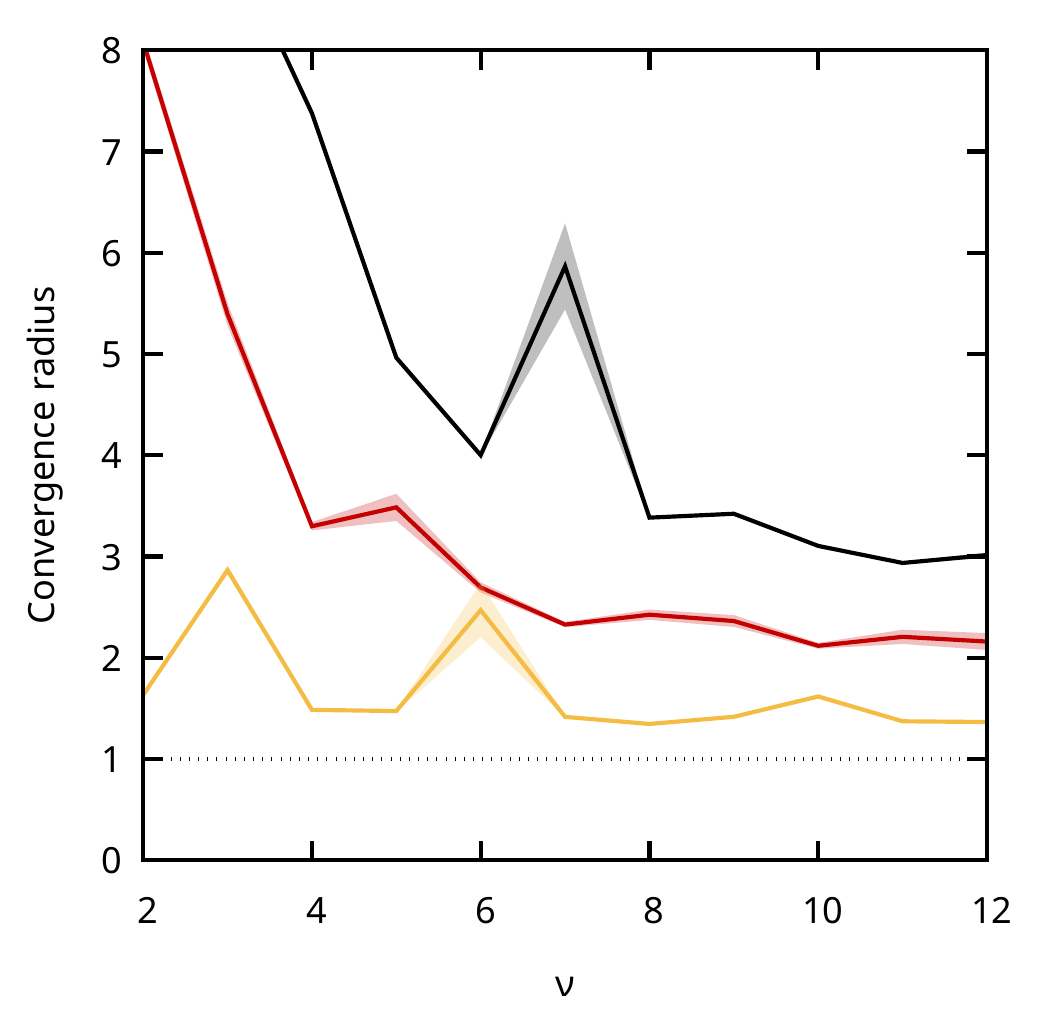}
    \caption{\label{fig:poland2_rs} Dependence of the radius of convergence estimate of a Taylor series around $\xi_\textnormal{T}=0$ as a function of the polynomial order $\nu$ [cf.~Eq.~(\ref{eq:radius_root})] for the unpolarized UEG with $N=8$ particles at $\Theta=0.5$. Results for $r_s=10$ (dashed black), $r_s=3.23$ (dashed red), and $r_s=0.5$ (dashed golden).}
\end{figure}

Let us next consider the dependence of these findings on the number of simulated electrons $N$. In the top panel of Fig.~\ref{fig:poland}, we show the $\xi$-dependence of the potential energy per particle, with a particular focus on the location of the "pseudo-poles". Interestingly, the latter only appears in our PIMC results for $N=4$ (dashed gold) and $N=8$ (solid black), but not for $N=6$ (dotted red). This is further substantiated by the average sign $S$ shown in the bottom panel of Fig.~\ref{fig:poland}. Evidently, $S$ vanishes for all depicted $N$ within the given Monte Carlo error bars around $|\xi|=1.5$, except for $N=6$. This is likely a consequence of some special symmetry for this particular case; otherwise, the occurrence and the location of the "pseudo-pole" appear to be a general feature of the UEG, and likely of Fermi-Dirac systems overall. In Fig.~\ref{fig:poland2}, we show the corresponding convergence radii as a function of the polynomial degree $\nu$.
Overall, these curves are consistently above unity, substantiating the principal possibility of a Taylor extrapolation to the fermionic limit. 

As the final question, we investigate the dependence of these findings on the density parameter in Fig.~\ref{fig:poland_rs} for $N=8$ and $\Theta=0.5$. The black curve has been obtained for $r_s=10$, which is on the margin of the strongly coupled electron liquid regime~\cite{dornheim_dynamic,dornheim_electron_liquid,Takada_PRB_2016,quantum_theory}, and the red curve corresponds to $r_s=3.23$, i.e., the electron number density of solid density hydrogen that is realized, e.g., in experiments with hydrogen jets~\cite{Fletcher_Frontiers_2022,zastrau,Hamann_PRR_2023,Bonitz_POP_2024}; the yellow curve represents $r_s=0.5$, the weakly coupled high-density regime that we have considered so far. Interestingly, we only find the "pseudo-poles" for $r_s=3.23$ and $r_s=0.5$, whereas it is absent for $r_s=10$. Indeed, the sign remains finite and can be resolved within the given error bars over the entire $\xi$-range. Secondly, we find that the position of the "pseudo-poles" moves towards smaller $|\xi|$ with decreasing coupling strength and, thus, with increasing degree of quantum degeneracy. This is also directly reflected by the convergence radii of the Taylor expansion around $\xi_\textnormal{T}=0$ shown in Fig.~\ref{fig:poland2_rs}.

Let us conclude this section with an attempted practical interpretation of the reported observations. Empirically, we find that the radius of convergence of the Taylor expansion around the Maxwell-Boltzmann system is directly connected to the (negative) value of $\xi$ at which the average sign and, hence, the fermionic partition function vanishes. For the physically relevant domain of $\xi\in[-1,1]$, the sign is expected to vanish only in the ground-state limit of $\beta\to\infty$ ($\Theta=0$). We, therefore, expect Taylor extrapolations to the fermionic limit to converge at any finite temperature, even though, as we shall see below, the estimation of all significant coefficients might be unfeasible in practice.

\subsection{PIMC results for $\xi$-derivatives\label{sec:derivatives_results}}

\begin{figure}
    \centering
    \includegraphics[width=0.44\textwidth]{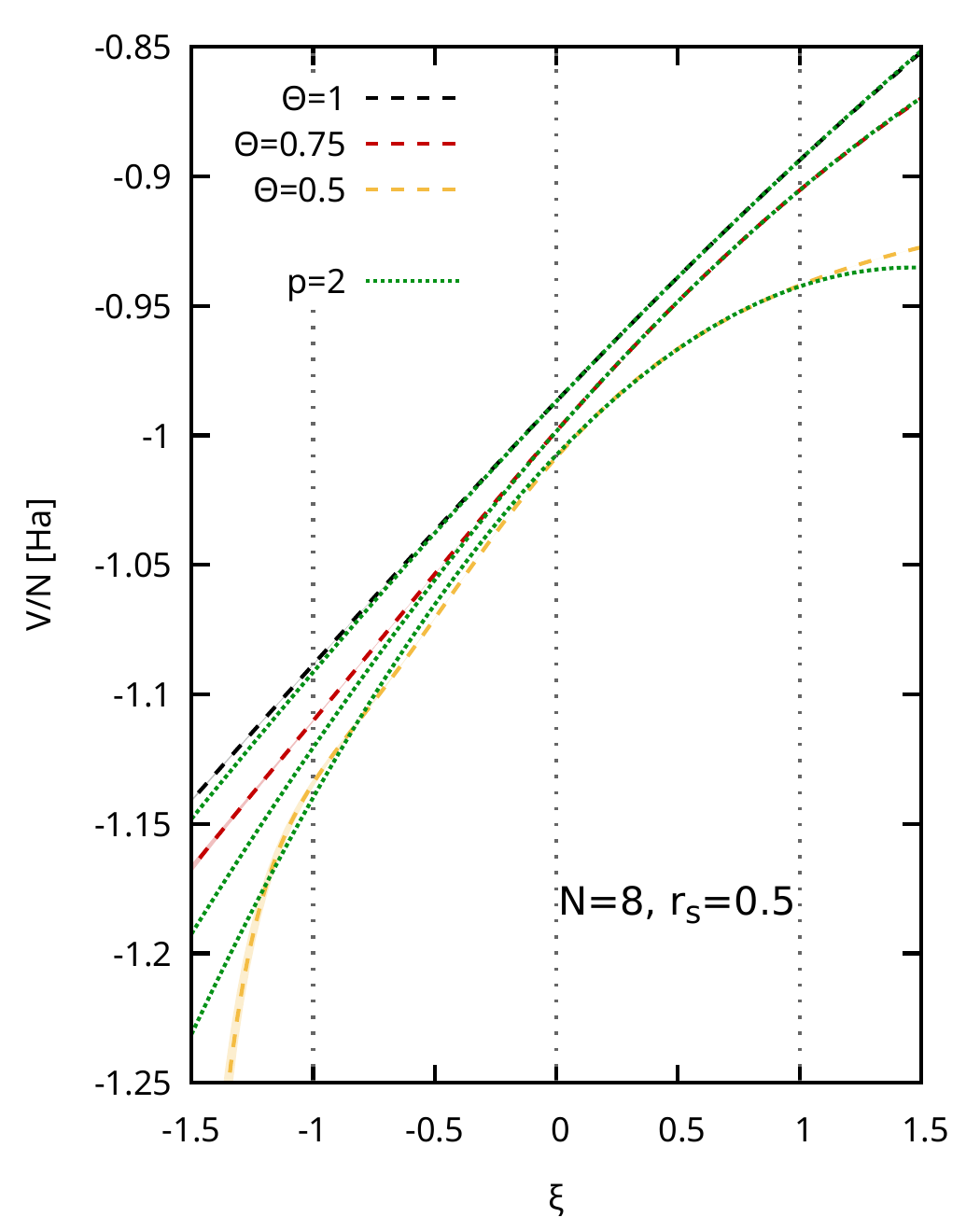}
    \caption{\label{fig:poly_fit} Dependence of the interaction energy per particle on the fictitious quantum statistics variable $\xi$ for the unpolarized UEG with $N=8$ particles at $r_s=0.5$ and $\Theta=1$ (black), $\Theta=0.75$ (red), $\Theta=0.5$ (gold). The green dots correspond to bosonic sector (within the fitting range of $\xi\in[0,1]$) polynomial fits of the order of $p=2$, cf.~Eq.~(\ref{eq:canonical_polynomial}).}
\end{figure}

\begin{figure}
    \centering
    \includegraphics[width=0.44\textwidth]{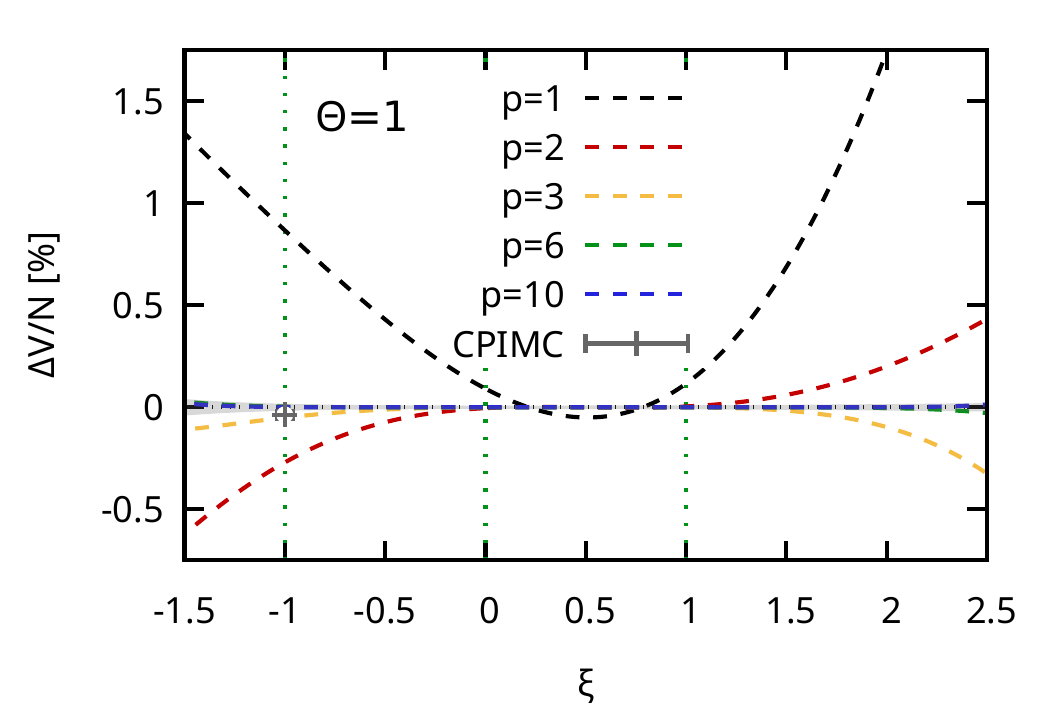}\\\vspace*{-1.15cm}\includegraphics[width=0.44\textwidth]{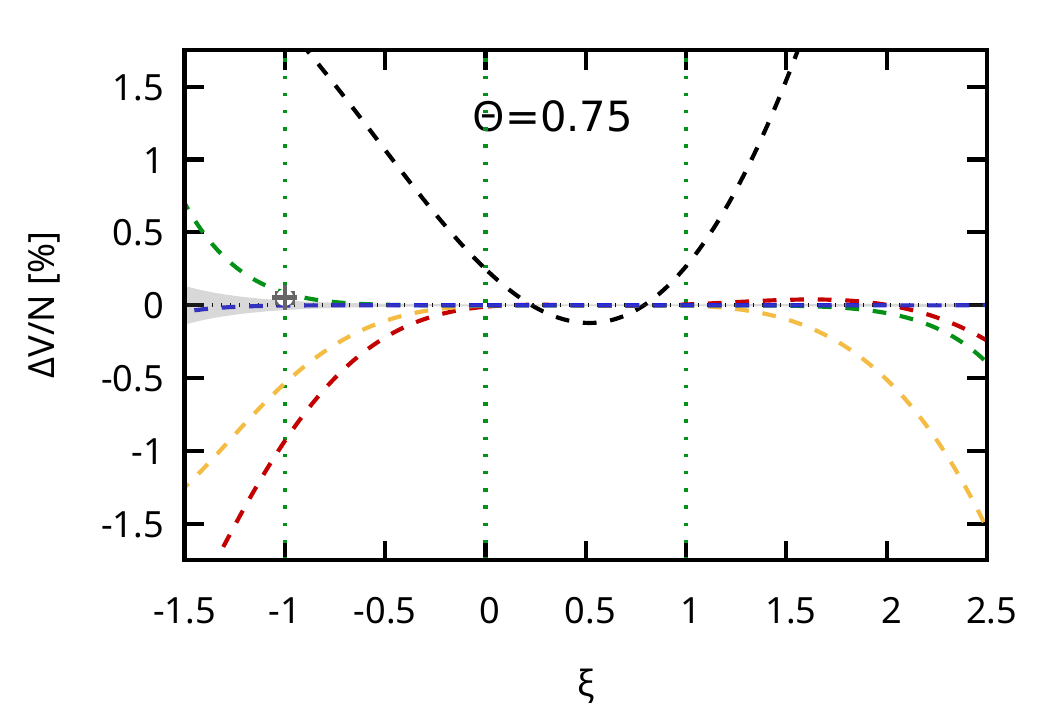}\\\vspace*{-1.15cm}\includegraphics[width=0.44\textwidth]{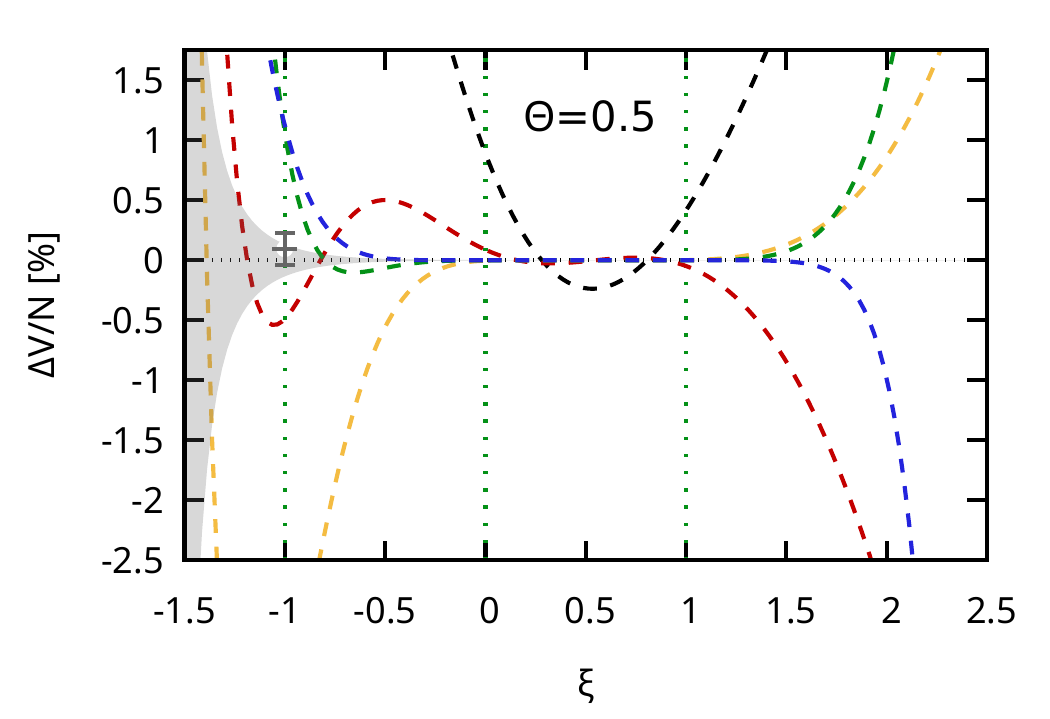}
    \caption{\label{fig:delta_poly_fit} Relative deviation of polynomial fits of different order $p$ [cf.~Eq.~(\ref{eq:canonical_polynomial})] for the interaction energy per particle shown in Fig.~\ref{fig:poly_fit}. The shaded grey areas show the statistical error in the PIMC estimates for $V(\xi)$. Also shown are comparisons to the CPIMC reference data in the fermionic limit of $\xi=-1$ (grey symbols).
    }
\end{figure}

\begin{figure*}
    \centering
    \includegraphics[width=0.32\textwidth]{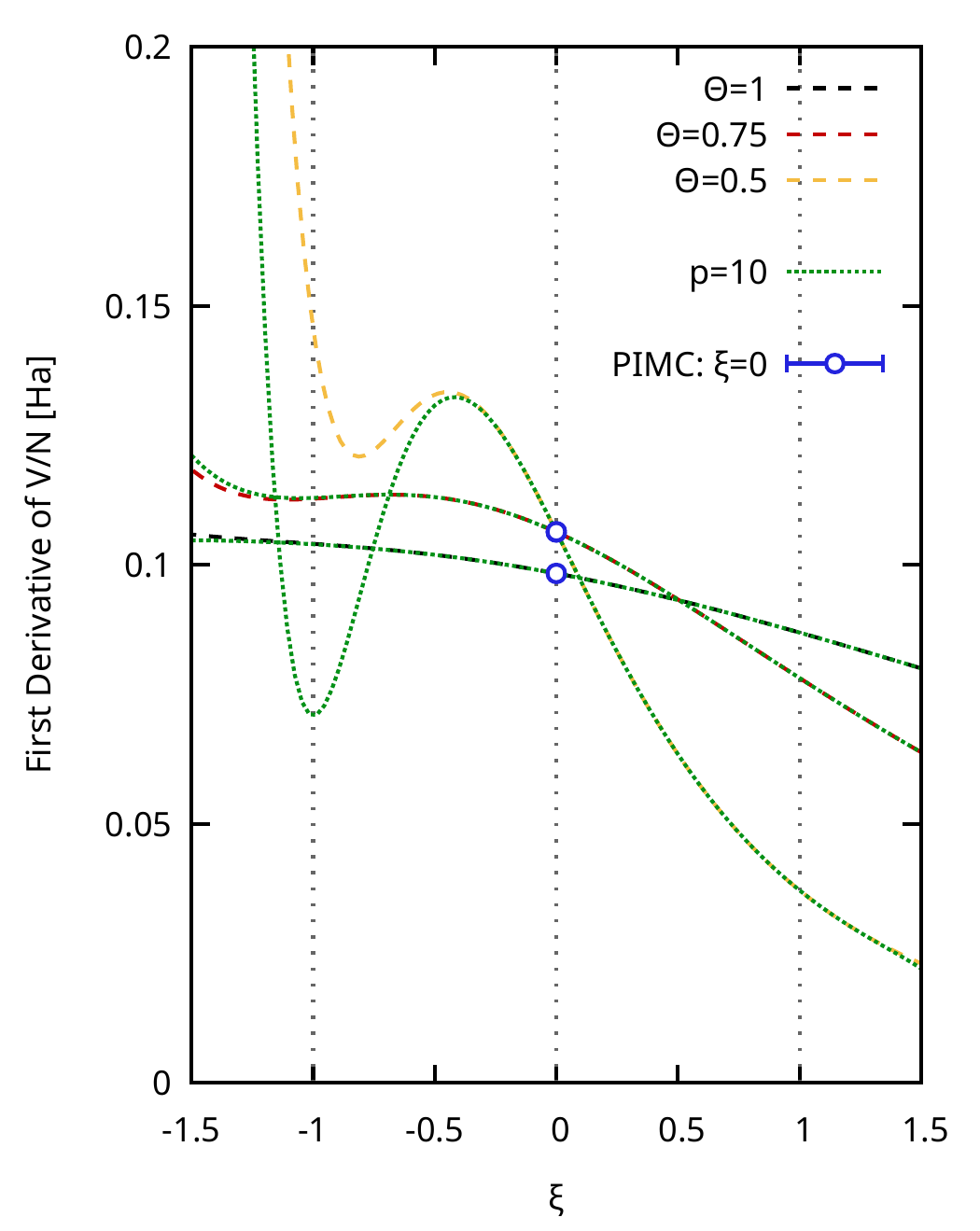} \includegraphics[width=0.32\textwidth]{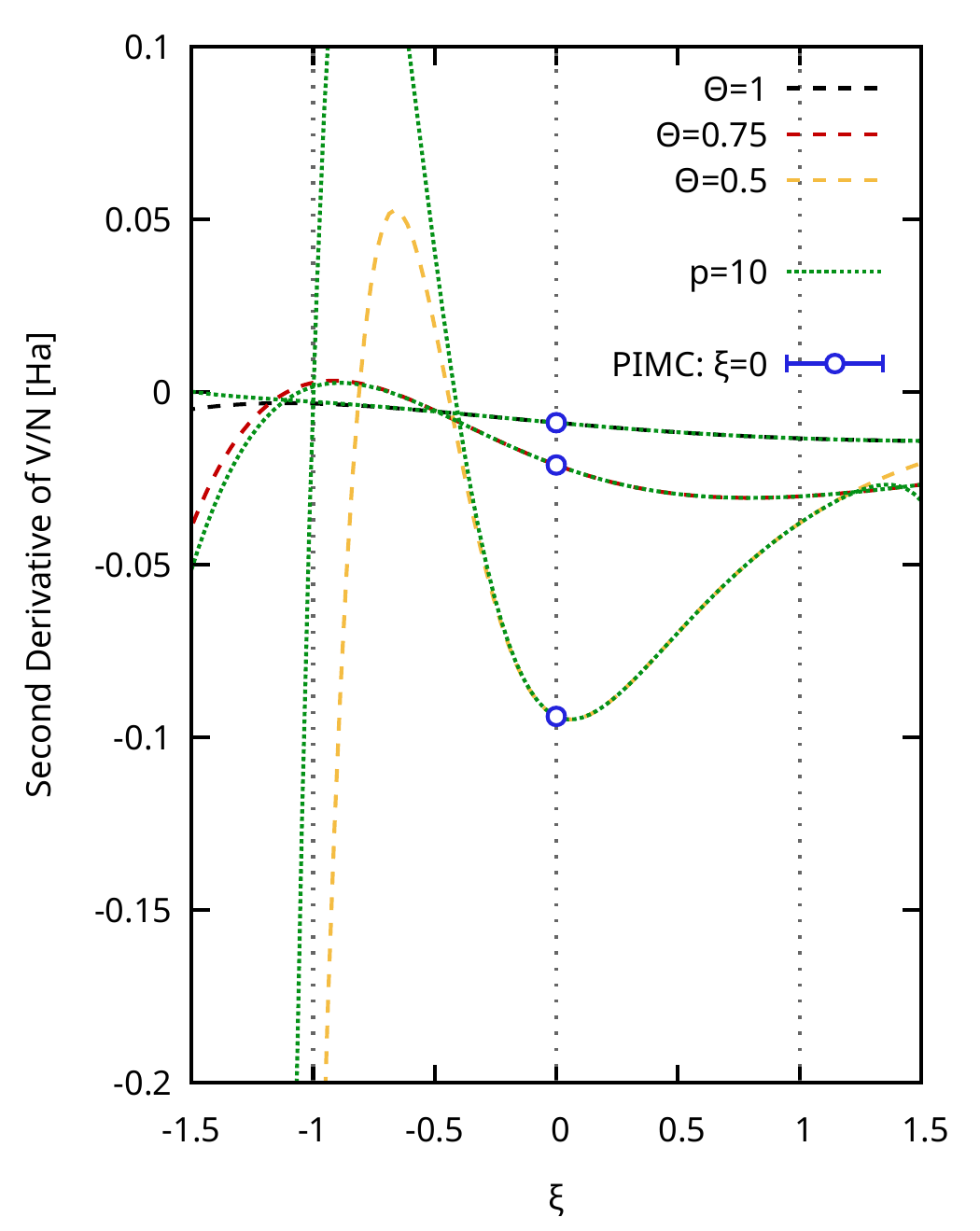} \includegraphics[width=0.32\textwidth]{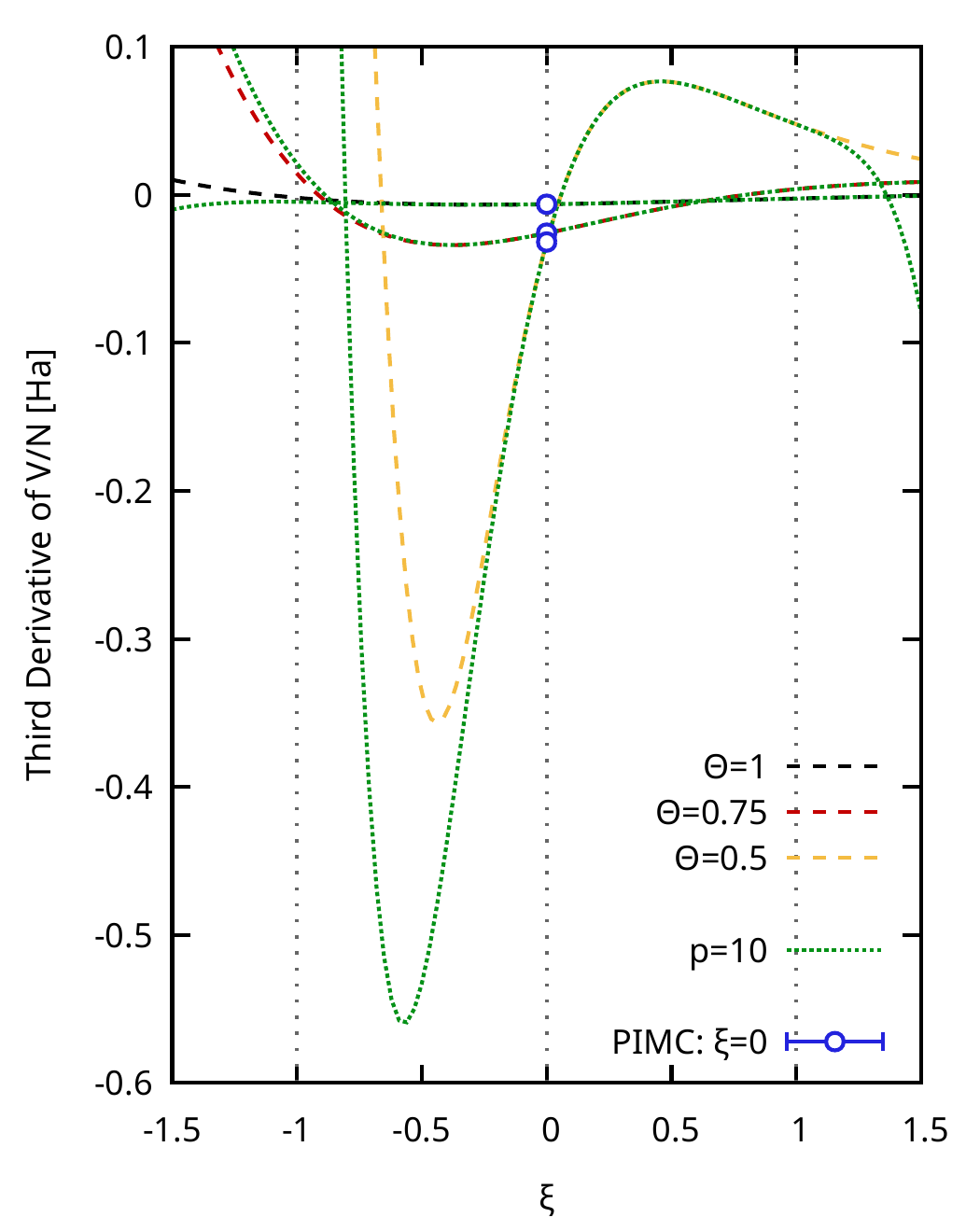}
    \caption{\label{fig:derivatives} Dependence of the first (left), second (center) and third (right) order $\xi-$derivatives of the interaction energy per particle on the fictitious quantum statistics variable $\xi$ for the unpolarized UEG with $N=8$ particles at $r_s=0.5$. The dashed black, red, and golden curves have been computed using Eqs.~(\ref{eq:derivative}), (\ref{eq:derivative2}) and (\ref{eq:derivative3}) for $\Theta=1$, $\Theta=0.75$ and $\Theta=0.5$, respectively. The dotted green curves show the corresponding derivatives of the polynomial fits to $V(\xi)$ with $p=10$, see Figs.~\ref{fig:poly_fit} and \ref{fig:delta_poly_fit}. The blue dots correspond to PIMC results for the $\xi$-derivatives at $\xi=0$ computed via Eq.~(\ref{eq:quotient_rule}).
    }
\end{figure*}

Let us next turn our attention to the derivatives of expectation values with respect to $\xi$, which play a central role for our Taylor series perspective. As an example, we consider the interaction energy per particle $V/N$, which is shown in Fig.~\ref{fig:poly_fit} for $N=8$ at $r_s=0.5$ and three temperatures. In addition, we have performed polynomial fits of the order $p=2$ according to Eq.~(\ref{eq:canonical_polynomial}) above for $\xi\in[0,1]$, corresponding to the usual isothermal $\xi$-extrapolation carried out in previous works~\cite{Xiong_JCP_2022,Dornheim_JCP_xi_2023,Dornheim_JCP_2024,Dornheim_JPCL_2024}.
For $\Theta=1$ (black curve), the dependence of $V$ on $\xi$ is almost linear for $\xi\in[-1,1]$ and the fermionic limit of $\xi=-1$ is accurately reproduced by the parabolic extrapolation. For $\Theta=0.75$ (red curve), the dependence of $V$ on $\xi$ is noticeably less linear. As a result, the parabolic polynomial fit does reproduce its input data in the fitting interval of $\xi\in[0,1]$, but the gap to the true fermionic limit widens compared to the higher temperature. Finally, we find a substantially less trivial dependence of $V$ on $\xi$ for $\Theta=0.5$ (golden curve), with a pronounced curvature for $\xi\in[0,1]$ and a significantly more linear progression for $\xi\in[-1,0]$. The parabolic extrapolation is not capable of reproducing this intricate $\xi$-dependence in the signful domain, and the relatively good agreement at $\xi=-1$ has to be viewed as coincidental.

A more systematic analysis of the polynomial fitting and extrapolation is presented in Fig.~\ref{fig:delta_poly_fit} for different polynomial degrees $p$. For $\Theta=1$ (top panel), only the linear fit ($p=1$, black) is incapable of reproducing the interaction energy in the fitting range of $\xi\in[0,1]$ with deviations of $\Delta V/V\lesssim0.1\%$, resulting in an extrapolation to the fermionic limit with an error of $\sim0.8\%$. Interestingly, this error can be reduced with increasing $p$ beyond $p=2$, although this strategy is only expected to work when the accuracy of the underlying PIMC data is high enough to directly resolve the fermionic limit. The parabolic fit gives a systematic error of $\lesssim0.3\%$ for $\xi=-1$, which is consistent with previous works~\cite{Dornheim_JCP_xi_2023}. Let us next turn to $\Theta=0.75$, for which an analogous analysis is shown in the center panel of Fig.~\ref{fig:delta_poly_fit}. Overall, we observe the same qualitative trends as for $\Theta=1$, albeit with somewhat larger systematic errors compared to $\Theta=1$. Specifically, we find a systematic error of $1\%$ for the standard $\xi$-extrapolation with $p=2$ and even for $p=6$ a small yet significant deviation of $\sim0.1\%$ can be resolved with the given accuracy. The bottom panel of Fig.~\ref{fig:delta_poly_fit} shows results for $\Theta=0.5$, for which the situation again becomes noticeably more complex. First, we note that $p=2$ is no longer sufficient to accurately reproduce the interaction energy in the fitting range of $\xi\in[0,1]$; $p\geq3$ is required. Second, none of the depicted polynomial degrees is sufficient to reproduce the fermionic limit and the good agreement at $p=2$ is decisively confirmed as being coincidental. For completeness, we have also included a comparison to highly accurate CPIMC reference data for $\xi=-1$ as the grey symbols for all three temperatures in Fig.~\ref{fig:delta_poly_fit}. We find excellent agreement within the given error bars for all three cases, amounting to an agreement on the level of $\sim0.01\%$ for $\Theta=1$ and $\Theta=0.75$ and $\sim0.1\%$ for $\Theta=0.5$ due to the more severe sign problem in our direct PIMC calculations.

Let us proceed with the topic at hand, which is the investigation of the derivatives of $V(\xi)$ with respect to $\xi$ shown in Fig.~\ref{fig:derivatives}. The left panel corresponds to the first derivative and the dashed curves show our direct PIMC results for $\xi\neq0$ evaluated from Eq.~(\ref{eq:derivative}). The blue dots have been computed for $\xi=0$ via Eq.~(\ref{eq:quotient_rule}) and are in perfect agreement with the former data sets for all temperatures.  The intersection of the results for $\Theta=0.5$ and $\Theta=0.75$ for $\xi=0$ is likely a coincidence. Finally, the dotted green curves have been obtained by taking the derivative of the polynomial fits with $p=10$ (cf.~Fig.~\ref{fig:delta_poly_fit}). They are in perfect agreement with the direct PIMC results within the fitting range of $\xi\in[0,1]$, which further substantiates the correctness of our implementation. For $\Theta=1$, we find a smooth curve without any marked features, and which is reproduced well by the polynomial fit over the entire depicted $\xi$-range. For $\Theta=0.75$, the first derivative with respect to $\xi$ is substantially more complex than it was evident from $V$ itself (cf.~Fig.~\ref{fig:poly_fit}) and we find a shallow local maximum followed by a shallow local minimum around $\xi=-0.5$ and $\xi=-1$, respectively. Both features are well reproduced by the polynomial fit with $p=10$.
For $\Theta=0.5$, these features become substantially more pronounced, indicating the onset of strong quantum degeneracy effects. The tenth-order polynomial fit qualitatively captures the existence of both maximum and minimum, but fails to describe in particular the latter accurately.

The center and right panels of Fig.~\ref{fig:derivatives} show analogous investigations of the second and third derivatives of $V$ with respect to $\xi$, with overall similar trends.
Most importantly, we again find excellent agreement between the different PIMC estimators and the polynomials (within their fitting interval). In addition, our new PIMC results reveal further intricacies of the impact of quantum statistics in both bosonic and fermionic sectors. While the present study is focused on the Taylor series estimation of the fermionic limit, future works might apply similar techniques also to gain further insights into the role and manifestation of Bose-Einstein statistics for a variety of systems, e.g., in the context of superfluidity~\cite{Sindzingre_PRL_1989,Kwon_PRB_2006,Dornheim_PRA_2020,Dornheim_NJP_2022,Yan_Blume_PRL_2014,Saccani_Supersolid_PRL_2012,Boninsegni_RMP_2012,Filinov_PRA_2012}.

\subsection{Taylor series extrapolation\label{sec:taylor_results}}

\begin{figure*}
    \centering
    \includegraphics[width=0.44\textwidth]{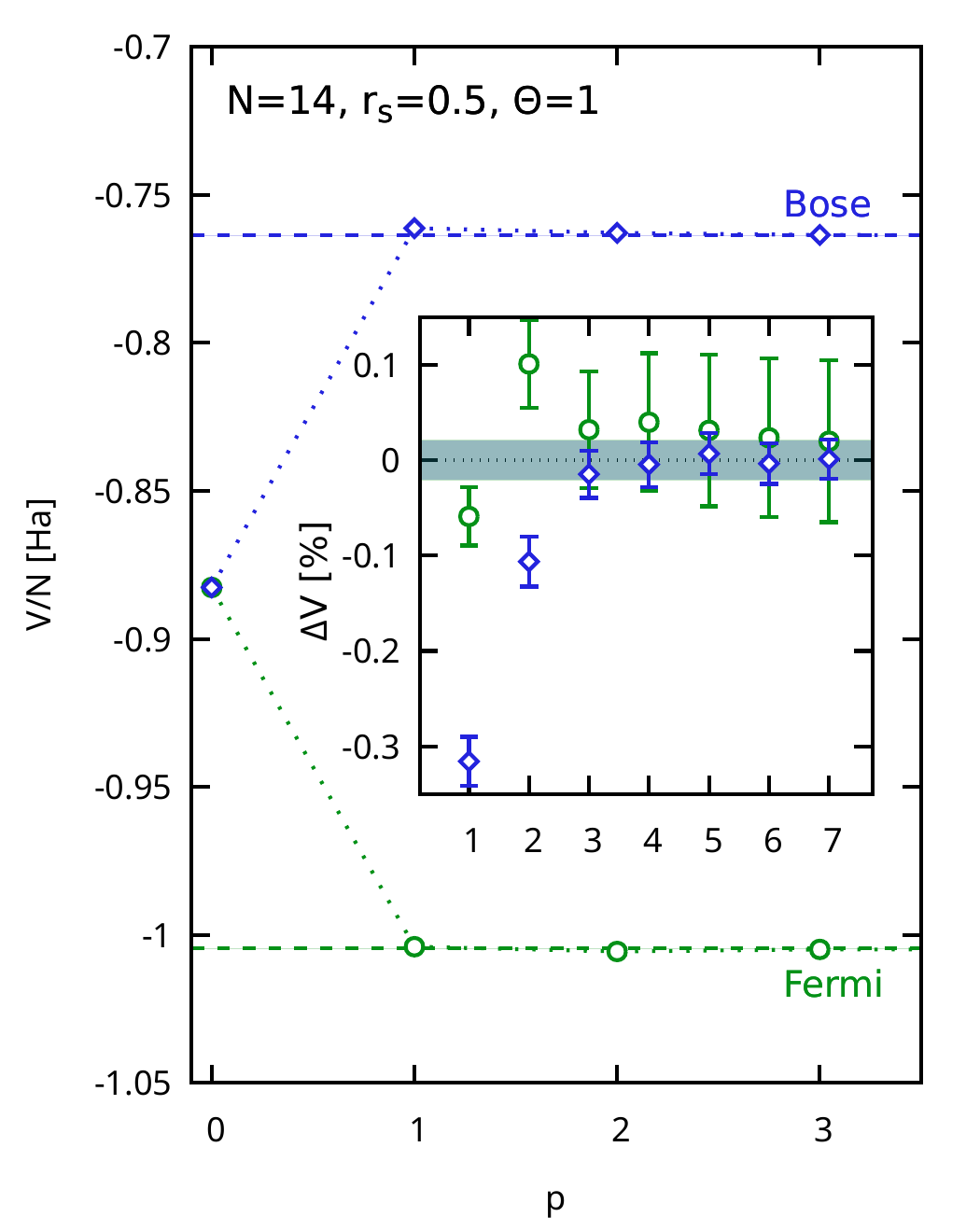}
   \includegraphics[width=0.44\textwidth]{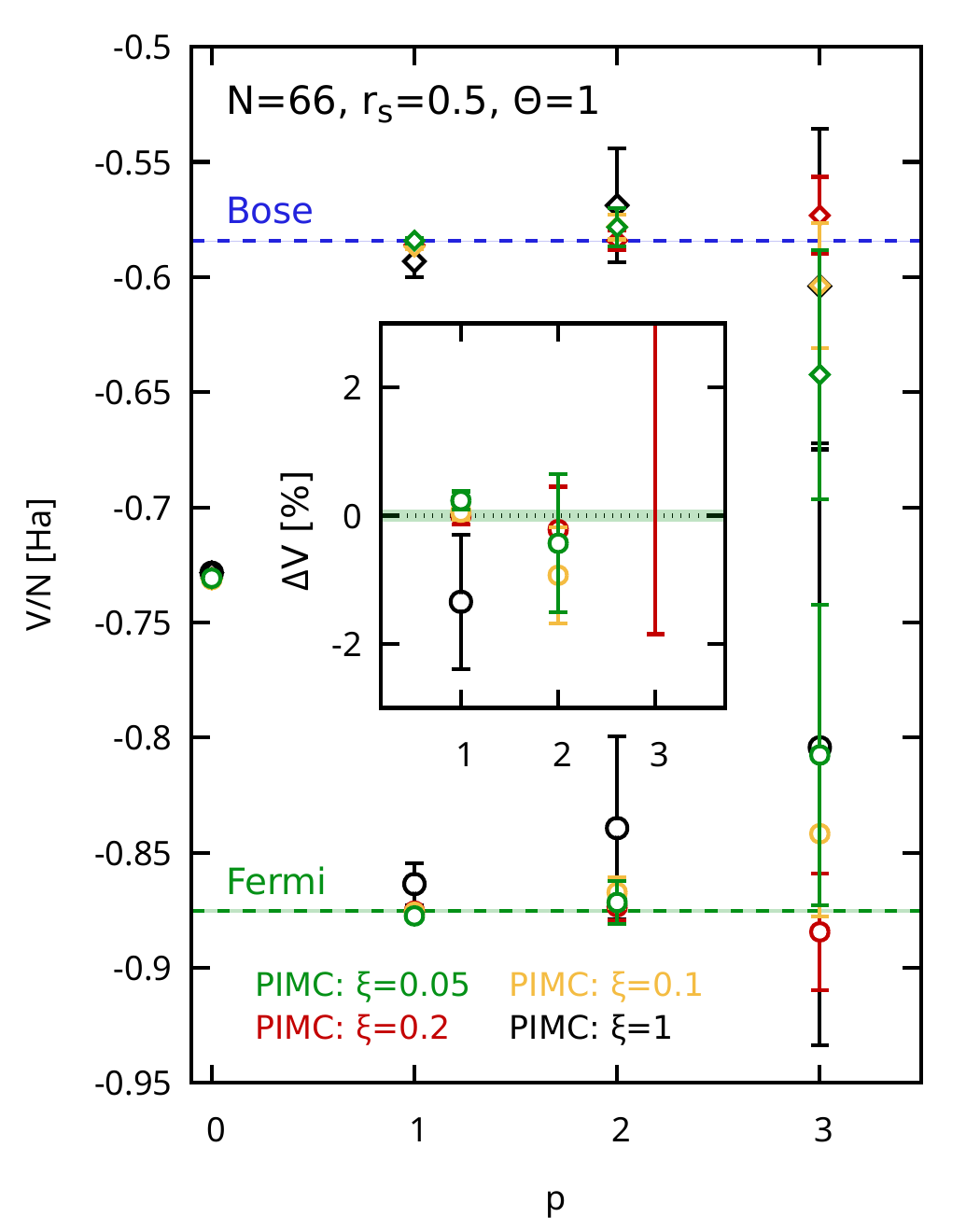} 
    \caption{\label{fig:N14} Taylor series extrapolation of the interaction energy per particle for the unpolarized UEG at $r_s=0.5$, $\Theta=1$ to the bosonic limit of $\xi=1$ (blue) and the fermionic limit of $\xi=-1$ (green) as a function of the Taylor order $p$ [cf.~Eq.~(\ref{eq:taylor_truncated})]. Results for $N=14$ [left] and $N=66$ [right]. The fermionic reference data (horizontal dashed green) have been obtained with CPIMC. For $N=14$, all results have been obtained for $\xi_\textnormal{ref}=1$. For $N=66$, we compare results for different $\xi_\textnormal{ref}$, see the color code. The insets show the relative deviation to the correct bosonic and fermionic limits ($N=14$) and the relative deviation to the correct fermionic limit for different $\xi_\textnormal{ref}$ ($N=66$).
    }
\end{figure*}

\begin{figure}
    \centering
    \includegraphics[width=0.44\textwidth]{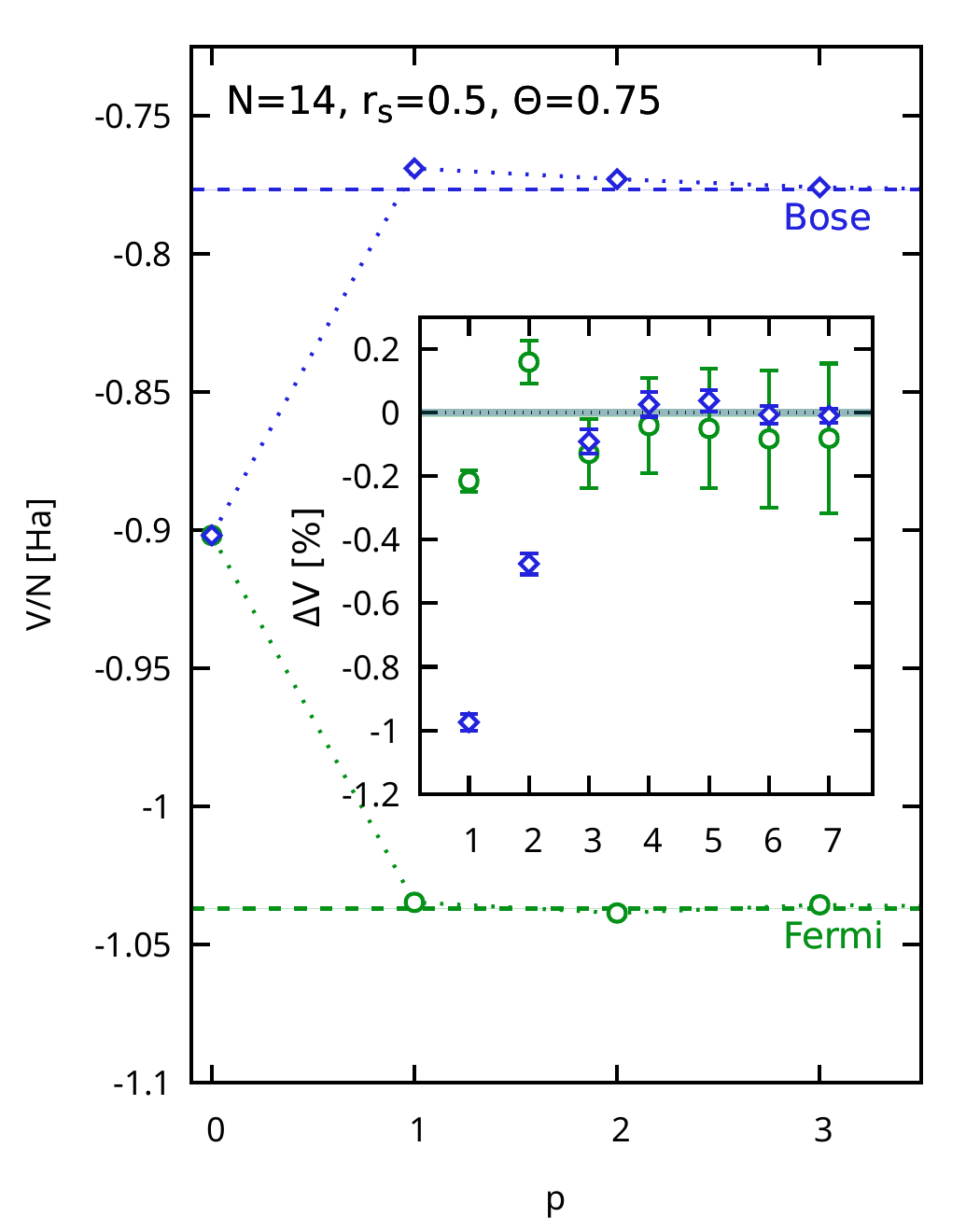}
    \caption{\label{fig:0p75} Taylor series extrapolation of the interaction energy per particle for the unpolarized UEG with $N=14$ at $r_s=0.5$ and $\Theta=0.75$ to the bosonic limit of $\xi=1$ (blue) and the fermionic limit of $\xi=-1$ (green) as a function of the Taylor order $p$ [cf.~Eq.~(\ref{eq:taylor_truncated})]. The fermionic reference data (horizontal dashed green) have been obtained with CPIMC. All results have been obtained for $\xi_\textnormal{ref}=1$. The inset shows the relative deviation to the correct bosonic and fermionic limits.}
\end{figure}

As the capstone of our work, we utilize our new PIMC results for the $\xi-$derivatives around $\xi=0$ for a controlled Taylor series extrapolation to the fermionic limit of $\xi=-1$ and the bosonic limit of $\xi=1$. In the left panel of Fig.~\ref{fig:N14}, we analyze the convergence of Eq.~(\ref{eq:canonical_polynomial}) with respect to the Taylor series order $p$ at $r_s=0.5$, $\Theta=1$ with $N=14$ particles. The green and blue symbols correspond to fermions and bosons, respectively. CPIMC results are employed as a reference for fermions and direct PIMC results are employed as a reference for bosons. As evident from Fig.~\ref{fig:poly_fit}, the bosonic and fermionic limits are nearly equidistant from the boltzmannon results that correspond to $p=0$. Both data sets rapidly converge with $p$. This can be discerned particularly well in the inset that shows deviations from the respective exact limits. Interestingly, truncating the Taylor series at $p=1$ works better for fermions ($\Delta V\sim0.1\%$) than bosons ($\Delta V\sim 0.3\%$), although both errors are small and of the same order of magnitude.

In practice, a successfull application of the original $\xi$-extrapolation method has been its utilization for larger systems, whose PIMC simulations are otherwise computationally unfeasible (even in case of a moderate degree of quantum degeneracy) due to the exponential increase in compute time with $N$ that is inherent to the sign problem~\cite{Dornheim_JCP_xi_2023}. As an example, we consider $N=66$ at the same density and temperature in the right panel of Fig.~\ref{fig:N14}. The green and blue horizontal lines still correspond to the correct fermionic (CPIMC) and bosonic (direct PIMC) limits, while the colored symbols now depict the PIMC results for different values of $\xi_\textnormal{ref}$, cf.~Eq.~(\ref{eq:final}). It can be deduced from Eqs.~(\ref{eq:wroclaw}) and (\ref{eq:quotient_rule}) [see Eqs.~(\ref{eq:I1}-\ref{eq:I3}) for an explicit form for $\nu=1,2,3$] that the Taylor series around $\xi_\textnormal{T}=0$ truncated at an order $p$ only contains measurements with a number of pair permutations up to $N_\textnormal{pp}\leq p$. However, such configurations are rare for larger systems even for weak to moderate degrees of quantum degeneracy. It is straightforward that even if the number of pair permutations per particle is low, there will be more such exchanges when there are more particles. Consequently, configurations with $N_\textnormal{pp}\ll{N}$ will be underexplored and the corresponding $\xi$-derivatives around $\xi=0$ will suffer from comparably bad statistics for $\xi_\textnormal{ref}=1$. This causes the relatively large error bars of the black symbols in the right panel of Fig.~\ref{fig:N14}. As a simple workaround, we have also performed PIMC calculations with $\xi_\textnormal{ref}=0.2$ (red), $\xi_\textnormal{ref}=0.1$ (gold) and $\xi_\textnormal{ref}=0.05$ (green). In these calculations, the number of pair permutations is artificially suppressed (see also the discussion in Ref.~\cite{dornheim2025reweightingestimatorabinitio}), giving us a higher accuracy in the relevant sectors. Indeed, we obtain error bars that are smaller by two orders of magnitude, in particular for $\xi_\textnormal{ref}=0.1$ and $\xi_\textnormal{ref}=0.2$. In both cases, we find perfect agreement with the true fermionic limit within the associated statistical error bars of $<0.15\%$ for $p=1$, which surpasses the accuracy attained in previous studies for the same parameters using the standard $\xi$-extrapolation method. We stress that a direct PIMC simulation would give an average sign of $S\sim10^{-6}$ for $\xi=-1$, making this fundamentally impossible. The current scheme gives us a speed-up by twelve orders of magnitude \emph{from a single PIMC simulation}, e.g., at $\xi_\textnormal{ref}=0.2$.

The final question concerns how to decide if the Taylor series has indeed converged at a given $p$ and with what accuracy. The easiest option would be to rely on PIMC simulations of a smaller system. Indeed, the case of $N=14$ depicted in the left panel of Fig.~\ref{fig:N14} nicely confirms that the systematic error should be $\lesssim0.1\%$ for $p=1$. Alternatively, we can check a larger $p$ for the larger system of interest. For the present case, considering $p=2$ would then confirm convergence within an error bar of $\sim0.6\%$, comparable to the accuracy level in the two previous investigations~\cite{Dornheim_JCP_xi_2023,dornheim2025reweightingestimatorabinitio}, where such a size-consistent check was not carried out.

In Fig.~\ref{fig:0p75}, we analyze the convergence of the truncated Taylor series with $N=14$, at $r_s=0.5$ and the somewhat lower temperature of $\Theta=0.75$ (for $\xi_\textnormal{ref}=1$). Overall, we find the same trends as for $\Theta=1$, although there appear systematic errors of $\sim0.2\%$ for $p=1$ and $p=2$ in the extrapolation to the fermionic limit. This is often sufficient for practical applications~\cite{groth_prl,review,Militzer_PRE_2021,vorberger2025roadmapwarmdensematter}, in particular when it, at the same time, facilitates the simulation of larger systems and thus helps mitigate possible finite-size errors~\cite{Chiesa_PRL_2006,Drummond_PRB_2008,Holzmann_PRB_2016,Dornheim_JCP_2021}.
Interestingly, we again observe a slower convergence for bosons compared to fermions.

\begin{figure*}
    \centering
    \includegraphics[width=0.32\textwidth]{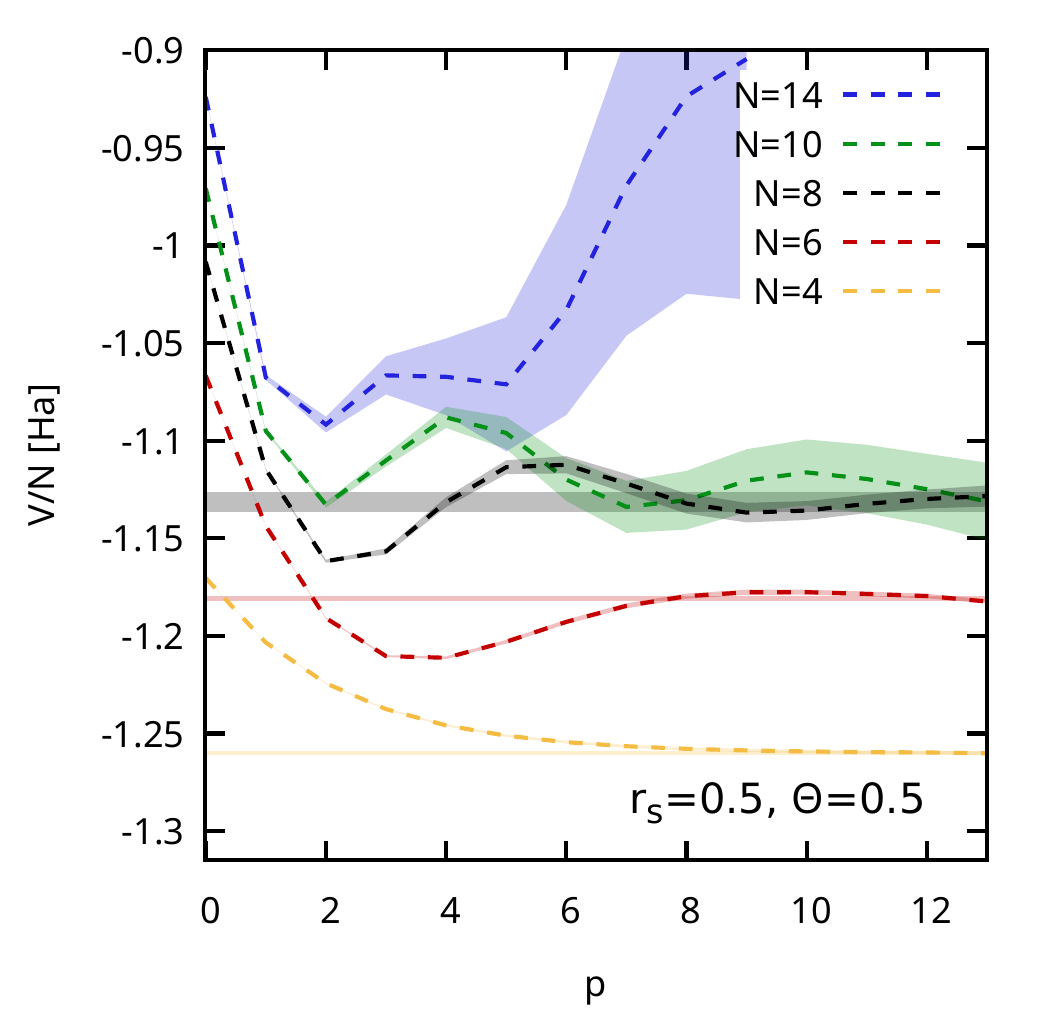}\includegraphics[width=0.32\textwidth]{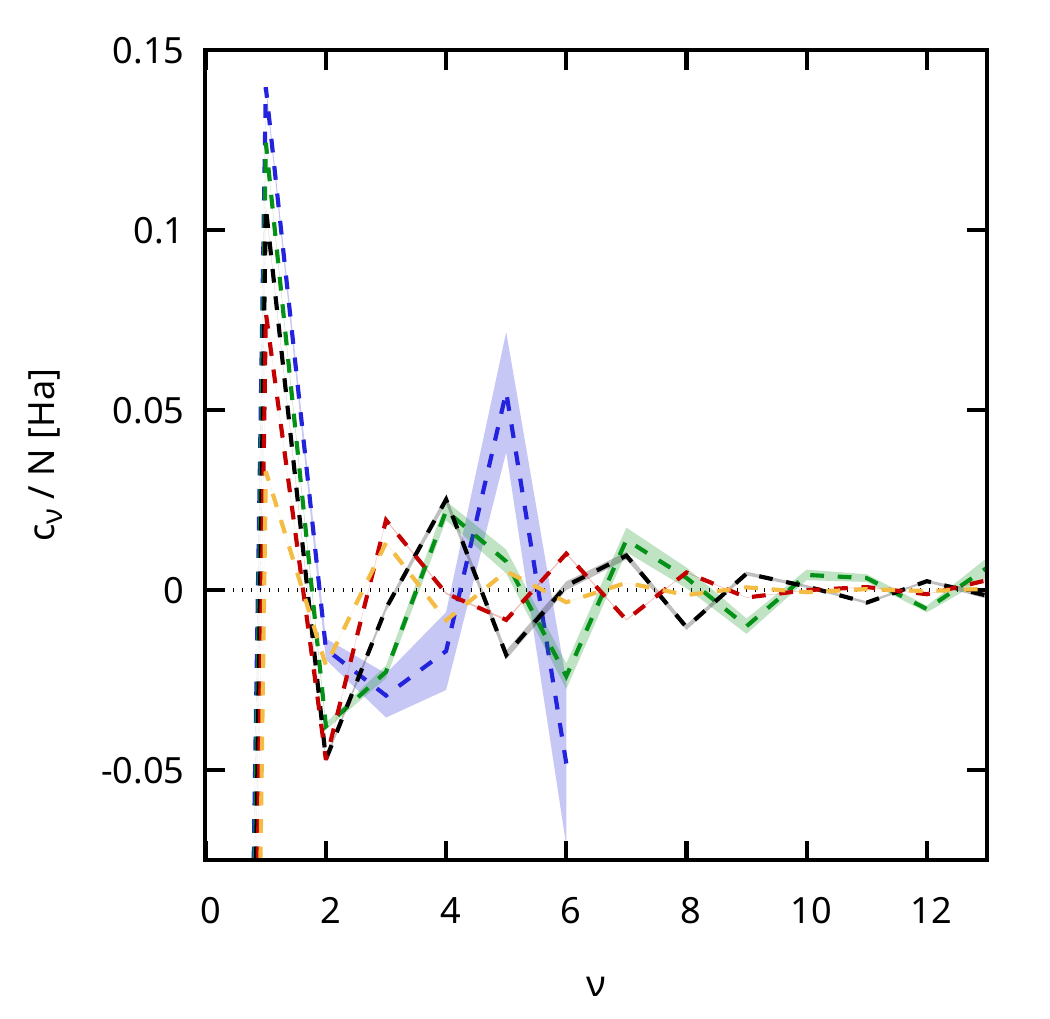}
\includegraphics[width=0.32\textwidth]{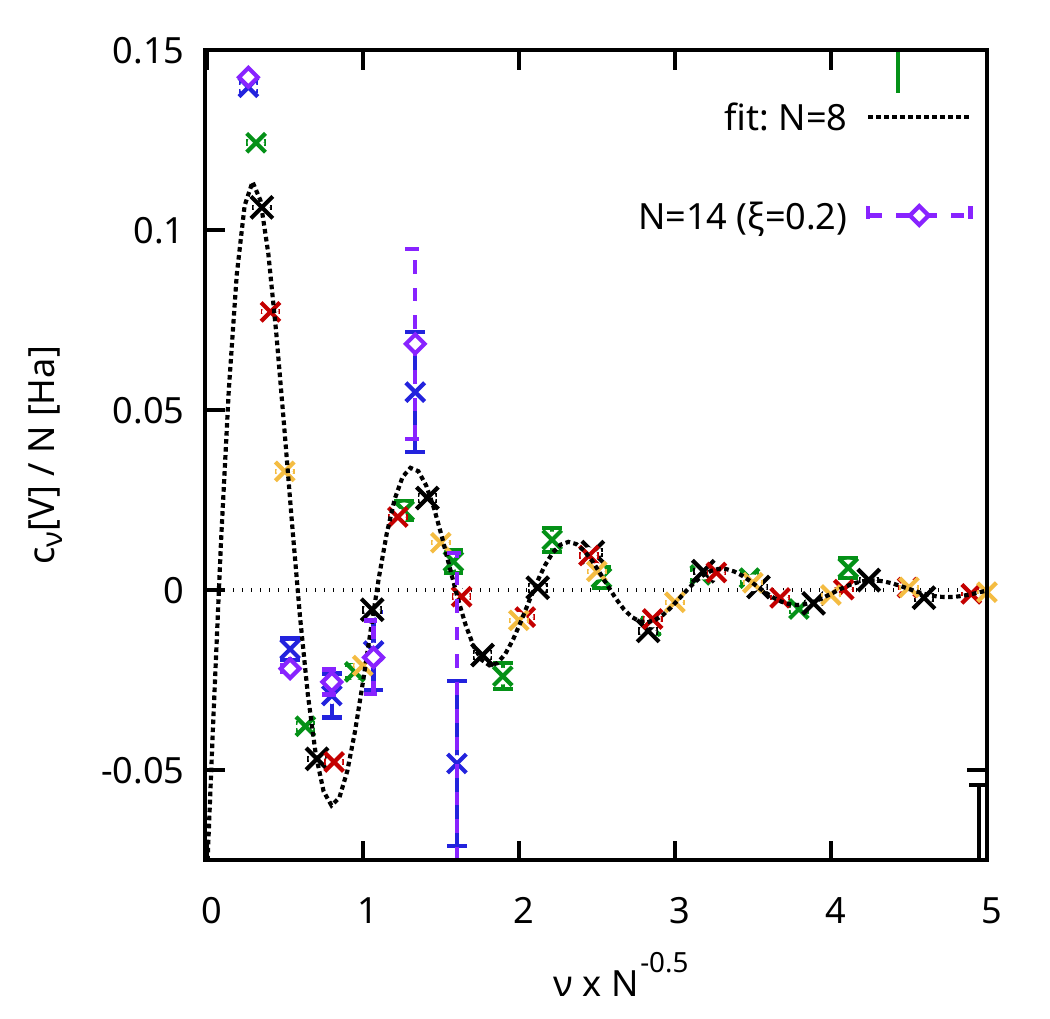}\\
\includegraphics[width=0.32\textwidth]{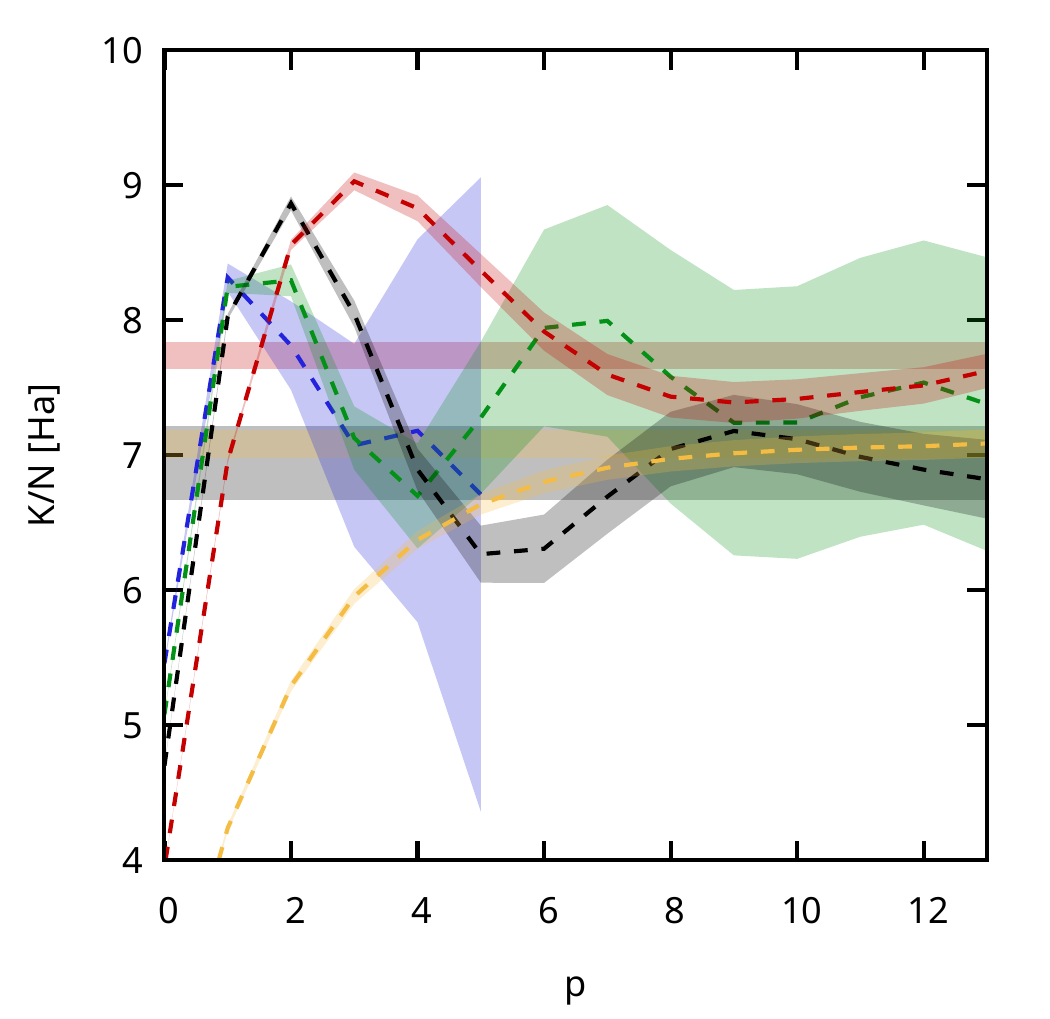}\includegraphics[width=0.32\textwidth]{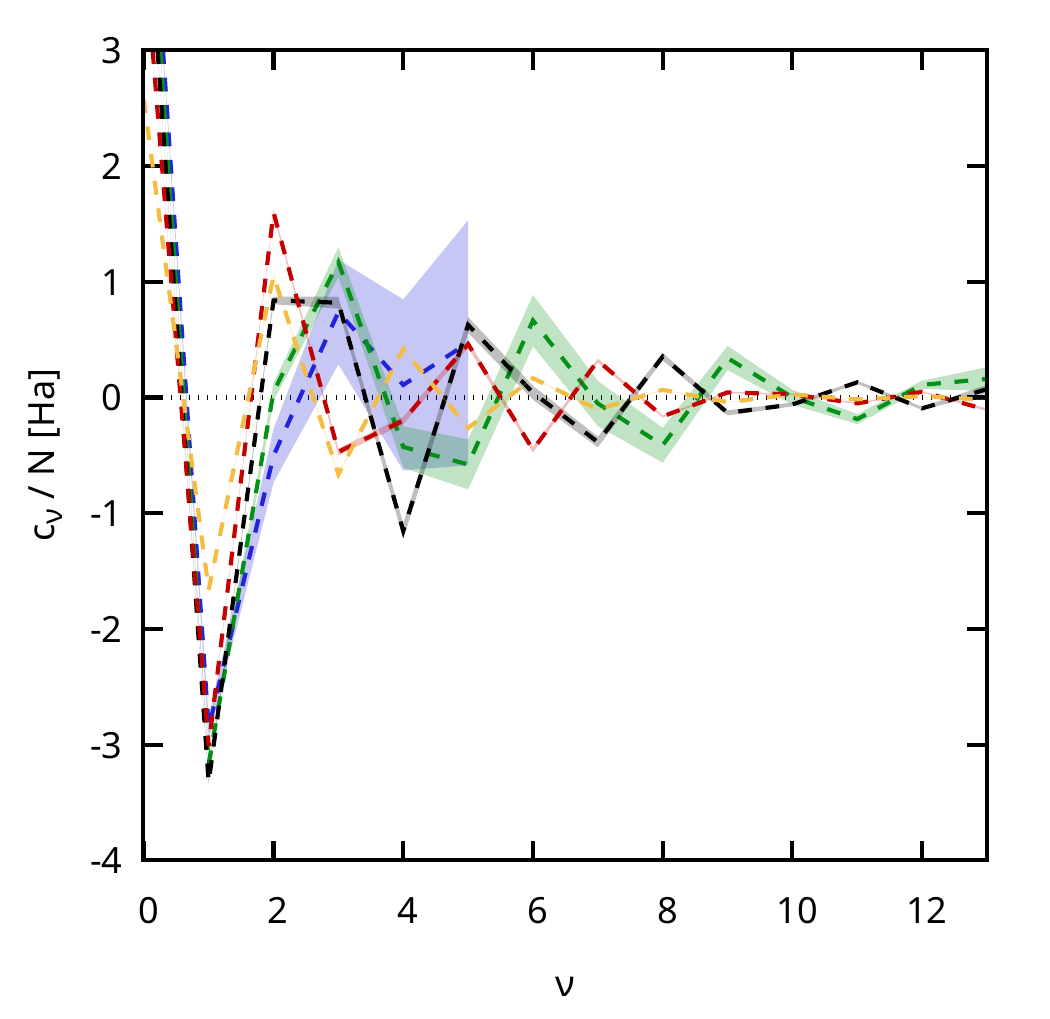}
\includegraphics[width=0.32\textwidth]{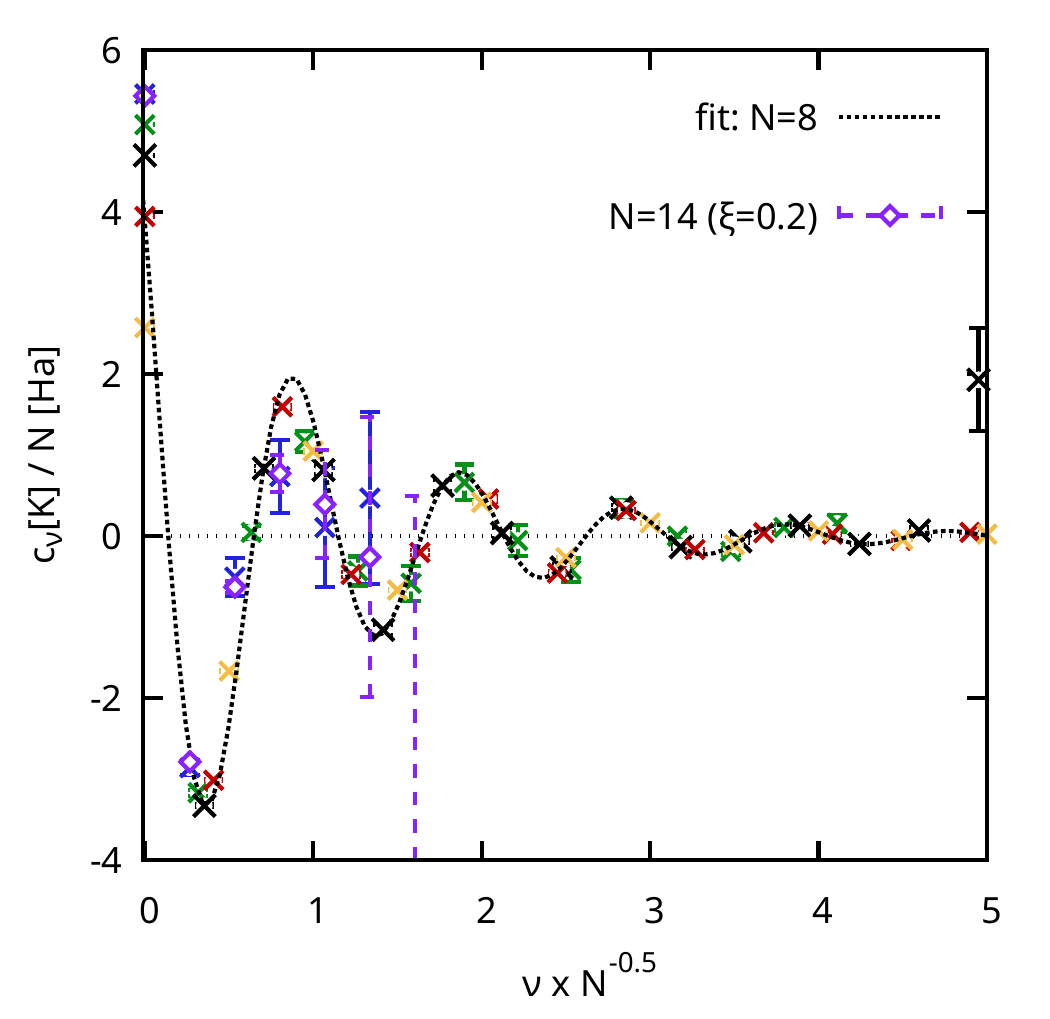}
    \caption{\label{fig:six}  Left: convergence of the Taylor series extrapolation to the fermionic limit of $\xi=-1$ around $\xi_\textnormal{T}=0$ for the potential (top row) and kinetic (bottom row) energy per particle for the unpolarized UEG at $r_s=0.5$ and $\Theta=0.5$ with different $N$. The horizontal shaded areas indicate the full direct PIMC results for $\xi=-1$ with the corresponding uncertainty interval for $N=4,6,8$. Center: convergence of the extracted expansion coefficients $c_\nu$ [cf.~Eq.~(\ref{eq:canonical_polynomial})]. Right: same as center panel, but the $x$-axis has been re-scaled by a factor of $1/\sqrt{N}$. The dotted black lines show empirical fits according to Eq.~(\ref{eq:simple_fit}) to the $N=8$ data set obtained within the fitting interval of $\nu\in[1,11]$.}
\end{figure*}

As the final and most difficult example, we consider the lower temperature of $\Theta=0.5$ in Fig.~\ref{fig:six}. Since for $N=14$ we cannot reach the fermionic limit of $\xi=-1$ due to the fermion sign problem, we consider a range of particle numbers $N=4,\dots,14$, and the results are depicted by the differently colored curves. The top row corresponds to the interaction energy per particle and the left panel shows the convergence for $\xi=-1$ with respect to the Taylor order $p$. Overall, we find a smooth and monotonic convergence for $N=4$ (gold), whereas the curves start to increasingly oscillate for larger numbers of electrons. These oscillations have amplitudes of the order of a few percent, and the resulting non-monotonicity makes any kind of controlled truncation or extrapolation non-trivial in practice.

The kinetic energy per particle shown in the bottom row of Fig.~\ref{fig:six} overall exhibits the same qualitative trends, albeit with larger statistical error bars due to the thermodynamic PIMC estimator~\cite{Janke_JCP_1997}. A second difference with respect to the interaction energy is that the true kinetic energy is underestimated in the limit of $p=0$, whereas the (negative) interaction is overestimated in the boltzmannon systems compared to proper Fermi-Dirac statistics.

Let us next analyze the behavior of the Taylor expansion coefficients $c_\nu$, which are shown in the central column of Fig.~\ref{fig:six}. In this representation, it is easy to see that (i) the amplitude of the oscillations in the $c_\nu$ decays with increasing $\nu$, which is expected as $\xi=\pm1$ lies within the radius of convergence, cf.~Fig.~\ref{fig:poland2} above; (ii) the wavelength of the oscillations increases with the number of particles. Empirically, we find that a re-scaling of the expansion order $\nu$ by a factor of $1/\sqrt{N}$ leads to a remarkable invariance of the oscillations in the kinetic and potential energy coefficients with respect to $N$, see the right column of Fig.~\ref{fig:six}. In practice, the $\nu$-dependence of the Taylor coefficients of an observable $O\in\{V,K\}$ is well reproduced by 
\begin{eqnarray}\label{eq:simple_fit}
    c_\nu[\hat{O}] &=& a_O\ e^{-b_O \nu}\ \textnormal{sin}\left(2\pi\ c_O\ \nu\right)\\\nonumber & & + d_O\ e^{-b_O\nu/2}\ \textnormal{sin}\left(2\pi\ c_O\ \nu + \pi/2\right)\quad ,
\end{eqnarray}
with $a_O$, $b_O$, $c_O$, and $d_O$ the four free parameters. The results are shown as the dotted black lines in the right column of Fig.~\ref{fig:six} and fit all data sets remarkably well. A possible exception is given by $N=14$, which is afflicted with substantially higher error bars. It is also pointed out that the re-scaling of the $x$-axis by $1/\sqrt{N}$ means that we need larger $\nu$ for larger systems to reach the same characteristic oscillation decay on the re-scaled scale compared to smaller systems. For completeness, we have also performed additional PIMC simulations with $\xi_\textnormal{ref}=0.2$, resulting in a higher accuracy for small $\nu$; see the lilac diamonds in the right column. While indeed offering an improvement for small $\nu$, this advantage is not decisive.

\begin{figure}
    \centering
    \includegraphics[width=0.44\textwidth]{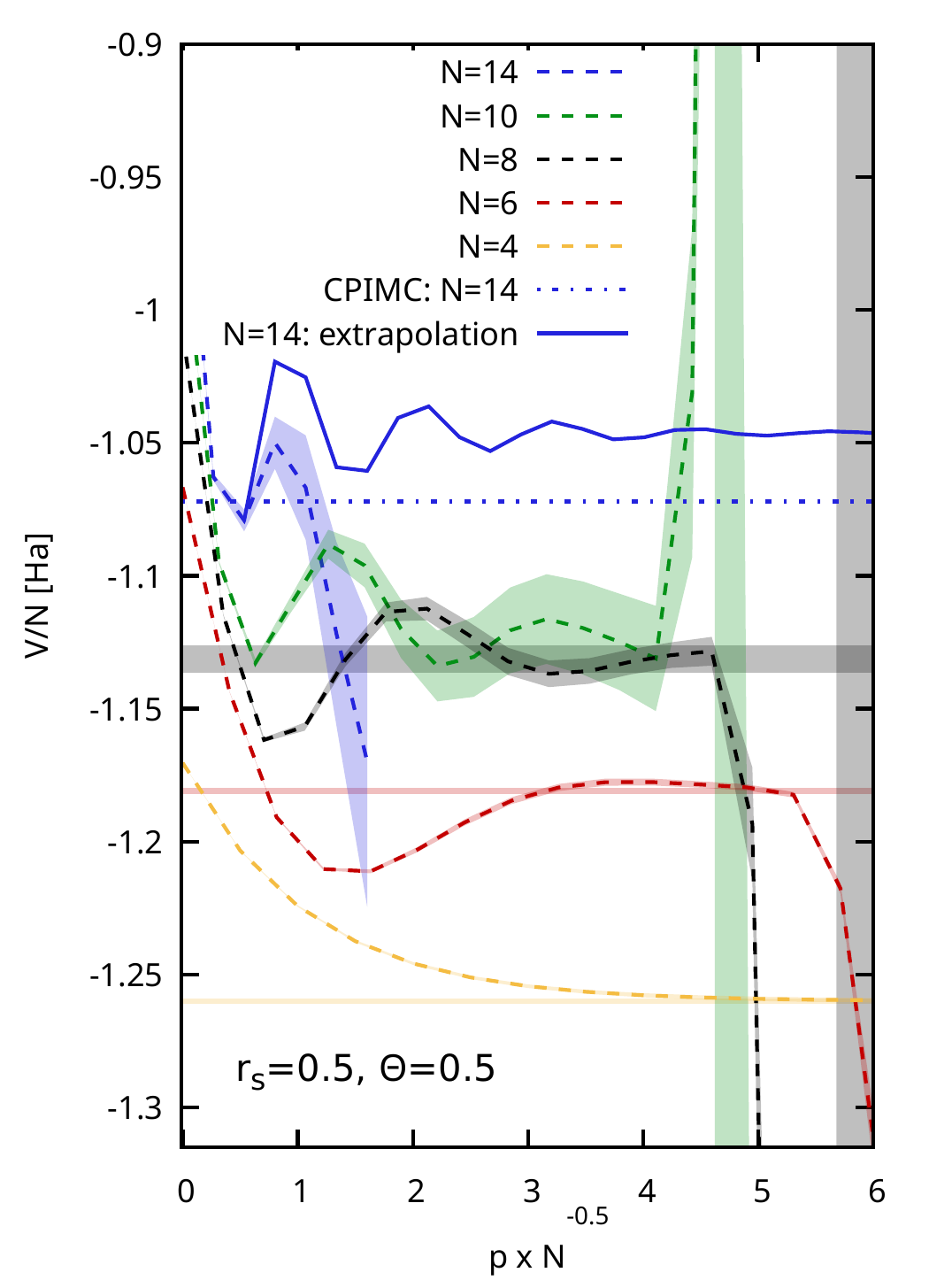}
\caption{\label{fig:rescale_coefficients_interaction} Convergence of the Taylor series extrapolation around $\xi_\textnormal{T}=0$ to the fermionic limit of $\xi=-1$ for the interaction energy per particle for the same conditions as in Fig.~\ref{fig:six}, but with the $x$-axis having been re-scaled by a factor of $1/\sqrt{N}$. The horizontal shaded areas show direct PIMC results for $\xi=-1$ with the associated uncertainty interval for $N=4,6,8$. The dashed blue line shows CPIMC reference data for $N=14$ and the solid blue line has been obtained by re-computing the Taylor coefficients for $\nu>2$ with the fit function Eq.~(\ref{eq:simple_fit}) determined for $N=8$.}
\end{figure}

\begin{figure}
    \centering
    \includegraphics[width=0.44\textwidth]{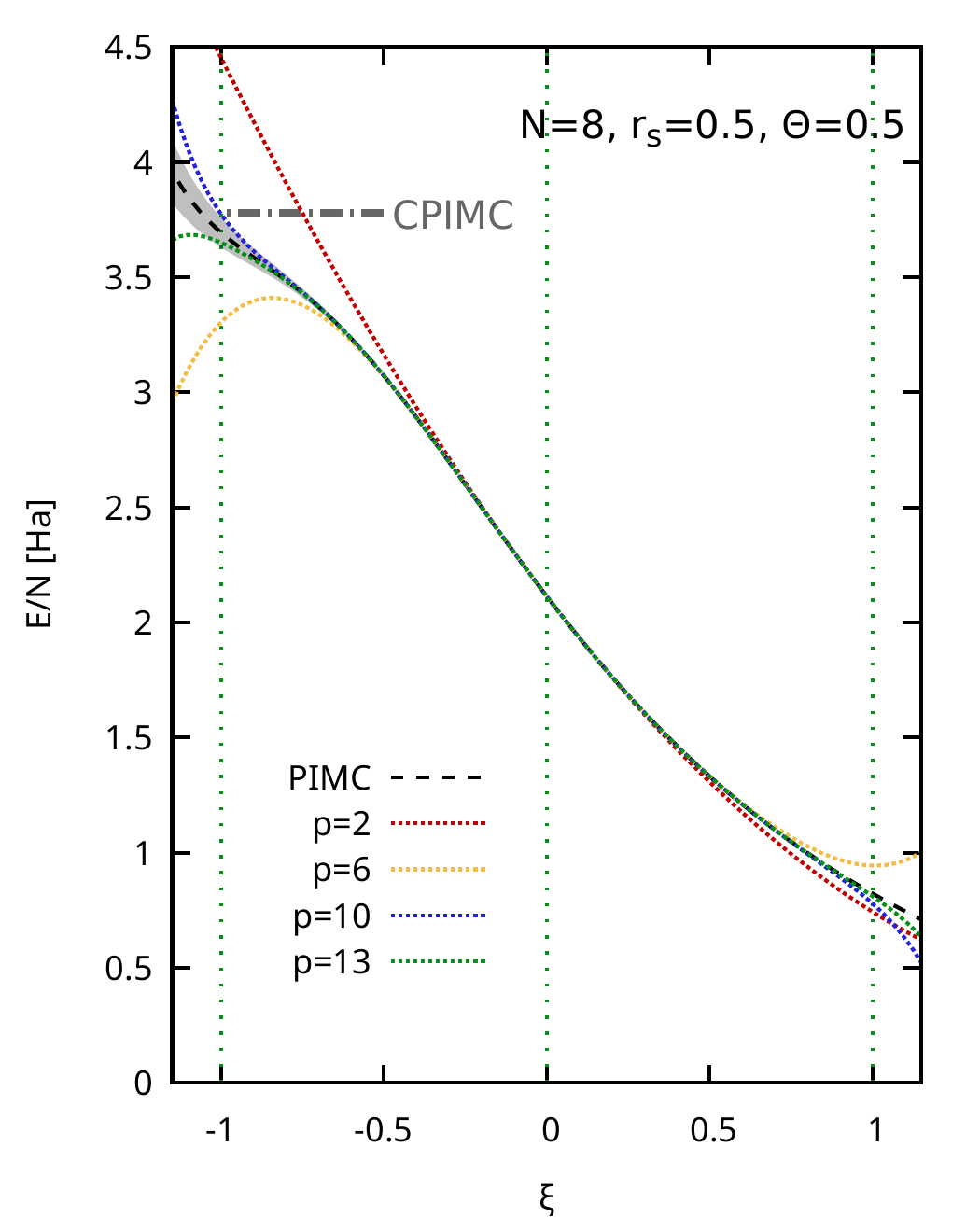}
\caption{\label{fig:das_letzte} Dependence of the total energy per particle on the fictitious quantum statistics variable $\xi$ for $N=8$, $r_s=0.5$, and $\Theta=0.5$. Dashed black: full PIMC results obtained from a single simulation at $\xi_\textnormal{ref}=1$ via Eq.~(\ref{eq:final}). The shaded gray area indicates the associated uncertainty interval. The horizontal dash-dotted gray line indicates the highly accurate CPIMC result at the $\xi=-1$ fermionic limit. The dotted curves have been computed by combining the $\xi$-derivatives around $\xi=0$ [Eq.~(\ref{eq:quotient_rule})] with the corresponding Taylor series expansion [Eq.~(\ref{eq:canonical_polynomial})] truncated at different polynomial degrees $p$.
    }
\end{figure}

In Fig.~\ref{fig:rescale_coefficients_interaction}, we show the convergence of the Taylor series extrapolation to $\xi=-1$, but with the $x$-axis again being re-scaled by a factor of $1/\sqrt{N}$. In this representation, it appears that all curves converge towards the true fermionic limit around $p/\sqrt{N}\approx5$. Unfortunately, this limit cannot be reached for $N=14$ at these conditions, as discussed above. In light of the remarkable $N$-invariance observed in Fig.~\ref{fig:six}, it makes sense to ponder if it can be exploited to model the convergence of a given larger $N$ based on a fit following Eq.~(\ref{eq:simple_fit}) for a smaller $N$; here, we again choose $N=8$ for the latter. Following this procedure to obtain the Taylor coefficients for $\nu>2$ with $N=14$ results in the solid blue curve in Fig.~\ref{fig:rescale_coefficients_interaction}. Evidently, these data exhibit even faster oscillations compared to the other data sets. This is a consequence of the denser $\nu$-grid on this re-scaled representation. In addition, the blue curve shows a nice and controlled convergence; yet, its $p\to\infty$ limit differs from the true fermionic interaction energy (dotted blue line obtained from CPIMC) by $\sim3\%$, which is even worse than a simple truncation at $p=2$ for this case. Thus, we have to conclude that there presently does not exist a reliable strategy to converge the Taylor series to the true fermionic limit for strong degrees of quantum degeneracy.

Let us finish our investigation with an analysis of the capability of the Taylor series expansion around $\xi_\textnormal{T}=0$ to describe the actual $\xi$-dependence of an observable over the entire relevant $\xi$-range. To this end, we show the total energy per particle for $N=8$, $r_s=0.5$, and $\Theta=0.5$ in Fig.~\ref{fig:das_letzte}. The dashed black line shows our direct PIMC results, which exhibit a complicated behavior in particular around $\xi\sim-0.5$ already hinted at in Ref.~\cite{Dornheim_JCP_xi_2023}. 
The dash-dotted horizontal grey bar indicates a highly accurate CPIMC reference result for the fermionic limit, which agrees with our direct PIMC calculations within the given error interval (shaded grey area), as it is expected. We note that we find an average sign of $S\approx0.0036$, making our simulations computationally very demanding despite the comparably small system size.
The dotted colored curves correspond to Taylor series expansions around $\xi_\textnormal{T}=0$ evaluated via Eq.~(\ref{eq:quotient_rule}) and truncated at different polynomial degrees $p$.
At a first glance, the extrapolation to the fermionic limit appears to be more challenging than to the bosonic limit. This is somewhat unsurprising as the $\xi$-dependence in the bosonic sector of $\xi>0$ seems less complicated than in the fermionic domain of $\xi<0$.
Then again, $p=6$ leads to worse results for $\xi=1$ compared to $p=2$, indicating a non-monotonic convergence.
A key question is if the Taylor series around $\xi_\textnormal{T}=0$ is capable of capturing the full complex $\xi$-dependence in the fermionic sector. 
This is indeed the case for $p=10$ (blue) and $p=13$ (green), leading us to the following conclusion: empirically, the polynomial ansatz Eq.~(\ref{eq:canonical_polynomial}) is well justified by the present Taylor series perspective even for strong levels of quantum degeneracy. The practical challenge will be to find suitable truncation schemes that avoid the evaluation of coefficients involving larger numbers of pair exchanges $N_\textnormal{pp}$, or to find alternative analytical simplifications.

\section{Summary and Outlook\label{sec:outlook}}

In this work, we investigated the \emph{ab initio} PIMC simulation of quantum degenerate fermions ($\xi=-1$) [and bosons ($\xi=1$)] from the perspective of a Taylor expansion around the limit of distinguishable quantum particles, i.e., boltzmannons ($\xi=0)$. To this end, we derived new PIMC estimators to evaluate the arbitrary-order derivatives of arbitrary observables $\hat{O}$ with respect to the continuous quantum statistics variable $\xi$ for finite values of $\xi$ and the particularly important case of $\xi=0$. This allowed us to evaluate a Taylor series representation of $O(\xi)$ up to, in principle, arbitrary order $p$, which includes the previous empirical parabolic isothermal $\xi$-extrapolation method~\cite{Xiong_JCP_2022,Dornheim_JCP_xi_2023,Dornheim_JCP_2024,Dornheim_JPCL_2024} as a special case.

In the first part of our investigation, we analyzed the radius of convergence of the Taylor series, which, in all considered cases, includes the bosonic and fermionic limits of $\xi=\pm1$. Interestingly, for some parameters, we observed "pseudo-poles" for $\xi<-1$, which would be consistent with the recent study of Lee-Yang zeros by He \emph{et al.}~\cite{he2025revisitingfermionsignproblem}. As the next step, we explicitly investigated the derivatives of an observable with respect to $\xi$, which are difficult to capture within polynomial extrapolation ansätze based on fits in a limiting fitting range such as $\xi\in[0,1]$. Finally, we investigated the convergence of the Taylor extrapolation to the fermionic and bosonic limits for a range of conditions. Remarkably, we find that accurate estimates of fermionic properties with an accuracy of $\Delta O\sim0.1\%$ are possible even after truncating at $p=1$ for $\Theta\geq0.75$. This indicates that, while not necessarily small, the influence of quantum statistics remains approximately linear in this regime.
In practice, this has allowed us to compute highly accurate results for the UEG with $N=66$, $r_s=0.5$ and $\Theta=1$ from a single PIMC simulation with, e.g., $\xi_\textnormal{ref}=0.2$. For $\Theta=0.5$, on the other hand, the situation is considerably more complex. In particular, the convergence of the Taylor series is non-monotonic, with some characteristic damped oscillations of the Taylor coefficients amenable to near $N$-invariance after re-scaling the coefficient order $\nu$ by an empirical factor of $1/\sqrt{N}$. Unfortunately, these commonalities between different $N$ do not allow us to estimate the converged fermionic limit for, say, $N=14$ without a systematic error of $\sim3\%$. Conversely, we found that more than ten Taylor coefficients are required to converge to the fermionic limit for $\Theta=0.5$, which is challenging in practice.

We expect our results to be useful for a variety of PIMC simulations over a range of research fields. First and foremost, the fermion sign problem remains a central bottleneck in physics, quantum chemistry and related disciplines. The present Taylor series perspective put the empirical isothermal $\xi$-extrapolation technique in a more firm theoretical ground and makes it more controlled by providing the possibility for convergence analysis. In addition, we have shown that one can evaluate, in principle, the entire Taylor series from a single simulation at a given reference value $\xi_\textnormal{ref}$ (also, without any re-weighting~\cite{dornheim2025reweightingestimatorabinitio}), which is computationally efficient. We note that the estimation of the derivatives around $\xi=0$ simply involve the estimation of expectation values for different numbers of pair permutations $N_\textnormal{pp}$, which can be easily incorporated into existing PIMC codes.

Future works might focus on the dedicated analysis of the Taylor coefficients for strong quantum degeneracy, and on the development of improved truncation schemes. In addition, we note that the presented estimators for the $\xi$-derivatives might be used to improve polynomial fitting schemes, and thus improve previously used $\xi$-extrapolation ansätze. A particularly important application will be the Taylor series analysis of real warm dense matter systems that include both electrons and positively charged nuclei~\cite{Bohme_PRL_2022,Bohme_PRE_2023,Bonitz_POP_2024,Filinov_PRE_2023,Dornheim_JCP_2024,Dornheim_MRE_2024,Dornheim_NatComm_2025}, which will be pursued in a dedicated future study.
Finally, we note that the gradual activation of quantum statistics in terms of the fictitious quantum statistics parameter $\xi$ is interesting in itself, and might also open up new perspectives onto important physical effects in Bose-Einstein systems such as superfluidity.

\begin{acknowledgements}

\noindent This work was partially supported by the Center for Advanced Systems Understanding (CASUS), financed by Germany’s Federal Ministry of Education and Research and the Saxon state government out of the State budget approved by the Saxon State Parliament. This work has received funding from the European Research Council (ERC) under the European Union’s Horizon 2022 research and innovation programme (Grant agreement No. 101076233, "PREXTREME"). 
Views and opinions expressed are however those of the authors only and do not necessarily reflect those of the European Union or the European Research Council Executive Agency. Neither the European Union nor the granting authority can be held responsible for them.
This work has received funding from the European Union's Just Transition Fund (JTF) within the project \emph{R\"ontgenlaser-Optimierung der Laserfusion} (ROLF), contract number 5086999001, co-financed by the Saxon state government out of the State budget approved by the Saxon State Parliament. Computations were performed on a Bull Cluster at the Center for Information Services and High-Performance Computing (ZIH) at Technische Universit\"at Dresden and at the Norddeutscher Verbund f\"ur Hoch- und H\"ochstleistungsrechnen (HLRN) under grant mvp00024.

\end{acknowledgements}

\bibliography{bibliography}

\begin{thebibliography}{213}%
\makeatletter
\providecommand \@ifxundefined [1]{%
 \@ifx{#1\undefined}
}%
\providecommand \@ifnum [1]{%
 \ifnum #1\expandafter \@firstoftwo
 \else \expandafter \@secondoftwo
 \fi
}%
\providecommand \@ifx [1]{%
 \ifx #1\expandafter \@firstoftwo
 \else \expandafter \@secondoftwo
 \fi
}%
\providecommand \natexlab [1]{#1}%
\providecommand \enquote  [1]{``#1''}%
\providecommand \bibnamefont  [1]{#1}%
\providecommand \bibfnamefont [1]{#1}%
\providecommand \citenamefont [1]{#1}%
\providecommand \href@noop [0]{\@secondoftwo}%
\providecommand \href [0]{\begingroup \@sanitize@url \@href}%
\providecommand \@href[1]{\@@startlink{#1}\@@href}%
\providecommand \@@href[1]{\endgroup#1\@@endlink}%
\providecommand \@sanitize@url [0]{\catcode `\\12\catcode `\$12\catcode `\&12\catcode `\#12\catcode `\^12\catcode `\_12\catcode `\%12\relax}%
\providecommand \@@startlink[1]{}%
\providecommand \@@endlink[0]{}%
\providecommand \url  [0]{\begingroup\@sanitize@url \@url }%
\providecommand \@url [1]{\endgroup\@href {#1}{\urlprefix }}%
\providecommand \urlprefix  [0]{URL }%
\providecommand \Eprint [0]{\href }%
\providecommand \doibase [0]{http://dx.doi.org/}%
\providecommand \selectlanguage [0]{\@gobble}%
\providecommand \bibinfo  [0]{\@secondoftwo}%
\providecommand \bibfield  [0]{\@secondoftwo}%
\providecommand \translation [1]{[#1]}%
\providecommand \BibitemOpen [0]{}%
\providecommand \bibitemStop [0]{}%
\providecommand \bibitemNoStop [0]{.\EOS\space}%
\providecommand \EOS [0]{\spacefactor3000\relax}%
\providecommand \BibitemShut  [1]{\csname bibitem#1\endcsname}%
\let\auto@bib@innerbib\@empty
\bibitem [{\citenamefont {Anderson}(2007)}]{anderson2007quantum}%
  \BibitemOpen
  \bibfield  {author} {\bibinfo {author} {\bibfnamefont {J.B.}\ \bibnamefont {Anderson}},\ }\href {https://books.google.de/books?id=\_QUSDAAAQBAJ} {\emph {\bibinfo {title} {Quantum Monte Carlo: Origins, Development, Applications}}}\ (\bibinfo  {publisher} {Oxford University Press, USA},\ \bibinfo {year} {2007})\BibitemShut {NoStop}%
\bibitem [{\citenamefont {Foulkes}\ \emph {et~al.}(2001)\citenamefont {Foulkes}, \citenamefont {Mitas}, \citenamefont {Needs},\ and\ \citenamefont {Rajagopal}}]{Foulkes_RMP_2001}%
  \BibitemOpen
  \bibfield  {author} {\bibinfo {author} {\bibfnamefont {W.~M.~C.}\ \bibnamefont {Foulkes}}, \bibinfo {author} {\bibfnamefont {L.}~\bibnamefont {Mitas}}, \bibinfo {author} {\bibfnamefont {R.~J.}\ \bibnamefont {Needs}}, \ and\ \bibinfo {author} {\bibfnamefont {G.}~\bibnamefont {Rajagopal}},\ }\bibfield  {title} {\enquote {\bibinfo {title} {Quantum monte carlo simulations of solids},}\ }\href {\doibase 10.1103/RevModPhys.73.33} {\bibfield  {journal} {\bibinfo  {journal} {Rev. Mod. Phys.}\ }\textbf {\bibinfo {volume} {73}},\ \bibinfo {pages} {33--83} (\bibinfo {year} {2001})}\BibitemShut {NoStop}%
\bibitem [{\citenamefont {Ceperley}(1995)}]{cep}%
  \BibitemOpen
  \bibfield  {author} {\bibinfo {author} {\bibfnamefont {D.~M.}\ \bibnamefont {Ceperley}},\ }\bibfield  {title} {\enquote {\bibinfo {title} {Path integrals in the theory of condensed helium},}\ }\href {https://journals.aps.org/rmp/abstract/10.1103/RevModPhys.67.279} {\bibfield  {journal} {\bibinfo  {journal} {Rev. Mod. Phys}\ }\textbf {\bibinfo {volume} {67}},\ \bibinfo {pages} {279} (\bibinfo {year} {1995})}\BibitemShut {NoStop}%
\bibitem [{\citenamefont {Pollet}(2012)}]{Pollet_2012}%
  \BibitemOpen
  \bibfield  {author} {\bibinfo {author} {\bibfnamefont {Lode}\ \bibnamefont {Pollet}},\ }\bibfield  {title} {\enquote {\bibinfo {title} {Recent developments in quantum monte carlo simulations with applications for cold gases},}\ }\href {\doibase 10.1088/0034-4885/75/9/094501} {\bibfield  {journal} {\bibinfo  {journal} {Reports on Progress in Physics}\ }\textbf {\bibinfo {volume} {75}},\ \bibinfo {pages} {094501} (\bibinfo {year} {2012})}\BibitemShut {NoStop}%
\bibitem [{\citenamefont {Booth}\ \emph {et~al.}(2013)\citenamefont {Booth}, \citenamefont {Gr{\"u}neis}, \citenamefont {Kresse},\ and\ \citenamefont {Alavi}}]{Booth_Nature_2013}%
  \BibitemOpen
  \bibfield  {author} {\bibinfo {author} {\bibfnamefont {George~H.}\ \bibnamefont {Booth}}, \bibinfo {author} {\bibfnamefont {Andreas}\ \bibnamefont {Gr{\"u}neis}}, \bibinfo {author} {\bibfnamefont {Georg}\ \bibnamefont {Kresse}}, \ and\ \bibinfo {author} {\bibfnamefont {Ali}\ \bibnamefont {Alavi}},\ }\bibfield  {title} {\enquote {\bibinfo {title} {Towards an exact description of electronic wavefunctions in real solids},}\ }\href {\doibase 10.1038/nature11770} {\bibfield  {journal} {\bibinfo  {journal} {Nature}\ }\textbf {\bibinfo {volume} {493}},\ \bibinfo {pages} {365--370} (\bibinfo {year} {2013})}\BibitemShut {NoStop}%
\bibitem [{\citenamefont {Ceperley}\ and\ \citenamefont {Alder}(1980)}]{Ceperley_Alder_PRL_1980}%
  \BibitemOpen
  \bibfield  {author} {\bibinfo {author} {\bibfnamefont {D.~M.}\ \bibnamefont {Ceperley}}\ and\ \bibinfo {author} {\bibfnamefont {B.~J.}\ \bibnamefont {Alder}},\ }\bibfield  {title} {\enquote {\bibinfo {title} {Ground state of the electron gas by a stochastic method},}\ }\href {\doibase 10.1103/PhysRevLett.45.566} {\bibfield  {journal} {\bibinfo  {journal} {Phys. Rev. Lett.}\ }\textbf {\bibinfo {volume} {45}},\ \bibinfo {pages} {566--569} (\bibinfo {year} {1980})}\BibitemShut {NoStop}%
\bibitem [{\citenamefont {Booth}\ \emph {et~al.}(2009)\citenamefont {Booth}, \citenamefont {Thom},\ and\ \citenamefont {Alavi}}]{Booth_JCP_2009}%
  \BibitemOpen
  \bibfield  {author} {\bibinfo {author} {\bibfnamefont {George~H.}\ \bibnamefont {Booth}}, \bibinfo {author} {\bibfnamefont {Alex J.~W.}\ \bibnamefont {Thom}}, \ and\ \bibinfo {author} {\bibfnamefont {Ali}\ \bibnamefont {Alavi}},\ }\bibfield  {title} {\enquote {\bibinfo {title} {Fermion monte carlo without fixed nodes: A game of life, death, and annihilation in slater determinant space},}\ }\href {\doibase 10.1063/1.3193710} {\bibfield  {journal} {\bibinfo  {journal} {The Journal of Chemical Physics}\ }\textbf {\bibinfo {volume} {131}},\ \bibinfo {pages} {054106} (\bibinfo {year} {2009})}\BibitemShut {NoStop}%
\bibitem [{\citenamefont {Pfau}\ \emph {et~al.}(2020)\citenamefont {Pfau}, \citenamefont {Spencer}, \citenamefont {Matthews},\ and\ \citenamefont {Foulkes}}]{Pfau_PRR_2020}%
  \BibitemOpen
  \bibfield  {author} {\bibinfo {author} {\bibfnamefont {David}\ \bibnamefont {Pfau}}, \bibinfo {author} {\bibfnamefont {James~S.}\ \bibnamefont {Spencer}}, \bibinfo {author} {\bibfnamefont {Alexander G. D.~G.}\ \bibnamefont {Matthews}}, \ and\ \bibinfo {author} {\bibfnamefont {W.~M.~C.}\ \bibnamefont {Foulkes}},\ }\bibfield  {title} {\enquote {\bibinfo {title} {Ab initio solution of the many-electron schr\"odinger equation with deep neural networks},}\ }\href {\doibase 10.1103/PhysRevResearch.2.033429} {\bibfield  {journal} {\bibinfo  {journal} {Phys. Rev. Res.}\ }\textbf {\bibinfo {volume} {2}},\ \bibinfo {pages} {033429} (\bibinfo {year} {2020})}\BibitemShut {NoStop}%
\bibitem [{\citenamefont {Van~Houcke}\ \emph {et~al.}(2006)\citenamefont {Van~Houcke}, \citenamefont {Rombouts},\ and\ \citenamefont {Pollet}}]{PhysRevE.73.056703}%
  \BibitemOpen
  \bibfield  {author} {\bibinfo {author} {\bibfnamefont {K.}~\bibnamefont {Van~Houcke}}, \bibinfo {author} {\bibfnamefont {S.~M.~A.}\ \bibnamefont {Rombouts}}, \ and\ \bibinfo {author} {\bibfnamefont {L.}~\bibnamefont {Pollet}},\ }\bibfield  {title} {\enquote {\bibinfo {title} {Quantum monte carlo simulation in the canonical ensemble at finite temperature},}\ }\href {\doibase 10.1103/PhysRevE.73.056703} {\bibfield  {journal} {\bibinfo  {journal} {Phys. Rev. E}\ }\textbf {\bibinfo {volume} {73}},\ \bibinfo {pages} {056703} (\bibinfo {year} {2006})}\BibitemShut {NoStop}%
\bibitem [{\citenamefont {Ceperley}(1992)}]{Ceperley_PRL_1992}%
  \BibitemOpen
  \bibfield  {author} {\bibinfo {author} {\bibfnamefont {D.~M.}\ \bibnamefont {Ceperley}},\ }\bibfield  {title} {\enquote {\bibinfo {title} {Path-integral calculations of normal liquid $^{3}\mathrm{He}$},}\ }\href {\doibase 10.1103/PhysRevLett.69.331} {\bibfield  {journal} {\bibinfo  {journal} {Phys. Rev. Lett.}\ }\textbf {\bibinfo {volume} {69}},\ \bibinfo {pages} {331--334} (\bibinfo {year} {1992})}\BibitemShut {NoStop}%
\bibitem [{\citenamefont {Dornheim}\ \emph {et~al.}(2017{\natexlab{a}})\citenamefont {Dornheim}, \citenamefont {Groth}, \citenamefont {Malone}, \citenamefont {Schoof}, \citenamefont {Sjostrom}, \citenamefont {Foulkes},\ and\ \citenamefont {Bonitz}}]{Dornheim_POP_2017}%
  \BibitemOpen
  \bibfield  {author} {\bibinfo {author} {\bibfnamefont {Tobias}\ \bibnamefont {Dornheim}}, \bibinfo {author} {\bibfnamefont {Simon}\ \bibnamefont {Groth}}, \bibinfo {author} {\bibfnamefont {Fionn~D.}\ \bibnamefont {Malone}}, \bibinfo {author} {\bibfnamefont {Tim}\ \bibnamefont {Schoof}}, \bibinfo {author} {\bibfnamefont {Travis}\ \bibnamefont {Sjostrom}}, \bibinfo {author} {\bibfnamefont {W.~M.~C.}\ \bibnamefont {Foulkes}}, \ and\ \bibinfo {author} {\bibfnamefont {Michael}\ \bibnamefont {Bonitz}},\ }\bibfield  {title} {\enquote {\bibinfo {title} {Ab initio quantum monte carlo simulation of the warm dense electron gas},}\ }\href {\doibase 10.1063/1.4977920} {\bibfield  {journal} {\bibinfo  {journal} {Phys. Plasmas}\ }\textbf {\bibinfo {volume} {24}},\ \bibinfo {pages} {056303} (\bibinfo {year} {2017}{\natexlab{a}})}\BibitemShut {NoStop}%
\bibitem [{\citenamefont {Schoof}\ \emph {et~al.}(2015{\natexlab{a}})\citenamefont {Schoof}, \citenamefont {Groth}, \citenamefont {Vorberger},\ and\ \citenamefont {Bonitz}}]{Schoof_PRL_2015}%
  \BibitemOpen
  \bibfield  {author} {\bibinfo {author} {\bibfnamefont {T.}~\bibnamefont {Schoof}}, \bibinfo {author} {\bibfnamefont {S.}~\bibnamefont {Groth}}, \bibinfo {author} {\bibfnamefont {J.}~\bibnamefont {Vorberger}}, \ and\ \bibinfo {author} {\bibfnamefont {M.}~\bibnamefont {Bonitz}},\ }\bibfield  {title} {\enquote {\bibinfo {title} {Ab initio thermodynamic results for the degenerate electron gas at finite temperature},}\ }\href {\doibase 10.1103/PhysRevLett.115.130402} {\bibfield  {journal} {\bibinfo  {journal} {Phys. Rev. Lett.}\ }\textbf {\bibinfo {volume} {115}},\ \bibinfo {pages} {130402} (\bibinfo {year} {2015}{\natexlab{a}})}\BibitemShut {NoStop}%
\bibitem [{\citenamefont {Blunt}\ \emph {et~al.}(2014)\citenamefont {Blunt}, \citenamefont {Rogers}, \citenamefont {Spencer},\ and\ \citenamefont {Foulkes}}]{Blunt_PRB_2014}%
  \BibitemOpen
  \bibfield  {author} {\bibinfo {author} {\bibfnamefont {N.~S.}\ \bibnamefont {Blunt}}, \bibinfo {author} {\bibfnamefont {T.~W.}\ \bibnamefont {Rogers}}, \bibinfo {author} {\bibfnamefont {J.~S.}\ \bibnamefont {Spencer}}, \ and\ \bibinfo {author} {\bibfnamefont {W.~M.~C.}\ \bibnamefont {Foulkes}},\ }\bibfield  {title} {\enquote {\bibinfo {title} {Density-matrix quantum monte carlo method},}\ }\href {\doibase 10.1103/PhysRevB.89.245124} {\bibfield  {journal} {\bibinfo  {journal} {Phys. Rev. B}\ }\textbf {\bibinfo {volume} {89}},\ \bibinfo {pages} {245124} (\bibinfo {year} {2014})}\BibitemShut {NoStop}%
\bibitem [{\citenamefont {Malone}\ \emph {et~al.}(2015)\citenamefont {Malone}, \citenamefont {Blunt}, \citenamefont {Shepherd}, \citenamefont {Lee}, \citenamefont {Spencer},\ and\ \citenamefont {Foulkes}}]{Malone_JCP_2015}%
  \BibitemOpen
  \bibfield  {author} {\bibinfo {author} {\bibfnamefont {Fionn~D.}\ \bibnamefont {Malone}}, \bibinfo {author} {\bibfnamefont {N.~S.}\ \bibnamefont {Blunt}}, \bibinfo {author} {\bibfnamefont {James~J.}\ \bibnamefont {Shepherd}}, \bibinfo {author} {\bibfnamefont {D.~K.~K.}\ \bibnamefont {Lee}}, \bibinfo {author} {\bibfnamefont {J.~S.}\ \bibnamefont {Spencer}}, \ and\ \bibinfo {author} {\bibfnamefont {W.~M.~C.}\ \bibnamefont {Foulkes}},\ }\bibfield  {title} {\enquote {\bibinfo {title} {Interaction picture density matrix quantum monte carlo},}\ }\href {\doibase 10.1063/1.4927434} {\bibfield  {journal} {\bibinfo  {journal} {The Journal of Chemical Physics}\ }\textbf {\bibinfo {volume} {143}},\ \bibinfo {pages} {044116} (\bibinfo {year} {2015})}\BibitemShut {NoStop}%
\bibitem [{\citenamefont {Driver}\ and\ \citenamefont {Militzer}(2012)}]{Driver_PRL_2012}%
  \BibitemOpen
  \bibfield  {author} {\bibinfo {author} {\bibfnamefont {K.~P.}\ \bibnamefont {Driver}}\ and\ \bibinfo {author} {\bibfnamefont {B.}~\bibnamefont {Militzer}},\ }\bibfield  {title} {\enquote {\bibinfo {title} {All-electron path integral monte carlo simulations of warm dense matter: Application to water and carbon plasmas},}\ }\href {\doibase 10.1103/PhysRevLett.108.115502} {\bibfield  {journal} {\bibinfo  {journal} {Phys. Rev. Lett.}\ }\textbf {\bibinfo {volume} {108}},\ \bibinfo {pages} {115502} (\bibinfo {year} {2012})}\BibitemShut {NoStop}%
\bibitem [{\citenamefont {Boninsegni}\ \emph {et~al.}(2006{\natexlab{a}})\citenamefont {Boninsegni}, \citenamefont {Prokofev},\ and\ \citenamefont {Svistunov}}]{boninsegni1}%
  \BibitemOpen
  \bibfield  {author} {\bibinfo {author} {\bibfnamefont {M.}~\bibnamefont {Boninsegni}}, \bibinfo {author} {\bibfnamefont {N.~V.}\ \bibnamefont {Prokofev}}, \ and\ \bibinfo {author} {\bibfnamefont {B.~V.}\ \bibnamefont {Svistunov}},\ }\bibfield  {title} {\enquote {\bibinfo {title} {Worm algorithm and diagrammatic {M}onte {C}arlo: A new approach to continuous-space path integral {M}onte {C}arlo simulations},}\ }\href {https://journals.aps.org/pre/abstract/10.1103/PhysRevE.74.036701} {\bibfield  {journal} {\bibinfo  {journal} {Phys. Rev. E}\ }\textbf {\bibinfo {volume} {74}},\ \bibinfo {pages} {036701} (\bibinfo {year} {2006}{\natexlab{a}})}\BibitemShut {NoStop}%
\bibitem [{\citenamefont {Saccani}\ \emph {et~al.}(2012)\citenamefont {Saccani}, \citenamefont {Moroni},\ and\ \citenamefont {Boninsegni}}]{Saccani_Supersolid_PRL_2012}%
  \BibitemOpen
  \bibfield  {author} {\bibinfo {author} {\bibfnamefont {S.}~\bibnamefont {Saccani}}, \bibinfo {author} {\bibfnamefont {S.}~\bibnamefont {Moroni}}, \ and\ \bibinfo {author} {\bibfnamefont {M.}~\bibnamefont {Boninsegni}},\ }\bibfield  {title} {\enquote {\bibinfo {title} {Excitation spectrum of a supersolid},}\ }\href {\doibase 10.1103/PhysRevLett.108.175301} {\bibfield  {journal} {\bibinfo  {journal} {Phys. Rev. Lett.}\ }\textbf {\bibinfo {volume} {108}},\ \bibinfo {pages} {175301} (\bibinfo {year} {2012})}\BibitemShut {NoStop}%
\bibitem [{\citenamefont {Kawashima}\ and\ \citenamefont {Harada}(2004)}]{kawashima2004recent}%
  \BibitemOpen
  \bibfield  {author} {\bibinfo {author} {\bibfnamefont {Naoki}\ \bibnamefont {Kawashima}}\ and\ \bibinfo {author} {\bibfnamefont {Kenji}\ \bibnamefont {Harada}},\ }\bibfield  {title} {\enquote {\bibinfo {title} {Recent developments of world-line monte carlo methods},}\ }\href@noop {} {\bibfield  {journal} {\bibinfo  {journal} {Journal of the Physical Society of Japan}\ }\textbf {\bibinfo {volume} {73}},\ \bibinfo {pages} {1379--1414} (\bibinfo {year} {2004})}\BibitemShut {NoStop}%
\bibitem [{\citenamefont {Lee}\ \emph {et~al.}(2021)\citenamefont {Lee}, \citenamefont {Morales},\ and\ \citenamefont {Malone}}]{Joonho_JCP_2021}%
  \BibitemOpen
  \bibfield  {author} {\bibinfo {author} {\bibfnamefont {Joonho}\ \bibnamefont {Lee}}, \bibinfo {author} {\bibfnamefont {Miguel~A.}\ \bibnamefont {Morales}}, \ and\ \bibinfo {author} {\bibfnamefont {Fionn~D.}\ \bibnamefont {Malone}},\ }\bibfield  {title} {\enquote {\bibinfo {title} {A phaseless auxiliary-field quantum monte carlo perspective on the uniform electron gas at finite temperatures: Issues, observations, and benchmark study},}\ }\href {\doibase 10.1063/5.0041378} {\bibfield  {journal} {\bibinfo  {journal} {J. Chem. Phys.}\ }\textbf {\bibinfo {volume} {154}},\ \bibinfo {pages} {064109} (\bibinfo {year} {2021})}\BibitemShut {NoStop}%
\bibitem [{\citenamefont {Filinov}\ and\ \citenamefont {Bonitz}(2012)}]{Filinov_PRA_2012}%
  \BibitemOpen
  \bibfield  {author} {\bibinfo {author} {\bibfnamefont {A.}~\bibnamefont {Filinov}}\ and\ \bibinfo {author} {\bibfnamefont {M.}~\bibnamefont {Bonitz}},\ }\bibfield  {title} {\enquote {\bibinfo {title} {Collective and single-particle excitations in two-dimensional dipolar bose gases},}\ }\href {\doibase 10.1103/PhysRevA.86.043628} {\bibfield  {journal} {\bibinfo  {journal} {Phys. Rev. A}\ }\textbf {\bibinfo {volume} {86}},\ \bibinfo {pages} {043628} (\bibinfo {year} {2012})}\BibitemShut {NoStop}%
\bibitem [{\citenamefont {Metropolis}\ \emph {et~al.}(1953)\citenamefont {Metropolis}, \citenamefont {Rosenbluth}, \citenamefont {Rosenbluth}, \citenamefont {Teller},\ and\ \citenamefont {Teller}}]{metropolis}%
  \BibitemOpen
  \bibfield  {author} {\bibinfo {author} {\bibfnamefont {Nicholas}\ \bibnamefont {Metropolis}}, \bibinfo {author} {\bibfnamefont {Arianna~W.}\ \bibnamefont {Rosenbluth}}, \bibinfo {author} {\bibfnamefont {Marshall~N.}\ \bibnamefont {Rosenbluth}}, \bibinfo {author} {\bibfnamefont {Augusta~H.}\ \bibnamefont {Teller}}, \ and\ \bibinfo {author} {\bibfnamefont {Edward}\ \bibnamefont {Teller}},\ }\bibfield  {title} {\enquote {\bibinfo {title} {Equation of state calculations by fast computing machines},}\ }\href {\doibase 10.1063/1.1699114} {\bibfield  {journal} {\bibinfo  {journal} {J. Chem. Phys.}\ }\textbf {\bibinfo {volume} {21}},\ \bibinfo {pages} {1087--1092} (\bibinfo {year} {1953})}\BibitemShut {NoStop}%
\bibitem [{\citenamefont {Pollock}\ and\ \citenamefont {Ceperley}(1984)}]{Pollock_PRB_1984}%
  \BibitemOpen
  \bibfield  {author} {\bibinfo {author} {\bibfnamefont {E.~L.}\ \bibnamefont {Pollock}}\ and\ \bibinfo {author} {\bibfnamefont {D.~M.}\ \bibnamefont {Ceperley}},\ }\bibfield  {title} {\enquote {\bibinfo {title} {Simulation of quantum many-body systems by path-integral methods},}\ }\href {\doibase 10.1103/PhysRevB.30.2555} {\bibfield  {journal} {\bibinfo  {journal} {Phys. Rev. B}\ }\textbf {\bibinfo {volume} {30}},\ \bibinfo {pages} {2555--2568} (\bibinfo {year} {1984})}\BibitemShut {NoStop}%
\bibitem [{\citenamefont {Herman}\ \emph {et~al.}(1982)\citenamefont {Herman}, \citenamefont {Bruskin},\ and\ \citenamefont {Berne}}]{Berne_JCP_1982}%
  \BibitemOpen
  \bibfield  {author} {\bibinfo {author} {\bibfnamefont {M.~F.}\ \bibnamefont {Herman}}, \bibinfo {author} {\bibfnamefont {E.~J.}\ \bibnamefont {Bruskin}}, \ and\ \bibinfo {author} {\bibfnamefont {B.~J.}\ \bibnamefont {Berne}},\ }\bibfield  {title} {\enquote {\bibinfo {title} {On path integral monte carlo simulations},}\ }\href {\doibase 10.1063/1.442815} {\bibfield  {journal} {\bibinfo  {journal} {The Journal of Chemical Physics}\ }\textbf {\bibinfo {volume} {76}},\ \bibinfo {pages} {5150--5155} (\bibinfo {year} {1982})}\BibitemShut {NoStop}%
\bibitem [{\citenamefont {Takahashi}\ and\ \citenamefont {Imada}(1984)}]{Takahashi_Imada_PIMC_1984}%
  \BibitemOpen
  \bibfield  {author} {\bibinfo {author} {\bibfnamefont {Minoru}\ \bibnamefont {Takahashi}}\ and\ \bibinfo {author} {\bibfnamefont {Masatoshi}\ \bibnamefont {Imada}},\ }\bibfield  {title} {\enquote {\bibinfo {title} {Monte carlo calculation of quantum systems},}\ }\href@noop {} {\bibfield  {journal} {\bibinfo  {journal} {Journal of the Physical Society of Japan}\ }\textbf {\bibinfo {volume} {53}},\ \bibinfo {pages} {963--974} (\bibinfo {year} {1984})}\BibitemShut {NoStop}%
\bibitem [{\citenamefont {Chandler}\ and\ \citenamefont {Wolynes}(1981)}]{Chandler_JCP_1981}%
  \BibitemOpen
  \bibfield  {author} {\bibinfo {author} {\bibfnamefont {David}\ \bibnamefont {Chandler}}\ and\ \bibinfo {author} {\bibfnamefont {Peter~G.}\ \bibnamefont {Wolynes}},\ }\bibfield  {title} {\enquote {\bibinfo {title} {Exploiting the isomorphism between quantum theory and classical statistical mechanics of polyatomic fluids},}\ }\href {\doibase 10.1063/1.441588} {\bibfield  {journal} {\bibinfo  {journal} {The Journal of Chemical Physics}\ }\textbf {\bibinfo {volume} {74}},\ \bibinfo {pages} {4078--4095} (\bibinfo {year} {1981})}\BibitemShut {NoStop}%
\bibitem [{\citenamefont {Jordan}\ and\ \citenamefont {Fosdick}(1968)}]{Jordan_PR_1968}%
  \BibitemOpen
  \bibfield  {author} {\bibinfo {author} {\bibfnamefont {Harry~F.}\ \bibnamefont {Jordan}}\ and\ \bibinfo {author} {\bibfnamefont {Lloyd~D.}\ \bibnamefont {Fosdick}},\ }\bibfield  {title} {\enquote {\bibinfo {title} {Three-particle effects in the pair distribution function for ${\mathrm{he}}^{4}$ gas},}\ }\href {\doibase 10.1103/PhysRev.171.128} {\bibfield  {journal} {\bibinfo  {journal} {Phys. Rev.}\ }\textbf {\bibinfo {volume} {171}},\ \bibinfo {pages} {128--149} (\bibinfo {year} {1968})}\BibitemShut {NoStop}%
\bibitem [{\citenamefont {Fosdick}\ and\ \citenamefont {Jordan}(1966)}]{Fosdick_PR_1966}%
  \BibitemOpen
  \bibfield  {author} {\bibinfo {author} {\bibfnamefont {Lloyd~D.}\ \bibnamefont {Fosdick}}\ and\ \bibinfo {author} {\bibfnamefont {Harry~F.}\ \bibnamefont {Jordan}},\ }\bibfield  {title} {\enquote {\bibinfo {title} {{Path-Integral Calculation of the Two-Particle Slater Sum for ${\mathrm{He}}^{4}$}},}\ }\href {\doibase 10.1103/PhysRev.143.58} {\bibfield  {journal} {\bibinfo  {journal} {Phys. Rev.}\ }\textbf {\bibinfo {volume} {143}},\ \bibinfo {pages} {58--66} (\bibinfo {year} {1966})}\BibitemShut {NoStop}%
\bibitem [{\citenamefont {Sindzingre}\ \emph {et~al.}(1989)\citenamefont {Sindzingre}, \citenamefont {Klein},\ and\ \citenamefont {Ceperley}}]{Sindzingre_PRL_1989}%
  \BibitemOpen
  \bibfield  {author} {\bibinfo {author} {\bibfnamefont {Philippe}\ \bibnamefont {Sindzingre}}, \bibinfo {author} {\bibfnamefont {Michael~L.}\ \bibnamefont {Klein}}, \ and\ \bibinfo {author} {\bibfnamefont {David~M.}\ \bibnamefont {Ceperley}},\ }\bibfield  {title} {\enquote {\bibinfo {title} {Path-integral monte carlo study of low-temperature $^{4}\mathrm{He}$ clusters},}\ }\href {\doibase 10.1103/PhysRevLett.63.1601} {\bibfield  {journal} {\bibinfo  {journal} {Phys. Rev. Lett.}\ }\textbf {\bibinfo {volume} {63}},\ \bibinfo {pages} {1601--1604} (\bibinfo {year} {1989})}\BibitemShut {NoStop}%
\bibitem [{\citenamefont {Filinov}\ \emph {et~al.}(2010)\citenamefont {Filinov}, \citenamefont {Prokof'ev},\ and\ \citenamefont {Bonitz}}]{Filinov_PRL_2010}%
  \BibitemOpen
  \bibfield  {author} {\bibinfo {author} {\bibfnamefont {A.}~\bibnamefont {Filinov}}, \bibinfo {author} {\bibfnamefont {N.~V.}\ \bibnamefont {Prokof'ev}}, \ and\ \bibinfo {author} {\bibfnamefont {M.}~\bibnamefont {Bonitz}},\ }\bibfield  {title} {\enquote {\bibinfo {title} {Berezinskii-kosterlitz-thouless transition in two-dimensional dipole systems},}\ }\href {\doibase 10.1103/PhysRevLett.105.070401} {\bibfield  {journal} {\bibinfo  {journal} {Phys. Rev. Lett.}\ }\textbf {\bibinfo {volume} {105}},\ \bibinfo {pages} {070401} (\bibinfo {year} {2010})}\BibitemShut {NoStop}%
\bibitem [{\citenamefont {Dornheim}\ \emph {et~al.}(2015{\natexlab{a}})\citenamefont {Dornheim}, \citenamefont {Filinov},\ and\ \citenamefont {Bonitz}}]{Dornheim_PRB_2015}%
  \BibitemOpen
  \bibfield  {author} {\bibinfo {author} {\bibfnamefont {T.}~\bibnamefont {Dornheim}}, \bibinfo {author} {\bibfnamefont {A.}~\bibnamefont {Filinov}}, \ and\ \bibinfo {author} {\bibfnamefont {M.}~\bibnamefont {Bonitz}},\ }\bibfield  {title} {\enquote {\bibinfo {title} {Superfluidity of strongly correlated bosons in two- and three-dimensional traps},}\ }\href {\doibase 10.1103/PhysRevB.91.054503} {\bibfield  {journal} {\bibinfo  {journal} {Phys. Rev. B}\ }\textbf {\bibinfo {volume} {91}},\ \bibinfo {pages} {054503} (\bibinfo {year} {2015}{\natexlab{a}})}\BibitemShut {NoStop}%
\bibitem [{\citenamefont {Pollock}\ and\ \citenamefont {Ceperley}(1987)}]{Pollock_PRB_1987}%
  \BibitemOpen
  \bibfield  {author} {\bibinfo {author} {\bibfnamefont {E.~L.}\ \bibnamefont {Pollock}}\ and\ \bibinfo {author} {\bibfnamefont {D.~M.}\ \bibnamefont {Ceperley}},\ }\bibfield  {title} {\enquote {\bibinfo {title} {Path-integral computation of superfluid densities},}\ }\href {\doibase 10.1103/PhysRevB.36.8343} {\bibfield  {journal} {\bibinfo  {journal} {Phys. Rev. B}\ }\textbf {\bibinfo {volume} {36}},\ \bibinfo {pages} {8343--8352} (\bibinfo {year} {1987})}\BibitemShut {NoStop}%
\bibitem [{\citenamefont {Ferr\'e}\ and\ \citenamefont {Boronat}(2016)}]{Ferre_PRB_2016}%
  \BibitemOpen
  \bibfield  {author} {\bibinfo {author} {\bibfnamefont {G.}~\bibnamefont {Ferr\'e}}\ and\ \bibinfo {author} {\bibfnamefont {J.}~\bibnamefont {Boronat}},\ }\bibfield  {title} {\enquote {\bibinfo {title} {Dynamic structure factor of liquid $^{4}\mathrm{He}$ across the normal-superfluid transition},}\ }\href {\doibase 10.1103/PhysRevB.93.104510} {\bibfield  {journal} {\bibinfo  {journal} {Phys. Rev. B}\ }\textbf {\bibinfo {volume} {93}},\ \bibinfo {pages} {104510} (\bibinfo {year} {2016})}\BibitemShut {NoStop}%
\bibitem [{\citenamefont {Dornheim}\ \emph {et~al.}(2022{\natexlab{a}})\citenamefont {Dornheim}, \citenamefont {Moldabekov}, \citenamefont {Vorberger},\ and\ \citenamefont {Militzer}}]{Dornheim_SciRep_2022}%
  \BibitemOpen
  \bibfield  {author} {\bibinfo {author} {\bibfnamefont {Tobias}\ \bibnamefont {Dornheim}}, \bibinfo {author} {\bibfnamefont {Zhandos~A.}\ \bibnamefont {Moldabekov}}, \bibinfo {author} {\bibfnamefont {Jan}\ \bibnamefont {Vorberger}}, \ and\ \bibinfo {author} {\bibfnamefont {Burkhard}\ \bibnamefont {Militzer}},\ }\bibfield  {title} {\enquote {\bibinfo {title} {Path integral monte carlo approach to the structural properties and collective excitations of liquid $^3$he without fixed nodes},}\ }\href {\doibase 10.1038/s41598-021-04355-9} {\bibfield  {journal} {\bibinfo  {journal} {Sci. Rep.}\ }\textbf {\bibinfo {volume} {12}},\ \bibinfo {pages} {708} (\bibinfo {year} {2022}{\natexlab{a}})}\BibitemShut {NoStop}%
\bibitem [{\citenamefont {Dornheim}\ \emph {et~al.}(2018{\natexlab{a}})\citenamefont {Dornheim}, \citenamefont {Groth}, \citenamefont {Vorberger},\ and\ \citenamefont {Bonitz}}]{dornheim_dynamic}%
  \BibitemOpen
  \bibfield  {author} {\bibinfo {author} {\bibfnamefont {T.}~\bibnamefont {Dornheim}}, \bibinfo {author} {\bibfnamefont {S.}~\bibnamefont {Groth}}, \bibinfo {author} {\bibfnamefont {J.}~\bibnamefont {Vorberger}}, \ and\ \bibinfo {author} {\bibfnamefont {M.}~\bibnamefont {Bonitz}},\ }\bibfield  {title} {\enquote {\bibinfo {title} {Ab initio path integral {M}onte {C}arlo results for the dynamic structure factor of correlated electrons: From the electron liquid to warm dense matter},}\ }\href {https://journals.aps.org/prl/abstract/10.1103/PhysRevLett.121.255001} {\bibfield  {journal} {\bibinfo  {journal} {Phys. Rev. Lett.}\ }\textbf {\bibinfo {volume} {121}},\ \bibinfo {pages} {255001} (\bibinfo {year} {2018}{\natexlab{a}})}\BibitemShut {NoStop}%
\bibitem [{\citenamefont {Vitali}\ \emph {et~al.}(2010)\citenamefont {Vitali}, \citenamefont {Rossi}, \citenamefont {Reatto},\ and\ \citenamefont {Galli}}]{Vitali_PRB_2010}%
  \BibitemOpen
  \bibfield  {author} {\bibinfo {author} {\bibfnamefont {E.}~\bibnamefont {Vitali}}, \bibinfo {author} {\bibfnamefont {M.}~\bibnamefont {Rossi}}, \bibinfo {author} {\bibfnamefont {L.}~\bibnamefont {Reatto}}, \ and\ \bibinfo {author} {\bibfnamefont {D.~E.}\ \bibnamefont {Galli}},\ }\bibfield  {title} {\enquote {\bibinfo {title} {Ab initio low-energy dynamics of superfluid and solid $^{4}\textnormal{H}\textnormal{e}$},}\ }\href {\doibase 10.1103/PhysRevB.82.174510} {\bibfield  {journal} {\bibinfo  {journal} {Phys. Rev. B}\ }\textbf {\bibinfo {volume} {82}},\ \bibinfo {pages} {174510} (\bibinfo {year} {2010})}\BibitemShut {NoStop}%
\bibitem [{\citenamefont {Kora}\ and\ \citenamefont {Boninsegni}(2018)}]{Boninsegni_maximum_entropy}%
  \BibitemOpen
  \bibfield  {author} {\bibinfo {author} {\bibfnamefont {Youssef}\ \bibnamefont {Kora}}\ and\ \bibinfo {author} {\bibfnamefont {Massimo}\ \bibnamefont {Boninsegni}},\ }\bibfield  {title} {\enquote {\bibinfo {title} {Dynamic structure factor of superfluid $^{4}\mathrm{He}$ from quantum monte carlo: Maximum entropy revisited},}\ }\href {\doibase 10.1103/PhysRevB.98.134509} {\bibfield  {journal} {\bibinfo  {journal} {Phys. Rev. B}\ }\textbf {\bibinfo {volume} {98}},\ \bibinfo {pages} {134509} (\bibinfo {year} {2018})}\BibitemShut {NoStop}%
\bibitem [{\citenamefont {Clark}\ \emph {et~al.}(2009)\citenamefont {Clark}, \citenamefont {Casula},\ and\ \citenamefont {Ceperley}}]{Clark_PRL_2009}%
  \BibitemOpen
  \bibfield  {author} {\bibinfo {author} {\bibfnamefont {Bryan~K.}\ \bibnamefont {Clark}}, \bibinfo {author} {\bibfnamefont {Michele}\ \bibnamefont {Casula}}, \ and\ \bibinfo {author} {\bibfnamefont {D.~M.}\ \bibnamefont {Ceperley}},\ }\bibfield  {title} {\enquote {\bibinfo {title} {Hexatic and mesoscopic phases in a 2d quantum coulomb system},}\ }\href {\doibase 10.1103/PhysRevLett.103.055701} {\bibfield  {journal} {\bibinfo  {journal} {Phys. Rev. Lett.}\ }\textbf {\bibinfo {volume} {103}},\ \bibinfo {pages} {055701} (\bibinfo {year} {2009})}\BibitemShut {NoStop}%
\bibitem [{\citenamefont {Filinov}\ \emph {et~al.}(2001)\citenamefont {Filinov}, \citenamefont {Bonitz},\ and\ \citenamefont {Lozovik}}]{Filinov_PRL_2001}%
  \BibitemOpen
  \bibfield  {author} {\bibinfo {author} {\bibfnamefont {A.~V.}\ \bibnamefont {Filinov}}, \bibinfo {author} {\bibfnamefont {M.}~\bibnamefont {Bonitz}}, \ and\ \bibinfo {author} {\bibfnamefont {Yu.~E.}\ \bibnamefont {Lozovik}},\ }\bibfield  {title} {\enquote {\bibinfo {title} {Wigner crystallization in mesoscopic 2d electron systems},}\ }\href {\doibase 10.1103/PhysRevLett.86.3851} {\bibfield  {journal} {\bibinfo  {journal} {Phys. Rev. Lett.}\ }\textbf {\bibinfo {volume} {86}},\ \bibinfo {pages} {3851--3854} (\bibinfo {year} {2001})}\BibitemShut {NoStop}%
\bibitem [{\citenamefont {M\"user}\ \emph {et~al.}(1995)\citenamefont {M\"user}, \citenamefont {Nielaba},\ and\ \citenamefont {Binder}}]{PhysRevB.51.2723}%
  \BibitemOpen
  \bibfield  {author} {\bibinfo {author} {\bibfnamefont {M.~H.}\ \bibnamefont {M\"user}}, \bibinfo {author} {\bibfnamefont {P.}~\bibnamefont {Nielaba}}, \ and\ \bibinfo {author} {\bibfnamefont {K.}~\bibnamefont {Binder}},\ }\bibfield  {title} {\enquote {\bibinfo {title} {Path-integral monte carlo study of crystalline lennard-jones systems},}\ }\href {\doibase 10.1103/PhysRevB.51.2723} {\bibfield  {journal} {\bibinfo  {journal} {Phys. Rev. B}\ }\textbf {\bibinfo {volume} {51}},\ \bibinfo {pages} {2723--2731} (\bibinfo {year} {1995})}\BibitemShut {NoStop}%
\bibitem [{\citenamefont {Bhattacharya}\ \emph {et~al.}(2016)\citenamefont {Bhattacharya}, \citenamefont {Filinov}, \citenamefont {Ghosal},\ and\ \citenamefont {Bonitz}}]{Bhattacharya2016}%
  \BibitemOpen
  \bibfield  {author} {\bibinfo {author} {\bibfnamefont {Dyuti}\ \bibnamefont {Bhattacharya}}, \bibinfo {author} {\bibfnamefont {Alexei~V.}\ \bibnamefont {Filinov}}, \bibinfo {author} {\bibfnamefont {Amit}\ \bibnamefont {Ghosal}}, \ and\ \bibinfo {author} {\bibfnamefont {Michael}\ \bibnamefont {Bonitz}},\ }\bibfield  {title} {\enquote {\bibinfo {title} {Role of confinements on the melting of wigner molecules in quantum dots},}\ }\href {\doibase 10.1140/epjb/e2016-60448-5} {\bibfield  {journal} {\bibinfo  {journal} {The European Physical Journal B}\ }\textbf {\bibinfo {volume} {89}},\ \bibinfo {pages} {60} (\bibinfo {year} {2016})}\BibitemShut {NoStop}%
\bibitem [{\citenamefont {Boninsegni}\ \emph {et~al.}(2006{\natexlab{b}})\citenamefont {Boninsegni}, \citenamefont {Prokofev},\ and\ \citenamefont {Svistunov}}]{boninsegni2}%
  \BibitemOpen
  \bibfield  {author} {\bibinfo {author} {\bibfnamefont {M.}~\bibnamefont {Boninsegni}}, \bibinfo {author} {\bibfnamefont {N.~V.}\ \bibnamefont {Prokofev}}, \ and\ \bibinfo {author} {\bibfnamefont {B.~V.}\ \bibnamefont {Svistunov}},\ }\bibfield  {title} {\enquote {\bibinfo {title} {Worm algorithm for continuous-space path integral {M}onte {C}arlo simulations},}\ }\href {https://journals.aps.org/prl/abstract/10.1103/PhysRevLett.96.070601} {\bibfield  {journal} {\bibinfo  {journal} {Phys. Rev. Lett}\ }\textbf {\bibinfo {volume} {96}},\ \bibinfo {pages} {070601} (\bibinfo {year} {2006}{\natexlab{b}})}\BibitemShut {NoStop}%
\bibitem [{\citenamefont {Troyer}\ and\ \citenamefont {Wiese}(2005)}]{troyer}%
  \BibitemOpen
  \bibfield  {author} {\bibinfo {author} {\bibfnamefont {M.}~\bibnamefont {Troyer}}\ and\ \bibinfo {author} {\bibfnamefont {U.~J.}\ \bibnamefont {Wiese}},\ }\bibfield  {title} {\enquote {\bibinfo {title} {Computational complexity and fundamental limitations to fermionic quantum {M}onte {C}arlo simulations},}\ }\href {http://link.aps.org/doi/10.1103/PhysRevLett.94.170201} {\bibfield  {journal} {\bibinfo  {journal} {Phys. Rev. Lett}\ }\textbf {\bibinfo {volume} {94}},\ \bibinfo {pages} {170201} (\bibinfo {year} {2005})}\BibitemShut {NoStop}%
\bibitem [{\citenamefont {Loh}\ \emph {et~al.}(1990)\citenamefont {Loh}, \citenamefont {Gubernatis}, \citenamefont {Scalettar}, \citenamefont {White}, \citenamefont {Scalapino},\ and\ \citenamefont {Sugar}}]{Loh_PRB_1990}%
  \BibitemOpen
  \bibfield  {author} {\bibinfo {author} {\bibfnamefont {E.~Y.}\ \bibnamefont {Loh}}, \bibinfo {author} {\bibfnamefont {J.~E.}\ \bibnamefont {Gubernatis}}, \bibinfo {author} {\bibfnamefont {R.~T.}\ \bibnamefont {Scalettar}}, \bibinfo {author} {\bibfnamefont {S.~R.}\ \bibnamefont {White}}, \bibinfo {author} {\bibfnamefont {D.~J.}\ \bibnamefont {Scalapino}}, \ and\ \bibinfo {author} {\bibfnamefont {R.~L.}\ \bibnamefont {Sugar}},\ }\bibfield  {title} {\enquote {\bibinfo {title} {Sign problem in the numerical simulation of many-electron systems},}\ }\href {\doibase 10.1103/PhysRevB.41.9301} {\bibfield  {journal} {\bibinfo  {journal} {Phys. Rev. B}\ }\textbf {\bibinfo {volume} {41}},\ \bibinfo {pages} {9301--9307} (\bibinfo {year} {1990})}\BibitemShut {NoStop}%
\bibitem [{\citenamefont {Dornheim}(2019)}]{dornheim_sign_problem}%
  \BibitemOpen
  \bibfield  {author} {\bibinfo {author} {\bibfnamefont {T.}~\bibnamefont {Dornheim}},\ }\bibfield  {title} {\enquote {\bibinfo {title} {Fermion sign problem in path integral {M}onte {C}arlo simulations: Quantum dots, ultracold atoms, and warm dense matter},}\ }\href {https://journals.aps.org/pre/abstract/10.1103/PhysRevE.100.023307} {\bibfield  {journal} {\bibinfo  {journal} {Phys. Rev. E}\ }\textbf {\bibinfo {volume} {100}},\ \bibinfo {pages} {023307} (\bibinfo {year} {2019})}\BibitemShut {NoStop}%
\bibitem [{\citenamefont {Dornheim}(2021)}]{Dornheim_JPA_2021}%
  \BibitemOpen
  \bibfield  {author} {\bibinfo {author} {\bibfnamefont {Tobias}\ \bibnamefont {Dornheim}},\ }\bibfield  {title} {\enquote {\bibinfo {title} {Fermion sign problem in path integral monte carlo simulations: grand-canonical ensemble},}\ }\href {\doibase 10.1088/1751-8121/ac1481} {\bibfield  {journal} {\bibinfo  {journal} {Journal of Physics A: Mathematical and Theoretical}\ }\textbf {\bibinfo {volume} {54}},\ \bibinfo {pages} {335001} (\bibinfo {year} {2021})}\BibitemShut {NoStop}%
\bibitem [{\citenamefont {Kagan}\ and\ \citenamefont {Turlapov}(2019)}]{Kagan_2019}%
  \BibitemOpen
  \bibfield  {author} {\bibinfo {author} {\bibfnamefont {M~Yu}\ \bibnamefont {Kagan}}\ and\ \bibinfo {author} {\bibfnamefont {A~V}\ \bibnamefont {Turlapov}},\ }\bibfield  {title} {\enquote {\bibinfo {title} {Bcs – bec crossover, collective excitations, and hydrodynamics of superfluid quantum liquids and gases},}\ }\href {\doibase 10.3367/UFNe.2018.10.038471} {\bibfield  {journal} {\bibinfo  {journal} {Physics-Uspekhi}\ }\textbf {\bibinfo {volume} {62}},\ \bibinfo {pages} {215} (\bibinfo {year} {2019})}\BibitemShut {NoStop}%
\bibitem [{\citenamefont {Godfrin}\ \emph {et~al.}(2012)\citenamefont {Godfrin}, \citenamefont {Meschke}, \citenamefont {Lauter}, \citenamefont {Sultan}, \citenamefont {B{\"o}hm}, \citenamefont {Krotscheck},\ and\ \citenamefont {Panholzer}}]{Godfrin2012}%
  \BibitemOpen
  \bibfield  {author} {\bibinfo {author} {\bibfnamefont {Henri}\ \bibnamefont {Godfrin}}, \bibinfo {author} {\bibfnamefont {Matthias}\ \bibnamefont {Meschke}}, \bibinfo {author} {\bibfnamefont {Hans-Jochen}\ \bibnamefont {Lauter}}, \bibinfo {author} {\bibfnamefont {Ahmad}\ \bibnamefont {Sultan}}, \bibinfo {author} {\bibfnamefont {Helga~M.}\ \bibnamefont {B{\"o}hm}}, \bibinfo {author} {\bibfnamefont {Eckhard}\ \bibnamefont {Krotscheck}}, \ and\ \bibinfo {author} {\bibfnamefont {Martin}\ \bibnamefont {Panholzer}},\ }\bibfield  {title} {\enquote {\bibinfo {title} {Observation of a roton collective mode in a two-dimensional fermi liquid},}\ }\href {\doibase 10.1038/nature10919} {\bibfield  {journal} {\bibinfo  {journal} {Nature}\ }\textbf {\bibinfo {volume} {483}},\ \bibinfo {pages} {576--579} (\bibinfo {year} {2012})}\BibitemShut {NoStop}%
\bibitem [{\citenamefont {Dornheim}\ \emph {et~al.}(2022{\natexlab{b}})\citenamefont {Dornheim}, \citenamefont {Moldabekov}, \citenamefont {Vorberger}, \citenamefont {K{\"a}hlert},\ and\ \citenamefont {Bonitz}}]{Dornheim_Nature_2022}%
  \BibitemOpen
  \bibfield  {author} {\bibinfo {author} {\bibfnamefont {Tobias}\ \bibnamefont {Dornheim}}, \bibinfo {author} {\bibfnamefont {Zhandos}\ \bibnamefont {Moldabekov}}, \bibinfo {author} {\bibfnamefont {Jan}\ \bibnamefont {Vorberger}}, \bibinfo {author} {\bibfnamefont {Hanno}\ \bibnamefont {K{\"a}hlert}}, \ and\ \bibinfo {author} {\bibfnamefont {Michael}\ \bibnamefont {Bonitz}},\ }\bibfield  {title} {\enquote {\bibinfo {title} {Electronic pair alignment and roton feature in the warm dense electron gas},}\ }\href {\doibase 10.1038/s42005-022-01078-9} {\bibfield  {journal} {\bibinfo  {journal} {Communications Physics}\ }\textbf {\bibinfo {volume} {5}},\ \bibinfo {pages} {304} (\bibinfo {year} {2022}{\natexlab{b}})}\BibitemShut {NoStop}%
\bibitem [{\citenamefont {Hamann}\ \emph {et~al.}(2023)\citenamefont {Hamann}, \citenamefont {Kordts}, \citenamefont {Filinov}, \citenamefont {Bonitz}, \citenamefont {Dornheim},\ and\ \citenamefont {Vorberger}}]{Hamann_PRR_2023}%
  \BibitemOpen
  \bibfield  {author} {\bibinfo {author} {\bibfnamefont {Paul}\ \bibnamefont {Hamann}}, \bibinfo {author} {\bibfnamefont {Linda}\ \bibnamefont {Kordts}}, \bibinfo {author} {\bibfnamefont {Alexey}\ \bibnamefont {Filinov}}, \bibinfo {author} {\bibfnamefont {Michael}\ \bibnamefont {Bonitz}}, \bibinfo {author} {\bibfnamefont {Tobias}\ \bibnamefont {Dornheim}}, \ and\ \bibinfo {author} {\bibfnamefont {Jan}\ \bibnamefont {Vorberger}},\ }\bibfield  {title} {\enquote {\bibinfo {title} {Prediction of a roton-type feature in warm dense hydrogen},}\ }\href {\doibase 10.1103/PhysRevResearch.5.033039} {\bibfield  {journal} {\bibinfo  {journal} {Phys. Rev. Res.}\ }\textbf {\bibinfo {volume} {5}},\ \bibinfo {pages} {033039} (\bibinfo {year} {2023})}\BibitemShut {NoStop}%
\bibitem [{\citenamefont {Chuna}\ \emph {et~al.}(2025{\natexlab{a}})\citenamefont {Chuna}, \citenamefont {Vorberger}, \citenamefont {Tolias}, \citenamefont {Benedix~Robles}, \citenamefont {Hecht}, \citenamefont {Hofmann}, \citenamefont {Moldabekov},\ and\ \citenamefont {Dornheim}}]{Chuna_JCP_2025}%
  \BibitemOpen
  \bibfield  {author} {\bibinfo {author} {\bibfnamefont {Thomas~M.}\ \bibnamefont {Chuna}}, \bibinfo {author} {\bibfnamefont {Jan}\ \bibnamefont {Vorberger}}, \bibinfo {author} {\bibfnamefont {Panagiotis}\ \bibnamefont {Tolias}}, \bibinfo {author} {\bibfnamefont {Alexander}\ \bibnamefont {Benedix~Robles}}, \bibinfo {author} {\bibfnamefont {Michael}\ \bibnamefont {Hecht}}, \bibinfo {author} {\bibfnamefont {Phil-Alexander}\ \bibnamefont {Hofmann}}, \bibinfo {author} {\bibfnamefont {Zhandos~A.}\ \bibnamefont {Moldabekov}}, \ and\ \bibinfo {author} {\bibfnamefont {Tobias}\ \bibnamefont {Dornheim}},\ }\bibfield  {title} {\enquote {\bibinfo {title} {Second roton feature in the strongly coupled electron liquid},}\ }\href {\doibase 10.1063/5.0281085} {\bibfield  {journal} {\bibinfo  {journal} {The Journal of Chemical Physics}\ }\textbf {\bibinfo {volume} {163}},\ \bibinfo {pages} {034117} (\bibinfo {year} {2025}{\natexlab{a}})}\BibitemShut {NoStop}%
\bibitem [{\citenamefont {Bobrov}\ \emph {et~al.}(2016)\citenamefont {Bobrov}, \citenamefont {Trigger},\ and\ \citenamefont {Litinski}}]{Trigger}%
  \BibitemOpen
  \bibfield  {author} {\bibinfo {author} {\bibfnamefont {Viktor}\ \bibnamefont {Bobrov}}, \bibinfo {author} {\bibfnamefont {Sergey}\ \bibnamefont {Trigger}}, \ and\ \bibinfo {author} {\bibfnamefont {Daniel}\ \bibnamefont {Litinski}},\ }\bibfield  {title} {\enquote {\bibinfo {title} {Universality of the phonon–roton spectrum in liquids and superfluidity of 4he},}\ }\href {\doibase doi:10.1515/zna-2015-0397} {\bibfield  {journal} {\bibinfo  {journal} {Z. Naturforsch. A}\ }\textbf {\bibinfo {volume} {71}},\ \bibinfo {pages} {565--575} (\bibinfo {year} {2016})}\BibitemShut {NoStop}%
\bibitem [{\citenamefont {Azadi}\ and\ \citenamefont {Drummond}(2022)}]{Azadi_PRB_2022}%
  \BibitemOpen
  \bibfield  {author} {\bibinfo {author} {\bibfnamefont {Sam}\ \bibnamefont {Azadi}}\ and\ \bibinfo {author} {\bibfnamefont {N.~D.}\ \bibnamefont {Drummond}},\ }\bibfield  {title} {\enquote {\bibinfo {title} {Low-density phase diagram of the three-dimensional electron gas},}\ }\href {\doibase 10.1103/PhysRevB.105.245135} {\bibfield  {journal} {\bibinfo  {journal} {Phys. Rev. B}\ }\textbf {\bibinfo {volume} {105}},\ \bibinfo {pages} {245135} (\bibinfo {year} {2022})}\BibitemShut {NoStop}%
\bibitem [{\citenamefont {Drummond}\ \emph {et~al.}(2004)\citenamefont {Drummond}, \citenamefont {Radnai}, \citenamefont {Trail}, \citenamefont {Towler},\ and\ \citenamefont {Needs}}]{Drummond_PRB_Wigner_2004}%
  \BibitemOpen
  \bibfield  {author} {\bibinfo {author} {\bibfnamefont {N.~D.}\ \bibnamefont {Drummond}}, \bibinfo {author} {\bibfnamefont {Z.}~\bibnamefont {Radnai}}, \bibinfo {author} {\bibfnamefont {J.~R.}\ \bibnamefont {Trail}}, \bibinfo {author} {\bibfnamefont {M.~D.}\ \bibnamefont {Towler}}, \ and\ \bibinfo {author} {\bibfnamefont {R.~J.}\ \bibnamefont {Needs}},\ }\bibfield  {title} {\enquote {\bibinfo {title} {Diffusion quantum monte carlo study of three-dimensional wigner crystals},}\ }\href {\doibase 10.1103/PhysRevB.69.085116} {\bibfield  {journal} {\bibinfo  {journal} {Phys. Rev. B}\ }\textbf {\bibinfo {volume} {69}},\ \bibinfo {pages} {085116} (\bibinfo {year} {2004})}\BibitemShut {NoStop}%
\bibitem [{\citenamefont {Egger}\ \emph {et~al.}(1999)\citenamefont {Egger}, \citenamefont {H\"ausler}, \citenamefont {Mak},\ and\ \citenamefont {Grabert}}]{PhysRevLett.82.3320}%
  \BibitemOpen
  \bibfield  {author} {\bibinfo {author} {\bibfnamefont {R.}~\bibnamefont {Egger}}, \bibinfo {author} {\bibfnamefont {W.}~\bibnamefont {H\"ausler}}, \bibinfo {author} {\bibfnamefont {C.~H.}\ \bibnamefont {Mak}}, \ and\ \bibinfo {author} {\bibfnamefont {H.}~\bibnamefont {Grabert}},\ }\bibfield  {title} {\enquote {\bibinfo {title} {Crossover from fermi liquid to wigner molecule behavior in quantum dots},}\ }\href {\doibase 10.1103/PhysRevLett.82.3320} {\bibfield  {journal} {\bibinfo  {journal} {Phys. Rev. Lett.}\ }\textbf {\bibinfo {volume} {82}},\ \bibinfo {pages} {3320--3323} (\bibinfo {year} {1999})}\BibitemShut {NoStop}%
\bibitem [{\citenamefont {Vorberger}\ \emph {et~al.}(2025)\citenamefont {Vorberger}, \citenamefont {Graziani}, \citenamefont {Riley}, \citenamefont {Baczewski}, \citenamefont {Baraffe}, \citenamefont {Bethkenhagen}, \citenamefont {Blouin}, \citenamefont {Böhme}, \citenamefont {Bonitz}, \citenamefont {Bussmann}, \citenamefont {Casner}, \citenamefont {Cayzac}, \citenamefont {Celliers}, \citenamefont {Chabrier}, \citenamefont {Chamel}, \citenamefont {Chapman}, \citenamefont {Chen}, \citenamefont {Clérouin}, \citenamefont {Collins}, \citenamefont {Coppari}, \citenamefont {Döppner}, \citenamefont {Dornheim}, \citenamefont {Fletcher}, \citenamefont {Gericke}, \citenamefont {Glenzer}, \citenamefont {Goncharov}, \citenamefont {Gregori}, \citenamefont {Hamel}, \citenamefont {Hansen}, \citenamefont {Hartley}, \citenamefont {Hu}, \citenamefont {Hurricane}, \citenamefont {Karasiev}, \citenamefont {Kas}, \citenamefont {Kettle}, \citenamefont {Kluge}, \citenamefont {Knudson}, \citenamefont {Kononov}, \citenamefont {á},
  \citenamefont {Kraus}, \citenamefont {Kritcher}, \citenamefont {Malko}, \citenamefont {Massacrier}, \citenamefont {Militzer}, \citenamefont {Moldabekov}, \citenamefont {Murillo}, \citenamefont {Nagler}, \citenamefont {Nettelmann}, \citenamefont {Neumayer}, \citenamefont {Ofori-Okai}, \citenamefont {Oleynik}, \citenamefont {Preising}, \citenamefont {Pribram-Jones}, \citenamefont {Ramazanov}, \citenamefont {Ravasio}, \citenamefont {Redmer}, \citenamefont {Rethfeld}, \citenamefont {Robinson}, \citenamefont {Röpke}, \citenamefont {Soubiran}, \citenamefont {Starrett}, \citenamefont {Steinle-Neumann}, \citenamefont {Sterne}, \citenamefont {Tanaka}, \citenamefont {Thompson}, \citenamefont {Trickey}, \citenamefont {Vinci}, \citenamefont {Vinko}, \citenamefont {Wang}, \citenamefont {White}, \citenamefont {White}, \citenamefont {Zastrau}, \citenamefont {Zurek},\ and\ \citenamefont {Tolias}}]{vorberger2025roadmapwarmdensematter}%
  \BibitemOpen
  \bibfield  {author} {\bibinfo {author} {\bibfnamefont {Jan}\ \bibnamefont {Vorberger}}, \bibinfo {author} {\bibfnamefont {Frank}\ \bibnamefont {Graziani}}, \bibinfo {author} {\bibfnamefont {David}\ \bibnamefont {Riley}}, \bibinfo {author} {\bibfnamefont {Andrew~D.}\ \bibnamefont {Baczewski}}, \bibinfo {author} {\bibfnamefont {Isabelle}\ \bibnamefont {Baraffe}}, \bibinfo {author} {\bibfnamefont {Mandy}\ \bibnamefont {Bethkenhagen}}, \bibinfo {author} {\bibfnamefont {Simon}\ \bibnamefont {Blouin}}, \bibinfo {author} {\bibfnamefont {Maximilian~P.}\ \bibnamefont {Böhme}}, \bibinfo {author} {\bibfnamefont {Michael}\ \bibnamefont {Bonitz}}, \bibinfo {author} {\bibfnamefont {Michael}\ \bibnamefont {Bussmann}}, \bibinfo {author} {\bibfnamefont {Alexis}\ \bibnamefont {Casner}}, \bibinfo {author} {\bibfnamefont {Witold}\ \bibnamefont {Cayzac}}, \bibinfo {author} {\bibfnamefont {Peter}\ \bibnamefont {Celliers}}, \bibinfo {author} {\bibfnamefont {Gilles}\ \bibnamefont {Chabrier}}, \bibinfo {author} {\bibfnamefont
  {Nicolas}\ \bibnamefont {Chamel}}, \bibinfo {author} {\bibfnamefont {Dave}\ \bibnamefont {Chapman}}, \bibinfo {author} {\bibfnamefont {Mohan}\ \bibnamefont {Chen}}, \bibinfo {author} {\bibfnamefont {Jean}\ \bibnamefont {Clérouin}}, \bibinfo {author} {\bibfnamefont {Gilbert}\ \bibnamefont {Collins}}, \bibinfo {author} {\bibfnamefont {Federica}\ \bibnamefont {Coppari}}, \bibinfo {author} {\bibfnamefont {Tilo}\ \bibnamefont {Döppner}}, \bibinfo {author} {\bibfnamefont {Tobias}\ \bibnamefont {Dornheim}}, \bibinfo {author} {\bibfnamefont {Luke~B.}\ \bibnamefont {Fletcher}}, \bibinfo {author} {\bibfnamefont {Dirk~O.}\ \bibnamefont {Gericke}}, \bibinfo {author} {\bibfnamefont {Siegfried}\ \bibnamefont {Glenzer}}, \bibinfo {author} {\bibfnamefont {Alexander~F.}\ \bibnamefont {Goncharov}}, \bibinfo {author} {\bibfnamefont {Gianluca}\ \bibnamefont {Gregori}}, \bibinfo {author} {\bibfnamefont {Sebastien}\ \bibnamefont {Hamel}}, \bibinfo {author} {\bibfnamefont {Stephanie~B.}\ \bibnamefont {Hansen}}, \bibinfo
  {author} {\bibfnamefont {Nicholas~J.}\ \bibnamefont {Hartley}}, \bibinfo {author} {\bibfnamefont {Suxing}\ \bibnamefont {Hu}}, \bibinfo {author} {\bibfnamefont {Omar~A.}\ \bibnamefont {Hurricane}}, \bibinfo {author} {\bibfnamefont {Valentin~V.}\ \bibnamefont {Karasiev}}, \bibinfo {author} {\bibfnamefont {Joshua~J.}\ \bibnamefont {Kas}}, \bibinfo {author} {\bibfnamefont {Brendan}\ \bibnamefont {Kettle}}, \bibinfo {author} {\bibfnamefont {Thomas}\ \bibnamefont {Kluge}}, \bibinfo {author} {\bibfnamefont {Marcus~D.}\ \bibnamefont {Knudson}}, \bibinfo {author} {\bibfnamefont {Alina}\ \bibnamefont {Kononov}}, \bibinfo {author} {\bibfnamefont {Zuzana~Konôpkov}\ \bibnamefont {á}}, \bibinfo {author} {\bibfnamefont {Dominik}\ \bibnamefont {Kraus}}, \bibinfo {author} {\bibfnamefont {Andrea}\ \bibnamefont {Kritcher}}, \bibinfo {author} {\bibfnamefont {Sophia}\ \bibnamefont {Malko}}, \bibinfo {author} {\bibfnamefont {Gérard}\ \bibnamefont {Massacrier}}, \bibinfo {author} {\bibfnamefont {Burkhard}\ \bibnamefont
  {Militzer}}, \bibinfo {author} {\bibfnamefont {Zhandos~A.}\ \bibnamefont {Moldabekov}}, \bibinfo {author} {\bibfnamefont {Michael~S.}\ \bibnamefont {Murillo}}, \bibinfo {author} {\bibfnamefont {Bob}\ \bibnamefont {Nagler}}, \bibinfo {author} {\bibfnamefont {Nadine}\ \bibnamefont {Nettelmann}}, \bibinfo {author} {\bibfnamefont {Paul}\ \bibnamefont {Neumayer}}, \bibinfo {author} {\bibfnamefont {Benjamin~K.}\ \bibnamefont {Ofori-Okai}}, \bibinfo {author} {\bibfnamefont {Ivan~I.}\ \bibnamefont {Oleynik}}, \bibinfo {author} {\bibfnamefont {Martin}\ \bibnamefont {Preising}}, \bibinfo {author} {\bibfnamefont {Aurora}\ \bibnamefont {Pribram-Jones}}, \bibinfo {author} {\bibfnamefont {Tlekkabul}\ \bibnamefont {Ramazanov}}, \bibinfo {author} {\bibfnamefont {Alessandra}\ \bibnamefont {Ravasio}}, \bibinfo {author} {\bibfnamefont {Ronald}\ \bibnamefont {Redmer}}, \bibinfo {author} {\bibfnamefont {Baerbel}\ \bibnamefont {Rethfeld}}, \bibinfo {author} {\bibfnamefont {Alex P.~L.}\ \bibnamefont {Robinson}}, \bibinfo {author}
  {\bibfnamefont {Gerd}\ \bibnamefont {Röpke}}, \bibinfo {author} {\bibfnamefont {François}\ \bibnamefont {Soubiran}}, \bibinfo {author} {\bibfnamefont {Charles~E.}\ \bibnamefont {Starrett}}, \bibinfo {author} {\bibfnamefont {Gerd}\ \bibnamefont {Steinle-Neumann}}, \bibinfo {author} {\bibfnamefont {Phillip~A.}\ \bibnamefont {Sterne}}, \bibinfo {author} {\bibfnamefont {Shigenori}\ \bibnamefont {Tanaka}}, \bibinfo {author} {\bibfnamefont {Aidan~P.}\ \bibnamefont {Thompson}}, \bibinfo {author} {\bibfnamefont {Samuel~B.}\ \bibnamefont {Trickey}}, \bibinfo {author} {\bibfnamefont {Tommaso}\ \bibnamefont {Vinci}}, \bibinfo {author} {\bibfnamefont {Sam~M.}\ \bibnamefont {Vinko}}, \bibinfo {author} {\bibfnamefont {Lei}\ \bibnamefont {Wang}}, \bibinfo {author} {\bibfnamefont {Alexander~J.}\ \bibnamefont {White}}, \bibinfo {author} {\bibfnamefont {Thomas~G.}\ \bibnamefont {White}}, \bibinfo {author} {\bibfnamefont {Ulf}\ \bibnamefont {Zastrau}}, \bibinfo {author} {\bibfnamefont {Eva}\ \bibnamefont {Zurek}}, \ and\
  \bibinfo {author} {\bibfnamefont {Panagiotis}\ \bibnamefont {Tolias}},\ }\href {https://arxiv.org/abs/2505.02494} {\enquote {\bibinfo {title} {Roadmap for warm dense matter physics},}\ } (\bibinfo {year} {2025}),\ \Eprint {http://arxiv.org/abs/2505.02494} {arXiv:2505.02494 [physics.plasm-ph]} \BibitemShut {NoStop}%
\bibitem [{\citenamefont {Graziani}\ \emph {et~al.}(2014)\citenamefont {Graziani}, \citenamefont {Desjarlais}, \citenamefont {Redmer},\ and\ \citenamefont {Trickey}}]{wdm_book}%
  \BibitemOpen
  \bibinfo {editor} {\bibfnamefont {F.}~\bibnamefont {Graziani}}, \bibinfo {editor} {\bibfnamefont {M.~P.}\ \bibnamefont {Desjarlais}}, \bibinfo {editor} {\bibfnamefont {R.}~\bibnamefont {Redmer}}, \ and\ \bibinfo {editor} {\bibfnamefont {S.~B.}\ \bibnamefont {Trickey}},\ eds.,\ \href@noop {} {\emph {\bibinfo {title} {Frontiers and Challenges in Warm Dense Matter}}}\ (\bibinfo  {publisher} {Springer},\ \bibinfo {address} {International Publishing},\ \bibinfo {year} {2014})\BibitemShut {NoStop}%
\bibitem [{\citenamefont {Drake}(2018)}]{drake2018high}%
  \BibitemOpen
  \bibfield  {author} {\bibinfo {author} {\bibfnamefont {R.P.}\ \bibnamefont {Drake}},\ }\href {https://books.google.de/books?id=1AoZtAEACAAJ} {\emph {\bibinfo {title} {High-Energy-Density Physics: Foundation of Inertial Fusion and Experimental Astrophysics}}},\ Graduate Texts in Physics\ (\bibinfo  {publisher} {Springer International Publishing},\ \bibinfo {year} {2018})\BibitemShut {NoStop}%
\bibitem [{\citenamefont {Guillot}\ \emph {et~al.}(2018)\citenamefont {Guillot}, \citenamefont {Miguel}, \citenamefont {Militzer}, \citenamefont {Hubbard}, \citenamefont {Kaspi}, \citenamefont {Galanti}, \citenamefont {Cao}, \citenamefont {Helled}, \citenamefont {Wahl}, \citenamefont {Iess}, \citenamefont {Folkner}, \citenamefont {Stevenson}, \citenamefont {Lunine}, \citenamefont {Reese}, \citenamefont {Biekman}, \citenamefont {Parisi}, \citenamefont {Durante}, \citenamefont {Connerney}, \citenamefont {Levin},\ and\ \citenamefont {Bolton}}]{Guillot2018}%
  \BibitemOpen
  \bibfield  {author} {\bibinfo {author} {\bibfnamefont {T.}~\bibnamefont {Guillot}}, \bibinfo {author} {\bibfnamefont {Y.}~\bibnamefont {Miguel}}, \bibinfo {author} {\bibfnamefont {B.}~\bibnamefont {Militzer}}, \bibinfo {author} {\bibfnamefont {W.~B.}\ \bibnamefont {Hubbard}}, \bibinfo {author} {\bibfnamefont {Y.}~\bibnamefont {Kaspi}}, \bibinfo {author} {\bibfnamefont {E.}~\bibnamefont {Galanti}}, \bibinfo {author} {\bibfnamefont {H.}~\bibnamefont {Cao}}, \bibinfo {author} {\bibfnamefont {R.}~\bibnamefont {Helled}}, \bibinfo {author} {\bibfnamefont {S.~M.}\ \bibnamefont {Wahl}}, \bibinfo {author} {\bibfnamefont {L.}~\bibnamefont {Iess}}, \bibinfo {author} {\bibfnamefont {W.~M.}\ \bibnamefont {Folkner}}, \bibinfo {author} {\bibfnamefont {D.~J.}\ \bibnamefont {Stevenson}}, \bibinfo {author} {\bibfnamefont {J.~I.}\ \bibnamefont {Lunine}}, \bibinfo {author} {\bibfnamefont {D.~R.}\ \bibnamefont {Reese}}, \bibinfo {author} {\bibfnamefont {A.}~\bibnamefont {Biekman}}, \bibinfo {author} {\bibfnamefont
  {M.}~\bibnamefont {Parisi}}, \bibinfo {author} {\bibfnamefont {D.}~\bibnamefont {Durante}}, \bibinfo {author} {\bibfnamefont {J.~E.~P.}\ \bibnamefont {Connerney}}, \bibinfo {author} {\bibfnamefont {S.~M.}\ \bibnamefont {Levin}}, \ and\ \bibinfo {author} {\bibfnamefont {S.~J.}\ \bibnamefont {Bolton}},\ }\bibfield  {title} {\enquote {\bibinfo {title} {A suppression of differential rotation in jupiter's deep interior},}\ }\href {\doibase 10.1038/nature25775} {\bibfield  {journal} {\bibinfo  {journal} {Nature}\ }\textbf {\bibinfo {volume} {555}},\ \bibinfo {pages} {227--230} (\bibinfo {year} {2018})}\BibitemShut {NoStop}%
\bibitem [{\citenamefont {Saumon}\ \emph {et~al.}(2022)\citenamefont {Saumon}, \citenamefont {Blouin},\ and\ \citenamefont {Tremblay}}]{SAUMON20221}%
  \BibitemOpen
  \bibfield  {author} {\bibinfo {author} {\bibfnamefont {Didier}\ \bibnamefont {Saumon}}, \bibinfo {author} {\bibfnamefont {Simon}\ \bibnamefont {Blouin}}, \ and\ \bibinfo {author} {\bibfnamefont {Pier-Emmanuel}\ \bibnamefont {Tremblay}},\ }\bibfield  {title} {\enquote {\bibinfo {title} {Current challenges in the physics of white dwarf stars},}\ }\href {\doibase https://doi.org/10.1016/j.physrep.2022.09.001} {\bibfield  {journal} {\bibinfo  {journal} {Phys. Rep.}\ }\textbf {\bibinfo {volume} {988}},\ \bibinfo {pages} {1--63} (\bibinfo {year} {2022})},\ \bibinfo {note} {current Challenges in the Physics of White Dwarf Stars}\BibitemShut {NoStop}%
\bibitem [{\citenamefont {Kraus}\ \emph {et~al.}(2016)\citenamefont {Kraus}, \citenamefont {Ravasio}, \citenamefont {Gauthier}, \citenamefont {Gericke}, \citenamefont {Vorberger}, \citenamefont {Frydrych}, \citenamefont {Helfrich}, \citenamefont {Fletcher}, \citenamefont {Schaumann}, \citenamefont {Nagler}, \citenamefont {Barbrel}, \citenamefont {Bachmann}, \citenamefont {Gamboa}, \citenamefont {G{\"o}de}, \citenamefont {Granados}, \citenamefont {Gregori}, \citenamefont {Lee}, \citenamefont {Neumayer}, \citenamefont {Schumaker}, \citenamefont {D{\"o}ppner}, \citenamefont {Falcone}, \citenamefont {Glenzer},\ and\ \citenamefont {Roth}}]{Kraus2016}%
  \BibitemOpen
  \bibfield  {author} {\bibinfo {author} {\bibfnamefont {D.}~\bibnamefont {Kraus}}, \bibinfo {author} {\bibfnamefont {A.}~\bibnamefont {Ravasio}}, \bibinfo {author} {\bibfnamefont {M.}~\bibnamefont {Gauthier}}, \bibinfo {author} {\bibfnamefont {D.~O.}\ \bibnamefont {Gericke}}, \bibinfo {author} {\bibfnamefont {J.}~\bibnamefont {Vorberger}}, \bibinfo {author} {\bibfnamefont {S.}~\bibnamefont {Frydrych}}, \bibinfo {author} {\bibfnamefont {J.}~\bibnamefont {Helfrich}}, \bibinfo {author} {\bibfnamefont {L.~B.}\ \bibnamefont {Fletcher}}, \bibinfo {author} {\bibfnamefont {G.}~\bibnamefont {Schaumann}}, \bibinfo {author} {\bibfnamefont {B.}~\bibnamefont {Nagler}}, \bibinfo {author} {\bibfnamefont {B.}~\bibnamefont {Barbrel}}, \bibinfo {author} {\bibfnamefont {B.}~\bibnamefont {Bachmann}}, \bibinfo {author} {\bibfnamefont {E.~J.}\ \bibnamefont {Gamboa}}, \bibinfo {author} {\bibfnamefont {S.}~\bibnamefont {G{\"o}de}}, \bibinfo {author} {\bibfnamefont {E.}~\bibnamefont {Granados}}, \bibinfo {author} {\bibfnamefont
  {G.}~\bibnamefont {Gregori}}, \bibinfo {author} {\bibfnamefont {H.~J.}\ \bibnamefont {Lee}}, \bibinfo {author} {\bibfnamefont {P.}~\bibnamefont {Neumayer}}, \bibinfo {author} {\bibfnamefont {W.}~\bibnamefont {Schumaker}}, \bibinfo {author} {\bibfnamefont {T.}~\bibnamefont {D{\"o}ppner}}, \bibinfo {author} {\bibfnamefont {R.~W.}\ \bibnamefont {Falcone}}, \bibinfo {author} {\bibfnamefont {S.~H.}\ \bibnamefont {Glenzer}}, \ and\ \bibinfo {author} {\bibfnamefont {M.}~\bibnamefont {Roth}},\ }\bibfield  {title} {\enquote {\bibinfo {title} {Nanosecond formation of diamond and lonsdaleite by shock compression of graphite},}\ }\href {\doibase 10.1038/ncomms10970} {\bibfield  {journal} {\bibinfo  {journal} {Nature Communications}\ }\textbf {\bibinfo {volume} {7}},\ \bibinfo {pages} {10970} (\bibinfo {year} {2016})}\BibitemShut {NoStop}%
\bibitem [{\citenamefont {Kraus}\ \emph {et~al.}(2017)\citenamefont {Kraus}, \citenamefont {Vorberger}, \citenamefont {Pak}, \citenamefont {Hartley}, \citenamefont {Fletcher}, \citenamefont {Frydrych}, \citenamefont {Galtier}, \citenamefont {Gamboa}, \citenamefont {Gericke}, \citenamefont {Glenzer}, \citenamefont {Granados}, \citenamefont {MacDonald}, \citenamefont {MacKinnon}, \citenamefont {McBride}, \citenamefont {Nam}, \citenamefont {Neumayer}, \citenamefont {Roth}, \citenamefont {Saunders}, \citenamefont {Schuster}, \citenamefont {Sun}, \citenamefont {van Driel}, \citenamefont {D{\"o}ppner},\ and\ \citenamefont {Falcone}}]{Kraus2017}%
  \BibitemOpen
  \bibfield  {author} {\bibinfo {author} {\bibfnamefont {D.}~\bibnamefont {Kraus}}, \bibinfo {author} {\bibfnamefont {J.}~\bibnamefont {Vorberger}}, \bibinfo {author} {\bibfnamefont {A.}~\bibnamefont {Pak}}, \bibinfo {author} {\bibfnamefont {N.~J.}\ \bibnamefont {Hartley}}, \bibinfo {author} {\bibfnamefont {L.~B.}\ \bibnamefont {Fletcher}}, \bibinfo {author} {\bibfnamefont {S.}~\bibnamefont {Frydrych}}, \bibinfo {author} {\bibfnamefont {E.}~\bibnamefont {Galtier}}, \bibinfo {author} {\bibfnamefont {E.~J.}\ \bibnamefont {Gamboa}}, \bibinfo {author} {\bibfnamefont {D.~O.}\ \bibnamefont {Gericke}}, \bibinfo {author} {\bibfnamefont {S.~H.}\ \bibnamefont {Glenzer}}, \bibinfo {author} {\bibfnamefont {E.}~\bibnamefont {Granados}}, \bibinfo {author} {\bibfnamefont {M.~J.}\ \bibnamefont {MacDonald}}, \bibinfo {author} {\bibfnamefont {A.~J.}\ \bibnamefont {MacKinnon}}, \bibinfo {author} {\bibfnamefont {E.~E.}\ \bibnamefont {McBride}}, \bibinfo {author} {\bibfnamefont {I.}~\bibnamefont {Nam}}, \bibinfo {author}
  {\bibfnamefont {P.}~\bibnamefont {Neumayer}}, \bibinfo {author} {\bibfnamefont {M.}~\bibnamefont {Roth}}, \bibinfo {author} {\bibfnamefont {A.~M.}\ \bibnamefont {Saunders}}, \bibinfo {author} {\bibfnamefont {A.~K.}\ \bibnamefont {Schuster}}, \bibinfo {author} {\bibfnamefont {P.}~\bibnamefont {Sun}}, \bibinfo {author} {\bibfnamefont {T.}~\bibnamefont {van Driel}}, \bibinfo {author} {\bibfnamefont {T.}~\bibnamefont {D{\"o}ppner}}, \ and\ \bibinfo {author} {\bibfnamefont {R.~W.}\ \bibnamefont {Falcone}},\ }\bibfield  {title} {\enquote {\bibinfo {title} {Formation of diamonds in laser-compressed hydrocarbons at planetary interior conditions},}\ }\href {\doibase 10.1038/s41550-017-0219-9} {\bibfield  {journal} {\bibinfo  {journal} {Nature Astronomy}\ }\textbf {\bibinfo {volume} {1}},\ \bibinfo {pages} {606--611} (\bibinfo {year} {2017})}\BibitemShut {NoStop}%
\bibitem [{\citenamefont {Lazicki}\ \emph {et~al.}(2021)\citenamefont {Lazicki}, \citenamefont {McGonegle}, \citenamefont {Rygg}, \citenamefont {Braun}, \citenamefont {Swift}, \citenamefont {Gorman}, \citenamefont {Smith}, \citenamefont {Heighway}, \citenamefont {Higginbotham}, \citenamefont {Suggit}, \citenamefont {Fratanduono}, \citenamefont {Coppari}, \citenamefont {Wehrenberg}, \citenamefont {Kraus}, \citenamefont {Erskine}, \citenamefont {Bernier}, \citenamefont {McNaney}, \citenamefont {Rudd}, \citenamefont {Collins}, \citenamefont {Eggert},\ and\ \citenamefont {Wark}}]{Lazicki2021}%
  \BibitemOpen
  \bibfield  {author} {\bibinfo {author} {\bibfnamefont {A.}~\bibnamefont {Lazicki}}, \bibinfo {author} {\bibfnamefont {D.}~\bibnamefont {McGonegle}}, \bibinfo {author} {\bibfnamefont {J.~R.}\ \bibnamefont {Rygg}}, \bibinfo {author} {\bibfnamefont {D.~G.}\ \bibnamefont {Braun}}, \bibinfo {author} {\bibfnamefont {D.~C.}\ \bibnamefont {Swift}}, \bibinfo {author} {\bibfnamefont {M.~G.}\ \bibnamefont {Gorman}}, \bibinfo {author} {\bibfnamefont {R.~F.}\ \bibnamefont {Smith}}, \bibinfo {author} {\bibfnamefont {P.~G.}\ \bibnamefont {Heighway}}, \bibinfo {author} {\bibfnamefont {A.}~\bibnamefont {Higginbotham}}, \bibinfo {author} {\bibfnamefont {M.~J.}\ \bibnamefont {Suggit}}, \bibinfo {author} {\bibfnamefont {D.~E.}\ \bibnamefont {Fratanduono}}, \bibinfo {author} {\bibfnamefont {F.}~\bibnamefont {Coppari}}, \bibinfo {author} {\bibfnamefont {C.~E.}\ \bibnamefont {Wehrenberg}}, \bibinfo {author} {\bibfnamefont {R.~G.}\ \bibnamefont {Kraus}}, \bibinfo {author} {\bibfnamefont {D.}~\bibnamefont {Erskine}}, \bibinfo
  {author} {\bibfnamefont {J.~V.}\ \bibnamefont {Bernier}}, \bibinfo {author} {\bibfnamefont {J.~M.}\ \bibnamefont {McNaney}}, \bibinfo {author} {\bibfnamefont {R.~E.}\ \bibnamefont {Rudd}}, \bibinfo {author} {\bibfnamefont {G.~W.}\ \bibnamefont {Collins}}, \bibinfo {author} {\bibfnamefont {J.~H.}\ \bibnamefont {Eggert}}, \ and\ \bibinfo {author} {\bibfnamefont {J.~S.}\ \bibnamefont {Wark}},\ }\bibfield  {title} {\enquote {\bibinfo {title} {Metastability of diamond ramp-compressed to 2 terapascals},}\ }\href {\doibase 10.1038/s41586-020-03140-4} {\bibfield  {journal} {\bibinfo  {journal} {Nature}\ }\textbf {\bibinfo {volume} {589}},\ \bibinfo {pages} {532--535} (\bibinfo {year} {2021})}\BibitemShut {NoStop}%
\bibitem [{\citenamefont {Hu}\ \emph {et~al.}(2011)\citenamefont {Hu}, \citenamefont {Militzer}, \citenamefont {Goncharov},\ and\ \citenamefont {Skupsky}}]{hu_ICF}%
  \BibitemOpen
  \bibfield  {author} {\bibinfo {author} {\bibfnamefont {S.~X.}\ \bibnamefont {Hu}}, \bibinfo {author} {\bibfnamefont {B.}~\bibnamefont {Militzer}}, \bibinfo {author} {\bibfnamefont {V.~N.}\ \bibnamefont {Goncharov}}, \ and\ \bibinfo {author} {\bibfnamefont {S.}~\bibnamefont {Skupsky}},\ }\bibfield  {title} {\enquote {\bibinfo {title} {First-principles equation-of-state table of deuterium for inertial confinement fusion applications},}\ }\href {https://journals.aps.org/prb/abstract/10.1103/PhysRevB.84.224109} {\bibfield  {journal} {\bibinfo  {journal} {Phys. Rev. B}\ }\textbf {\bibinfo {volume} {84}},\ \bibinfo {pages} {224109} (\bibinfo {year} {2011})}\BibitemShut {NoStop}%
\bibitem [{\citenamefont {Batani}\ \emph {et~al.}(2023)\citenamefont {Batani}, \citenamefont {Colaïtis}, \citenamefont {Consoli}, \citenamefont {Danson}, \citenamefont {Gizzi}, \citenamefont {Honrubia}, \citenamefont {Kühl}, \citenamefont {Le~Pape}, \citenamefont {Miquel}, \citenamefont {Perlado},\ and\ \citenamefont {et~al.}}]{Batani_roadmap}%
  \BibitemOpen
  \bibfield  {author} {\bibinfo {author} {\bibfnamefont {Dimitri}\ \bibnamefont {Batani}}, \bibinfo {author} {\bibfnamefont {Arnaud}\ \bibnamefont {Colaïtis}}, \bibinfo {author} {\bibfnamefont {Fabrizio}\ \bibnamefont {Consoli}}, \bibinfo {author} {\bibfnamefont {Colin~N.}\ \bibnamefont {Danson}}, \bibinfo {author} {\bibfnamefont {Leonida~Antonio}\ \bibnamefont {Gizzi}}, \bibinfo {author} {\bibfnamefont {Javier}\ \bibnamefont {Honrubia}}, \bibinfo {author} {\bibfnamefont {Thomas}\ \bibnamefont {Kühl}}, \bibinfo {author} {\bibfnamefont {Sebastien}\ \bibnamefont {Le~Pape}}, \bibinfo {author} {\bibfnamefont {Jean-Luc}\ \bibnamefont {Miquel}}, \bibinfo {author} {\bibfnamefont {Jose~Manuel}\ \bibnamefont {Perlado}}, \ and\ \bibinfo {author} {\bibnamefont {et~al.}},\ }\bibfield  {title} {\enquote {\bibinfo {title} {Future for inertial-fusion energy in europe: a roadmap},}\ }\href {\doibase 10.1017/hpl.2023.80} {\bibfield  {journal} {\bibinfo  {journal} {High Power Laser Science and Engineering}\ }\textbf
  {\bibinfo {volume} {11}},\ \bibinfo {pages} {e83} (\bibinfo {year} {2023})}\BibitemShut {NoStop}%
\bibitem [{\citenamefont {Hurricane}\ \emph {et~al.}(2023)\citenamefont {Hurricane}, \citenamefont {Patel}, \citenamefont {Betti}, \citenamefont {Froula}, \citenamefont {Regan}, \citenamefont {Slutz}, \citenamefont {Gomez},\ and\ \citenamefont {Sweeney}}]{Hurricane_RevModPhys_2023}%
  \BibitemOpen
  \bibfield  {author} {\bibinfo {author} {\bibfnamefont {O.~A.}\ \bibnamefont {Hurricane}}, \bibinfo {author} {\bibfnamefont {P.~K.}\ \bibnamefont {Patel}}, \bibinfo {author} {\bibfnamefont {R.}~\bibnamefont {Betti}}, \bibinfo {author} {\bibfnamefont {D.~H.}\ \bibnamefont {Froula}}, \bibinfo {author} {\bibfnamefont {S.~P.}\ \bibnamefont {Regan}}, \bibinfo {author} {\bibfnamefont {S.~A.}\ \bibnamefont {Slutz}}, \bibinfo {author} {\bibfnamefont {M.~R.}\ \bibnamefont {Gomez}}, \ and\ \bibinfo {author} {\bibfnamefont {M.~A.}\ \bibnamefont {Sweeney}},\ }\bibfield  {title} {\enquote {\bibinfo {title} {Physics principles of inertial confinement fusion and u.s. program overview},}\ }\href {\doibase 10.1103/RevModPhys.95.025005} {\bibfield  {journal} {\bibinfo  {journal} {Rev. Mod. Phys.}\ }\textbf {\bibinfo {volume} {95}},\ \bibinfo {pages} {025005} (\bibinfo {year} {2023})}\BibitemShut {NoStop}%
\bibitem [{\citenamefont {Abu-Shawareb}\ \emph {et~al.}(2024)\citenamefont {Abu-Shawareb}, \citenamefont {Acree}, \citenamefont {Adams}, \citenamefont {Adams}, \citenamefont {Addis}, \citenamefont {Aden}, \citenamefont {Adrian}, \citenamefont {Afeyan}, \citenamefont {Aggleton}, \citenamefont {Aghaian}, \citenamefont {Aguirre}, \citenamefont {Aikens}, \citenamefont {Akre}, \citenamefont {Albert} \emph {et~al.}}]{AbuShawareb_PRL_2024}%
  \BibitemOpen
  \bibfield  {author} {\bibinfo {author} {\bibfnamefont {H.}~\bibnamefont {Abu-Shawareb}}, \bibinfo {author} {\bibfnamefont {R.}~\bibnamefont {Acree}}, \bibinfo {author} {\bibfnamefont {P.}~\bibnamefont {Adams}}, \bibinfo {author} {\bibfnamefont {J.}~\bibnamefont {Adams}}, \bibinfo {author} {\bibfnamefont {B.}~\bibnamefont {Addis}}, \bibinfo {author} {\bibfnamefont {R.}~\bibnamefont {Aden}}, \bibinfo {author} {\bibfnamefont {P.}~\bibnamefont {Adrian}}, \bibinfo {author} {\bibfnamefont {B.~B.}\ \bibnamefont {Afeyan}}, \bibinfo {author} {\bibfnamefont {M.}~\bibnamefont {Aggleton}}, \bibinfo {author} {\bibfnamefont {L.}~\bibnamefont {Aghaian}}, \bibinfo {author} {\bibfnamefont {A.}~\bibnamefont {Aguirre}}, \bibinfo {author} {\bibfnamefont {D.}~\bibnamefont {Aikens}}, \bibinfo {author} {\bibfnamefont {J.}~\bibnamefont {Akre}}, \bibinfo {author} {\bibfnamefont {F.}~\bibnamefont {Albert}},  \emph {et~al.} (\bibinfo {collaboration} {The Indirect Drive ICF Collaboration}),\ }\bibfield  {title} {\enquote {\bibinfo
  {title} {Achievement of target gain larger than unity in an inertial fusion experiment},}\ }\href {\doibase 10.1103/PhysRevLett.132.065102} {\bibfield  {journal} {\bibinfo  {journal} {Phys. Rev. Lett.}\ }\textbf {\bibinfo {volume} {132}},\ \bibinfo {pages} {065102} (\bibinfo {year} {2024})}\BibitemShut {NoStop}%
\bibitem [{\citenamefont {Bonitz}\ \emph {et~al.}(2020)\citenamefont {Bonitz}, \citenamefont {Dornheim}, \citenamefont {Moldabekov}, \citenamefont {Zhang}, \citenamefont {Hamann}, \citenamefont {Kählert}, \citenamefont {Filinov}, \citenamefont {Ramakrishna},\ and\ \citenamefont {Vorberger}}]{new_POP}%
  \BibitemOpen
  \bibfield  {author} {\bibinfo {author} {\bibfnamefont {M.}~\bibnamefont {Bonitz}}, \bibinfo {author} {\bibfnamefont {T.}~\bibnamefont {Dornheim}}, \bibinfo {author} {\bibfnamefont {Zh.~A.}\ \bibnamefont {Moldabekov}}, \bibinfo {author} {\bibfnamefont {S.}~\bibnamefont {Zhang}}, \bibinfo {author} {\bibfnamefont {P.}~\bibnamefont {Hamann}}, \bibinfo {author} {\bibfnamefont {H.}~\bibnamefont {Kählert}}, \bibinfo {author} {\bibfnamefont {A.}~\bibnamefont {Filinov}}, \bibinfo {author} {\bibfnamefont {K.}~\bibnamefont {Ramakrishna}}, \ and\ \bibinfo {author} {\bibfnamefont {J.}~\bibnamefont {Vorberger}},\ }\bibfield  {title} {\enquote {\bibinfo {title} {Ab initio simulation of warm dense matter},}\ }\href {\doibase 10.1063/1.5143225} {\bibfield  {journal} {\bibinfo  {journal} {Phys. Plasmas}\ }\textbf {\bibinfo {volume} {27}},\ \bibinfo {pages} {042710} (\bibinfo {year} {2020})}\BibitemShut {NoStop}%
\bibitem [{\citenamefont {Dornheim}\ \emph {et~al.}(2023{\natexlab{a}})\citenamefont {Dornheim}, \citenamefont {Moldabekov}, \citenamefont {Ramakrishna}, \citenamefont {Tolias}, \citenamefont {Baczewski}, \citenamefont {Kraus}, \citenamefont {Preston}, \citenamefont {Chapman}, \citenamefont {Böhme}, \citenamefont {Döppner}, \citenamefont {Graziani}, \citenamefont {Bonitz}, \citenamefont {Cangi},\ and\ \citenamefont {Vorberger}}]{Dornheim_review}%
  \BibitemOpen
  \bibfield  {author} {\bibinfo {author} {\bibfnamefont {Tobias}\ \bibnamefont {Dornheim}}, \bibinfo {author} {\bibfnamefont {Zhandos~A.}\ \bibnamefont {Moldabekov}}, \bibinfo {author} {\bibfnamefont {Kushal}\ \bibnamefont {Ramakrishna}}, \bibinfo {author} {\bibfnamefont {Panagiotis}\ \bibnamefont {Tolias}}, \bibinfo {author} {\bibfnamefont {Andrew~D.}\ \bibnamefont {Baczewski}}, \bibinfo {author} {\bibfnamefont {Dominik}\ \bibnamefont {Kraus}}, \bibinfo {author} {\bibfnamefont {Thomas~R.}\ \bibnamefont {Preston}}, \bibinfo {author} {\bibfnamefont {David~A.}\ \bibnamefont {Chapman}}, \bibinfo {author} {\bibfnamefont {Maximilian~P.}\ \bibnamefont {Böhme}}, \bibinfo {author} {\bibfnamefont {Tilo}\ \bibnamefont {Döppner}}, \bibinfo {author} {\bibfnamefont {Frank}\ \bibnamefont {Graziani}}, \bibinfo {author} {\bibfnamefont {Michael}\ \bibnamefont {Bonitz}}, \bibinfo {author} {\bibfnamefont {Attila}\ \bibnamefont {Cangi}}, \ and\ \bibinfo {author} {\bibfnamefont {Jan}\ \bibnamefont {Vorberger}},\ }\bibfield
  {title} {\enquote {\bibinfo {title} {Electronic density response of warm dense matter},}\ }\href {\doibase 10.1063/5.0138955} {\bibfield  {journal} {\bibinfo  {journal} {Phys. Plasmas}\ }\textbf {\bibinfo {volume} {30}},\ \bibinfo {pages} {032705} (\bibinfo {year} {2023}{\natexlab{a}})}\BibitemShut {NoStop}%
\bibitem [{\citenamefont {Bonitz}\ \emph {et~al.}(2024)\citenamefont {Bonitz}, \citenamefont {Vorberger}, \citenamefont {Bethkenhagen}, \citenamefont {Böhme}, \citenamefont {Ceperley}, \citenamefont {Filinov}, \citenamefont {Gawne}, \citenamefont {Graziani}, \citenamefont {Gregori}, \citenamefont {Hamann}, \citenamefont {Hansen}, \citenamefont {Holzmann}, \citenamefont {Hu}, \citenamefont {Kählert}, \citenamefont {Karasiev}, \citenamefont {Kleinschmidt}, \citenamefont {Kordts}, \citenamefont {Makait}, \citenamefont {Militzer}, \citenamefont {Moldabekov}, \citenamefont {Pierleoni}, \citenamefont {Preising}, \citenamefont {Ramakrishna}, \citenamefont {Redmer}, \citenamefont {Schwalbe}, \citenamefont {Svensson},\ and\ \citenamefont {Dornheim}}]{Bonitz_POP_2024}%
  \BibitemOpen
  \bibfield  {author} {\bibinfo {author} {\bibfnamefont {Michael}\ \bibnamefont {Bonitz}}, \bibinfo {author} {\bibfnamefont {Jan}\ \bibnamefont {Vorberger}}, \bibinfo {author} {\bibfnamefont {Mandy}\ \bibnamefont {Bethkenhagen}}, \bibinfo {author} {\bibfnamefont {Maximilian~P.}\ \bibnamefont {Böhme}}, \bibinfo {author} {\bibfnamefont {David~M.}\ \bibnamefont {Ceperley}}, \bibinfo {author} {\bibfnamefont {Alexey}\ \bibnamefont {Filinov}}, \bibinfo {author} {\bibfnamefont {Thomas}\ \bibnamefont {Gawne}}, \bibinfo {author} {\bibfnamefont {Frank}\ \bibnamefont {Graziani}}, \bibinfo {author} {\bibfnamefont {Gianluca}\ \bibnamefont {Gregori}}, \bibinfo {author} {\bibfnamefont {Paul}\ \bibnamefont {Hamann}}, \bibinfo {author} {\bibfnamefont {Stephanie~B.}\ \bibnamefont {Hansen}}, \bibinfo {author} {\bibfnamefont {Markus}\ \bibnamefont {Holzmann}}, \bibinfo {author} {\bibfnamefont {S.~X.}\ \bibnamefont {Hu}}, \bibinfo {author} {\bibfnamefont {Hanno}\ \bibnamefont {Kählert}}, \bibinfo {author} {\bibfnamefont
  {Valentin~V.}\ \bibnamefont {Karasiev}}, \bibinfo {author} {\bibfnamefont {Uwe}\ \bibnamefont {Kleinschmidt}}, \bibinfo {author} {\bibfnamefont {Linda}\ \bibnamefont {Kordts}}, \bibinfo {author} {\bibfnamefont {Christopher}\ \bibnamefont {Makait}}, \bibinfo {author} {\bibfnamefont {Burkhard}\ \bibnamefont {Militzer}}, \bibinfo {author} {\bibfnamefont {Zhandos~A.}\ \bibnamefont {Moldabekov}}, \bibinfo {author} {\bibfnamefont {Carlo}\ \bibnamefont {Pierleoni}}, \bibinfo {author} {\bibfnamefont {Martin}\ \bibnamefont {Preising}}, \bibinfo {author} {\bibfnamefont {Kushal}\ \bibnamefont {Ramakrishna}}, \bibinfo {author} {\bibfnamefont {Ronald}\ \bibnamefont {Redmer}}, \bibinfo {author} {\bibfnamefont {Sebastian}\ \bibnamefont {Schwalbe}}, \bibinfo {author} {\bibfnamefont {Pontus}\ \bibnamefont {Svensson}}, \ and\ \bibinfo {author} {\bibfnamefont {Tobias}\ \bibnamefont {Dornheim}},\ }\bibfield  {title} {\enquote {\bibinfo {title} {Toward first principles-based simulations of dense hydrogen},}\ }\href {\doibase
  10.1063/5.0219405} {\bibfield  {journal} {\bibinfo  {journal} {Physics of Plasmas}\ }\textbf {\bibinfo {volume} {31}},\ \bibinfo {pages} {110501} (\bibinfo {year} {2024})}\BibitemShut {NoStop}%
\bibitem [{\citenamefont {Schoof}\ \emph {et~al.}(2015{\natexlab{b}})\citenamefont {Schoof}, \citenamefont {Groth},\ and\ \citenamefont {Bonitz}}]{Schoof_CPP_2015}%
  \BibitemOpen
  \bibfield  {author} {\bibinfo {author} {\bibfnamefont {T.}~\bibnamefont {Schoof}}, \bibinfo {author} {\bibfnamefont {S.}~\bibnamefont {Groth}}, \ and\ \bibinfo {author} {\bibfnamefont {M.}~\bibnamefont {Bonitz}},\ }\bibfield  {title} {\enquote {\bibinfo {title} {Towards ab initio thermodynamics of the electron gas at strong degeneracy},}\ }\href {\doibase https://doi.org/10.1002/ctpp.201400072} {\bibfield  {journal} {\bibinfo  {journal} {Contributions to Plasma Physics}\ }\textbf {\bibinfo {volume} {55}},\ \bibinfo {pages} {136--143} (\bibinfo {year} {2015}{\natexlab{b}})}\BibitemShut {NoStop}%
\bibitem [{\citenamefont {Groth}\ \emph {et~al.}(2016)\citenamefont {Groth}, \citenamefont {Schoof}, \citenamefont {Dornheim},\ and\ \citenamefont {Bonitz}}]{Groth_PRB_2016}%
  \BibitemOpen
  \bibfield  {author} {\bibinfo {author} {\bibfnamefont {S.}~\bibnamefont {Groth}}, \bibinfo {author} {\bibfnamefont {T.}~\bibnamefont {Schoof}}, \bibinfo {author} {\bibfnamefont {T.}~\bibnamefont {Dornheim}}, \ and\ \bibinfo {author} {\bibfnamefont {M.}~\bibnamefont {Bonitz}},\ }\bibfield  {title} {\enquote {\bibinfo {title} {Ab initio quantum monte carlo simulations of the uniform electron gas without fixed nodes},}\ }\href {\doibase 10.1103/PhysRevB.93.085102} {\bibfield  {journal} {\bibinfo  {journal} {Phys. Rev. B}\ }\textbf {\bibinfo {volume} {93}},\ \bibinfo {pages} {085102} (\bibinfo {year} {2016})}\BibitemShut {NoStop}%
\bibitem [{\citenamefont {Chin}(2015)}]{Chin_PRE_2015}%
  \BibitemOpen
  \bibfield  {author} {\bibinfo {author} {\bibfnamefont {Siu~A.}\ \bibnamefont {Chin}},\ }\bibfield  {title} {\enquote {\bibinfo {title} {High-order path-integral monte carlo methods for solving quantum dot problems},}\ }\href {\doibase 10.1103/PhysRevE.91.031301} {\bibfield  {journal} {\bibinfo  {journal} {Phys. Rev. E}\ }\textbf {\bibinfo {volume} {91}},\ \bibinfo {pages} {031301} (\bibinfo {year} {2015})}\BibitemShut {NoStop}%
\bibitem [{\citenamefont {Dornheim}\ \emph {et~al.}(2015{\natexlab{b}})\citenamefont {Dornheim}, \citenamefont {Groth}, \citenamefont {Filinov},\ and\ \citenamefont {Bonitz}}]{Dornheim_NJP_2015}%
  \BibitemOpen
  \bibfield  {author} {\bibinfo {author} {\bibfnamefont {Tobias}\ \bibnamefont {Dornheim}}, \bibinfo {author} {\bibfnamefont {Simon}\ \bibnamefont {Groth}}, \bibinfo {author} {\bibfnamefont {Alexey}\ \bibnamefont {Filinov}}, \ and\ \bibinfo {author} {\bibfnamefont {Michael}\ \bibnamefont {Bonitz}},\ }\bibfield  {title} {\enquote {\bibinfo {title} {Permutation blocking path integral monte carlo: a highly efficient approach to the simulation of strongly degenerate non-ideal fermions},}\ }\href {\doibase 10.1088/1367-2630/17/7/073017} {\bibfield  {journal} {\bibinfo  {journal} {New Journal of Physics}\ }\textbf {\bibinfo {volume} {17}},\ \bibinfo {pages} {073017} (\bibinfo {year} {2015}{\natexlab{b}})}\BibitemShut {NoStop}%
\bibitem [{\citenamefont {Malone}\ \emph {et~al.}(2016)\citenamefont {Malone}, \citenamefont {Blunt}, \citenamefont {Brown}, \citenamefont {Lee}, \citenamefont {Spencer}, \citenamefont {Foulkes},\ and\ \citenamefont {Shepherd}}]{Malone_PRL_2016}%
  \BibitemOpen
  \bibfield  {author} {\bibinfo {author} {\bibfnamefont {Fionn~D.}\ \bibnamefont {Malone}}, \bibinfo {author} {\bibfnamefont {N.~S.}\ \bibnamefont {Blunt}}, \bibinfo {author} {\bibfnamefont {Ethan~W.}\ \bibnamefont {Brown}}, \bibinfo {author} {\bibfnamefont {D.~K.~K.}\ \bibnamefont {Lee}}, \bibinfo {author} {\bibfnamefont {J.~S.}\ \bibnamefont {Spencer}}, \bibinfo {author} {\bibfnamefont {W.~M.~C.}\ \bibnamefont {Foulkes}}, \ and\ \bibinfo {author} {\bibfnamefont {James~J.}\ \bibnamefont {Shepherd}},\ }\bibfield  {title} {\enquote {\bibinfo {title} {Accurate exchange-correlation energies for the warm dense electron gas},}\ }\href {\doibase 10.1103/PhysRevLett.117.115701} {\bibfield  {journal} {\bibinfo  {journal} {Phys. Rev. Lett.}\ }\textbf {\bibinfo {volume} {117}},\ \bibinfo {pages} {115701} (\bibinfo {year} {2016})}\BibitemShut {NoStop}%
\bibitem [{\citenamefont {Shen}\ \emph {et~al.}(2020)\citenamefont {Shen}, \citenamefont {Liu}, \citenamefont {Yu},\ and\ \citenamefont {Rubenstein}}]{Rubenstein_JCTC_2020}%
  \BibitemOpen
  \bibfield  {author} {\bibinfo {author} {\bibfnamefont {Tong}\ \bibnamefont {Shen}}, \bibinfo {author} {\bibfnamefont {Yuan}\ \bibnamefont {Liu}}, \bibinfo {author} {\bibfnamefont {Yang}\ \bibnamefont {Yu}}, \ and\ \bibinfo {author} {\bibfnamefont {Brenda~M.}\ \bibnamefont {Rubenstein}},\ }\bibfield  {title} {\enquote {\bibinfo {title} {Finite temperature auxiliary field quantum monte carlo in the canonical ensemble},}\ }\href {\doibase 10.1063/5.0026606} {\bibfield  {journal} {\bibinfo  {journal} {The Journal of Chemical Physics}\ }\textbf {\bibinfo {volume} {153}},\ \bibinfo {pages} {204108} (\bibinfo {year} {2020})}\BibitemShut {NoStop}%
\bibitem [{\citenamefont {Brown}\ \emph {et~al.}(2013{\natexlab{a}})\citenamefont {Brown}, \citenamefont {Clark}, \citenamefont {DuBois},\ and\ \citenamefont {Ceperley}}]{Brown_PRL_2013}%
  \BibitemOpen
  \bibfield  {author} {\bibinfo {author} {\bibfnamefont {Ethan~W.}\ \bibnamefont {Brown}}, \bibinfo {author} {\bibfnamefont {Bryan~K.}\ \bibnamefont {Clark}}, \bibinfo {author} {\bibfnamefont {Jonathan~L.}\ \bibnamefont {DuBois}}, \ and\ \bibinfo {author} {\bibfnamefont {David~M.}\ \bibnamefont {Ceperley}},\ }\bibfield  {title} {\enquote {\bibinfo {title} {Path-integral monte carlo simulation of the warm dense homogeneous electron gas},}\ }\href {\doibase 10.1103/PhysRevLett.110.146405} {\bibfield  {journal} {\bibinfo  {journal} {Phys. Rev. Lett.}\ }\textbf {\bibinfo {volume} {110}},\ \bibinfo {pages} {146405} (\bibinfo {year} {2013}{\natexlab{a}})}\BibitemShut {NoStop}%
\bibitem [{\citenamefont {Xiong}\ and\ \citenamefont {Xiong}(2022)}]{Xiong_JCP_2022}%
  \BibitemOpen
  \bibfield  {author} {\bibinfo {author} {\bibfnamefont {Yunuo}\ \bibnamefont {Xiong}}\ and\ \bibinfo {author} {\bibfnamefont {Hongwei}\ \bibnamefont {Xiong}},\ }\bibfield  {title} {\enquote {\bibinfo {title} {On the thermodynamic properties of fictitious identical particles and the application to fermion sign problem},}\ }\href {https://doi.org/10.1063/5.0106067} {\bibfield  {journal} {\bibinfo  {journal} {The Journal of Chemical Physics}\ }\textbf {\bibinfo {volume} {157}},\ \bibinfo {pages} {094112} (\bibinfo {year} {2022})}\BibitemShut {NoStop}%
\bibitem [{\citenamefont {Xiong}\ \emph {et~al.}(2024)\citenamefont {Xiong}, \citenamefont {Liu},\ and\ \citenamefont {Xiong}}]{Xiong_PRE_2024}%
  \BibitemOpen
  \bibfield  {author} {\bibinfo {author} {\bibfnamefont {Yunuo}\ \bibnamefont {Xiong}}, \bibinfo {author} {\bibfnamefont {Shujuan}\ \bibnamefont {Liu}}, \ and\ \bibinfo {author} {\bibfnamefont {Hongwei}\ \bibnamefont {Xiong}},\ }\bibfield  {title} {\enquote {\bibinfo {title} {Quadratic scaling path integral molecular dynamics for fictitious identical particles and its application to fermion systems},}\ }\href {\doibase 10.1103/PhysRevE.110.065303} {\bibfield  {journal} {\bibinfo  {journal} {Phys. Rev. E}\ }\textbf {\bibinfo {volume} {110}},\ \bibinfo {pages} {065303} (\bibinfo {year} {2024})}\BibitemShut {NoStop}%
\bibitem [{\citenamefont {Hirshberg}\ \emph {et~al.}(2020)\citenamefont {Hirshberg}, \citenamefont {Invernizzi},\ and\ \citenamefont {Parrinello}}]{Hirshberg_JCP_2020}%
  \BibitemOpen
  \bibfield  {author} {\bibinfo {author} {\bibfnamefont {Barak}\ \bibnamefont {Hirshberg}}, \bibinfo {author} {\bibfnamefont {Michele}\ \bibnamefont {Invernizzi}}, \ and\ \bibinfo {author} {\bibfnamefont {Michele}\ \bibnamefont {Parrinello}},\ }\bibfield  {title} {\enquote {\bibinfo {title} {Path integral molecular dynamics for fermions: Alleviating the sign problem with the bogoliubov inequality},}\ }\href {\doibase 10.1063/5.0008720} {\bibfield  {journal} {\bibinfo  {journal} {The Journal of Chemical Physics}\ }\textbf {\bibinfo {volume} {152}},\ \bibinfo {pages} {171102} (\bibinfo {year} {2020})}\BibitemShut {NoStop}%
\bibitem [{\citenamefont {Dornheim}\ \emph {et~al.}(2020{\natexlab{a}})\citenamefont {Dornheim}, \citenamefont {Invernizzi}, \citenamefont {Vorberger},\ and\ \citenamefont {Hirshberg}}]{Dornheim_JCP_2020}%
  \BibitemOpen
  \bibfield  {author} {\bibinfo {author} {\bibfnamefont {Tobias}\ \bibnamefont {Dornheim}}, \bibinfo {author} {\bibfnamefont {Michele}\ \bibnamefont {Invernizzi}}, \bibinfo {author} {\bibfnamefont {Jan}\ \bibnamefont {Vorberger}}, \ and\ \bibinfo {author} {\bibfnamefont {Barak}\ \bibnamefont {Hirshberg}},\ }\bibfield  {title} {\enquote {\bibinfo {title} {{Attenuating the fermion sign problem in path integral Monte Carlo simulations using the Bogoliubov inequality and thermodynamic integration}},}\ }\href {\doibase 10.1063/5.0030760} {\bibfield  {journal} {\bibinfo  {journal} {The Journal of Chemical Physics}\ }\textbf {\bibinfo {volume} {153}},\ \bibinfo {pages} {234104} (\bibinfo {year} {2020}{\natexlab{a}})}\BibitemShut {NoStop}%
\bibitem [{\citenamefont {Dornheim}\ \emph {et~al.}(2023{\natexlab{b}})\citenamefont {Dornheim}, \citenamefont {Tolias}, \citenamefont {Groth}, \citenamefont {Moldabekov}, \citenamefont {Vorberger},\ and\ \citenamefont {Hirshberg}}]{Dornheim_JCP_xi_2023}%
  \BibitemOpen
  \bibfield  {author} {\bibinfo {author} {\bibfnamefont {Tobias}\ \bibnamefont {Dornheim}}, \bibinfo {author} {\bibfnamefont {Panagiotis}\ \bibnamefont {Tolias}}, \bibinfo {author} {\bibfnamefont {Simon}\ \bibnamefont {Groth}}, \bibinfo {author} {\bibfnamefont {Zhandos~A.}\ \bibnamefont {Moldabekov}}, \bibinfo {author} {\bibfnamefont {Jan}\ \bibnamefont {Vorberger}}, \ and\ \bibinfo {author} {\bibfnamefont {Barak}\ \bibnamefont {Hirshberg}},\ }\bibfield  {title} {\enquote {\bibinfo {title} {{Fermionic physics from ab initio path integral Monte Carlo simulations of fictitious identical particles}},}\ }\href {\doibase 10.1063/5.0171930} {\bibfield  {journal} {\bibinfo  {journal} {The Journal of Chemical Physics}\ }\textbf {\bibinfo {volume} {159}},\ \bibinfo {pages} {164113} (\bibinfo {year} {2023}{\natexlab{b}})}\BibitemShut {NoStop}%
\bibitem [{\citenamefont {Yilmaz}\ \emph {et~al.}(2020)\citenamefont {Yilmaz}, \citenamefont {Hunger}, \citenamefont {Dornheim}, \citenamefont {Groth},\ and\ \citenamefont {Bonitz}}]{Yilmaz_JCP_2020}%
  \BibitemOpen
  \bibfield  {author} {\bibinfo {author} {\bibfnamefont {A.}~\bibnamefont {Yilmaz}}, \bibinfo {author} {\bibfnamefont {K.}~\bibnamefont {Hunger}}, \bibinfo {author} {\bibfnamefont {T.}~\bibnamefont {Dornheim}}, \bibinfo {author} {\bibfnamefont {S.}~\bibnamefont {Groth}}, \ and\ \bibinfo {author} {\bibfnamefont {M.}~\bibnamefont {Bonitz}},\ }\bibfield  {title} {\enquote {\bibinfo {title} {Restricted configuration path integral monte carlo},}\ }\href {\doibase 10.1063/5.0022800} {\bibfield  {journal} {\bibinfo  {journal} {The Journal of Chemical Physics}\ }\textbf {\bibinfo {volume} {153}},\ \bibinfo {pages} {124114} (\bibinfo {year} {2020})}\BibitemShut {NoStop}%
\bibitem [{\citenamefont {Dornheim}\ \emph {et~al.}(2024{\natexlab{a}})\citenamefont {Dornheim}, \citenamefont {Schwalbe}, \citenamefont {Böhme}, \citenamefont {Moldabekov}, \citenamefont {Vorberger},\ and\ \citenamefont {Tolias}}]{Dornheim_JCP_2024}%
  \BibitemOpen
  \bibfield  {author} {\bibinfo {author} {\bibfnamefont {Tobias}\ \bibnamefont {Dornheim}}, \bibinfo {author} {\bibfnamefont {Sebastian}\ \bibnamefont {Schwalbe}}, \bibinfo {author} {\bibfnamefont {Maximilian~P.}\ \bibnamefont {Böhme}}, \bibinfo {author} {\bibfnamefont {Zhandos~A.}\ \bibnamefont {Moldabekov}}, \bibinfo {author} {\bibfnamefont {Jan}\ \bibnamefont {Vorberger}}, \ and\ \bibinfo {author} {\bibfnamefont {Panagiotis}\ \bibnamefont {Tolias}},\ }\bibfield  {title} {\enquote {\bibinfo {title} {{Ab initio path integral Monte Carlo simulations of warm dense two-component systems without fixed nodes: Structural properties}},}\ }\href {\doibase 10.1063/5.0206787} {\bibfield  {journal} {\bibinfo  {journal} {J. Chem. Phys.}\ }\textbf {\bibinfo {volume} {160}},\ \bibinfo {pages} {164111} (\bibinfo {year} {2024}{\natexlab{a}})}\BibitemShut {NoStop}%
\bibitem [{\citenamefont {Dornheim}\ \emph {et~al.}(2025{\natexlab{a}})\citenamefont {Dornheim}, \citenamefont {D{\"o}ppner}, \citenamefont {Tolias}, \citenamefont {B{\"o}hme}, \citenamefont {Fletcher}, \citenamefont {Gawne}, \citenamefont {Graziani}, \citenamefont {Kraus}, \citenamefont {MacDonald}, \citenamefont {Moldabekov}, \citenamefont {Schwalbe}, \citenamefont {Gericke},\ and\ \citenamefont {Vorberger}}]{Dornheim_NatComm_2025}%
  \BibitemOpen
  \bibfield  {author} {\bibinfo {author} {\bibfnamefont {Tobias}\ \bibnamefont {Dornheim}}, \bibinfo {author} {\bibfnamefont {Tilo}\ \bibnamefont {D{\"o}ppner}}, \bibinfo {author} {\bibfnamefont {Panagiotis}\ \bibnamefont {Tolias}}, \bibinfo {author} {\bibfnamefont {Maximilian~P.}\ \bibnamefont {B{\"o}hme}}, \bibinfo {author} {\bibfnamefont {Luke~B.}\ \bibnamefont {Fletcher}}, \bibinfo {author} {\bibfnamefont {Thomas}\ \bibnamefont {Gawne}}, \bibinfo {author} {\bibfnamefont {Frank~R.}\ \bibnamefont {Graziani}}, \bibinfo {author} {\bibfnamefont {Dominik}\ \bibnamefont {Kraus}}, \bibinfo {author} {\bibfnamefont {Michael~J.}\ \bibnamefont {MacDonald}}, \bibinfo {author} {\bibfnamefont {Zhandos~A.}\ \bibnamefont {Moldabekov}}, \bibinfo {author} {\bibfnamefont {Sebastian}\ \bibnamefont {Schwalbe}}, \bibinfo {author} {\bibfnamefont {Dirk~O.}\ \bibnamefont {Gericke}}, \ and\ \bibinfo {author} {\bibfnamefont {Jan}\ \bibnamefont {Vorberger}},\ }\bibfield  {title} {\enquote {\bibinfo {title} {Unraveling electronic
  correlations in warm dense quantum plasmas},}\ }\href {\doibase 10.1038/s41467-025-60278-3} {\bibfield  {journal} {\bibinfo  {journal} {Nature Communications}\ }\textbf {\bibinfo {volume} {16}},\ \bibinfo {pages} {5103} (\bibinfo {year} {2025}{\natexlab{a}})}\BibitemShut {NoStop}%
\bibitem [{\citenamefont {Filinov}\ \emph {et~al.}(2020)\citenamefont {Filinov}, \citenamefont {Larkin},\ and\ \citenamefont {Levashov}}]{Filinov_PRE_2020}%
  \BibitemOpen
  \bibfield  {author} {\bibinfo {author} {\bibfnamefont {V.~S.}\ \bibnamefont {Filinov}}, \bibinfo {author} {\bibfnamefont {A.~S.}\ \bibnamefont {Larkin}}, \ and\ \bibinfo {author} {\bibfnamefont {P.~R.}\ \bibnamefont {Levashov}},\ }\bibfield  {title} {\enquote {\bibinfo {title} {Uniform electron gas at finite temperature by fermionic-path-integral monte carlo simulations},}\ }\href {\doibase 10.1103/PhysRevE.102.033203} {\bibfield  {journal} {\bibinfo  {journal} {Phys. Rev. E}\ }\textbf {\bibinfo {volume} {102}},\ \bibinfo {pages} {033203} (\bibinfo {year} {2020})}\BibitemShut {NoStop}%
\bibitem [{\citenamefont {Filinov}\ \emph {et~al.}(2025)\citenamefont {Filinov}, \citenamefont {Levashov},\ and\ \citenamefont {Larkin}}]{FILINOV2025130542}%
  \BibitemOpen
  \bibfield  {author} {\bibinfo {author} {\bibfnamefont {Vladimir}\ \bibnamefont {Filinov}}, \bibinfo {author} {\bibfnamefont {Pavel}\ \bibnamefont {Levashov}}, \ and\ \bibinfo {author} {\bibfnamefont {Alexander}\ \bibnamefont {Larkin}},\ }\bibfield  {title} {\enquote {\bibinfo {title} {Density response and correlation functions in the wigner path integral representation. monte carlo simulations},}\ }\href {\doibase https://doi.org/10.1016/j.physleta.2025.130542} {\bibfield  {journal} {\bibinfo  {journal} {Physics Letters A}\ }\textbf {\bibinfo {volume} {548}},\ \bibinfo {pages} {130542} (\bibinfo {year} {2025})}\BibitemShut {NoStop}%
\bibitem [{\citenamefont {Xiong}\ and\ \citenamefont {Xiong}(2025)}]{Xiong:2025dwx}%
  \BibitemOpen
  \bibfield  {author} {\bibinfo {author} {\bibfnamefont {Yunuo}\ \bibnamefont {Xiong}}\ and\ \bibinfo {author} {\bibfnamefont {Hongwei}\ \bibnamefont {Xiong}},\ }\bibfield  {title} {\enquote {\bibinfo {title} {{A Pseudo-Fermion Propagator Approach to the Fermion Sign Problem}},}\ }\href@noop {} {\  (\bibinfo {year} {2025})},\ \Eprint {http://arxiv.org/abs/2508.09557} {arXiv:2508.09557 [physics.comp-ph]} \BibitemShut {NoStop}%
\bibitem [{\citenamefont {Dornheim}\ \emph {et~al.}(2025{\natexlab{b}})\citenamefont {Dornheim}, \citenamefont {Svensson}, \citenamefont {Hamann}, \citenamefont {Schwalbe}, \citenamefont {Moldabekov}, \citenamefont {Tolias},\ and\ \citenamefont {Vorberger}}]{dornheim2025reweightingestimatorabinitio}%
  \BibitemOpen
  \bibfield  {author} {\bibinfo {author} {\bibfnamefont {Tobias}\ \bibnamefont {Dornheim}}, \bibinfo {author} {\bibfnamefont {Pontus}\ \bibnamefont {Svensson}}, \bibinfo {author} {\bibfnamefont {Paul}\ \bibnamefont {Hamann}}, \bibinfo {author} {\bibfnamefont {Sebastian}\ \bibnamefont {Schwalbe}}, \bibinfo {author} {\bibfnamefont {Zhandos}\ \bibnamefont {Moldabekov}}, \bibinfo {author} {\bibfnamefont {Panagiotis}\ \bibnamefont {Tolias}}, \ and\ \bibinfo {author} {\bibfnamefont {Jan}\ \bibnamefont {Vorberger}},\ }\href {https://arxiv.org/abs/2508.12323} {\enquote {\bibinfo {title} {Re-weighting estimator for ab initio path integral monte carlo simulations of fictitious identical particles},}\ } (\bibinfo {year} {2025}{\natexlab{b}}),\ \Eprint {http://arxiv.org/abs/2508.12323} {arXiv:2508.12323 [physics.chem-ph]} \BibitemShut {NoStop}%
\bibitem [{\citenamefont {Xiong}\ and\ \citenamefont {Xiong}(2023)}]{Xiong_PRE_2023}%
  \BibitemOpen
  \bibfield  {author} {\bibinfo {author} {\bibfnamefont {Yunuo}\ \bibnamefont {Xiong}}\ and\ \bibinfo {author} {\bibfnamefont {Hongwei}\ \bibnamefont {Xiong}},\ }\bibfield  {title} {\enquote {\bibinfo {title} {Thermodynamics of fermions at any temperature based on parametrized partition function},}\ }\href {\doibase 10.1103/PhysRevE.107.055308} {\bibfield  {journal} {\bibinfo  {journal} {Phys. Rev. E}\ }\textbf {\bibinfo {volume} {107}},\ \bibinfo {pages} {055308} (\bibinfo {year} {2023})}\BibitemShut {NoStop}%
\bibitem [{\citenamefont {Dornheim}\ \emph {et~al.}(2024{\natexlab{b}})\citenamefont {Dornheim}, \citenamefont {Schwalbe}, \citenamefont {Moldabekov}, \citenamefont {Vorberger},\ and\ \citenamefont {Tolias}}]{Dornheim_JPCL_2024}%
  \BibitemOpen
  \bibfield  {author} {\bibinfo {author} {\bibfnamefont {Tobias}\ \bibnamefont {Dornheim}}, \bibinfo {author} {\bibfnamefont {Sebastian}\ \bibnamefont {Schwalbe}}, \bibinfo {author} {\bibfnamefont {Zhandos~A.}\ \bibnamefont {Moldabekov}}, \bibinfo {author} {\bibfnamefont {Jan}\ \bibnamefont {Vorberger}}, \ and\ \bibinfo {author} {\bibfnamefont {Panagiotis}\ \bibnamefont {Tolias}},\ }\bibfield  {title} {\enquote {\bibinfo {title} {Ab initio path integral {Monte Carlo} simulations of the uniform electron gas on large length scales},}\ }\href {\doibase 10.1021/acs.jpclett.3c03193} {\bibfield  {journal} {\bibinfo  {journal} {J. Phys. Chem. Lett.}\ }\textbf {\bibinfo {volume} {15}},\ \bibinfo {pages} {1305--1313} (\bibinfo {year} {2024}{\natexlab{b}})}\BibitemShut {NoStop}%
\bibitem [{\citenamefont {Dornheim}\ \emph {et~al.}(2024{\natexlab{c}})\citenamefont {Dornheim}, \citenamefont {Schwalbe}, \citenamefont {Tolias}, \citenamefont {Böhme}, \citenamefont {Moldabekov},\ and\ \citenamefont {Vorberger}}]{Dornheim_MRE_2024}%
  \BibitemOpen
  \bibfield  {author} {\bibinfo {author} {\bibfnamefont {Tobias}\ \bibnamefont {Dornheim}}, \bibinfo {author} {\bibfnamefont {Sebastian}\ \bibnamefont {Schwalbe}}, \bibinfo {author} {\bibfnamefont {Panagiotis}\ \bibnamefont {Tolias}}, \bibinfo {author} {\bibfnamefont {Maximilian~P.}\ \bibnamefont {Böhme}}, \bibinfo {author} {\bibfnamefont {Zhandos~A.}\ \bibnamefont {Moldabekov}}, \ and\ \bibinfo {author} {\bibfnamefont {Jan}\ \bibnamefont {Vorberger}},\ }\bibfield  {title} {\enquote {\bibinfo {title} {{Ab initio density response and local field factor of warm dense hydrogen}},}\ }\href {\doibase 10.1063/5.0211407} {\bibfield  {journal} {\bibinfo  {journal} {Matter Radiat. Extrem.}\ }\textbf {\bibinfo {volume} {9}},\ \bibinfo {pages} {057401} (\bibinfo {year} {2024}{\natexlab{c}})}\BibitemShut {NoStop}%
\bibitem [{\citenamefont {Schwalbe}\ \emph {et~al.}(2025)\citenamefont {Schwalbe}, \citenamefont {Bellenbaum}, \citenamefont {Döppner}, \citenamefont {Böhme}, \citenamefont {Gawne}, \citenamefont {Kraus}, \citenamefont {MacDonald}, \citenamefont {Moldabekov}, \citenamefont {Tolias}, \citenamefont {Vorberger},\ and\ \citenamefont {Dornheim}}]{schwalbe2025staticlineardensityresponse}%
  \BibitemOpen
  \bibfield  {author} {\bibinfo {author} {\bibfnamefont {Sebastian}\ \bibnamefont {Schwalbe}}, \bibinfo {author} {\bibfnamefont {Hannah}\ \bibnamefont {Bellenbaum}}, \bibinfo {author} {\bibfnamefont {Tilo}\ \bibnamefont {Döppner}}, \bibinfo {author} {\bibfnamefont {Maximilian}\ \bibnamefont {Böhme}}, \bibinfo {author} {\bibfnamefont {Thomas}\ \bibnamefont {Gawne}}, \bibinfo {author} {\bibfnamefont {Dominik}\ \bibnamefont {Kraus}}, \bibinfo {author} {\bibfnamefont {Michael~J.}\ \bibnamefont {MacDonald}}, \bibinfo {author} {\bibfnamefont {Zhandos}\ \bibnamefont {Moldabekov}}, \bibinfo {author} {\bibfnamefont {Panagiotis}\ \bibnamefont {Tolias}}, \bibinfo {author} {\bibfnamefont {Jan}\ \bibnamefont {Vorberger}}, \ and\ \bibinfo {author} {\bibfnamefont {Tobias}\ \bibnamefont {Dornheim}},\ }\href {https://arxiv.org/abs/2504.13611} {\enquote {\bibinfo {title} {Static linear density response from x-ray thomson scattering measurements: a case study of warm dense beryllium},}\ } (\bibinfo {year} {2025}),\ \Eprint
  {http://arxiv.org/abs/2504.13611} {arXiv:2504.13611 [physics.plasm-ph]} \BibitemShut {NoStop}%
\bibitem [{\citenamefont {Dornheim}\ \emph {et~al.}(2025{\natexlab{c}})\citenamefont {Dornheim}, \citenamefont {Moldabekov}, \citenamefont {Schwalbe}, \citenamefont {Tolias},\ and\ \citenamefont {Vorberger}}]{Dornheim_JCTC_2025}%
  \BibitemOpen
  \bibfield  {author} {\bibinfo {author} {\bibfnamefont {Tobias}\ \bibnamefont {Dornheim}}, \bibinfo {author} {\bibfnamefont {Zhandos}\ \bibnamefont {Moldabekov}}, \bibinfo {author} {\bibfnamefont {Sebastian}\ \bibnamefont {Schwalbe}}, \bibinfo {author} {\bibfnamefont {Panagiotis}\ \bibnamefont {Tolias}}, \ and\ \bibinfo {author} {\bibfnamefont {Jan}\ \bibnamefont {Vorberger}},\ }\bibfield  {title} {\enquote {\bibinfo {title} {Fermionic free energies from ab initio path integral monte carlo simulations of fictitious identical particles},}\ }\href {\doibase 10.1021/acs.jctc.5c00301} {\bibfield  {journal} {\bibinfo  {journal} {Journal of Chemical Theory and Computation}\ }\textbf {\bibinfo {volume} {21}},\ \bibinfo {pages} {7290--7303} (\bibinfo {year} {2025}{\natexlab{c}})}\BibitemShut {NoStop}%
\bibitem [{\citenamefont {Svensson}\ \emph {et~al.}(2025)\citenamefont {Svensson}, \citenamefont {Kalkavouras}, \citenamefont {Acosta}, \citenamefont {Moldabekov}, \citenamefont {Tolias}, \citenamefont {Vorberger},\ and\ \citenamefont {Dornheim}}]{svensson2025acceleratedfreeenergyestimation}%
  \BibitemOpen
  \bibfield  {author} {\bibinfo {author} {\bibfnamefont {Pontus}\ \bibnamefont {Svensson}}, \bibinfo {author} {\bibfnamefont {Fotios}\ \bibnamefont {Kalkavouras}}, \bibinfo {author} {\bibfnamefont {Uwe~Hernandez}\ \bibnamefont {Acosta}}, \bibinfo {author} {\bibfnamefont {Zhandos~A.}\ \bibnamefont {Moldabekov}}, \bibinfo {author} {\bibfnamefont {Panagiotis}\ \bibnamefont {Tolias}}, \bibinfo {author} {\bibfnamefont {Jan}\ \bibnamefont {Vorberger}}, \ and\ \bibinfo {author} {\bibfnamefont {Tobias}\ \bibnamefont {Dornheim}},\ }\href {https://arxiv.org/abs/2507.12960} {\enquote {\bibinfo {title} {Accelerated free energy estimation in ab initio path integral monte carlo simulations},}\ } (\bibinfo {year} {2025}),\ \Eprint {http://arxiv.org/abs/2507.12960} {arXiv:2507.12960 [physics.chem-ph]} \BibitemShut {NoStop}%
\bibitem [{\citenamefont {Morresi}\ \emph {et~al.}(2025)\citenamefont {Morresi}, \citenamefont {Garberoglio}, \citenamefont {Xiong},\ and\ \citenamefont {Xiong}}]{morresi2025studyuniformelectrongas}%
  \BibitemOpen
  \bibfield  {author} {\bibinfo {author} {\bibfnamefont {Tommaso}\ \bibnamefont {Morresi}}, \bibinfo {author} {\bibfnamefont {Giovanni}\ \bibnamefont {Garberoglio}}, \bibinfo {author} {\bibfnamefont {Hongwei}\ \bibnamefont {Xiong}}, \ and\ \bibinfo {author} {\bibfnamefont {Yunuo}\ \bibnamefont {Xiong}},\ }\href {https://arxiv.org/abs/2506.10113} {\enquote {\bibinfo {title} {Study of the uniform electron gas through parametrized partition functions},}\ } (\bibinfo {year} {2025}),\ \Eprint {http://arxiv.org/abs/2506.10113} {arXiv:2506.10113 [cond-mat.mtrl-sci]} \BibitemShut {NoStop}%
\bibitem [{\citenamefont {Morresi}\ and\ \citenamefont {Garberoglio}(2025)}]{Morresi_PRB_2025}%
  \BibitemOpen
  \bibfield  {author} {\bibinfo {author} {\bibfnamefont {Tommaso}\ \bibnamefont {Morresi}}\ and\ \bibinfo {author} {\bibfnamefont {Giovanni}\ \bibnamefont {Garberoglio}},\ }\bibfield  {title} {\enquote {\bibinfo {title} {Normal liquid $^{3}\mathrm{He}$ studied by path-integral monte carlo with a parametrized partition function},}\ }\href {\doibase 10.1103/PhysRevB.111.014521} {\bibfield  {journal} {\bibinfo  {journal} {Phys. Rev. B}\ }\textbf {\bibinfo {volume} {111}},\ \bibinfo {pages} {014521} (\bibinfo {year} {2025})}\BibitemShut {NoStop}%
\bibitem [{\citenamefont {Yang}\ \emph {et~al.}(2025)\citenamefont {Yang}, \citenamefont {Yu}, \citenamefont {Liu},\ and\ \citenamefont {Zhu}}]{Yang_Entropy_2025}%
  \BibitemOpen
  \bibfield  {author} {\bibinfo {author} {\bibfnamefont {Bo}~\bibnamefont {Yang}}, \bibinfo {author} {\bibfnamefont {Hongsheng}\ \bibnamefont {Yu}}, \bibinfo {author} {\bibfnamefont {Shujuan}\ \bibnamefont {Liu}}, \ and\ \bibinfo {author} {\bibfnamefont {Fengzheng}\ \bibnamefont {Zhu}},\ }\bibfield  {title} {\enquote {\bibinfo {title} {Density distribution of strongly quantum degenerate fermi systems simulated by fictitious identical particle thermodynamics},}\ }\href {\doibase 10.3390/e27050458} {\bibfield  {journal} {\bibinfo  {journal} {Entropy}\ }\textbf {\bibinfo {volume} {27}} (\bibinfo {year} {2025}),\ 10.3390/e27050458}\BibitemShut {NoStop}%
\bibitem [{\citenamefont {Dornheim}\ \emph {et~al.}(2025{\natexlab{d}})\citenamefont {Dornheim}, \citenamefont {Bellenbaum}, \citenamefont {Bethkenhagen}, \citenamefont {Hansen}, \citenamefont {Böhme}, \citenamefont {Döppner}, \citenamefont {Fletcher}, \citenamefont {Gawne}, \citenamefont {Gericke}, \citenamefont {Hamel}, \citenamefont {Kraus}, \citenamefont {MacDonald}, \citenamefont {Moldabekov}, \citenamefont {Preston}, \citenamefont {Redmer}, \citenamefont {Schörner}, \citenamefont {Schwalbe}, \citenamefont {Tolias},\ and\ \citenamefont {Vorberger}}]{Dornheim_POP_2025}%
  \BibitemOpen
  \bibfield  {author} {\bibinfo {author} {\bibfnamefont {T.}~\bibnamefont {Dornheim}}, \bibinfo {author} {\bibfnamefont {H.~M.}\ \bibnamefont {Bellenbaum}}, \bibinfo {author} {\bibfnamefont {M.}~\bibnamefont {Bethkenhagen}}, \bibinfo {author} {\bibfnamefont {S.~B.}\ \bibnamefont {Hansen}}, \bibinfo {author} {\bibfnamefont {M.~P.}\ \bibnamefont {Böhme}}, \bibinfo {author} {\bibfnamefont {T.}~\bibnamefont {Döppner}}, \bibinfo {author} {\bibfnamefont {L.~B.}\ \bibnamefont {Fletcher}}, \bibinfo {author} {\bibfnamefont {T.}~\bibnamefont {Gawne}}, \bibinfo {author} {\bibfnamefont {D.~O.}\ \bibnamefont {Gericke}}, \bibinfo {author} {\bibfnamefont {S.}~\bibnamefont {Hamel}}, \bibinfo {author} {\bibfnamefont {D.}~\bibnamefont {Kraus}}, \bibinfo {author} {\bibfnamefont {M.~J.}\ \bibnamefont {MacDonald}}, \bibinfo {author} {\bibfnamefont {Zh.~A.}\ \bibnamefont {Moldabekov}}, \bibinfo {author} {\bibfnamefont {T.~R.}\ \bibnamefont {Preston}}, \bibinfo {author} {\bibfnamefont {R.}~\bibnamefont {Redmer}}, \bibinfo
  {author} {\bibfnamefont {M.}~\bibnamefont {Schörner}}, \bibinfo {author} {\bibfnamefont {S.}~\bibnamefont {Schwalbe}}, \bibinfo {author} {\bibfnamefont {P.}~\bibnamefont {Tolias}}, \ and\ \bibinfo {author} {\bibfnamefont {J.}~\bibnamefont {Vorberger}},\ }\bibfield  {title} {\enquote {\bibinfo {title} {Model-free rayleigh weight from x-ray thomson scattering measurements},}\ }\href {\doibase 10.1063/5.0238630} {\bibfield  {journal} {\bibinfo  {journal} {Physics of Plasmas}\ }\textbf {\bibinfo {volume} {32}},\ \bibinfo {pages} {052712} (\bibinfo {year} {2025}{\natexlab{d}})}\BibitemShut {NoStop}%
\bibitem [{\citenamefont {Dornheim}\ \emph {et~al.}(2021{\natexlab{a}})\citenamefont {Dornheim}, \citenamefont {Moldabekov},\ and\ \citenamefont {Vorberger}}]{Dornheim_JCP_ITCF_2021}%
  \BibitemOpen
  \bibfield  {author} {\bibinfo {author} {\bibfnamefont {Tobias}\ \bibnamefont {Dornheim}}, \bibinfo {author} {\bibfnamefont {Zhandos~A.}\ \bibnamefont {Moldabekov}}, \ and\ \bibinfo {author} {\bibfnamefont {Jan}\ \bibnamefont {Vorberger}},\ }\bibfield  {title} {\enquote {\bibinfo {title} {Nonlinear density response from imaginary-time correlation functions: Ab initio path integral monte carlo simulations of the warm dense electron gas},}\ }\href {\doibase 10.1063/5.0058988} {\bibfield  {journal} {\bibinfo  {journal} {The Journal of Chemical Physics}\ }\textbf {\bibinfo {volume} {155}},\ \bibinfo {pages} {054110} (\bibinfo {year} {2021}{\natexlab{a}})}\BibitemShut {NoStop}%
\bibitem [{\citenamefont {Boninsegni}\ and\ \citenamefont {Ceperley}(1996)}]{Boninsegni1996}%
  \BibitemOpen
  \bibfield  {author} {\bibinfo {author} {\bibfnamefont {Massimo}\ \bibnamefont {Boninsegni}}\ and\ \bibinfo {author} {\bibfnamefont {David~M.}\ \bibnamefont {Ceperley}},\ }\bibfield  {title} {\enquote {\bibinfo {title} {{Density fluctuations in liquid $^4$He. Path integrals and maximum entropy}},}\ }\href {\doibase 10.1007/BF00751861} {\bibfield  {journal} {\bibinfo  {journal} {J. Low Temp. Phys.}\ }\textbf {\bibinfo {volume} {104}},\ \bibinfo {pages} {339--357} (\bibinfo {year} {1996})}\BibitemShut {NoStop}%
\bibitem [{\citenamefont {Thirumalai}\ and\ \citenamefont {Berne}(1983)}]{Berne_JCP_1983}%
  \BibitemOpen
  \bibfield  {author} {\bibinfo {author} {\bibfnamefont {Devarajan}\ \bibnamefont {Thirumalai}}\ and\ \bibinfo {author} {\bibfnamefont {Bruce~J.}\ \bibnamefont {Berne}},\ }\bibfield  {title} {\enquote {\bibinfo {title} {On the calculation of time correlation functions in quantum systems: Path integral techniquesa)},}\ }\href {\doibase 10.1063/1.445597} {\bibfield  {journal} {\bibinfo  {journal} {The Journal of Chemical Physics}\ }\textbf {\bibinfo {volume} {79}},\ \bibinfo {pages} {5029--5033} (\bibinfo {year} {1983})}\BibitemShut {NoStop}%
\bibitem [{\citenamefont {Dornheim}\ \emph {et~al.}(2023{\natexlab{c}})\citenamefont {Dornheim}, \citenamefont {Moldabekov}, \citenamefont {Tolias}, \citenamefont {Böhme},\ and\ \citenamefont {Vorberger}}]{Dornheim_MRE_2023}%
  \BibitemOpen
  \bibfield  {author} {\bibinfo {author} {\bibfnamefont {Tobias}\ \bibnamefont {Dornheim}}, \bibinfo {author} {\bibfnamefont {Zhandos}\ \bibnamefont {Moldabekov}}, \bibinfo {author} {\bibfnamefont {Panagiotis}\ \bibnamefont {Tolias}}, \bibinfo {author} {\bibfnamefont {Maximilian}\ \bibnamefont {Böhme}}, \ and\ \bibinfo {author} {\bibfnamefont {Jan}\ \bibnamefont {Vorberger}},\ }\bibfield  {title} {\enquote {\bibinfo {title} {Physical insights from imaginary-time density--density correlation functions},}\ }\href {\doibase 10.1063/5.0149638} {\bibfield  {journal} {\bibinfo  {journal} {Matter and Radiation at Extremes}\ }\textbf {\bibinfo {volume} {8}},\ \bibinfo {pages} {056601} (\bibinfo {year} {2023}{\natexlab{c}})}\BibitemShut {NoStop}%
\bibitem [{\citenamefont {Dornheim}\ \emph {et~al.}(2023{\natexlab{d}})\citenamefont {Dornheim}, \citenamefont {Vorberger}, \citenamefont {Moldabekov},\ and\ \citenamefont {Böhme}}]{Dornheim_PTR_2023}%
  \BibitemOpen
  \bibfield  {author} {\bibinfo {author} {\bibfnamefont {Tobias}\ \bibnamefont {Dornheim}}, \bibinfo {author} {\bibfnamefont {Jan}\ \bibnamefont {Vorberger}}, \bibinfo {author} {\bibfnamefont {Zhandos~A.}\ \bibnamefont {Moldabekov}}, \ and\ \bibinfo {author} {\bibfnamefont {Maximilian}\ \bibnamefont {Böhme}},\ }\bibfield  {title} {\enquote {\bibinfo {title} {Analysing the dynamic structure of warm dense matter in the imaginary-time domain: theoretical models and simulations},}\ }\href {\doibase 10.1098/rsta.2022.0217} {\bibfield  {journal} {\bibinfo  {journal} {Philosophical Transactions of the Royal Society A: Mathematical, Physical and Engineering Sciences}\ }\textbf {\bibinfo {volume} {381}},\ \bibinfo {pages} {20220217} (\bibinfo {year} {2023}{\natexlab{d}})}\BibitemShut {NoStop}%
\bibitem [{\citenamefont {Ceperley}(1991)}]{Ceperley1991}%
  \BibitemOpen
  \bibfield  {author} {\bibinfo {author} {\bibfnamefont {D.~M.}\ \bibnamefont {Ceperley}},\ }\bibfield  {title} {\enquote {\bibinfo {title} {Fermion nodes},}\ }\href {\doibase 10.1007/BF01030009} {\bibfield  {journal} {\bibinfo  {journal} {Journal of Statistical Physics}\ }\textbf {\bibinfo {volume} {63}},\ \bibinfo {pages} {1237--1267} (\bibinfo {year} {1991})}\BibitemShut {NoStop}%
\bibitem [{\citenamefont {Dornheim}\ \emph {et~al.}(2018{\natexlab{b}})\citenamefont {Dornheim}, \citenamefont {Groth},\ and\ \citenamefont {Bonitz}}]{review}%
  \BibitemOpen
  \bibfield  {author} {\bibinfo {author} {\bibfnamefont {T.}~\bibnamefont {Dornheim}}, \bibinfo {author} {\bibfnamefont {S.}~\bibnamefont {Groth}}, \ and\ \bibinfo {author} {\bibfnamefont {M.}~\bibnamefont {Bonitz}},\ }\bibfield  {title} {\enquote {\bibinfo {title} {The uniform electron gas at warm dense matter conditions},}\ }\href {https://www.sciencedirect.com/science/article/abs/pii/S0370157318300516} {\bibfield  {journal} {\bibinfo  {journal} {Phys. Reports}\ }\textbf {\bibinfo {volume} {744}},\ \bibinfo {pages} {1--86} (\bibinfo {year} {2018}{\natexlab{b}})}\BibitemShut {NoStop}%
\bibitem [{\citenamefont {Loos}\ and\ \citenamefont {Gill}(2016)}]{loos}%
  \BibitemOpen
  \bibfield  {author} {\bibinfo {author} {\bibfnamefont {P.-F.}\ \bibnamefont {Loos}}\ and\ \bibinfo {author} {\bibfnamefont {P.~M.~W.}\ \bibnamefont {Gill}},\ }\bibfield  {title} {\enquote {\bibinfo {title} {The uniform electron gas},}\ }\href {http://onlinelibrary.wiley.com/doi/10.1002/wcms.1257/abstract} {\bibfield  {journal} {\bibinfo  {journal} {Comput. Mol. Sci}\ }\textbf {\bibinfo {volume} {6}},\ \bibinfo {pages} {410--429} (\bibinfo {year} {2016})}\BibitemShut {NoStop}%
\bibitem [{\citenamefont {Giuliani}\ and\ \citenamefont {Vignale}(2008)}]{quantum_theory}%
  \BibitemOpen
  \bibfield  {author} {\bibinfo {author} {\bibfnamefont {G.}~\bibnamefont {Giuliani}}\ and\ \bibinfo {author} {\bibfnamefont {G.}~\bibnamefont {Vignale}},\ }\href@noop {} {\emph {\bibinfo {title} {Quantum Theory of the Electron Liquid}}}\ (\bibinfo  {publisher} {Cambridge University Press},\ \bibinfo {address} {Cambridge},\ \bibinfo {year} {2008})\BibitemShut {NoStop}%
\bibitem [{\citenamefont {Brown}\ \emph {et~al.}(2013{\natexlab{b}})\citenamefont {Brown}, \citenamefont {DuBois}, \citenamefont {Holzmann},\ and\ \citenamefont {Ceperley}}]{Brown_PRB_2013}%
  \BibitemOpen
  \bibfield  {author} {\bibinfo {author} {\bibfnamefont {Ethan~W.}\ \bibnamefont {Brown}}, \bibinfo {author} {\bibfnamefont {Jonathan~L.}\ \bibnamefont {DuBois}}, \bibinfo {author} {\bibfnamefont {Markus}\ \bibnamefont {Holzmann}}, \ and\ \bibinfo {author} {\bibfnamefont {David~M.}\ \bibnamefont {Ceperley}},\ }\bibfield  {title} {\enquote {\bibinfo {title} {Exchange-correlation energy for the three-dimensional homogeneous electron gas at arbitrary temperature},}\ }\href {\doibase 10.1103/PhysRevB.88.081102} {\bibfield  {journal} {\bibinfo  {journal} {Phys. Rev. B}\ }\textbf {\bibinfo {volume} {88}},\ \bibinfo {pages} {081102} (\bibinfo {year} {2013}{\natexlab{b}})}\BibitemShut {NoStop}%
\bibitem [{\citenamefont {Karasiev}\ \emph {et~al.}(2014)\citenamefont {Karasiev}, \citenamefont {Sjostrom}, \citenamefont {Dufty},\ and\ \citenamefont {Trickey}}]{ksdt}%
  \BibitemOpen
  \bibfield  {author} {\bibinfo {author} {\bibfnamefont {Valentin~V.}\ \bibnamefont {Karasiev}}, \bibinfo {author} {\bibfnamefont {Travis}\ \bibnamefont {Sjostrom}}, \bibinfo {author} {\bibfnamefont {James}\ \bibnamefont {Dufty}}, \ and\ \bibinfo {author} {\bibfnamefont {S.~B.}\ \bibnamefont {Trickey}},\ }\bibfield  {title} {\enquote {\bibinfo {title} {Accurate homogeneous electron gas exchange-correlation free energy for local spin-density calculations},}\ }\href {\doibase 10.1103/PhysRevLett.112.076403} {\bibfield  {journal} {\bibinfo  {journal} {Phys. Rev. Lett.}\ }\textbf {\bibinfo {volume} {112}},\ \bibinfo {pages} {076403} (\bibinfo {year} {2014})}\BibitemShut {NoStop}%
\bibitem [{\citenamefont {Karasiev}\ \emph {et~al.}(2019)\citenamefont {Karasiev}, \citenamefont {Trickey},\ and\ \citenamefont {Dufty}}]{Karasiev_status_2019}%
  \BibitemOpen
  \bibfield  {author} {\bibinfo {author} {\bibfnamefont {Valentin~V.}\ \bibnamefont {Karasiev}}, \bibinfo {author} {\bibfnamefont {S.~B.}\ \bibnamefont {Trickey}}, \ and\ \bibinfo {author} {\bibfnamefont {James~W.}\ \bibnamefont {Dufty}},\ }\bibfield  {title} {\enquote {\bibinfo {title} {Status of free-energy representations for the homogeneous electron gas},}\ }\href {\doibase 10.1103/PhysRevB.99.195134} {\bibfield  {journal} {\bibinfo  {journal} {Phys. Rev. B}\ }\textbf {\bibinfo {volume} {99}},\ \bibinfo {pages} {195134} (\bibinfo {year} {2019})}\BibitemShut {NoStop}%
\bibitem [{\citenamefont {Dornheim}\ \emph {et~al.}(2016{\natexlab{a}})\citenamefont {Dornheim}, \citenamefont {Groth}, \citenamefont {Sjostrom}, \citenamefont {Malone}, \citenamefont {Foulkes},\ and\ \citenamefont {Bonitz}}]{dornheim_prl}%
  \BibitemOpen
  \bibfield  {author} {\bibinfo {author} {\bibfnamefont {T.}~\bibnamefont {Dornheim}}, \bibinfo {author} {\bibfnamefont {S.}~\bibnamefont {Groth}}, \bibinfo {author} {\bibfnamefont {T.}~\bibnamefont {Sjostrom}}, \bibinfo {author} {\bibfnamefont {F.~D.}\ \bibnamefont {Malone}}, \bibinfo {author} {\bibfnamefont {W.~M.~C.}\ \bibnamefont {Foulkes}}, \ and\ \bibinfo {author} {\bibfnamefont {M.}~\bibnamefont {Bonitz}},\ }\bibfield  {title} {\enquote {\bibinfo {title} {Ab initio quantum {M}onte {C}arlo simulation of the warm dense electron gas in the thermodynamic limit},}\ }\href {http://link.aps.org/doi/10.1103/PhysRevLett.117.156403} {\bibfield  {journal} {\bibinfo  {journal} {Phys. Rev. Lett.}\ }\textbf {\bibinfo {volume} {117}},\ \bibinfo {pages} {156403} (\bibinfo {year} {2016}{\natexlab{a}})}\BibitemShut {NoStop}%
\bibitem [{\citenamefont {Groth}\ \emph {et~al.}(2017{\natexlab{a}})\citenamefont {Groth}, \citenamefont {Dornheim}, \citenamefont {Sjostrom}, \citenamefont {Malone}, \citenamefont {Foulkes},\ and\ \citenamefont {Bonitz}}]{groth_prl}%
  \BibitemOpen
  \bibfield  {author} {\bibinfo {author} {\bibfnamefont {S.}~\bibnamefont {Groth}}, \bibinfo {author} {\bibfnamefont {T.}~\bibnamefont {Dornheim}}, \bibinfo {author} {\bibfnamefont {T.}~\bibnamefont {Sjostrom}}, \bibinfo {author} {\bibfnamefont {F.~D.}\ \bibnamefont {Malone}}, \bibinfo {author} {\bibfnamefont {W.~M.~C.}\ \bibnamefont {Foulkes}}, \ and\ \bibinfo {author} {\bibfnamefont {M.}~\bibnamefont {Bonitz}},\ }\bibfield  {title} {\enquote {\bibinfo {title} {Ab initio exchange--correlation free energy of the uniform electron gas at warm dense matter conditions},}\ }\href {https://journals.aps.org/prl/abstract/10.1103/PhysRevLett.119.135001} {\bibfield  {journal} {\bibinfo  {journal} {Phys. Rev. Lett.}\ }\textbf {\bibinfo {volume} {119}},\ \bibinfo {pages} {135001} (\bibinfo {year} {2017}{\natexlab{a}})}\BibitemShut {NoStop}%
\bibitem [{\citenamefont {Hou}\ \emph {et~al.}(2022)\citenamefont {Hou}, \citenamefont {Wang}, \citenamefont {Haule}, \citenamefont {Deng},\ and\ \citenamefont {Chen}}]{Hou_PRB_2022}%
  \BibitemOpen
  \bibfield  {author} {\bibinfo {author} {\bibfnamefont {Peng-Cheng}\ \bibnamefont {Hou}}, \bibinfo {author} {\bibfnamefont {Bao-Zong}\ \bibnamefont {Wang}}, \bibinfo {author} {\bibfnamefont {Kristjan}\ \bibnamefont {Haule}}, \bibinfo {author} {\bibfnamefont {Youjin}\ \bibnamefont {Deng}}, \ and\ \bibinfo {author} {\bibfnamefont {Kun}\ \bibnamefont {Chen}},\ }\bibfield  {title} {\enquote {\bibinfo {title} {Exchange-correlation effect in the charge response of a warm dense electron gas},}\ }\href {\doibase 10.1103/PhysRevB.106.L081126} {\bibfield  {journal} {\bibinfo  {journal} {Phys. Rev. B}\ }\textbf {\bibinfo {volume} {106}},\ \bibinfo {pages} {L081126} (\bibinfo {year} {2022})}\BibitemShut {NoStop}%
\bibitem [{\citenamefont {Dornheim}\ \emph {et~al.}(2019{\natexlab{a}})\citenamefont {Dornheim}, \citenamefont {Vorberger}, \citenamefont {Groth}, \citenamefont {Hoffmann}, \citenamefont {Moldabekov},\ and\ \citenamefont {Bonitz}}]{dornheim_ML}%
  \BibitemOpen
  \bibfield  {author} {\bibinfo {author} {\bibfnamefont {T.}~\bibnamefont {Dornheim}}, \bibinfo {author} {\bibfnamefont {J.}~\bibnamefont {Vorberger}}, \bibinfo {author} {\bibfnamefont {S.}~\bibnamefont {Groth}}, \bibinfo {author} {\bibfnamefont {N.}~\bibnamefont {Hoffmann}}, \bibinfo {author} {\bibfnamefont {Zh.A.}\ \bibnamefont {Moldabekov}}, \ and\ \bibinfo {author} {\bibfnamefont {M.}~\bibnamefont {Bonitz}},\ }\bibfield  {title} {\enquote {\bibinfo {title} {The static local field correction of the warm dense electron gas: An ab initio path integral {M}onte {C}arlo study and machine learning representation},}\ }\href {https://aip.scitation.org/doi/full/10.1063/1.5123013} {\bibfield  {journal} {\bibinfo  {journal} {J. Chem. Phys}\ }\textbf {\bibinfo {volume} {151}},\ \bibinfo {pages} {194104} (\bibinfo {year} {2019}{\natexlab{a}})}\BibitemShut {NoStop}%
\bibitem [{\citenamefont {Dornheim}\ \emph {et~al.}(2020{\natexlab{b}})\citenamefont {Dornheim}, \citenamefont {Vorberger},\ and\ \citenamefont {Bonitz}}]{Dornheim_PRL_2020}%
  \BibitemOpen
  \bibfield  {author} {\bibinfo {author} {\bibfnamefont {Tobias}\ \bibnamefont {Dornheim}}, \bibinfo {author} {\bibfnamefont {Jan}\ \bibnamefont {Vorberger}}, \ and\ \bibinfo {author} {\bibfnamefont {Michael}\ \bibnamefont {Bonitz}},\ }\bibfield  {title} {\enquote {\bibinfo {title} {Nonlinear electronic density response in warm dense matter},}\ }\href {\doibase 10.1103/PhysRevLett.125.085001} {\bibfield  {journal} {\bibinfo  {journal} {Phys. Rev. Lett.}\ }\textbf {\bibinfo {volume} {125}},\ \bibinfo {pages} {085001} (\bibinfo {year} {2020}{\natexlab{b}})}\BibitemShut {NoStop}%
\bibitem [{\citenamefont {Dornheim}\ \emph {et~al.}(2020{\natexlab{c}})\citenamefont {Dornheim}, \citenamefont {Cangi}, \citenamefont {Ramakrishna}, \citenamefont {B\"ohme}, \citenamefont {Tanaka},\ and\ \citenamefont {Vorberger}}]{Dornheim_PRL_2020_ESA}%
  \BibitemOpen
  \bibfield  {author} {\bibinfo {author} {\bibfnamefont {Tobias}\ \bibnamefont {Dornheim}}, \bibinfo {author} {\bibfnamefont {Attila}\ \bibnamefont {Cangi}}, \bibinfo {author} {\bibfnamefont {Kushal}\ \bibnamefont {Ramakrishna}}, \bibinfo {author} {\bibfnamefont {Maximilian}\ \bibnamefont {B\"ohme}}, \bibinfo {author} {\bibfnamefont {Shigenori}\ \bibnamefont {Tanaka}}, \ and\ \bibinfo {author} {\bibfnamefont {Jan}\ \bibnamefont {Vorberger}},\ }\bibfield  {title} {\enquote {\bibinfo {title} {Effective static approximation: A fast and reliable tool for warm-dense matter theory},}\ }\href {\doibase 10.1103/PhysRevLett.125.235001} {\bibfield  {journal} {\bibinfo  {journal} {Phys. Rev. Lett.}\ }\textbf {\bibinfo {volume} {125}},\ \bibinfo {pages} {235001} (\bibinfo {year} {2020}{\natexlab{c}})}\BibitemShut {NoStop}%
\bibitem [{\citenamefont {Hunger}\ \emph {et~al.}(2021)\citenamefont {Hunger}, \citenamefont {Schoof}, \citenamefont {Dornheim}, \citenamefont {Bonitz},\ and\ \citenamefont {Filinov}}]{Hunger_PRE_2021}%
  \BibitemOpen
  \bibfield  {author} {\bibinfo {author} {\bibfnamefont {Kai}\ \bibnamefont {Hunger}}, \bibinfo {author} {\bibfnamefont {Tim}\ \bibnamefont {Schoof}}, \bibinfo {author} {\bibfnamefont {Tobias}\ \bibnamefont {Dornheim}}, \bibinfo {author} {\bibfnamefont {Michael}\ \bibnamefont {Bonitz}}, \ and\ \bibinfo {author} {\bibfnamefont {Alexey}\ \bibnamefont {Filinov}},\ }\bibfield  {title} {\enquote {\bibinfo {title} {Momentum distribution function and short-range correlations of the warm dense electron gas: Ab initio quantum monte carlo results},}\ }\href {https://link.aps.org/doi/10.1103/PhysRevE.103.053204} {\bibfield  {journal} {\bibinfo  {journal} {Phys. Rev. E}\ }\textbf {\bibinfo {volume} {103}},\ \bibinfo {pages} {053204} (\bibinfo {year} {2021})}\BibitemShut {NoStop}%
\bibitem [{\citenamefont {Tanaka}(2016)}]{tanaka_hnc}%
  \BibitemOpen
  \bibfield  {author} {\bibinfo {author} {\bibfnamefont {S.}~\bibnamefont {Tanaka}},\ }\bibfield  {title} {\enquote {\bibinfo {title} {Correlational and thermodynamic properties of finite-temperature electron liquids in the hypernetted-chain approximation},}\ }\href {https://aip.scitation.org/doi/abs/10.1063/1.4969071} {\bibfield  {journal} {\bibinfo  {journal} {J. Chem. Phys}\ }\textbf {\bibinfo {volume} {145}},\ \bibinfo {pages} {214104} (\bibinfo {year} {2016})}\BibitemShut {NoStop}%
\bibitem [{\citenamefont {Tanaka}(2017)}]{Tanaka_CPP_2017}%
  \BibitemOpen
  \bibfield  {author} {\bibinfo {author} {\bibfnamefont {Shigenori}\ \bibnamefont {Tanaka}},\ }\bibfield  {title} {\enquote {\bibinfo {title} {Improved equation of state for finite-temperature spin-polarized electron liquids on the basis of singwi–tosi–land–sjölander approximation},}\ }\href {\doibase https://doi.org/10.1002/ctpp.201600096} {\bibfield  {journal} {\bibinfo  {journal} {Contrib. Plasma Phys.}\ }\textbf {\bibinfo {volume} {57}},\ \bibinfo {pages} {126--136} (\bibinfo {year} {2017})}\BibitemShut {NoStop}%
\bibitem [{\citenamefont {Arora}\ \emph {et~al.}(2017)\citenamefont {Arora}, \citenamefont {Kumar},\ and\ \citenamefont {Moudgil}}]{arora}%
  \BibitemOpen
  \bibfield  {author} {\bibinfo {author} {\bibfnamefont {P.}~\bibnamefont {Arora}}, \bibinfo {author} {\bibfnamefont {K.}~\bibnamefont {Kumar}}, \ and\ \bibinfo {author} {\bibfnamefont {R.~K.}\ \bibnamefont {Moudgil}},\ }\bibfield  {title} {\enquote {\bibinfo {title} {Spin-resolved correlations in the warm-dense homogeneous electron gas},}\ }\href {https://link.springer.com/article/10.1140/epjb/e2017-70532-y} {\bibfield  {journal} {\bibinfo  {journal} {Eur. Phys. J. B}\ }\textbf {\bibinfo {volume} {90}},\ \bibinfo {pages} {76} (\bibinfo {year} {2017})}\BibitemShut {NoStop}%
\bibitem [{\citenamefont {Tolias}\ \emph {et~al.}(2021)\citenamefont {Tolias}, \citenamefont {Lucco~Castello},\ and\ \citenamefont {Dornheim}}]{Tolias_JCP_2021}%
  \BibitemOpen
  \bibfield  {author} {\bibinfo {author} {\bibfnamefont {P.}~\bibnamefont {Tolias}}, \bibinfo {author} {\bibfnamefont {F.}~\bibnamefont {Lucco~Castello}}, \ and\ \bibinfo {author} {\bibfnamefont {T.}~\bibnamefont {Dornheim}},\ }\bibfield  {title} {\enquote {\bibinfo {title} {Integral equation theory based dielectric scheme for strongly coupled electron liquids},}\ }\href {\doibase 10.1063/5.0065988} {\bibfield  {journal} {\bibinfo  {journal} {J. Chem. Phys.}\ }\textbf {\bibinfo {volume} {155}},\ \bibinfo {pages} {134115} (\bibinfo {year} {2021})}\BibitemShut {NoStop}%
\bibitem [{\citenamefont {Tolias}\ \emph {et~al.}(2023{\natexlab{a}})\citenamefont {Tolias}, \citenamefont {Lucco~Castello},\ and\ \citenamefont {Dornheim}}]{Tolias_JCP_2023}%
  \BibitemOpen
  \bibfield  {author} {\bibinfo {author} {\bibfnamefont {Panagiotis}\ \bibnamefont {Tolias}}, \bibinfo {author} {\bibfnamefont {Federico}\ \bibnamefont {Lucco~Castello}}, \ and\ \bibinfo {author} {\bibfnamefont {Tobias}\ \bibnamefont {Dornheim}},\ }\bibfield  {title} {\enquote {\bibinfo {title} {{Quantum version of the integral equation theory-based dielectric scheme for strongly coupled electron liquids}},}\ }\href {\doibase 10.1063/5.0145687} {\bibfield  {journal} {\bibinfo  {journal} {J. Chem. Phys.}\ }\textbf {\bibinfo {volume} {158}},\ \bibinfo {pages} {141102} (\bibinfo {year} {2023}{\natexlab{a}})}\BibitemShut {NoStop}%
\bibitem [{\citenamefont {Dornheim}\ \emph {et~al.}(2024{\natexlab{d}})\citenamefont {Dornheim}, \citenamefont {Tolias}, \citenamefont {Kalkavouras}, \citenamefont {Moldabekov},\ and\ \citenamefont {Vorberger}}]{Dornheim_PRB_2024}%
  \BibitemOpen
  \bibfield  {author} {\bibinfo {author} {\bibfnamefont {Tobias}\ \bibnamefont {Dornheim}}, \bibinfo {author} {\bibfnamefont {Panagiotis}\ \bibnamefont {Tolias}}, \bibinfo {author} {\bibfnamefont {Fotios}\ \bibnamefont {Kalkavouras}}, \bibinfo {author} {\bibfnamefont {Zhandos~A.}\ \bibnamefont {Moldabekov}}, \ and\ \bibinfo {author} {\bibfnamefont {Jan}\ \bibnamefont {Vorberger}},\ }\bibfield  {title} {\enquote {\bibinfo {title} {Dynamic exchange correlation effects in the strongly coupled electron liquid},}\ }\href {\doibase 10.1103/PhysRevB.110.075137} {\bibfield  {journal} {\bibinfo  {journal} {Phys. Rev. B}\ }\textbf {\bibinfo {volume} {110}},\ \bibinfo {pages} {075137} (\bibinfo {year} {2024}{\natexlab{d}})}\BibitemShut {NoStop}%
\bibitem [{\citenamefont {Tolias}\ \emph {et~al.}(2024)\citenamefont {Tolias}, \citenamefont {Lucco~Castello}, \citenamefont {Kalkavouras},\ and\ \citenamefont {Dornheim}}]{Tolias_PRB_2024}%
  \BibitemOpen
  \bibfield  {author} {\bibinfo {author} {\bibfnamefont {Panagiotis}\ \bibnamefont {Tolias}}, \bibinfo {author} {\bibfnamefont {Federico}\ \bibnamefont {Lucco~Castello}}, \bibinfo {author} {\bibfnamefont {Fotios}\ \bibnamefont {Kalkavouras}}, \ and\ \bibinfo {author} {\bibfnamefont {Tobias}\ \bibnamefont {Dornheim}},\ }\bibfield  {title} {\enquote {\bibinfo {title} {Revisiting the vashishta-singwi dielectric scheme for the warm dense uniform electron fluid},}\ }\href {\doibase 10.1103/PhysRevB.109.125134} {\bibfield  {journal} {\bibinfo  {journal} {Phys. Rev. B}\ }\textbf {\bibinfo {volume} {109}},\ \bibinfo {pages} {125134} (\bibinfo {year} {2024})}\BibitemShut {NoStop}%
\bibitem [{\citenamefont {Sjostrom}\ and\ \citenamefont {Dufty}(2013)}]{stls2}%
  \BibitemOpen
  \bibfield  {author} {\bibinfo {author} {\bibfnamefont {T.}~\bibnamefont {Sjostrom}}\ and\ \bibinfo {author} {\bibfnamefont {J.}~\bibnamefont {Dufty}},\ }\bibfield  {title} {\enquote {\bibinfo {title} {Uniform electron gas at finite temperatures},}\ }\href {http://link.aps.org/doi/10.1103/PhysRevB.88.115123} {\bibfield  {journal} {\bibinfo  {journal} {Phys. Rev. B}\ }\textbf {\bibinfo {volume} {88}},\ \bibinfo {pages} {115123} (\bibinfo {year} {2013})}\BibitemShut {NoStop}%
\bibitem [{\citenamefont {Karasiev}\ \emph {et~al.}(2016)\citenamefont {Karasiev}, \citenamefont {Calderin},\ and\ \citenamefont {Trickey}}]{karasiev_importance}%
  \BibitemOpen
  \bibfield  {author} {\bibinfo {author} {\bibfnamefont {V.~V.}\ \bibnamefont {Karasiev}}, \bibinfo {author} {\bibfnamefont {L.}~\bibnamefont {Calderin}}, \ and\ \bibinfo {author} {\bibfnamefont {S.~B.}\ \bibnamefont {Trickey}},\ }\bibfield  {title} {\enquote {\bibinfo {title} {Importance of finite-temperature exchange correlation for warm dense matter calculations},}\ }\href {https://journals.aps.org/pre/abstract/10.1103/PhysRevE.93.063207} {\bibfield  {journal} {\bibinfo  {journal} {Phys. Rev. E}\ }\textbf {\bibinfo {volume} {93}},\ \bibinfo {pages} {063207} (\bibinfo {year} {2016})}\BibitemShut {NoStop}%
\bibitem [{\citenamefont {Ramakrishna}\ \emph {et~al.}(2020)\citenamefont {Ramakrishna}, \citenamefont {Dornheim},\ and\ \citenamefont {Vorberger}}]{kushal}%
  \BibitemOpen
  \bibfield  {author} {\bibinfo {author} {\bibfnamefont {Kushal}\ \bibnamefont {Ramakrishna}}, \bibinfo {author} {\bibfnamefont {Tobias}\ \bibnamefont {Dornheim}}, \ and\ \bibinfo {author} {\bibfnamefont {Jan}\ \bibnamefont {Vorberger}},\ }\bibfield  {title} {\enquote {\bibinfo {title} {Influence of finite temperature exchange-correlation effects in hydrogen},}\ }\href {\doibase 10.1103/PhysRevB.101.195129} {\bibfield  {journal} {\bibinfo  {journal} {Phys. Rev. B}\ }\textbf {\bibinfo {volume} {101}},\ \bibinfo {pages} {195129} (\bibinfo {year} {2020})}\BibitemShut {NoStop}%
\bibitem [{\citenamefont {Karasiev}\ \emph {et~al.}(2018)\citenamefont {Karasiev}, \citenamefont {Dufty},\ and\ \citenamefont {Trickey}}]{Karasiev_PRL_2018}%
  \BibitemOpen
  \bibfield  {author} {\bibinfo {author} {\bibfnamefont {Valentin~V.}\ \bibnamefont {Karasiev}}, \bibinfo {author} {\bibfnamefont {James~W.}\ \bibnamefont {Dufty}}, \ and\ \bibinfo {author} {\bibfnamefont {S.~B.}\ \bibnamefont {Trickey}},\ }\bibfield  {title} {\enquote {\bibinfo {title} {Nonempirical semilocal free-energy density functional for matter under extreme conditions},}\ }\href {\doibase 10.1103/PhysRevLett.120.076401} {\bibfield  {journal} {\bibinfo  {journal} {Phys. Rev. Lett.}\ }\textbf {\bibinfo {volume} {120}},\ \bibinfo {pages} {076401} (\bibinfo {year} {2018})}\BibitemShut {NoStop}%
\bibitem [{\citenamefont {Karasiev}\ \emph {et~al.}(2022)\citenamefont {Karasiev}, \citenamefont {Mihaylov},\ and\ \citenamefont {Hu}}]{Karasiev_PRB_2022}%
  \BibitemOpen
  \bibfield  {author} {\bibinfo {author} {\bibfnamefont {Valentin~V.}\ \bibnamefont {Karasiev}}, \bibinfo {author} {\bibfnamefont {D.~I.}\ \bibnamefont {Mihaylov}}, \ and\ \bibinfo {author} {\bibfnamefont {S.~X.}\ \bibnamefont {Hu}},\ }\bibfield  {title} {\enquote {\bibinfo {title} {Meta-gga exchange-correlation free energy density functional to increase the accuracy of warm dense matter simulations},}\ }\href {\doibase 10.1103/PhysRevB.105.L081109} {\bibfield  {journal} {\bibinfo  {journal} {Phys. Rev. B}\ }\textbf {\bibinfo {volume} {105}},\ \bibinfo {pages} {L081109} (\bibinfo {year} {2022})}\BibitemShut {NoStop}%
\bibitem [{\citenamefont {Moldabekov}\ \emph {et~al.}(2025)\citenamefont {Moldabekov}, \citenamefont {Vorberger},\ and\ \citenamefont {Dornheim}}]{moldabekov2024density}%
  \BibitemOpen
  \bibfield  {author} {\bibinfo {author} {\bibfnamefont {Zhandos}\ \bibnamefont {Moldabekov}}, \bibinfo {author} {\bibfnamefont {Jan}\ \bibnamefont {Vorberger}}, \ and\ \bibinfo {author} {\bibfnamefont {Tobias}\ \bibnamefont {Dornheim}},\ }\bibfield  {title} {\enquote {\bibinfo {title} {From density response to energy functionals and back: An ab initio perspective on matter under extreme conditions},}\ }\href {\doibase https://doi.org/10.1016/j.ppnp.2024.104144} {\bibfield  {journal} {\bibinfo  {journal} {Progress in Particle and Nuclear Physics}\ }\textbf {\bibinfo {volume} {140}},\ \bibinfo {pages} {104144} (\bibinfo {year} {2025})}\BibitemShut {NoStop}%
\bibitem [{\citenamefont {Moldabekov}\ \emph {et~al.}(2024)\citenamefont {Moldabekov}, \citenamefont {Schwalbe}, \citenamefont {B{\"o}hme}, \citenamefont {Vorberger}, \citenamefont {Shao}, \citenamefont {Pavanello}, \citenamefont {Graziani},\ and\ \citenamefont {Dornheim}}]{Moldabekov_JCTC_2024}%
  \BibitemOpen
  \bibfield  {author} {\bibinfo {author} {\bibfnamefont {Zhandos}\ \bibnamefont {Moldabekov}}, \bibinfo {author} {\bibfnamefont {Sebastian}\ \bibnamefont {Schwalbe}}, \bibinfo {author} {\bibfnamefont {Maximilian~P.}\ \bibnamefont {B{\"o}hme}}, \bibinfo {author} {\bibfnamefont {Jan}\ \bibnamefont {Vorberger}}, \bibinfo {author} {\bibfnamefont {Xuecheng}\ \bibnamefont {Shao}}, \bibinfo {author} {\bibfnamefont {Michele}\ \bibnamefont {Pavanello}}, \bibinfo {author} {\bibfnamefont {Frank~R.}\ \bibnamefont {Graziani}}, \ and\ \bibinfo {author} {\bibfnamefont {Tobias}\ \bibnamefont {Dornheim}},\ }\bibfield  {title} {\enquote {\bibinfo {title} {Bound-state breaking and the importance of thermal exchange--correlation effects in warm dense hydrogen},}\ }\href {\doibase 10.1021/acs.jctc.3c00934} {\bibfield  {journal} {\bibinfo  {journal} {J. Chem. Theory Comput.}\ }\textbf {\bibinfo {volume} {20}},\ \bibinfo {pages} {68--78} (\bibinfo {year} {2024})}\BibitemShut {NoStop}%
\bibitem [{\citenamefont {Sjostrom}\ and\ \citenamefont {Daligault}(2014)}]{Sjostrom_PRB_2014}%
  \BibitemOpen
  \bibfield  {author} {\bibinfo {author} {\bibfnamefont {Travis}\ \bibnamefont {Sjostrom}}\ and\ \bibinfo {author} {\bibfnamefont {J\'er\^ome}\ \bibnamefont {Daligault}},\ }\bibfield  {title} {\enquote {\bibinfo {title} {Gradient corrections to the exchange-correlation free energy},}\ }\href {\doibase 10.1103/PhysRevB.90.155109} {\bibfield  {journal} {\bibinfo  {journal} {Phys. Rev. B}\ }\textbf {\bibinfo {volume} {90}},\ \bibinfo {pages} {155109} (\bibinfo {year} {2014})}\BibitemShut {NoStop}%
\bibitem [{\citenamefont {Baus}\ and\ \citenamefont {Hansen}(1980)}]{Baus_Hansen_OCP}%
  \BibitemOpen
  \bibfield  {author} {\bibinfo {author} {\bibfnamefont {Marc}\ \bibnamefont {Baus}}\ and\ \bibinfo {author} {\bibfnamefont {Jean-Pierre}\ \bibnamefont {Hansen}},\ }\bibfield  {title} {\enquote {\bibinfo {title} {Statistical mechanics of simple coulomb systems},}\ }\href {https://www.sciencedirect.com/science/article/pii/0370157380900228} {\bibfield  {journal} {\bibinfo  {journal} {Phys. Rep.}\ }\textbf {\bibinfo {volume} {59}},\ \bibinfo {pages} {1--94} (\bibinfo {year} {1980})}\BibitemShut {NoStop}%
\bibitem [{\citenamefont {Lucco~Castello}\ and\ \citenamefont {Tolias}(2022)}]{OCP_bridge_2022}%
  \BibitemOpen
  \bibfield  {author} {\bibinfo {author} {\bibfnamefont {F.}~\bibnamefont {Lucco~Castello}}\ and\ \bibinfo {author} {\bibfnamefont {P.}~\bibnamefont {Tolias}},\ }\bibfield  {title} {\enquote {\bibinfo {title} {Bridge functions of classical one-component plasmas},}\ }\href {\doibase 10.1103/PhysRevE.105.015208} {\bibfield  {journal} {\bibinfo  {journal} {Phys. Rev. E}\ }\textbf {\bibinfo {volume} {105}},\ \bibinfo {pages} {015208} (\bibinfo {year} {2022})}\BibitemShut {NoStop}%
\bibitem [{\citenamefont {Fraser}\ \emph {et~al.}(1996)\citenamefont {Fraser}, \citenamefont {Foulkes}, \citenamefont {Rajagopal}, \citenamefont {Needs}, \citenamefont {Kenny},\ and\ \citenamefont {Williamson}}]{Fraser_PRB_1996}%
  \BibitemOpen
  \bibfield  {author} {\bibinfo {author} {\bibfnamefont {Louisa~M.}\ \bibnamefont {Fraser}}, \bibinfo {author} {\bibfnamefont {W.~M.~C.}\ \bibnamefont {Foulkes}}, \bibinfo {author} {\bibfnamefont {G.}~\bibnamefont {Rajagopal}}, \bibinfo {author} {\bibfnamefont {R.~J.}\ \bibnamefont {Needs}}, \bibinfo {author} {\bibfnamefont {S.~D.}\ \bibnamefont {Kenny}}, \ and\ \bibinfo {author} {\bibfnamefont {A.~J.}\ \bibnamefont {Williamson}},\ }\bibfield  {title} {\enquote {\bibinfo {title} {Finite-size effects and coulomb interactions in quantum monte carlo calculations for homogeneous systems with periodic boundary conditions},}\ }\href {\doibase 10.1103/PhysRevB.53.1814} {\bibfield  {journal} {\bibinfo  {journal} {Phys. Rev. B}\ }\textbf {\bibinfo {volume} {53}},\ \bibinfo {pages} {1814--1832} (\bibinfo {year} {1996})}\BibitemShut {NoStop}%
\bibitem [{\citenamefont {Fukuda}\ and\ \citenamefont {Nakamura}(2012)}]{Fukuda2012}%
  \BibitemOpen
  \bibfield  {author} {\bibinfo {author} {\bibfnamefont {Ikuo}\ \bibnamefont {Fukuda}}\ and\ \bibinfo {author} {\bibfnamefont {Haruki}\ \bibnamefont {Nakamura}},\ }\bibfield  {title} {\enquote {\bibinfo {title} {Non-ewald methods: theory and applications to molecular systems},}\ }\href {\doibase 10.1007/s12551-012-0089-4} {\bibfield  {journal} {\bibinfo  {journal} {Biophysical Reviews}\ }\textbf {\bibinfo {volume} {4}},\ \bibinfo {pages} {161--170} (\bibinfo {year} {2012})}\BibitemShut {NoStop}%
\bibitem [{\citenamefont {Dornheim}\ \emph {et~al.}(2025{\natexlab{e}})\citenamefont {Dornheim}, \citenamefont {Chuna}, \citenamefont {Bellenbaum}, \citenamefont {Moldabekov}, \citenamefont {Tolias},\ and\ \citenamefont {Vorberger}}]{Dornheim_PRE_2025}%
  \BibitemOpen
  \bibfield  {author} {\bibinfo {author} {\bibfnamefont {Tobias}\ \bibnamefont {Dornheim}}, \bibinfo {author} {\bibfnamefont {Thomas~M.}\ \bibnamefont {Chuna}}, \bibinfo {author} {\bibfnamefont {Hannah~M.}\ \bibnamefont {Bellenbaum}}, \bibinfo {author} {\bibfnamefont {Zhandos~A.}\ \bibnamefont {Moldabekov}}, \bibinfo {author} {\bibfnamefont {Panagiotis}\ \bibnamefont {Tolias}}, \ and\ \bibinfo {author} {\bibfnamefont {Jan}\ \bibnamefont {Vorberger}},\ }\bibfield  {title} {\enquote {\bibinfo {title} {Application of a spherically averaged pair potential in ab initio path integral monte carlo simulations of a warm dense electron gas},}\ }\href {\doibase 10.1103/lj9c-bh48} {\bibfield  {journal} {\bibinfo  {journal} {Phys. Rev. E}\ }\textbf {\bibinfo {volume} {112}},\ \bibinfo {pages} {035203} (\bibinfo {year} {2025}{\natexlab{e}})}\BibitemShut {NoStop}%
\bibitem [{\citenamefont {Yakub}\ and\ \citenamefont {Ronchi}(2005)}]{Yakub2005}%
  \BibitemOpen
  \bibfield  {author} {\bibinfo {author} {\bibfnamefont {E.}~\bibnamefont {Yakub}}\ and\ \bibinfo {author} {\bibfnamefont {C.}~\bibnamefont {Ronchi}},\ }\bibfield  {title} {\enquote {\bibinfo {title} {A new method for computation of long ranged coulomb forces in computer simulation of disordered systems},}\ }\href {\doibase 10.1007/s10909-005-5451-5} {\bibfield  {journal} {\bibinfo  {journal} {Journal of Low Temperature Physics}\ }\textbf {\bibinfo {volume} {139}},\ \bibinfo {pages} {633--643} (\bibinfo {year} {2005})}\BibitemShut {NoStop}%
\bibitem [{\citenamefont {Yakub}\ and\ \citenamefont {Ronchi}(2003)}]{Yakub_JCP_2003}%
  \BibitemOpen
  \bibfield  {author} {\bibinfo {author} {\bibfnamefont {Eugene}\ \bibnamefont {Yakub}}\ and\ \bibinfo {author} {\bibfnamefont {Claudio}\ \bibnamefont {Ronchi}},\ }\bibfield  {title} {\enquote {\bibinfo {title} {An efficient method for computation of long-ranged coulomb forces in computer simulation of ionic fluids},}\ }\href {\doibase 10.1063/1.1624364} {\bibfield  {journal} {\bibinfo  {journal} {The Journal of Chemical Physics}\ }\textbf {\bibinfo {volume} {119}},\ \bibinfo {pages} {11556--11560} (\bibinfo {year} {2003})}\BibitemShut {NoStop}%
\bibitem [{\citenamefont {Filinov}\ and\ \citenamefont {Bonitz}(2023)}]{Filinov_PRE_2023}%
  \BibitemOpen
  \bibfield  {author} {\bibinfo {author} {\bibfnamefont {A.~V.}\ \bibnamefont {Filinov}}\ and\ \bibinfo {author} {\bibfnamefont {M.}~\bibnamefont {Bonitz}},\ }\bibfield  {title} {\enquote {\bibinfo {title} {Equation of state of partially ionized hydrogen and deuterium plasma revisited},}\ }\href {\doibase 10.1103/PhysRevE.108.055212} {\bibfield  {journal} {\bibinfo  {journal} {Phys. Rev. E}\ }\textbf {\bibinfo {volume} {108}},\ \bibinfo {pages} {055212} (\bibinfo {year} {2023})}\BibitemShut {NoStop}%
\bibitem [{\citenamefont {Demyanov}\ and\ \citenamefont {Levashov}(2022)}]{Demyanov_2022}%
  \BibitemOpen
  \bibfield  {author} {\bibinfo {author} {\bibfnamefont {G~S}\ \bibnamefont {Demyanov}}\ and\ \bibinfo {author} {\bibfnamefont {P~R}\ \bibnamefont {Levashov}},\ }\bibfield  {title} {\enquote {\bibinfo {title} {Systematic derivation of angular-averaged ewald potential},}\ }\href {\doibase 10.1088/1751-8121/ac870b} {\bibfield  {journal} {\bibinfo  {journal} {Journal of Physics A: Mathematical and Theoretical}\ }\textbf {\bibinfo {volume} {55}},\ \bibinfo {pages} {385202} (\bibinfo {year} {2022})}\BibitemShut {NoStop}%
\bibitem [{\citenamefont {Dornheim}\ \emph {et~al.}(2016{\natexlab{b}})\citenamefont {Dornheim}, \citenamefont {Groth}, \citenamefont {Schoof}, \citenamefont {Hann},\ and\ \citenamefont {Bonitz}}]{Dornheim_PRB_2016}%
  \BibitemOpen
  \bibfield  {author} {\bibinfo {author} {\bibfnamefont {T.}~\bibnamefont {Dornheim}}, \bibinfo {author} {\bibfnamefont {S.}~\bibnamefont {Groth}}, \bibinfo {author} {\bibfnamefont {T.}~\bibnamefont {Schoof}}, \bibinfo {author} {\bibfnamefont {C.}~\bibnamefont {Hann}}, \ and\ \bibinfo {author} {\bibfnamefont {M.}~\bibnamefont {Bonitz}},\ }\bibfield  {title} {\enquote {\bibinfo {title} {Ab initio quantum monte carlo simulations of the uniform electron gas without fixed nodes: The unpolarized case},}\ }\href {\doibase 10.1103/PhysRevB.93.205134} {\bibfield  {journal} {\bibinfo  {journal} {Phys. Rev. B}\ }\textbf {\bibinfo {volume} {93}},\ \bibinfo {pages} {205134} (\bibinfo {year} {2016}{\natexlab{b}})}\BibitemShut {NoStop}%
\bibitem [{\citenamefont {Ott}\ \emph {et~al.}(2018)\citenamefont {Ott}, \citenamefont {Thomsen}, \citenamefont {Abraham}, \citenamefont {Dornheim},\ and\ \citenamefont {Bonitz}}]{Ott2018}%
  \BibitemOpen
  \bibfield  {author} {\bibinfo {author} {\bibfnamefont {Torben}\ \bibnamefont {Ott}}, \bibinfo {author} {\bibfnamefont {Hauke}\ \bibnamefont {Thomsen}}, \bibinfo {author} {\bibfnamefont {Jan~Willem}\ \bibnamefont {Abraham}}, \bibinfo {author} {\bibfnamefont {Tobias}\ \bibnamefont {Dornheim}}, \ and\ \bibinfo {author} {\bibfnamefont {Michael}\ \bibnamefont {Bonitz}},\ }\bibfield  {title} {\enquote {\bibinfo {title} {Recent progress in the theory and simulation of strongly correlated plasmas: phase transitions, transport, quantum, and magnetic field effects},}\ }\href {\doibase 10.1140/epjd/e2018-80385-7} {\bibfield  {journal} {\bibinfo  {journal} {The European Physical Journal D}\ }\textbf {\bibinfo {volume} {72}},\ \bibinfo {pages} {84} (\bibinfo {year} {2018})}\BibitemShut {NoStop}%
\bibitem [{\citenamefont {Mahan}(1990)}]{mahan1990many}%
  \BibitemOpen
  \bibfield  {author} {\bibinfo {author} {\bibfnamefont {G.D.}\ \bibnamefont {Mahan}},\ }\href {https://books.google.de/books?id=v8du6cp0vUAC} {\emph {\bibinfo {title} {Many-Particle Physics}}},\ Physics of Solids and Liquids\ (\bibinfo  {publisher} {Springer US},\ \bibinfo {year} {1990})\BibitemShut {NoStop}%
\bibitem [{\citenamefont {Vosko}\ \emph {et~al.}(1980)\citenamefont {Vosko}, \citenamefont {Wilk},\ and\ \citenamefont {Nusair}}]{vwn}%
  \BibitemOpen
  \bibfield  {author} {\bibinfo {author} {\bibfnamefont {S.~H.}\ \bibnamefont {Vosko}}, \bibinfo {author} {\bibfnamefont {L.}~\bibnamefont {Wilk}}, \ and\ \bibinfo {author} {\bibfnamefont {M.}~\bibnamefont {Nusair}},\ }\bibfield  {title} {\enquote {\bibinfo {title} {Accurate spin-dependent electron liquid correlation energies for local spin density calculations: a critical analysis},}\ }\href {\doibase 10.1139/p80-159} {\bibfield  {journal} {\bibinfo  {journal} {Canadian Journal of Physics}\ }\textbf {\bibinfo {volume} {58}},\ \bibinfo {pages} {1200--1211} (\bibinfo {year} {1980})}\BibitemShut {NoStop}%
\bibitem [{\citenamefont {Perdew}\ and\ \citenamefont {Wang}(1992)}]{Perdew_Wang}%
  \BibitemOpen
  \bibfield  {author} {\bibinfo {author} {\bibfnamefont {John~P.}\ \bibnamefont {Perdew}}\ and\ \bibinfo {author} {\bibfnamefont {Yue}\ \bibnamefont {Wang}},\ }\bibfield  {title} {\enquote {\bibinfo {title} {Accurate and simple analytic representation of the electron-gas correlation energy},}\ }\href {\doibase 10.1103/PhysRevB.45.13244} {\bibfield  {journal} {\bibinfo  {journal} {Phys. Rev. B}\ }\textbf {\bibinfo {volume} {45}},\ \bibinfo {pages} {13244--13249} (\bibinfo {year} {1992})}\BibitemShut {NoStop}%
\bibitem [{\citenamefont {Perdew}\ and\ \citenamefont {Zunger}(1981)}]{Perdew_Zunger_PRB_1981}%
  \BibitemOpen
  \bibfield  {author} {\bibinfo {author} {\bibfnamefont {J.~P.}\ \bibnamefont {Perdew}}\ and\ \bibinfo {author} {\bibfnamefont {Alex}\ \bibnamefont {Zunger}},\ }\bibfield  {title} {\enquote {\bibinfo {title} {Self-interaction correction to density-functional approximations for many-electron systems},}\ }\href {\doibase 10.1103/PhysRevB.23.5048} {\bibfield  {journal} {\bibinfo  {journal} {Phys. Rev. B}\ }\textbf {\bibinfo {volume} {23}},\ \bibinfo {pages} {5048--5079} (\bibinfo {year} {1981})}\BibitemShut {NoStop}%
\bibitem [{\citenamefont {Corradini}\ \emph {et~al.}(1998)\citenamefont {Corradini}, \citenamefont {Sole}, \citenamefont {Onida},\ and\ \citenamefont {Palummo}}]{cdop}%
  \BibitemOpen
  \bibfield  {author} {\bibinfo {author} {\bibfnamefont {M.}~\bibnamefont {Corradini}}, \bibinfo {author} {\bibfnamefont {R.~Del}\ \bibnamefont {Sole}}, \bibinfo {author} {\bibfnamefont {G.}~\bibnamefont {Onida}}, \ and\ \bibinfo {author} {\bibfnamefont {M.}~\bibnamefont {Palummo}},\ }\bibfield  {title} {\enquote {\bibinfo {title} {Analytical expressions for the local-field factor $g(q)$ and the exchange-correlation kernel ${K}_{\mathrm{xc}}(r)$ of the homogeneous electron gas},}\ }\href {http://link.aps.org/doi/10.1103/PhysRevB.57.14569} {\bibfield  {journal} {\bibinfo  {journal} {Phys. Rev. B}\ }\textbf {\bibinfo {volume} {57}},\ \bibinfo {pages} {14569} (\bibinfo {year} {1998})}\BibitemShut {NoStop}%
\bibitem [{\citenamefont {Utsumi}\ and\ \citenamefont {Ichimaru}(1982)}]{Utsumi_PRA_1982}%
  \BibitemOpen
  \bibfield  {author} {\bibinfo {author} {\bibfnamefont {Kenichi}\ \bibnamefont {Utsumi}}\ and\ \bibinfo {author} {\bibfnamefont {Setsuo}\ \bibnamefont {Ichimaru}},\ }\bibfield  {title} {\enquote {\bibinfo {title} {Dielectric formulation of strongly coupled electron liquids at metallic densities. vi. analytic expression for the local-field correction},}\ }\href {\doibase 10.1103/PhysRevA.26.603} {\bibfield  {journal} {\bibinfo  {journal} {Phys. Rev. A}\ }\textbf {\bibinfo {volume} {26}},\ \bibinfo {pages} {603--610} (\bibinfo {year} {1982})}\BibitemShut {NoStop}%
\bibitem [{\citenamefont {Farid}\ \emph {et~al.}(1993)\citenamefont {Farid}, \citenamefont {Heine}, \citenamefont {Engel},\ and\ \citenamefont {Robertson}}]{farid}%
  \BibitemOpen
  \bibfield  {author} {\bibinfo {author} {\bibfnamefont {B.}~\bibnamefont {Farid}}, \bibinfo {author} {\bibfnamefont {V.}~\bibnamefont {Heine}}, \bibinfo {author} {\bibfnamefont {G.~E.}\ \bibnamefont {Engel}}, \ and\ \bibinfo {author} {\bibfnamefont {I.~J.}\ \bibnamefont {Robertson}},\ }\bibfield  {title} {\enquote {\bibinfo {title} {Extremal properties of the harris-foulkes functional and an improved screening calculation for the electron gas},}\ }\href {http://link.aps.org/doi/10.1103/PhysRevB.48.11602} {\bibfield  {journal} {\bibinfo  {journal} {Phys. Rev. B}\ }\textbf {\bibinfo {volume} {48}},\ \bibinfo {pages} {11602} (\bibinfo {year} {1993})}\BibitemShut {NoStop}%
\bibitem [{\citenamefont {Dornheim}\ \emph {et~al.}(2021{\natexlab{b}})\citenamefont {Dornheim}, \citenamefont {Moldabekov},\ and\ \citenamefont {Tolias}}]{Dornheim_PRB_ESA_2021}%
  \BibitemOpen
  \bibfield  {author} {\bibinfo {author} {\bibfnamefont {Tobias}\ \bibnamefont {Dornheim}}, \bibinfo {author} {\bibfnamefont {Zhandos~A.}\ \bibnamefont {Moldabekov}}, \ and\ \bibinfo {author} {\bibfnamefont {Panagiotis}\ \bibnamefont {Tolias}},\ }\bibfield  {title} {\enquote {\bibinfo {title} {Analytical representation of the local field correction of the uniform electron gas within the effective static approximation},}\ }\href {\doibase 10.1103/PhysRevB.103.165102} {\bibfield  {journal} {\bibinfo  {journal} {Phys. Rev. B}\ }\textbf {\bibinfo {volume} {103}},\ \bibinfo {pages} {165102} (\bibinfo {year} {2021}{\natexlab{b}})}\BibitemShut {NoStop}%
\bibitem [{\citenamefont {Ortiz}\ and\ \citenamefont {Ballone}(1994)}]{Ortiz_PRB_1994}%
  \BibitemOpen
  \bibfield  {author} {\bibinfo {author} {\bibfnamefont {G.}~\bibnamefont {Ortiz}}\ and\ \bibinfo {author} {\bibfnamefont {P.}~\bibnamefont {Ballone}},\ }\bibfield  {title} {\enquote {\bibinfo {title} {Correlation energy, structure factor, radial distribution function, and momentum distribution of the spin-polarized uniform electron gas},}\ }\href {\doibase 10.1103/PhysRevB.50.1391} {\bibfield  {journal} {\bibinfo  {journal} {Phys. Rev. B}\ }\textbf {\bibinfo {volume} {50}},\ \bibinfo {pages} {1391--1405} (\bibinfo {year} {1994})}\BibitemShut {NoStop}%
\bibitem [{\citenamefont {Ortiz}\ \emph {et~al.}(1999)\citenamefont {Ortiz}, \citenamefont {Harris},\ and\ \citenamefont {Ballone}}]{Ortiz_PRL_1999}%
  \BibitemOpen
  \bibfield  {author} {\bibinfo {author} {\bibfnamefont {G.}~\bibnamefont {Ortiz}}, \bibinfo {author} {\bibfnamefont {M.}~\bibnamefont {Harris}}, \ and\ \bibinfo {author} {\bibfnamefont {P.}~\bibnamefont {Ballone}},\ }\bibfield  {title} {\enquote {\bibinfo {title} {Zero temperature phases of the electron gas},}\ }\href {\doibase 10.1103/PhysRevLett.82.5317} {\bibfield  {journal} {\bibinfo  {journal} {Phys. Rev. Lett.}\ }\textbf {\bibinfo {volume} {82}},\ \bibinfo {pages} {5317--5320} (\bibinfo {year} {1999})}\BibitemShut {NoStop}%
\bibitem [{\citenamefont {Spink}\ \emph {et~al.}(2013)\citenamefont {Spink}, \citenamefont {Needs},\ and\ \citenamefont {Drummond}}]{Spink_PRB_2013}%
  \BibitemOpen
  \bibfield  {author} {\bibinfo {author} {\bibfnamefont {G.~G.}\ \bibnamefont {Spink}}, \bibinfo {author} {\bibfnamefont {R.~J.}\ \bibnamefont {Needs}}, \ and\ \bibinfo {author} {\bibfnamefont {N.~D.}\ \bibnamefont {Drummond}},\ }\bibfield  {title} {\enquote {\bibinfo {title} {Quantum monte carlo study of the three-dimensional spin-polarized homogeneous electron gas},}\ }\href {\doibase 10.1103/PhysRevB.88.085121} {\bibfield  {journal} {\bibinfo  {journal} {Phys. Rev. B}\ }\textbf {\bibinfo {volume} {88}},\ \bibinfo {pages} {085121} (\bibinfo {year} {2013})}\BibitemShut {NoStop}%
\bibitem [{\citenamefont {Groth}\ \emph {et~al.}(2019)\citenamefont {Groth}, \citenamefont {Dornheim},\ and\ \citenamefont {Vorberger}}]{dynamic_folgepaper}%
  \BibitemOpen
  \bibfield  {author} {\bibinfo {author} {\bibfnamefont {S.}~\bibnamefont {Groth}}, \bibinfo {author} {\bibfnamefont {T.}~\bibnamefont {Dornheim}}, \ and\ \bibinfo {author} {\bibfnamefont {J.}~\bibnamefont {Vorberger}},\ }\bibfield  {title} {\enquote {\bibinfo {title} {Ab initio path integral {M}onte {C}arlo approach to the static and dynamic density response of the uniform electron gas},}\ }\href {https://link.aps.org/doi/10.1103/PhysRevB.99.235122} {\bibfield  {journal} {\bibinfo  {journal} {Phys. Rev. B}\ }\textbf {\bibinfo {volume} {99}},\ \bibinfo {pages} {235122} (\bibinfo {year} {2019})}\BibitemShut {NoStop}%
\bibitem [{\citenamefont {Moroni}\ \emph {et~al.}(1992)\citenamefont {Moroni}, \citenamefont {Ceperley},\ and\ \citenamefont {Senatore}}]{moroni}%
  \BibitemOpen
  \bibfield  {author} {\bibinfo {author} {\bibfnamefont {S.}~\bibnamefont {Moroni}}, \bibinfo {author} {\bibfnamefont {D.~M.}\ \bibnamefont {Ceperley}}, \ and\ \bibinfo {author} {\bibfnamefont {G.}~\bibnamefont {Senatore}},\ }\bibfield  {title} {\enquote {\bibinfo {title} {Static response from quantum {M}onte {C}arlo calculations},}\ }\href {https://journals.aps.org/prl/abstract/10.1103/PhysRevLett.69.1837} {\bibfield  {journal} {\bibinfo  {journal} {Phys. Rev. Lett}\ }\textbf {\bibinfo {volume} {69}},\ \bibinfo {pages} {1837} (\bibinfo {year} {1992})}\BibitemShut {NoStop}%
\bibitem [{\citenamefont {Moroni}\ \emph {et~al.}(1995)\citenamefont {Moroni}, \citenamefont {Ceperley},\ and\ \citenamefont {Senatore}}]{moroni2}%
  \BibitemOpen
  \bibfield  {author} {\bibinfo {author} {\bibfnamefont {S.}~\bibnamefont {Moroni}}, \bibinfo {author} {\bibfnamefont {D.~M.}\ \bibnamefont {Ceperley}}, \ and\ \bibinfo {author} {\bibfnamefont {G.}~\bibnamefont {Senatore}},\ }\bibfield  {title} {\enquote {\bibinfo {title} {Static response and local field factor of the electron gas},}\ }\href {http://link.aps.org/doi/10.1103/PhysRevLett.75.689} {\bibfield  {journal} {\bibinfo  {journal} {Phys. Rev. Lett}\ }\textbf {\bibinfo {volume} {75}},\ \bibinfo {pages} {689} (\bibinfo {year} {1995})}\BibitemShut {NoStop}%
\bibitem [{\citenamefont {Jones}(2015)}]{Jones_RMP_2015}%
  \BibitemOpen
  \bibfield  {author} {\bibinfo {author} {\bibfnamefont {R.~O.}\ \bibnamefont {Jones}},\ }\bibfield  {title} {\enquote {\bibinfo {title} {Density functional theory: Its origins, rise to prominence, and future},}\ }\href {\doibase 10.1103/RevModPhys.87.897} {\bibfield  {journal} {\bibinfo  {journal} {Rev. Mod. Phys.}\ }\textbf {\bibinfo {volume} {87}},\ \bibinfo {pages} {897--923} (\bibinfo {year} {2015})}\BibitemShut {NoStop}%
\bibitem [{\citenamefont {Dornheim}\ \emph {et~al.}(2015{\natexlab{c}})\citenamefont {Dornheim}, \citenamefont {Schoof}, \citenamefont {Groth}, \citenamefont {Filinov},\ and\ \citenamefont {Bonitz}}]{Dornheim_JCP_2015}%
  \BibitemOpen
  \bibfield  {author} {\bibinfo {author} {\bibfnamefont {Tobias}\ \bibnamefont {Dornheim}}, \bibinfo {author} {\bibfnamefont {Tim}\ \bibnamefont {Schoof}}, \bibinfo {author} {\bibfnamefont {Simon}\ \bibnamefont {Groth}}, \bibinfo {author} {\bibfnamefont {Alexey}\ \bibnamefont {Filinov}}, \ and\ \bibinfo {author} {\bibfnamefont {Michael}\ \bibnamefont {Bonitz}},\ }\bibfield  {title} {\enquote {\bibinfo {title} {Permutation blocking path integral monte carlo approach to the uniform electron gas at finite temperature},}\ }\href {\doibase 10.1063/1.4936145} {\bibfield  {journal} {\bibinfo  {journal} {The Journal of Chemical Physics}\ }\textbf {\bibinfo {volume} {143}},\ \bibinfo {pages} {204101} (\bibinfo {year} {2015}{\natexlab{c}})}\BibitemShut {NoStop}%
\bibitem [{\citenamefont {Dornheim}\ \emph {et~al.}(2025{\natexlab{f}})\citenamefont {Dornheim}, \citenamefont {Moldabekov}, \citenamefont {Schwalbe},\ and\ \citenamefont {Vorberger}}]{Dornheim_PRB_2025}%
  \BibitemOpen
  \bibfield  {author} {\bibinfo {author} {\bibfnamefont {Tobias}\ \bibnamefont {Dornheim}}, \bibinfo {author} {\bibfnamefont {Zhandos~A.}\ \bibnamefont {Moldabekov}}, \bibinfo {author} {\bibfnamefont {Sebastian}\ \bibnamefont {Schwalbe}}, \ and\ \bibinfo {author} {\bibfnamefont {Jan}\ \bibnamefont {Vorberger}},\ }\bibfield  {title} {\enquote {\bibinfo {title} {Direct free energy calculation from ab initio path integral monte carlo simulations of warm dense matter},}\ }\href {\doibase 10.1103/PhysRevB.111.L041114} {\bibfield  {journal} {\bibinfo  {journal} {Phys. Rev. B}\ }\textbf {\bibinfo {volume} {111}},\ \bibinfo {pages} {L041114} (\bibinfo {year} {2025}{\natexlab{f}})}\BibitemShut {NoStop}%
\bibitem [{\citenamefont {Dornheim}\ \emph {et~al.}(2025{\natexlab{g}})\citenamefont {Dornheim}, \citenamefont {Tolias}, \citenamefont {Moldabekov},\ and\ \citenamefont {Vorberger}}]{Dornheim_PRR_2025}%
  \BibitemOpen
  \bibfield  {author} {\bibinfo {author} {\bibfnamefont {Tobias}\ \bibnamefont {Dornheim}}, \bibinfo {author} {\bibfnamefont {Panagiotis}\ \bibnamefont {Tolias}}, \bibinfo {author} {\bibfnamefont {Zhandos~A.}\ \bibnamefont {Moldabekov}}, \ and\ \bibinfo {author} {\bibfnamefont {Jan}\ \bibnamefont {Vorberger}},\ }\bibfield  {title} {\enquote {\bibinfo {title} {$\ensuremath{\eta}$-ensemble path integral monte carlo approach to the free energy of the warm dense electron gas and the uniform electron liquid},}\ }\href {\doibase 10.1103/4n7x-78fs} {\bibfield  {journal} {\bibinfo  {journal} {Phys. Rev. Res.}\ }\textbf {\bibinfo {volume} {7}},\ \bibinfo {pages} {023250} (\bibinfo {year} {2025}{\natexlab{g}})}\BibitemShut {NoStop}%
\bibitem [{\citenamefont {Dornheim}\ \emph {et~al.}(2025{\natexlab{h}})\citenamefont {Dornheim}, \citenamefont {Bonitz}, \citenamefont {Moldabekov}, \citenamefont {Schwalbe}, \citenamefont {Tolias},\ and\ \citenamefont {Vorberger}}]{Dornheim_PRB_ChemPot_2025}%
  \BibitemOpen
  \bibfield  {author} {\bibinfo {author} {\bibfnamefont {Tobias}\ \bibnamefont {Dornheim}}, \bibinfo {author} {\bibfnamefont {Michael}\ \bibnamefont {Bonitz}}, \bibinfo {author} {\bibfnamefont {Zhandos~A.}\ \bibnamefont {Moldabekov}}, \bibinfo {author} {\bibfnamefont {Sebastian}\ \bibnamefont {Schwalbe}}, \bibinfo {author} {\bibfnamefont {Panagiotis}\ \bibnamefont {Tolias}}, \ and\ \bibinfo {author} {\bibfnamefont {Jan}\ \bibnamefont {Vorberger}},\ }\bibfield  {title} {\enquote {\bibinfo {title} {Chemical potential of the warm dense electron gas from ab initio path integral monte carlo simulations},}\ }\href {\doibase 10.1103/PhysRevB.111.115149} {\bibfield  {journal} {\bibinfo  {journal} {Phys. Rev. B}\ }\textbf {\bibinfo {volume} {111}},\ \bibinfo {pages} {115149} (\bibinfo {year} {2025}{\natexlab{h}})}\BibitemShut {NoStop}%
\bibitem [{\citenamefont {Militzer}\ and\ \citenamefont {Pollock}(2002)}]{Militzer_Pollock_PRL_2002}%
  \BibitemOpen
  \bibfield  {author} {\bibinfo {author} {\bibfnamefont {Burkhard}\ \bibnamefont {Militzer}}\ and\ \bibinfo {author} {\bibfnamefont {E.~L.}\ \bibnamefont {Pollock}},\ }\bibfield  {title} {\enquote {\bibinfo {title} {Lowering of the kinetic energy in interacting quantum systems},}\ }\href {\doibase 10.1103/PhysRevLett.89.280401} {\bibfield  {journal} {\bibinfo  {journal} {Phys. Rev. Lett.}\ }\textbf {\bibinfo {volume} {89}},\ \bibinfo {pages} {280401} (\bibinfo {year} {2002})}\BibitemShut {NoStop}%
\bibitem [{\citenamefont {Dornheim}\ \emph {et~al.}(2021{\natexlab{c}})\citenamefont {Dornheim}, \citenamefont {B\"ohme}, \citenamefont {Militzer},\ and\ \citenamefont {Vorberger}}]{Dornheim_PRB_nk_2021}%
  \BibitemOpen
  \bibfield  {author} {\bibinfo {author} {\bibfnamefont {Tobias}\ \bibnamefont {Dornheim}}, \bibinfo {author} {\bibfnamefont {Maximilian}\ \bibnamefont {B\"ohme}}, \bibinfo {author} {\bibfnamefont {Burkhard}\ \bibnamefont {Militzer}}, \ and\ \bibinfo {author} {\bibfnamefont {Jan}\ \bibnamefont {Vorberger}},\ }\bibfield  {title} {\enquote {\bibinfo {title} {Ab initio path integral monte carlo approach to the momentum distribution of the uniform electron gas at finite temperature without fixed nodes},}\ }\href {\doibase 10.1103/PhysRevB.103.205142} {\bibfield  {journal} {\bibinfo  {journal} {Phys. Rev. B}\ }\textbf {\bibinfo {volume} {103}},\ \bibinfo {pages} {205142} (\bibinfo {year} {2021}{\natexlab{c}})}\BibitemShut {NoStop}%
\bibitem [{\citenamefont {Dornheim}\ \emph {et~al.}(2021{\natexlab{d}})\citenamefont {Dornheim}, \citenamefont {Vorberger}, \citenamefont {Militzer},\ and\ \citenamefont {Moldabekov}}]{Dornheim_PRE_2021}%
  \BibitemOpen
  \bibfield  {author} {\bibinfo {author} {\bibfnamefont {Tobias}\ \bibnamefont {Dornheim}}, \bibinfo {author} {\bibfnamefont {Jan}\ \bibnamefont {Vorberger}}, \bibinfo {author} {\bibfnamefont {Burkhard}\ \bibnamefont {Militzer}}, \ and\ \bibinfo {author} {\bibfnamefont {Zhandos~A.}\ \bibnamefont {Moldabekov}},\ }\bibfield  {title} {\enquote {\bibinfo {title} {Momentum distribution of the uniform electron gas at finite temperature: Effects of spin polarization},}\ }\href {\doibase 10.1103/PhysRevE.104.055206} {\bibfield  {journal} {\bibinfo  {journal} {Phys. Rev. E}\ }\textbf {\bibinfo {volume} {104}},\ \bibinfo {pages} {055206} (\bibinfo {year} {2021}{\natexlab{d}})}\BibitemShut {NoStop}%
\bibitem [{\citenamefont {Dornheim}\ \emph {et~al.}(2020{\natexlab{d}})\citenamefont {Dornheim}, \citenamefont {Sjostrom}, \citenamefont {Tanaka},\ and\ \citenamefont {Vorberger}}]{dornheim_electron_liquid}%
  \BibitemOpen
  \bibfield  {author} {\bibinfo {author} {\bibfnamefont {Tobias}\ \bibnamefont {Dornheim}}, \bibinfo {author} {\bibfnamefont {Travis}\ \bibnamefont {Sjostrom}}, \bibinfo {author} {\bibfnamefont {Shigenori}\ \bibnamefont {Tanaka}}, \ and\ \bibinfo {author} {\bibfnamefont {Jan}\ \bibnamefont {Vorberger}},\ }\bibfield  {title} {\enquote {\bibinfo {title} {Strongly coupled electron liquid: Ab initio path integral monte carlo simulations and dielectric theories},}\ }\href {\doibase 10.1103/PhysRevB.101.045129} {\bibfield  {journal} {\bibinfo  {journal} {Phys. Rev. B}\ }\textbf {\bibinfo {volume} {101}},\ \bibinfo {pages} {045129} (\bibinfo {year} {2020}{\natexlab{d}})}\BibitemShut {NoStop}%
\bibitem [{\citenamefont {Dornheim}\ \emph {et~al.}(2020{\natexlab{e}})\citenamefont {Dornheim}, \citenamefont {Moldabekov}, \citenamefont {Vorberger},\ and\ \citenamefont {Groth}}]{dornheim_HEDP}%
  \BibitemOpen
  \bibfield  {author} {\bibinfo {author} {\bibfnamefont {Tobias}\ \bibnamefont {Dornheim}}, \bibinfo {author} {\bibfnamefont {Zhandos~A}\ \bibnamefont {Moldabekov}}, \bibinfo {author} {\bibfnamefont {Jan}\ \bibnamefont {Vorberger}}, \ and\ \bibinfo {author} {\bibfnamefont {Simon}\ \bibnamefont {Groth}},\ }\bibfield  {title} {\enquote {\bibinfo {title} {Ab initio path integral monte carlo simulation of the uniform electron gas in the high energy density regime},}\ }\href {\doibase 10.1088/1361-6587/ab8bb4} {\bibfield  {journal} {\bibinfo  {journal} {Plasma Physics and Controlled Fusion}\ }\textbf {\bibinfo {volume} {62}},\ \bibinfo {pages} {075003} (\bibinfo {year} {2020}{\natexlab{e}})}\BibitemShut {NoStop}%
\bibitem [{\citenamefont {Dornheim}\ \emph {et~al.}(2022{\natexlab{c}})\citenamefont {Dornheim}, \citenamefont {Vorberger}, \citenamefont {Moldabekov}, \citenamefont {Röpke},\ and\ \citenamefont {Kraeft}}]{Dornheim_HEDP_2022}%
  \BibitemOpen
  \bibfield  {author} {\bibinfo {author} {\bibfnamefont {Tobias}\ \bibnamefont {Dornheim}}, \bibinfo {author} {\bibfnamefont {Jan}\ \bibnamefont {Vorberger}}, \bibinfo {author} {\bibfnamefont {Zhandos}\ \bibnamefont {Moldabekov}}, \bibinfo {author} {\bibfnamefont {Gerd}\ \bibnamefont {Röpke}}, \ and\ \bibinfo {author} {\bibfnamefont {Wolf-Dietrich}\ \bibnamefont {Kraeft}},\ }\bibfield  {title} {\enquote {\bibinfo {title} {The uniform electron gas at high temperatures: ab initio path integral monte carlo simulations and analytical theory},}\ }\href {\doibase https://doi.org/10.1016/j.hedp.2022.101015} {\bibfield  {journal} {\bibinfo  {journal} {High Energy Density Physics}\ }\textbf {\bibinfo {volume} {45}},\ \bibinfo {pages} {101015} (\bibinfo {year} {2022}{\natexlab{c}})}\BibitemShut {NoStop}%
\bibitem [{\citenamefont {Dornheim}\ \emph {et~al.}(2017{\natexlab{b}})\citenamefont {Dornheim}, \citenamefont {Groth}, \citenamefont {Vorberger},\ and\ \citenamefont {Bonitz}}]{dornheim_pre}%
  \BibitemOpen
  \bibfield  {author} {\bibinfo {author} {\bibfnamefont {T.}~\bibnamefont {Dornheim}}, \bibinfo {author} {\bibfnamefont {S.}~\bibnamefont {Groth}}, \bibinfo {author} {\bibfnamefont {J.}~\bibnamefont {Vorberger}}, \ and\ \bibinfo {author} {\bibfnamefont {M.}~\bibnamefont {Bonitz}},\ }\bibfield  {title} {\enquote {\bibinfo {title} {Permutation blocking path integral {M}onte {C}arlo approach to the static density response of the warm dense electron gas},}\ }\href {https://journals.aps.org/pre/abstract/10.1103/PhysRevE.96.023203} {\bibfield  {journal} {\bibinfo  {journal} {Phys. Rev. E}\ }\textbf {\bibinfo {volume} {96}},\ \bibinfo {pages} {023203} (\bibinfo {year} {2017}{\natexlab{b}})}\BibitemShut {NoStop}%
\bibitem [{\citenamefont {Groth}\ \emph {et~al.}(2017{\natexlab{b}})\citenamefont {Groth}, \citenamefont {Dornheim},\ and\ \citenamefont {Bonitz}}]{groth_jcp}%
  \BibitemOpen
  \bibfield  {author} {\bibinfo {author} {\bibfnamefont {S.}~\bibnamefont {Groth}}, \bibinfo {author} {\bibfnamefont {T.}~\bibnamefont {Dornheim}}, \ and\ \bibinfo {author} {\bibfnamefont {M.}~\bibnamefont {Bonitz}},\ }\bibfield  {title} {\enquote {\bibinfo {title} {Configuration path integral {M}onte {C}arlo approach to the static density response of the warm dense electron gas},}\ }\href {https://aip.scitation.org/doi/abs/10.1063/1.4999907} {\bibfield  {journal} {\bibinfo  {journal} {J. Chem. Phys}\ }\textbf {\bibinfo {volume} {147}},\ \bibinfo {pages} {164108} (\bibinfo {year} {2017}{\natexlab{b}})}\BibitemShut {NoStop}%
\bibitem [{\citenamefont {Dornheim}\ \emph {et~al.}(2021{\natexlab{e}})\citenamefont {Dornheim}, \citenamefont {B\"ohme}, \citenamefont {Moldabekov}, \citenamefont {Vorberger},\ and\ \citenamefont {Bonitz}}]{Dornheim_PRR_2021}%
  \BibitemOpen
  \bibfield  {author} {\bibinfo {author} {\bibfnamefont {Tobias}\ \bibnamefont {Dornheim}}, \bibinfo {author} {\bibfnamefont {Maximilian}\ \bibnamefont {B\"ohme}}, \bibinfo {author} {\bibfnamefont {Zhandos~A.}\ \bibnamefont {Moldabekov}}, \bibinfo {author} {\bibfnamefont {Jan}\ \bibnamefont {Vorberger}}, \ and\ \bibinfo {author} {\bibfnamefont {Michael}\ \bibnamefont {Bonitz}},\ }\bibfield  {title} {\enquote {\bibinfo {title} {Density response of the warm dense electron gas beyond linear response theory: Excitation of harmonics},}\ }\href {\doibase 10.1103/PhysRevResearch.3.033231} {\bibfield  {journal} {\bibinfo  {journal} {Phys. Rev. Research}\ }\textbf {\bibinfo {volume} {3}},\ \bibinfo {pages} {033231} (\bibinfo {year} {2021}{\natexlab{e}})}\BibitemShut {NoStop}%
\bibitem [{\citenamefont {Dornheim}\ \emph {et~al.}(2021{\natexlab{f}})\citenamefont {Dornheim}, \citenamefont {Vorberger},\ and\ \citenamefont {Moldabekov}}]{Dornheim_JPSJ_2021}%
  \BibitemOpen
  \bibfield  {author} {\bibinfo {author} {\bibfnamefont {Tobias}\ \bibnamefont {Dornheim}}, \bibinfo {author} {\bibfnamefont {Jan}\ \bibnamefont {Vorberger}}, \ and\ \bibinfo {author} {\bibfnamefont {Zhandos~A.}\ \bibnamefont {Moldabekov}},\ }\bibfield  {title} {\enquote {\bibinfo {title} {Nonlinear density response and higher order correlation functions in warm dense matter},}\ }\href {\doibase 10.7566/JPSJ.90.104002} {\bibfield  {journal} {\bibinfo  {journal} {Journal of the Physical Society of Japan}\ }\textbf {\bibinfo {volume} {90}},\ \bibinfo {pages} {104002} (\bibinfo {year} {2021}{\natexlab{f}})}\BibitemShut {NoStop}%
\bibitem [{\citenamefont {Dornheim}\ \emph {et~al.}(2021{\natexlab{g}})\citenamefont {Dornheim}, \citenamefont {Moldabekov},\ and\ \citenamefont {Vorberger}}]{Dornheim_CPP_2021}%
  \BibitemOpen
  \bibfield  {author} {\bibinfo {author} {\bibfnamefont {Tobias}\ \bibnamefont {Dornheim}}, \bibinfo {author} {\bibfnamefont {Zhandos~A.}\ \bibnamefont {Moldabekov}}, \ and\ \bibinfo {author} {\bibfnamefont {Jan}\ \bibnamefont {Vorberger}},\ }\bibfield  {title} {\enquote {\bibinfo {title} {Nonlinear electronic density response of the ferromagnetic uniform electron gas at warm dense matter conditions},}\ }\href {\doibase https://doi.org/10.1002/ctpp.202100098} {\bibfield  {journal} {\bibinfo  {journal} {Contributions to Plasma Physics}\ ,\ \bibinfo {pages} {e202100098}} (\bibinfo {year} {2021}{\natexlab{g}})}\BibitemShut {NoStop}%
\bibitem [{\citenamefont {Dornheim}\ \emph {et~al.}()\citenamefont {Dornheim}, \citenamefont {Vorberger}, \citenamefont {Moldabekov},\ and\ \citenamefont {Bonitz}}]{Dornheim_CPP_2022}%
  \BibitemOpen
  \bibfield  {author} {\bibinfo {author} {\bibfnamefont {Tobias}\ \bibnamefont {Dornheim}}, \bibinfo {author} {\bibfnamefont {Jan}\ \bibnamefont {Vorberger}}, \bibinfo {author} {\bibfnamefont {Zhandos~A.}\ \bibnamefont {Moldabekov}}, \ and\ \bibinfo {author} {\bibfnamefont {Michael}\ \bibnamefont {Bonitz}},\ }\bibfield  {title} {\enquote {\bibinfo {title} {Nonlinear interaction of external perturbations in warm dense matter},}\ }\href {\doibase https://doi.org/10.1002/ctpp.202100247} {\bibfield  {journal} {\bibinfo  {journal} {Contributions to Plasma Physics}\ }\textbf {\bibinfo {volume} {n/a}},\ \bibinfo {pages} {e202100247}}\BibitemShut {NoStop}%
\bibitem [{\citenamefont {Tolias}\ \emph {et~al.}(2023{\natexlab{b}})\citenamefont {Tolias}, \citenamefont {Dornheim}, \citenamefont {Moldabekov},\ and\ \citenamefont {Vorberger}}]{Tolias_EPL_2023}%
  \BibitemOpen
  \bibfield  {author} {\bibinfo {author} {\bibfnamefont {Panagiotis}\ \bibnamefont {Tolias}}, \bibinfo {author} {\bibfnamefont {Tobias}\ \bibnamefont {Dornheim}}, \bibinfo {author} {\bibfnamefont {Zhandos~A.}\ \bibnamefont {Moldabekov}}, \ and\ \bibinfo {author} {\bibfnamefont {Jan}\ \bibnamefont {Vorberger}},\ }\bibfield  {title} {\enquote {\bibinfo {title} {Unravelling the nonlinear ideal density response of many-body systems},}\ }\href {\doibase 10.1209/0295-5075/acd3a6} {\bibfield  {journal} {\bibinfo  {journal} {Europhysics Letters}\ }\textbf {\bibinfo {volume} {142}},\ \bibinfo {pages} {44001} (\bibinfo {year} {2023}{\natexlab{b}})}\BibitemShut {NoStop}%
\bibitem [{\citenamefont {Dornheim}\ and\ \citenamefont {Vorberger}(2020)}]{Dornheim_PRE_2020}%
  \BibitemOpen
  \bibfield  {author} {\bibinfo {author} {\bibfnamefont {Tobias}\ \bibnamefont {Dornheim}}\ and\ \bibinfo {author} {\bibfnamefont {Jan}\ \bibnamefont {Vorberger}},\ }\bibfield  {title} {\enquote {\bibinfo {title} {Finite-size effects in the reconstruction of dynamic properties from ab initio path integral monte carlo simulations},}\ }\href {\doibase 10.1103/PhysRevE.102.063301} {\bibfield  {journal} {\bibinfo  {journal} {Phys. Rev. E}\ }\textbf {\bibinfo {volume} {102}},\ \bibinfo {pages} {063301} (\bibinfo {year} {2020})}\BibitemShut {NoStop}%
\bibitem [{\citenamefont {Chuna}\ \emph {et~al.}(2025{\natexlab{b}})\citenamefont {Chuna}, \citenamefont {Barnfield}, \citenamefont {Vorberger}, \citenamefont {Friedlander}, \citenamefont {Hoheisel},\ and\ \citenamefont {Dornheim}}]{Chuna_PRB_2025}%
  \BibitemOpen
  \bibfield  {author} {\bibinfo {author} {\bibfnamefont {Thomas}\ \bibnamefont {Chuna}}, \bibinfo {author} {\bibfnamefont {Nicholas}\ \bibnamefont {Barnfield}}, \bibinfo {author} {\bibfnamefont {Jan}\ \bibnamefont {Vorberger}}, \bibinfo {author} {\bibfnamefont {Michael~P.}\ \bibnamefont {Friedlander}}, \bibinfo {author} {\bibfnamefont {Tim}\ \bibnamefont {Hoheisel}}, \ and\ \bibinfo {author} {\bibfnamefont {Tobias}\ \bibnamefont {Dornheim}},\ }\bibfield  {title} {\enquote {\bibinfo {title} {Estimates of the dynamic structure factor for the finite temperature electron liquid via analytic continuation of path integral monte carlo data},}\ }\href {\doibase 10.1103/4d4b-kgtk} {\bibfield  {journal} {\bibinfo  {journal} {Phys. Rev. B}\ }\textbf {\bibinfo {volume} {112}},\ \bibinfo {pages} {125112} (\bibinfo {year} {2025}{\natexlab{b}})}\BibitemShut {NoStop}%
\bibitem [{\citenamefont {Robles}\ \emph {et~al.}(2025)\citenamefont {Robles}, \citenamefont {Hofmann}, \citenamefont {Chuna}, \citenamefont {Dornheim},\ and\ \citenamefont {Hecht}}]{robles2025pylitreformulationimplementationanalytic}%
  \BibitemOpen
  \bibfield  {author} {\bibinfo {author} {\bibfnamefont {Alexander~Benedix}\ \bibnamefont {Robles}}, \bibinfo {author} {\bibfnamefont {Phil-Alexander}\ \bibnamefont {Hofmann}}, \bibinfo {author} {\bibfnamefont {Thomas}\ \bibnamefont {Chuna}}, \bibinfo {author} {\bibfnamefont {Tobias}\ \bibnamefont {Dornheim}}, \ and\ \bibinfo {author} {\bibfnamefont {Michael}\ \bibnamefont {Hecht}},\ }\href {https://arxiv.org/abs/2505.10211} {\enquote {\bibinfo {title} {Pylit: Reformulation and implementation of the analytic continuation problem using kernel representation methods},}\ } (\bibinfo {year} {2025}),\ \Eprint {http://arxiv.org/abs/2505.10211} {arXiv:2505.10211 [physics.comp-ph]} \BibitemShut {NoStop}%
\bibitem [{\citenamefont {Hamann}\ \emph {et~al.}(2020{\natexlab{a}})\citenamefont {Hamann}, \citenamefont {Vorberger}, \citenamefont {Dornheim}, \citenamefont {Moldabekov},\ and\ \citenamefont {Bonitz}}]{Hamann_CPP_2020}%
  \BibitemOpen
  \bibfield  {author} {\bibinfo {author} {\bibfnamefont {Paul}\ \bibnamefont {Hamann}}, \bibinfo {author} {\bibfnamefont {Jan}\ \bibnamefont {Vorberger}}, \bibinfo {author} {\bibfnamefont {Tobias}\ \bibnamefont {Dornheim}}, \bibinfo {author} {\bibfnamefont {Zhandos~A.}\ \bibnamefont {Moldabekov}}, \ and\ \bibinfo {author} {\bibfnamefont {Michael}\ \bibnamefont {Bonitz}},\ }\bibfield  {title} {\enquote {\bibinfo {title} {Ab initio results for the plasmon dispersion and damping of the warm dense electron gas},}\ }\href {\doibase https://doi.org/10.1002/ctpp.202000147} {\bibfield  {journal} {\bibinfo  {journal} {Contributions to Plasma Physics}\ }\textbf {\bibinfo {volume} {60}},\ \bibinfo {pages} {e202000147} (\bibinfo {year} {2020}{\natexlab{a}})}\BibitemShut {NoStop}%
\bibitem [{\citenamefont {Hamann}\ \emph {et~al.}(2020{\natexlab{b}})\citenamefont {Hamann}, \citenamefont {Dornheim}, \citenamefont {Vorberger}, \citenamefont {Moldabekov},\ and\ \citenamefont {Bonitz}}]{Hamann_PRB_2020}%
  \BibitemOpen
  \bibfield  {author} {\bibinfo {author} {\bibfnamefont {Paul}\ \bibnamefont {Hamann}}, \bibinfo {author} {\bibfnamefont {Tobias}\ \bibnamefont {Dornheim}}, \bibinfo {author} {\bibfnamefont {Jan}\ \bibnamefont {Vorberger}}, \bibinfo {author} {\bibfnamefont {Zhandos~A.}\ \bibnamefont {Moldabekov}}, \ and\ \bibinfo {author} {\bibfnamefont {Michael}\ \bibnamefont {Bonitz}},\ }\bibfield  {title} {\enquote {\bibinfo {title} {Dynamic properties of the warm dense electron gas based on $ab initio$ path integral monte carlo simulations},}\ }\href {\doibase 10.1103/PhysRevB.102.125150} {\bibfield  {journal} {\bibinfo  {journal} {Phys. Rev. B}\ }\textbf {\bibinfo {volume} {102}},\ \bibinfo {pages} {125150} (\bibinfo {year} {2020}{\natexlab{b}})}\BibitemShut {NoStop}%
\bibitem [{\citenamefont {Marienhagen}\ and\ \citenamefont {Meier}(2025)}]{Marienhagen_JCP_2025}%
  \BibitemOpen
  \bibfield  {author} {\bibinfo {author} {\bibfnamefont {Philipp}\ \bibnamefont {Marienhagen}}\ and\ \bibinfo {author} {\bibfnamefont {Karsten}\ \bibnamefont {Meier}},\ }\bibfield  {title} {\enquote {\bibinfo {title} {Calculation of thermodynamic properties using path integral monte carlo simulations in the canonical ensemble},}\ }\href {\doibase 10.1063/5.0282863} {\bibfield  {journal} {\bibinfo  {journal} {The Journal of Chemical Physics}\ }\textbf {\bibinfo {volume} {163}},\ \bibinfo {pages} {074116} (\bibinfo {year} {2025})}\BibitemShut {NoStop}%
\bibitem [{\citenamefont {Mezzacapo}\ and\ \citenamefont {Boninsegni}(2007)}]{mezza}%
  \BibitemOpen
  \bibfield  {author} {\bibinfo {author} {\bibfnamefont {F.}~\bibnamefont {Mezzacapo}}\ and\ \bibinfo {author} {\bibfnamefont {M.}~\bibnamefont {Boninsegni}},\ }\bibfield  {title} {\enquote {\bibinfo {title} {Structure, superfluidity, and quantum melting of hydrogen clusters},}\ }\href {https://journals.aps.org/pra/abstract/10.1103/PhysRevA.75.033201} {\bibfield  {journal} {\bibinfo  {journal} {Phys. Rev. A}\ }\textbf {\bibinfo {volume} {75}},\ \bibinfo {pages} {033201} (\bibinfo {year} {2007})}\BibitemShut {NoStop}%
\bibitem [{\citenamefont {Hirshberg}\ \emph {et~al.}(2019)\citenamefont {Hirshberg}, \citenamefont {Rizzi},\ and\ \citenamefont {Parrinello}}]{Hirshberg_PNAS_2019}%
  \BibitemOpen
  \bibfield  {author} {\bibinfo {author} {\bibfnamefont {Barak}\ \bibnamefont {Hirshberg}}, \bibinfo {author} {\bibfnamefont {Valerio}\ \bibnamefont {Rizzi}}, \ and\ \bibinfo {author} {\bibfnamefont {Michele}\ \bibnamefont {Parrinello}},\ }\bibfield  {title} {\enquote {\bibinfo {title} {Path integral molecular dynamics for bosons},}\ }\href {\doibase 10.1073/pnas.1913365116} {\bibfield  {journal} {\bibinfo  {journal} {Proceedings of the National Academy of Sciences}\ }\textbf {\bibinfo {volume} {116}},\ \bibinfo {pages} {21445--21449} (\bibinfo {year} {2019})}\BibitemShut {NoStop}%
\bibitem [{\citenamefont {Apostol}(1967)}]{apostol_book}%
  \BibitemOpen
  \bibfield  {author} {\bibinfo {author} {\bibfnamefont {T.~M.}\ \bibnamefont {Apostol}},\ }\href@noop {} {\emph {\bibinfo {title} {Calculus Volume I}}}\ (\bibinfo  {publisher} {John Wiley \& Sons, New York, US},\ \bibinfo {year} {1967})\BibitemShut {NoStop}%
\bibitem [{\citenamefont {Comtet}(1967)}]{comtet_book}%
  \BibitemOpen
  \bibfield  {author} {\bibinfo {author} {\bibfnamefont {L.}~\bibnamefont {Comtet}},\ }\href@noop {} {\emph {\bibinfo {title} {Advanced Combinatorics}}}\ (\bibinfo  {publisher} {D. Reidel Publishing Company, Dordrecht, Holland},\ \bibinfo {year} {1967})\BibitemShut {NoStop}%
\bibitem [{\citenamefont {Dornheim}\ \emph {et~al.}(2019{\natexlab{b}})\citenamefont {Dornheim}, \citenamefont {Groth}, \citenamefont {Filinov},\ and\ \citenamefont {Bonitz}}]{Dornheim_permutation_cycles}%
  \BibitemOpen
  \bibfield  {author} {\bibinfo {author} {\bibfnamefont {T.}~\bibnamefont {Dornheim}}, \bibinfo {author} {\bibfnamefont {S.}~\bibnamefont {Groth}}, \bibinfo {author} {\bibfnamefont {A.~V.}\ \bibnamefont {Filinov}}, \ and\ \bibinfo {author} {\bibfnamefont {M.}~\bibnamefont {Bonitz}},\ }\bibfield  {title} {\enquote {\bibinfo {title} {Path integral monte carlo simulation of degenerate electrons: Permutation-cycle properties},}\ }\href {\doibase 10.1063/1.5093171} {\bibfield  {journal} {\bibinfo  {journal} {J. Chem. Phys.}\ }\textbf {\bibinfo {volume} {151}},\ \bibinfo {pages} {014108} (\bibinfo {year} {2019}{\natexlab{b}})}\BibitemShut {NoStop}%
\bibitem [{\citenamefont {DuBois}\ \emph {et~al.}()\citenamefont {DuBois}, \citenamefont {Brown},\ and\ \citenamefont {Alder}}]{DuBois}%
  \BibitemOpen
  \bibfield  {author} {\bibinfo {author} {\bibfnamefont {Jonathan~L}\ \bibnamefont {DuBois}}, \bibinfo {author} {\bibfnamefont {Ethan~W.}\ \bibnamefont {Brown}}, \ and\ \bibinfo {author} {\bibfnamefont {Berni~J.}\ \bibnamefont {Alder}},\ }\enquote {\bibinfo {title} {Overcoming the fermion sign problem in homogeneous systems},}\ in\ \href {\doibase 10.1142/9789813209428_0013} {\emph {\bibinfo {booktitle} {Advances in the Computational Sciences}}},\ Chap.~\bibinfo {chapter} {13}, pp.\ \bibinfo {pages} {184--192}\BibitemShut {NoStop}%
\bibitem [{\citenamefont {Lyubartsev}\ and\ \citenamefont {Vorontsov-Velyaminov}(1993)}]{Vorontsov_PRA_1993}%
  \BibitemOpen
  \bibfield  {author} {\bibinfo {author} {\bibfnamefont {A.~P.}\ \bibnamefont {Lyubartsev}}\ and\ \bibinfo {author} {\bibfnamefont {P.~N.}\ \bibnamefont {Vorontsov-Velyaminov}},\ }\bibfield  {title} {\enquote {\bibinfo {title} {Path-integral monte carlo method in quantum statistics for a system of n identical fermions},}\ }\href {\doibase 10.1103/PhysRevA.48.4075} {\bibfield  {journal} {\bibinfo  {journal} {Phys. Rev. A}\ }\textbf {\bibinfo {volume} {48}},\ \bibinfo {pages} {4075--4083} (\bibinfo {year} {1993})}\BibitemShut {NoStop}%
\bibitem [{\citenamefont {Dornheim}\ \emph {et~al.}(2024{\natexlab{e}})\citenamefont {Dornheim}, \citenamefont {Böhme},\ and\ \citenamefont {Schwalbe}}]{ISHTAR}%
  \BibitemOpen
  \bibfield  {author} {\bibinfo {author} {\bibfnamefont {Tobias}\ \bibnamefont {Dornheim}}, \bibinfo {author} {\bibfnamefont {Maximilian}\ \bibnamefont {Böhme}}, \ and\ \bibinfo {author} {\bibfnamefont {Sebastian}\ \bibnamefont {Schwalbe}},\ }\href {\doibase 10.5281/zenodo.10497098} {\enquote {\bibinfo {title} {{ISHTAR - Imaginary-time Stochastic High- performance Tool for Ab initio Research}},}\ } (\bibinfo {year} {2024}{\natexlab{e}})\BibitemShut {NoStop}%
\bibitem [{rep()}]{repo}%
  \BibitemOpen
  \href@noop {} {}\bibinfo {note} {A link to a repository containing all PIMC results will be made available upon publication.}\BibitemShut {Stop}%
\bibitem [{\citenamefont {Hunger}\ \emph {et~al.}()\citenamefont {Hunger}, \citenamefont {Yilmaz},\ and\ \citenamefont {Hamann}}]{cpimc.jl}%
  \BibitemOpen
  \bibfield  {author} {\bibinfo {author} {\bibfnamefont {Kai}\ \bibnamefont {Hunger}}, \bibinfo {author} {\bibfnamefont {Arif}\ \bibnamefont {Yilmaz}}, \ and\ \bibinfo {author} {\bibfnamefont {Paul}\ \bibnamefont {Hamann}},\ }\href {https://github.com/CPIMC/CPIMC.jl/} {\enquote {\bibinfo {title} {{CPIMC.jl - Implementation of Configuration path-integral Monte Carlo (CPIMC) in Julia.}}}\ }\BibitemShut {NoStop}%
\bibitem [{\citenamefont {D{\"o}ppner}\ \emph {et~al.}(2023)\citenamefont {D{\"o}ppner}, \citenamefont {Bethkenhagen}, \citenamefont {Kraus}, \citenamefont {Neumayer}, \citenamefont {Chapman}, \citenamefont {Bachmann}, \citenamefont {Baggott}, \citenamefont {B{\"o}hme}, \citenamefont {Divol}, \citenamefont {Falcone}, \citenamefont {Fletcher}, \citenamefont {Landen}, \citenamefont {MacDonald}, \citenamefont {Saunders}, \citenamefont {Sch{\"o}rner}, \citenamefont {Sterne}, \citenamefont {Vorberger}, \citenamefont {Witte}, \citenamefont {Yi}, \citenamefont {Redmer}, \citenamefont {Glenzer},\ and\ \citenamefont {Gericke}}]{Tilo_Nature_2023}%
  \BibitemOpen
  \bibfield  {author} {\bibinfo {author} {\bibfnamefont {T.}~\bibnamefont {D{\"o}ppner}}, \bibinfo {author} {\bibfnamefont {M.}~\bibnamefont {Bethkenhagen}}, \bibinfo {author} {\bibfnamefont {D.}~\bibnamefont {Kraus}}, \bibinfo {author} {\bibfnamefont {P.}~\bibnamefont {Neumayer}}, \bibinfo {author} {\bibfnamefont {D.~A.}\ \bibnamefont {Chapman}}, \bibinfo {author} {\bibfnamefont {B.}~\bibnamefont {Bachmann}}, \bibinfo {author} {\bibfnamefont {R.~A.}\ \bibnamefont {Baggott}}, \bibinfo {author} {\bibfnamefont {M.~P.}\ \bibnamefont {B{\"o}hme}}, \bibinfo {author} {\bibfnamefont {L.}~\bibnamefont {Divol}}, \bibinfo {author} {\bibfnamefont {R.~W.}\ \bibnamefont {Falcone}}, \bibinfo {author} {\bibfnamefont {L.~B.}\ \bibnamefont {Fletcher}}, \bibinfo {author} {\bibfnamefont {O.~L.}\ \bibnamefont {Landen}}, \bibinfo {author} {\bibfnamefont {M.~J.}\ \bibnamefont {MacDonald}}, \bibinfo {author} {\bibfnamefont {A.~M.}\ \bibnamefont {Saunders}}, \bibinfo {author} {\bibfnamefont {M.}~\bibnamefont {Sch{\"o}rner}},
  \bibinfo {author} {\bibfnamefont {P.~A.}\ \bibnamefont {Sterne}}, \bibinfo {author} {\bibfnamefont {J.}~\bibnamefont {Vorberger}}, \bibinfo {author} {\bibfnamefont {B.~B.~L.}\ \bibnamefont {Witte}}, \bibinfo {author} {\bibfnamefont {A.}~\bibnamefont {Yi}}, \bibinfo {author} {\bibfnamefont {R.}~\bibnamefont {Redmer}}, \bibinfo {author} {\bibfnamefont {S.~H.}\ \bibnamefont {Glenzer}}, \ and\ \bibinfo {author} {\bibfnamefont {D.~O.}\ \bibnamefont {Gericke}},\ }\bibfield  {title} {\enquote {\bibinfo {title} {Observing the onset of pressure-driven k-shell delocalization},}\ }\href {\doibase 10.1038/s41586-023-05996-8} {\bibfield  {journal} {\bibinfo  {journal} {Nature}\ }\textbf {\bibinfo {volume} {618}},\ \bibinfo {pages} {270–275} (\bibinfo {year} {2023})}\BibitemShut {NoStop}%
\bibitem [{\citenamefont {Moses}\ \emph {et~al.}(2009)\citenamefont {Moses}, \citenamefont {Boyd}, \citenamefont {Remington}, \citenamefont {Keane},\ and\ \citenamefont {Al-Ayat}}]{Moses_NIF}%
  \BibitemOpen
  \bibfield  {author} {\bibinfo {author} {\bibfnamefont {E.~I.}\ \bibnamefont {Moses}}, \bibinfo {author} {\bibfnamefont {R.~N.}\ \bibnamefont {Boyd}}, \bibinfo {author} {\bibfnamefont {B.~A.}\ \bibnamefont {Remington}}, \bibinfo {author} {\bibfnamefont {C.~J.}\ \bibnamefont {Keane}}, \ and\ \bibinfo {author} {\bibfnamefont {R.}~\bibnamefont {Al-Ayat}},\ }\bibfield  {title} {\enquote {\bibinfo {title} {The national ignition facility: Ushering in a new age for high energy density science},}\ }\href {\doibase 10.1063/1.3116505} {\bibfield  {journal} {\bibinfo  {journal} {Phys. Plasmas}\ }\textbf {\bibinfo {volume} {16}},\ \bibinfo {pages} {041006} (\bibinfo {year} {2009})}\BibitemShut {NoStop}%
\bibitem [{\citenamefont {Hatano}(1994)}]{hatano1994data}%
  \BibitemOpen
  \bibfield  {author} {\bibinfo {author} {\bibfnamefont {Naomichi}\ \bibnamefont {Hatano}},\ }\bibfield  {title} {\enquote {\bibinfo {title} {Data analysis for quantum monte carlo simulations with the negative-sign problem},}\ }\href@noop {} {\bibfield  {journal} {\bibinfo  {journal} {Journal of the Physical Society of Japan}\ }\textbf {\bibinfo {volume} {63}},\ \bibinfo {pages} {1691--1697} (\bibinfo {year} {1994})}\BibitemShut {NoStop}%
\bibitem [{\citenamefont {Berg}(2004)}]{berg2004markov}%
  \BibitemOpen
  \bibfield  {author} {\bibinfo {author} {\bibfnamefont {Bernd~A}\ \bibnamefont {Berg}},\ }\href@noop {} {\emph {\bibinfo {title} {Markov chain Monte Carlo simulations and their statistical analysis: with web-based Fortran code}}}\ (\bibinfo  {publisher} {World Scientific Publishing Company},\ \bibinfo {year} {2004})\BibitemShut {NoStop}%
\bibitem [{\citenamefont {Janke}\ and\ \citenamefont {Sauer}(1997)}]{Janke_JCP_1997}%
  \BibitemOpen
  \bibfield  {author} {\bibinfo {author} {\bibfnamefont {Wolfhard}\ \bibnamefont {Janke}}\ and\ \bibinfo {author} {\bibfnamefont {Tilman}\ \bibnamefont {Sauer}},\ }\bibfield  {title} {\enquote {\bibinfo {title} {Optimal energy estimation in path-integral monte carlo simulations},}\ }\href {\doibase 10.1063/1.474309} {\bibfield  {journal} {\bibinfo  {journal} {The Journal of Chemical Physics}\ }\textbf {\bibinfo {volume} {107}},\ \bibinfo {pages} {5821--5839} (\bibinfo {year} {1997})}\BibitemShut {NoStop}%
\bibitem [{\citenamefont {He}\ \emph {et~al.}(2025)\citenamefont {He}, \citenamefont {Zeng}, \citenamefont {Yang}, \citenamefont {Wang}, \citenamefont {Ye},\ and\ \citenamefont {Li}}]{he2025revisitingfermionsignproblem}%
  \BibitemOpen
  \bibfield  {author} {\bibinfo {author} {\bibfnamefont {Ran-Chen}\ \bibnamefont {He}}, \bibinfo {author} {\bibfnamefont {Jia-Xi}\ \bibnamefont {Zeng}}, \bibinfo {author} {\bibfnamefont {Shu}\ \bibnamefont {Yang}}, \bibinfo {author} {\bibfnamefont {Cong}\ \bibnamefont {Wang}}, \bibinfo {author} {\bibfnamefont {Qi-Jun}\ \bibnamefont {Ye}}, \ and\ \bibinfo {author} {\bibfnamefont {Xin-Zheng}\ \bibnamefont {Li}},\ }\href {https://arxiv.org/abs/2507.22779} {\enquote {\bibinfo {title} {Revisiting the fermion sign problem from the structure of lee-yang zeros. i. the form of partition function for indistinguishable particles and its zeros at 0~k},}\ } (\bibinfo {year} {2025}),\ \Eprint {http://arxiv.org/abs/2507.22779} {arXiv:2507.22779 [cond-mat.stat-mech]} \BibitemShut {NoStop}%
\bibitem [{\citenamefont {Krauth}(2006)}]{krauth2006statistical}%
  \BibitemOpen
  \bibfield  {author} {\bibinfo {author} {\bibfnamefont {W.}~\bibnamefont {Krauth}},\ }\href {https://books.google.de/books?id=B3koVucDyKUC} {\emph {\bibinfo {title} {Statistical Mechanics: Algorithms and Computations}}},\ Oxford Master Series in Physics\ (\bibinfo  {publisher} {Oxford University Press, UK},\ \bibinfo {year} {2006})\BibitemShut {NoStop}%
\bibitem [{\citenamefont {Takada}(2016)}]{Takada_PRB_2016}%
  \BibitemOpen
  \bibfield  {author} {\bibinfo {author} {\bibfnamefont {Yasutami}\ \bibnamefont {Takada}},\ }\bibfield  {title} {\enquote {\bibinfo {title} {Emergence of an excitonic collective mode in the dilute electron gas},}\ }\href {\doibase 10.1103/PhysRevB.94.245106} {\bibfield  {journal} {\bibinfo  {journal} {Phys. Rev. B}\ }\textbf {\bibinfo {volume} {94}},\ \bibinfo {pages} {245106} (\bibinfo {year} {2016})}\BibitemShut {NoStop}%
\bibitem [{\citenamefont {Fletcher}\ \emph {et~al.}(2022)\citenamefont {Fletcher}, \citenamefont {Vorberger}, \citenamefont {Schumaker}, \citenamefont {Ruyer}, \citenamefont {Goede}, \citenamefont {Galtier}, \citenamefont {Zastrau}, \citenamefont {Alves}, \citenamefont {Baalrud}, \citenamefont {Baggott}, \citenamefont {Barbrel}, \citenamefont {Chen}, \citenamefont {Döppner}, \citenamefont {Gauthier}, \citenamefont {Granados}, \citenamefont {Kim}, \citenamefont {Kraus}, \citenamefont {Lee}, \citenamefont {MacDonald}, \citenamefont {Mishra}, \citenamefont {Pelka}, \citenamefont {Ravasio}, \citenamefont {Roedel}, \citenamefont {Fry}, \citenamefont {Redmer}, \citenamefont {Fiuza}, \citenamefont {Gericke},\ and\ \citenamefont {Glenzer}}]{Fletcher_Frontiers_2022}%
  \BibitemOpen
  \bibfield  {author} {\bibinfo {author} {\bibfnamefont {L.~B.}\ \bibnamefont {Fletcher}}, \bibinfo {author} {\bibfnamefont {J.}~\bibnamefont {Vorberger}}, \bibinfo {author} {\bibfnamefont {W.}~\bibnamefont {Schumaker}}, \bibinfo {author} {\bibfnamefont {C.}~\bibnamefont {Ruyer}}, \bibinfo {author} {\bibfnamefont {S.}~\bibnamefont {Goede}}, \bibinfo {author} {\bibfnamefont {E.}~\bibnamefont {Galtier}}, \bibinfo {author} {\bibfnamefont {U.}~\bibnamefont {Zastrau}}, \bibinfo {author} {\bibfnamefont {E.~P.}\ \bibnamefont {Alves}}, \bibinfo {author} {\bibfnamefont {S.~D.}\ \bibnamefont {Baalrud}}, \bibinfo {author} {\bibfnamefont {R.~A.}\ \bibnamefont {Baggott}}, \bibinfo {author} {\bibfnamefont {B.}~\bibnamefont {Barbrel}}, \bibinfo {author} {\bibfnamefont {Z.}~\bibnamefont {Chen}}, \bibinfo {author} {\bibfnamefont {T.}~\bibnamefont {Döppner}}, \bibinfo {author} {\bibfnamefont {M.}~\bibnamefont {Gauthier}}, \bibinfo {author} {\bibfnamefont {E.}~\bibnamefont {Granados}}, \bibinfo {author} {\bibfnamefont {J.~B.}\
  \bibnamefont {Kim}}, \bibinfo {author} {\bibfnamefont {D.}~\bibnamefont {Kraus}}, \bibinfo {author} {\bibfnamefont {H.~J.}\ \bibnamefont {Lee}}, \bibinfo {author} {\bibfnamefont {M.~J.}\ \bibnamefont {MacDonald}}, \bibinfo {author} {\bibfnamefont {R.}~\bibnamefont {Mishra}}, \bibinfo {author} {\bibfnamefont {A.}~\bibnamefont {Pelka}}, \bibinfo {author} {\bibfnamefont {A.}~\bibnamefont {Ravasio}}, \bibinfo {author} {\bibfnamefont {C.}~\bibnamefont {Roedel}}, \bibinfo {author} {\bibfnamefont {A.~R.}\ \bibnamefont {Fry}}, \bibinfo {author} {\bibfnamefont {R.}~\bibnamefont {Redmer}}, \bibinfo {author} {\bibfnamefont {F.}~\bibnamefont {Fiuza}}, \bibinfo {author} {\bibfnamefont {D.~O.}\ \bibnamefont {Gericke}}, \ and\ \bibinfo {author} {\bibfnamefont {S.~H.}\ \bibnamefont {Glenzer}},\ }\bibfield  {title} {\enquote {\bibinfo {title} {Electron-ion temperature relaxation in warm dense hydrogen observed with picosecond resolved x-ray scattering},}\ }\href {\doibase 10.3389/fphy.2022.838524} {\bibfield  {journal}
  {\bibinfo  {journal} {Frontiers in Physics}\ }\textbf {\bibinfo {volume} {10}} (\bibinfo {year} {2022}),\ 10.3389/fphy.2022.838524}\BibitemShut {NoStop}%
\bibitem [{\citenamefont {Zastrau}\ \emph {et~al.}(2014)\citenamefont {Zastrau}, \citenamefont {Sperling}, \citenamefont {Harmand}, \citenamefont {Becker}, \citenamefont {Bornath}, \citenamefont {Bredow}, \citenamefont {Dziarzhytski}, \citenamefont {Fennel}, \citenamefont {Fletcher}, \citenamefont {F\"orster}, \citenamefont {G\"ode}, \citenamefont {Gregori}, \citenamefont {Hilbert}, \citenamefont {Hochhaus}, \citenamefont {Holst}, \citenamefont {Laarmann}, \citenamefont {Lee}, \citenamefont {Ma}, \citenamefont {Mithen}, \citenamefont {Mitzner}, \citenamefont {Murphy}, \citenamefont {Nakatsutsumi}, \citenamefont {Neumayer}, \citenamefont {Przystawik}, \citenamefont {Roling}, \citenamefont {Schulz}, \citenamefont {Siemer}, \citenamefont {Skruszewicz}, \citenamefont {Tiggesb\"aumker}, \citenamefont {Toleikis}, \citenamefont {Tschentscher}, \citenamefont {White}, \citenamefont {W\"ostmann}, \citenamefont {Zacharias}, \citenamefont {D\"oppner}, \citenamefont {Glenzer},\ and\ \citenamefont {Redmer}}]{zastrau}%
  \BibitemOpen
  \bibfield  {author} {\bibinfo {author} {\bibfnamefont {U.}~\bibnamefont {Zastrau}}, \bibinfo {author} {\bibfnamefont {P.}~\bibnamefont {Sperling}}, \bibinfo {author} {\bibfnamefont {M.}~\bibnamefont {Harmand}}, \bibinfo {author} {\bibfnamefont {A.}~\bibnamefont {Becker}}, \bibinfo {author} {\bibfnamefont {T.}~\bibnamefont {Bornath}}, \bibinfo {author} {\bibfnamefont {R.}~\bibnamefont {Bredow}}, \bibinfo {author} {\bibfnamefont {S.}~\bibnamefont {Dziarzhytski}}, \bibinfo {author} {\bibfnamefont {T.}~\bibnamefont {Fennel}}, \bibinfo {author} {\bibfnamefont {L.~B.}\ \bibnamefont {Fletcher}}, \bibinfo {author} {\bibfnamefont {E.}~\bibnamefont {F\"orster}}, \bibinfo {author} {\bibfnamefont {S.}~\bibnamefont {G\"ode}}, \bibinfo {author} {\bibfnamefont {G.}~\bibnamefont {Gregori}}, \bibinfo {author} {\bibfnamefont {V.}~\bibnamefont {Hilbert}}, \bibinfo {author} {\bibfnamefont {D.}~\bibnamefont {Hochhaus}}, \bibinfo {author} {\bibfnamefont {B.}~\bibnamefont {Holst}}, \bibinfo {author} {\bibfnamefont
  {T.}~\bibnamefont {Laarmann}}, \bibinfo {author} {\bibfnamefont {H.~J.}\ \bibnamefont {Lee}}, \bibinfo {author} {\bibfnamefont {T.}~\bibnamefont {Ma}}, \bibinfo {author} {\bibfnamefont {J.~P.}\ \bibnamefont {Mithen}}, \bibinfo {author} {\bibfnamefont {R.}~\bibnamefont {Mitzner}}, \bibinfo {author} {\bibfnamefont {C.~D.}\ \bibnamefont {Murphy}}, \bibinfo {author} {\bibfnamefont {M.}~\bibnamefont {Nakatsutsumi}}, \bibinfo {author} {\bibfnamefont {P.}~\bibnamefont {Neumayer}}, \bibinfo {author} {\bibfnamefont {A.}~\bibnamefont {Przystawik}}, \bibinfo {author} {\bibfnamefont {S.}~\bibnamefont {Roling}}, \bibinfo {author} {\bibfnamefont {M.}~\bibnamefont {Schulz}}, \bibinfo {author} {\bibfnamefont {B.}~\bibnamefont {Siemer}}, \bibinfo {author} {\bibfnamefont {S.}~\bibnamefont {Skruszewicz}}, \bibinfo {author} {\bibfnamefont {J.}~\bibnamefont {Tiggesb\"aumker}}, \bibinfo {author} {\bibfnamefont {S.}~\bibnamefont {Toleikis}}, \bibinfo {author} {\bibfnamefont {T.}~\bibnamefont {Tschentscher}}, \bibinfo {author}
  {\bibfnamefont {T.}~\bibnamefont {White}}, \bibinfo {author} {\bibfnamefont {M.}~\bibnamefont {W\"ostmann}}, \bibinfo {author} {\bibfnamefont {H.}~\bibnamefont {Zacharias}}, \bibinfo {author} {\bibfnamefont {T.}~\bibnamefont {D\"oppner}}, \bibinfo {author} {\bibfnamefont {S.~H.}\ \bibnamefont {Glenzer}}, \ and\ \bibinfo {author} {\bibfnamefont {R.}~\bibnamefont {Redmer}},\ }\bibfield  {title} {\enquote {\bibinfo {title} {Resolving ultrafast heating of dense cryogenic hydrogen},}\ }\href {https://journals.aps.org/prl/abstract/10.1103/PhysRevLett.112.105002} {\bibfield  {journal} {\bibinfo  {journal} {Phys. Rev. Lett}\ }\textbf {\bibinfo {volume} {112}},\ \bibinfo {pages} {105002} (\bibinfo {year} {2014})}\BibitemShut {NoStop}%
\bibitem [{\citenamefont {Kwon}\ \emph {et~al.}(2006)\citenamefont {Kwon}, \citenamefont {Paesani},\ and\ \citenamefont {Whaley}}]{Kwon_PRB_2006}%
  \BibitemOpen
  \bibfield  {author} {\bibinfo {author} {\bibfnamefont {Yongkyung}\ \bibnamefont {Kwon}}, \bibinfo {author} {\bibfnamefont {Francesco}\ \bibnamefont {Paesani}}, \ and\ \bibinfo {author} {\bibfnamefont {K.~Birgitta}\ \bibnamefont {Whaley}},\ }\bibfield  {title} {\enquote {\bibinfo {title} {Local superfluidity in inhomogeneous quantum fluids},}\ }\href {\doibase 10.1103/PhysRevB.74.174522} {\bibfield  {journal} {\bibinfo  {journal} {Phys. Rev. B}\ }\textbf {\bibinfo {volume} {74}},\ \bibinfo {pages} {174522} (\bibinfo {year} {2006})}\BibitemShut {NoStop}%
\bibitem [{\citenamefont {Dornheim}(2020)}]{Dornheim_PRA_2020}%
  \BibitemOpen
  \bibfield  {author} {\bibinfo {author} {\bibfnamefont {Tobias}\ \bibnamefont {Dornheim}},\ }\bibfield  {title} {\enquote {\bibinfo {title} {Path-integral monte carlo simulations of quantum dipole systems in traps: Superfluidity, quantum statistics, and structural properties},}\ }\href {\doibase 10.1103/PhysRevA.102.023307} {\bibfield  {journal} {\bibinfo  {journal} {Phys. Rev. A}\ }\textbf {\bibinfo {volume} {102}},\ \bibinfo {pages} {023307} (\bibinfo {year} {2020})}\BibitemShut {NoStop}%
\bibitem [{\citenamefont {Dornheim}\ and\ \citenamefont {Yan}(2022)}]{Dornheim_NJP_2022}%
  \BibitemOpen
  \bibfield  {author} {\bibinfo {author} {\bibfnamefont {Tobias}\ \bibnamefont {Dornheim}}\ and\ \bibinfo {author} {\bibfnamefont {Yangqian}\ \bibnamefont {Yan}},\ }\bibfield  {title} {\enquote {\bibinfo {title} {Abnormal quantum moment of inertia and structural properties of electrons in 2d and 3d quantum dots: an ab initio path-integral monte carlo study},}\ }\href {\doibase 10.1088/1367-2630/ac9f29} {\bibfield  {journal} {\bibinfo  {journal} {New Journal of Physics}\ }\textbf {\bibinfo {volume} {24}},\ \bibinfo {pages} {113024} (\bibinfo {year} {2022})}\BibitemShut {NoStop}%
\bibitem [{\citenamefont {Yan}\ and\ \citenamefont {Blume}(2014)}]{Yan_Blume_PRL_2014}%
  \BibitemOpen
  \bibfield  {author} {\bibinfo {author} {\bibfnamefont {Yangqian}\ \bibnamefont {Yan}}\ and\ \bibinfo {author} {\bibfnamefont {D.}~\bibnamefont {Blume}},\ }\bibfield  {title} {\enquote {\bibinfo {title} {Abnormal superfluid fraction of harmonically trapped few-fermion systems},}\ }\href {\doibase 10.1103/PhysRevLett.112.235301} {\bibfield  {journal} {\bibinfo  {journal} {Phys. Rev. Lett.}\ }\textbf {\bibinfo {volume} {112}},\ \bibinfo {pages} {235301} (\bibinfo {year} {2014})}\BibitemShut {NoStop}%
\bibitem [{\citenamefont {Boninsegni}\ and\ \citenamefont {Prokof'ev}(2012)}]{Boninsegni_RMP_2012}%
  \BibitemOpen
  \bibfield  {author} {\bibinfo {author} {\bibfnamefont {Massimo}\ \bibnamefont {Boninsegni}}\ and\ \bibinfo {author} {\bibfnamefont {Nikolay~V.}\ \bibnamefont {Prokof'ev}},\ }\bibfield  {title} {\enquote {\bibinfo {title} {Colloquium: Supersolids: What and where are they?}}\ }\href {\doibase 10.1103/RevModPhys.84.759} {\bibfield  {journal} {\bibinfo  {journal} {Rev. Mod. Phys.}\ }\textbf {\bibinfo {volume} {84}},\ \bibinfo {pages} {759--776} (\bibinfo {year} {2012})}\BibitemShut {NoStop}%
\bibitem [{\citenamefont {Militzer}\ \emph {et~al.}(2021)\citenamefont {Militzer}, \citenamefont {Gonz\'alez-Cataldo}, \citenamefont {Zhang}, \citenamefont {Driver},\ and\ \citenamefont {Soubiran}}]{Militzer_PRE_2021}%
  \BibitemOpen
  \bibfield  {author} {\bibinfo {author} {\bibfnamefont {Burkhard}\ \bibnamefont {Militzer}}, \bibinfo {author} {\bibfnamefont {Felipe}\ \bibnamefont {Gonz\'alez-Cataldo}}, \bibinfo {author} {\bibfnamefont {Shuai}\ \bibnamefont {Zhang}}, \bibinfo {author} {\bibfnamefont {Kevin~P.}\ \bibnamefont {Driver}}, \ and\ \bibinfo {author} {\bibfnamefont {Fran\ifmmode \mbox{\c{c}}\else~\c{c}\fi{}ois}\ \bibnamefont {Soubiran}},\ }\bibfield  {title} {\enquote {\bibinfo {title} {First-principles equation of state database for warm dense matter computation},}\ }\href {\doibase 10.1103/PhysRevE.103.013203} {\bibfield  {journal} {\bibinfo  {journal} {Phys. Rev. E}\ }\textbf {\bibinfo {volume} {103}},\ \bibinfo {pages} {013203} (\bibinfo {year} {2021})}\BibitemShut {NoStop}%
\bibitem [{\citenamefont {Chiesa}\ \emph {et~al.}(2006)\citenamefont {Chiesa}, \citenamefont {Ceperley}, \citenamefont {Martin},\ and\ \citenamefont {Holzmann}}]{Chiesa_PRL_2006}%
  \BibitemOpen
  \bibfield  {author} {\bibinfo {author} {\bibfnamefont {Simone}\ \bibnamefont {Chiesa}}, \bibinfo {author} {\bibfnamefont {David~M.}\ \bibnamefont {Ceperley}}, \bibinfo {author} {\bibfnamefont {Richard~M.}\ \bibnamefont {Martin}}, \ and\ \bibinfo {author} {\bibfnamefont {Markus}\ \bibnamefont {Holzmann}},\ }\bibfield  {title} {\enquote {\bibinfo {title} {Finite-size error in many-body simulations with long-range interactions},}\ }\href {\doibase 10.1103/PhysRevLett.97.076404} {\bibfield  {journal} {\bibinfo  {journal} {Phys. Rev. Lett.}\ }\textbf {\bibinfo {volume} {97}},\ \bibinfo {pages} {076404} (\bibinfo {year} {2006})}\BibitemShut {NoStop}%
\bibitem [{\citenamefont {Drummond}\ \emph {et~al.}(2008)\citenamefont {Drummond}, \citenamefont {Needs}, \citenamefont {Sorouri},\ and\ \citenamefont {Foulkes}}]{Drummond_PRB_2008}%
  \BibitemOpen
  \bibfield  {author} {\bibinfo {author} {\bibfnamefont {N.~D.}\ \bibnamefont {Drummond}}, \bibinfo {author} {\bibfnamefont {R.~J.}\ \bibnamefont {Needs}}, \bibinfo {author} {\bibfnamefont {A.}~\bibnamefont {Sorouri}}, \ and\ \bibinfo {author} {\bibfnamefont {W.~M.~C.}\ \bibnamefont {Foulkes}},\ }\bibfield  {title} {\enquote {\bibinfo {title} {Finite-size errors in continuum quantum monte carlo calculations},}\ }\href {\doibase 10.1103/PhysRevB.78.125106} {\bibfield  {journal} {\bibinfo  {journal} {Phys. Rev. B}\ }\textbf {\bibinfo {volume} {78}},\ \bibinfo {pages} {125106} (\bibinfo {year} {2008})}\BibitemShut {NoStop}%
\bibitem [{\citenamefont {Holzmann}\ \emph {et~al.}(2016)\citenamefont {Holzmann}, \citenamefont {Clay}, \citenamefont {Morales}, \citenamefont {Tubman}, \citenamefont {Ceperley},\ and\ \citenamefont {Pierleoni}}]{Holzmann_PRB_2016}%
  \BibitemOpen
  \bibfield  {author} {\bibinfo {author} {\bibfnamefont {Markus}\ \bibnamefont {Holzmann}}, \bibinfo {author} {\bibfnamefont {Raymond~C.}\ \bibnamefont {Clay}}, \bibinfo {author} {\bibfnamefont {Miguel~A.}\ \bibnamefont {Morales}}, \bibinfo {author} {\bibfnamefont {Norm~M.}\ \bibnamefont {Tubman}}, \bibinfo {author} {\bibfnamefont {David~M.}\ \bibnamefont {Ceperley}}, \ and\ \bibinfo {author} {\bibfnamefont {Carlo}\ \bibnamefont {Pierleoni}},\ }\bibfield  {title} {\enquote {\bibinfo {title} {Theory of finite size effects for electronic quantum monte carlo calculations of liquids and solids},}\ }\href {\doibase 10.1103/PhysRevB.94.035126} {\bibfield  {journal} {\bibinfo  {journal} {Phys. Rev. B}\ }\textbf {\bibinfo {volume} {94}},\ \bibinfo {pages} {035126} (\bibinfo {year} {2016})}\BibitemShut {NoStop}%
\bibitem [{\citenamefont {Dornheim}\ and\ \citenamefont {Vorberger}(2021)}]{Dornheim_JCP_2021}%
  \BibitemOpen
  \bibfield  {author} {\bibinfo {author} {\bibfnamefont {Tobias}\ \bibnamefont {Dornheim}}\ and\ \bibinfo {author} {\bibfnamefont {Jan}\ \bibnamefont {Vorberger}},\ }\bibfield  {title} {\enquote {\bibinfo {title} {Overcoming finite-size effects in electronic structure simulations at extreme conditions},}\ }\href {\doibase 10.1063/5.0045634} {\bibfield  {journal} {\bibinfo  {journal} {J. Chem. Phys.}\ }\textbf {\bibinfo {volume} {154}},\ \bibinfo {pages} {144103} (\bibinfo {year} {2021})}\BibitemShut {NoStop}%
\bibitem [{\citenamefont {B\"ohme}\ \emph {et~al.}(2022)\citenamefont {B\"ohme}, \citenamefont {Moldabekov}, \citenamefont {Vorberger},\ and\ \citenamefont {Dornheim}}]{Bohme_PRL_2022}%
  \BibitemOpen
  \bibfield  {author} {\bibinfo {author} {\bibfnamefont {Maximilian}\ \bibnamefont {B\"ohme}}, \bibinfo {author} {\bibfnamefont {Zhandos~A.}\ \bibnamefont {Moldabekov}}, \bibinfo {author} {\bibfnamefont {Jan}\ \bibnamefont {Vorberger}}, \ and\ \bibinfo {author} {\bibfnamefont {Tobias}\ \bibnamefont {Dornheim}},\ }\bibfield  {title} {\enquote {\bibinfo {title} {Static electronic density response of warm dense hydrogen: Ab initio path integral monte carlo simulations},}\ }\href {\doibase 10.1103/PhysRevLett.129.066402} {\bibfield  {journal} {\bibinfo  {journal} {Phys. Rev. Lett.}\ }\textbf {\bibinfo {volume} {129}},\ \bibinfo {pages} {066402} (\bibinfo {year} {2022})}\BibitemShut {NoStop}%
\bibitem [{\citenamefont {B\"ohme}\ \emph {et~al.}(2023)\citenamefont {B\"ohme}, \citenamefont {Moldabekov}, \citenamefont {Vorberger},\ and\ \citenamefont {Dornheim}}]{Bohme_PRE_2023}%
  \BibitemOpen
  \bibfield  {author} {\bibinfo {author} {\bibfnamefont {Maximilian}\ \bibnamefont {B\"ohme}}, \bibinfo {author} {\bibfnamefont {Zhandos~A.}\ \bibnamefont {Moldabekov}}, \bibinfo {author} {\bibfnamefont {Jan}\ \bibnamefont {Vorberger}}, \ and\ \bibinfo {author} {\bibfnamefont {Tobias}\ \bibnamefont {Dornheim}},\ }\bibfield  {title} {\enquote {\bibinfo {title} {Ab initio path integral monte carlo simulations of hydrogen snapshots at warm dense matter conditions},}\ }\href {\doibase 10.1103/PhysRevE.107.015206} {\bibfield  {journal} {\bibinfo  {journal} {Phys. Rev. E}\ }\textbf {\bibinfo {volume} {107}},\ \bibinfo {pages} {015206} (\bibinfo {year} {2023})}\BibitemShut {NoStop}%
\end{thebibliography}%
\end{document}